\documentclass[twocolumn,twocolappendix]{aastex631}

\usepackage{amsmath,amssymb}

\usepackage{graphicx}
\usepackage{color}
\usepackage{natbib}
\usepackage[normalem]{ulem}
\usepackage[caption=false]{subfig}
\shorttitle{Spatially-Resolved Mid-IR Dual AGNs in DESI LS}
\shortauthors{Pfeifle et al.}
\graphicspath{{./}{figures/}}

\begin{document}

\title{The Hunt for Red \sout{October} Dual AGNs I: Spatially-Resolved Mid-IR Dual AGNs in the DESI Legacy Imaging Surveys}

\correspondingauthor{Ryan W. Pfeifle}
\email{ryan.w.pfeifle.civ@us.navy.mil}

\author[0000-0001-8640-8522]{Ryan W. Pfeifle}
\altaffiliation{NASA Postdoctoral Program Fellow}
\altaffiliation{Surname pronunciation: ``\textit{Fife-Lee}''}
\affiliation{U.S. Naval Observatory, 3450 Massachusetts Avenue NW, Washington, DC 20392, USA}
\affiliation{CRESST II, CSST, University of Maryland Baltimore County, 1000 Hilltop Cir, Baltimore, MD, 21250}
\affiliation{Oak Ridge Associated Universities, NASA NPP Program, Oak Ridge, TN 37831, USA}
\affiliation{X-ray Astrophysics Laboratory, NASA Goddard Space Flight Center, Code 662, Greenbelt, MD 20771, USA}

\author{Kimberly A. Weaver}
\affiliation{X-ray Astrophysics Laboratory, NASA Goddard Space Flight Center, Code 662, Greenbelt, MD 20771, USA}

\author[0000-0003-2283-2185]{Barry Rothberg}
\affiliation{U.S. Naval Observatory, 3450 Massachusetts Avenue NW, Washington, DC 20392, USA}
\affiliation{Department of Physics and Astronomy, George Mason University, 4400 University Drive, MSN 3F3, Fairfax, VA 22030, USA}

\author{Miranda McCarthy}
\affiliation{X-ray Astrophysics Laboratory, NASA Goddard Space Flight Center, Code 662, Greenbelt, MD 20771, USA}
\affiliation{Southeastern Universities Research Association, Washington DC 20005, USA}
\affiliation{Center for Research and Exploration in Space Science and Technology, NASA/GSFC, Greenbelt, MD 20771, USA}

\author[0000-0002-6454-861X]{Emma Schwartzman}
\affiliation{U.S. Naval Research Laboratory, 4555 Overlook Ave SW, Washington, DC 20375, USA}
\affiliation{Department of Physics and Astronomy, George Mason University, 4400 University Drive, MSN 3F3, Fairfax, VA 22030, USA}

\author[0000-0002-4902-8077]{Nathan J. Secrest}
\affiliation{U.S. Naval Observatory, 3450 Massachusetts Avenue NW, Washington, DC 20392, USA}

\author[0000-0001-9379-4716]{Peter~G.~Boorman}
\affiliation{Max-Planck-Institut f\"{u}r Extraterrestrische Physik, Gie{\ss}enbachstraße 1, 85748 Garching, Germany}
\affiliation{Cahill Center for Astronomy and Astrophysics, California Institute of Technology, 1216 E California Blvd, Pasadena, CA 91125, USA}

\author{Daniel Stern}
\affiliation{Cahill Center for Astronomy and Astrophysics, California Institute of Technology, 1216 E California Blvd, Pasadena, CA 91125, USA}
\affiliation{Jet Propulsion Laboratory, California Institute of Technology, Pasadena, CA 91109, USA}

\author{Joanna Piotrowska}
\affiliation{Cahill Center for Astronomy and Astrophysics, California Institute of Technology, 1216 E California Blvd, Pasadena, CA 91125, USA}

\author{Kevin McCarthy}
\altaffiliation{NASA Postdoctoral Program Fellow}
\affiliation{Jet Propulsion Laboratory, California Institute of Technology, Pasadena, CA 91109, USA}
\affiliation{Kavli Institute for the Physics and Mathematics of the Universe (WPI), University of Tokyo, Kashiwa, Chiba 277-8583, Japan}

\author[0000-0001-9793-5416,gname=Emily,sname=Moravec]{Emily Moravec}
\affiliation{Green Bank Observatory, P.O. Box 2, Green Bank, WV 24944}

\author{Jenna M. Cann}
\altaffiliation{NASA Postdoctoral Program Fellow}
\affiliation{CRESST II, CSST, University of Maryland Baltimore County, 1000 Hilltop Cir, Baltimore, MD, 21250}
\affiliation{Oak Ridge Associated Universities, NASA NPP Program, Oak Ridge, TN 37831, USA}
\affiliation{X-ray Astrophysics Laboratory, NASA Goddard Space Flight Center, Code 662, Greenbelt, MD 20771, USA}

\author{Kimberly Engle}
\affiliation{X-ray Astrophysics Laboratory, NASA Goddard Space Flight Center, Code 662, Greenbelt, MD 20771, USA}

\author{Kyla Mullaney}
\affiliation{Southeastern Universities Research Association, Washington DC 20005, USA}
\affiliation{Center for Research and Exploration in Space Science and Technology, NASA/GSFC, Greenbelt, MD 20771, USA}
\affiliation{X-ray Astrophysics Laboratory, NASA Goddard Space Flight Center, Code 662, Greenbelt, MD 20771, USA}

\author{Ryan Tanner}
\affiliation{X-ray Astrophysics Laboratory, NASA Goddard Space Flight Center, Code 662, Greenbelt, MD 20771, USA}

\author{Kelly Whalen}
\altaffiliation{NASA Postdoctoral Program Fellow}
\affiliation{Oak Ridge Associated Universities, NASA NPP Program, Oak Ridge, TN 37831, USA}
\affiliation{X-ray Astrophysics Laboratory, NASA Goddard Space Flight Center, Code 662, Greenbelt, MD 20771, USA}



\begin{abstract}
Theoretical studies predict that dual AGNs are a critical stage of galaxy merger-driven supermassive black hole growth. Systematic searches for dual AGNs typically target late-stage mergers ($\leq10$ kpc nuclear separations) and select AGNs based on optical diagnostics. Yet, simulations predict that obscuration can occur early in the merger sequence, and that a significant fraction of dual AGNs can be found beyond $10$ kpc. Here, we report on a new sample of {155} spatially resolved mid-IR dual AGNs candidates selected based upon their mid-IR $W1-W2$ colors from the Wide-Field Infrared Survey Explorer and optically classified as galaxy merger candidates using imaging from the DESI Legacy Imaging Surveys. Preliminary spectroscopic results are presented for approximately 2/3 of the sample. {78} candidates have been confirmed to reside in galaxy mergers; among these, 13 have been confirmed as bona fide mid-IR dual AGNs, while {65} represent strong dual AGN candidates that require further examination. {48} candidates have been rejected as non-merger contaminants (foreground-background AGNs, separations inconsistent with  interacting galaxies, etc.). {29} candidates still await spectroscopic coverage. The confirmed and high confidence dual AGN candidates exhibit separations of 14.5-129 kpc; $>50\%$ reside at separations $>50$\,kpc. Confirmed and high confidence candidates also exhibit a diversity of nuclear optical BPT classes. Seyfert-Seyferts and Seyfert-HIIs dominate the overall BPT pairings. 31\% of confirmed mid-IR dual AGNs reside in multi-mergers involving three or more galaxies. The diversity in AGN properties and environments identified in this work highlights the importance of multiwavelength selection strategies and analyses in the quest to holistically understand dual AGNs as a population.

\end{abstract}

\keywords{}


\section{Introduction} 
\label{sec:intro}

\subsection{Motivation}

Pairs of simultaneously accreting supermassive black holes (SMBHs) with nuclear separations on the order $\sim0.03-110$\,kpc, commonly referred to as dual active galactic nuclei \citep[dual AGNs,][]{derosa2019,pfeifle2025}, are a naturally expected byproduct of interactions and mergers of massive galaxies \citep{begelman1980,milo2001,yu2002,chen2023}. Interactions of galaxies can give rise to hydrodynamical and gravitational torques, the presence and strength of which depend upon the encounter dynamics and galaxy properties \citep{blumenthal2018}; these torques can excite a variety of gas inflow mechanisms \citep{barnes1996,cox2008,blumenthal2018,blecha2018} responsible for depositing gas reservoirs in the vicinity of the SMBHs thought to reside at the hearts of most massive galaxies \citep{kormendy1995}. Cosmological and closed-box simulations have both shown that, under the right encounter conditions, these processes can lead to repeated mass accretion onto the nuclear SMBHs, including periods of expeditious, heavily obscured SMBH growth, particularly in the later stages of the galaxy merger process \citep{vanwassenhove2012,capelo2015,capelo2017,blecha2018,chen2023}.

Cosmological simulations suggest that secular gas fueling processes cannot alone account for the occupation fraction of dual AGNs across cosmic history, which in turn suggests that dual AGNs represent environments characterized by significant merger-induced mass accretion onto the SMBHs \citep{foord2024}. Thus, dual AGNs may represent optimal environments to test the potential connection between galaxy mergers and SMBH growth and---by extension---the relative importance of mergers in the coevolution and establishment of the scaling relations between the SMBHs and their hosts \citep[e.g.,][]{ferrarese2000,gebhardt2000,gultekin2009,kormendy2013}, a long-standing and contentious topic still lacking an observational consensus \citep[see][and references therein]{ellison2019}. Furthermore, the mass, environmental, and dynamical evolution of kpc- and sub-kpc-scale dual AGNs govern the pairing rate and properties of the gravitationally-bound binary SMBH phase \citep[e.g.,][with {separations $\lesssim30$ pc, e.g., \citealp{liu2018,derosa2019,pfeifle2025}}]{callegari2011,chen2023,nianyichen2024} and are therefore, by extension, relevant to the production of gravitational waves. SMBH binaries in the continuous wave regime are one potential explanation for the gravitational wave background detected by Pulsar Timing Arrays \citep[][]{agazie2023a,agazie2023b}, and the eventual coalescence of a given SMBH binary may be detectable with the future Laser Interferometer Space Antenna \citep[or \textit{LISA},][]{amaro-seoane2023}. Taken all together, dual AGNs could embody a fundamental component of the current hierarchical mass assembly and structure formation paradigm.

\subsection{Prior Dual AGN Selection Approaches}

In practice, observational studies have yet to empirically ascertain the true, relative importance of dual AGNs in the context of SMBH and galaxy evolution due to the observationally demanding nature of robustly identifying and analyzing populations of dual AGNs, with only a few hundred known in the literature \citep[see][]{tonychen2022,pfeifle2025}. Optical spectroscopic diagnostics have been the most popular method by which dual AGNs and candidates have been selected {in the local universe} over the last two decades{, with the majority of dual AGN candidates in the local universe drawn from either (1) single fiber observations exhibiting double-peaked optical spectroscopic emission lines with AGN-like emission line ratios \citep[e.g.,][]{wang2009,liu2010a,smith2010,ge2012,fu2018}, or (2) spectroscopic fiber pairs \citep[e.g.,][]{liu2011b} or distinct emission line regions in integral field unit (IFU) observations exhibiting AGN-like emission line ratios \citep[e.g,][]{fu2018,steffen2023}}; see \citealp{pfeifle2025} for more details. As a result, the vast majority of follow-up works focusing on X-ray \citep[e.g.,][]{liu2013,comerford2015,derosa2018,hou2019,hou2020,derosa2023,he2023}, optical \citep[e.g,][]{liu2010b,fu2012,liu2013,comerford2015}, near-IR \citep[e.g.,][]{shen2011,fu2011a,mcgurk2011,rosario2011,mcgurk2015,johnstone2026}, and radio observations \citep[e.g.,][]{rosario2010,fu2011b,mullersanchez2015,liu2018,hou2023,johnstone2026} of dual AGNs have, by design, relied upon optical selection strategies as the foundation. However, optical diagnostics are at a particular disadvantage in the search for dual AGNs: large scale host galaxy dust attenuation can screen out not only AGN emission signatures but even entire galaxy nuclei in the optical \citep[and result in optically-elusive AGNs, e.g.,][]{koss2012,satyapal2017,koss2018,pfeifle2019a}. Moreover, dual AGNs are predicted to be preferentially obscured \citep{blecha2018}, with dust reddening and line-of-sight gas column densities expected to increase as the merger progresses and peak in the latest stages of the merger \citep{blecha2018,chen2023,barrows2023} -- the exact merger phase often targeted in systematic searches \citep[{separations $<10$\,kpc,} e.g.][]{wang2009,liu2010a,fu2012,fu2015a,comerford2015,satyapal2017,pfeifle2019a}. Notably, even the poster children of dual AGNs -- such as NGC 6240 \citep{komossa2003} and Mrk 463 \citep{bianchi2008} -- do not exhibit unambiguous signatures of dual AGNs in the optical. A statistically complete picture of dual AGN properties clearly demands a multiwavelength approach for selection, analysis, and confirmation. Indeed, small but no less important populations of dual AGNs and candidates have been recovered using techniques such as hard X-ray Burst Alert Telescope (BAT) preselection \citep{koss2012}, spatially-resolved radio imaging\footnote{\citet{zhang2021b} used radio imaging to preselect merging galaxies in a search for dual AGNs, but {optical emission line ratios indicative of AGN activity} \citep{baldwin1981} formed the basis for confirmation.} \citep{fu2015a,fu2015b,gross2019,gross2023radio}, and near-IR imaging \citep{imanishi2014,imanishi2020}, each of which are typically less reliant\footnote{\citet{koss2012} {relied upon optical emission line ratios indicative of AGN activity} \citep{baldwin1981} as one method in their confirmation strategy, but {this was }not {required for} all nuclei/AGNs; X-ray AGNs also served as evidence for dual AGNs.} upon optical spectroscopic evidence of AGNs (spectroscopic redshifts were often required, but optical AGNs were not a standard requirement) and less affected/unaffected by dust and gas attenuation. {Serendipitous findings of dual obscured/unobscured AGNs in recent years \citep[e.g.,][]{kim2019,glikman2023} reinforce the need for novel, multiwavelength strategies to pick out these missing dual AGN populations. }

\subsection{Mid-IR Selection for Dual AGNs: Prior Applications and Motivation for this Work}

With over a decade of all-sky mid-infrared (mid-IR) imaging from the Wide-field Infrared Survey Explorer \citep[WISE,][]{wright2010} and a variety of mid-IR AGN color selection criteria developed in the literature \citep{stern2012,jarrett2011,assef2018,mateos2012}, large-scale searches for dusty, mid-IR luminous dual AGNs can be readily performed. A particular benefit of mid-IR selection---like radio selection \citep[e.g.,][]{fu2015a}---is that it is less sensitive to line-of-sight obscuration and has been shown to reliably select both obscured and unobscured AGNs in the nearby Universe as well as out to higher redshifts \citep[assuming the AGN dominates over the host galaxy, e.g.,][{see also \citealp{boorman2025}, who found that mid- to far-IR selection criteria also identify unobscured and obscured AGNs equally well}]{stern2012,wu2012,assef2018,lamassa2017,glikman2018,glikman2022,glikman2023,petter2023}. Furthermore, dual AGNs are expected to be heavily obscured, dust enshrouded, and emitting intensely at mid-IR wavelengths \citep{blecha2018}; mid-IR color selection represents a natural -- and potentially optimal -- selection strategy. Excesses of mid-IR AGNs are consistently found in mergers relative to isolated mid-IR AGN control samples \citep{satyapal2014,weston2017,barrows2023}, and these AGN excesses increase significantly with decreasing pair separation, with the largest enhancements observed at pair separations below $20$\,kpc \citep[and most prominently observed in post-mergers,][]{satyapal2014,weston2017,barrows2023}; mid-IR AGN excesses in mergers are also distinctly larger than those found for optically-selected AGNs in mergers out to separations of $\sim40-50$\,kpc \citep{satyapal2014,barrows2023}. In fact, \citet{barrows2023} demonstrated that the AGN merger fraction increases significantly with increasing AGN dust extinction. If dual AGNs are indeed effective tracers of merger-driven SMBH growth, the excesses of mid-IR AGNs observed in mergers (relative to non-mergers and optical samples) suggest that mid-IR dual AGNs could represent a key population critical to better elucidating the AGN-merger connection.

The first systematic searches for dual AGNs in the mid-IR relied upon the color criterion proposed in \citet[][$W1-W2>0.5$]{blecha2018} to preselect late-stage ($<10$\,kpc) mergers; coupled with follow-up \textit{Chandra} X-ray imaging and ground-based spectroscopy, mid-IR preselection yielded a relatively high fraction of dual AGNs and candidates across a small pilot sample \citep[$\gtrsim40-50\%$,][]{satyapal2017,ellison2017,pfeifle2019a,pfeifle2019b}, but the majority of these mergers were spatially unresolved in WISE. Observational campaigns {(regardless of wavelength choice)} have {-- with some exceptions \citep[e.g.,][]{liu2011b,koss2012} --} primarily focused on late-stage mergers in search of dual AGNs \citep{liu2010a,liu2010b,liu2013,mcgurk2011,fu2011a,fu2011b,fu2012,comerford2012,comerford2015,mcgurk2015,fu2015a,satyapal2017,fu2018,pfeifle2019a,gross2019} due to predictions from closed-boxed simulations that dual AGNs undergo their most expeditious period of growth in advanced mergers when their separations fall consistently below 1\,kpc \citep[e.g.,][]{vanwassenhove2012,capelo2015,capelo2017,blecha2018}. Nevertheless, closed-box simulations have also demonstrated that gas inflow mechanisms can be triggered once the host galaxies have an encounter at separations of $\sim10-15$\,kpc \citep{blumenthal2018}---rather than the sub-kpc encounters required for inflows in earlier works \citep[e.g.,][]{barnes1996,blecha2018}---and that heavily obscured SMBH growth can occur even prior to the second pericenter passage depending on the geometry of the encounter and properties of the hosts \citep{blecha2018}. Coupled with predictions from recent cosmological simulations that dual AGNs flicker on and off throughout the merger sequence and that a significant fraction of all dual AGNs at a given snapshot in time have separations $>10$\,kpc \citep{chen2023}, it remains possible that dual AGNs in earlier-stage mergers could represent important sites of obscured SMBH growth. WISE selection then has a secondary benefit: with an angular resolution of $6.1''$, $6.4''$, and $6.5''$ across the $3.4\,\mu\rm{m}$, $4.6\,\mu\rm{m}$, and $12\,\mu\rm{m}$ bands, WISE is particularly well suited for identifying these spatially-resolved mid-IR dual AGNs in intermediate- and early-stage mergers ($\gtrsim10$\,kpc; at $z=0.1$, $6.1''$ corresponds to a separation of $11.2$\,kpc). \citet{barrows2023} most recently assembled a sample of $\sim200$ probabilistic WISE mid-IR dual AGN {candidates} with separations in the range of $10-100$\,kpc and, along with a large control sample of single mid-IR AGNs in mergers and isolated galaxies, found tentative evidence that the less massive SMBHs in dual AGNs are biased toward higher Eddington ratios---in line with prior predictions \citep[e.g.,][]{callegari2011,capelo2015,chen2023}. While the most significant excess (a factor of $\sim5\times$) of mid-IR dual AGN {candidates} was observed at separations below $20$\,kpc \citep[suggesting significant enhancements of obscured SMBH growth at smaller separations,][]{satyapal2014,barrows2023}, excesses on the order of $\sim2$ were observed for obscured mid-IR dual AGN candidates out to large projected separations \citep[][see Figure 15]{barrows2023}. A major next step in understanding populations of mid-IR dual AGNs and how they compare to samples selected at other wavelengths is to probe their multiwavelength properties and environments.

{Though the angular resolution of WISE largely limits the selection of spatially-resolved dual mid-IR AGNs to the local universe, WISE can nonetheless be used in concert with other multiwavelength diagnostics to select close pairs of dual AGNs at high redshift. For example, \citet{glikman2023} recently identified a strong candidate for an obscured/unobscured dual quasar at $z=1.889$; originally selected based as a Type 1 AGN with red WISE colors, J122016.9+112627.09 resolved into distinct blue and red quasars in follow-up Hubble Space Telescope imaging and spectroscopy. Efforts focusing on identifying further cases of obscured or obscured/unobscured dual AGNs at higher redshift are particularly important, as most recent high redshift dual AGN searches have relied upon optical imaging techniques such as varstrometry \citep{hwang2020,chentony2022,chentony2023,chentony2024,ishikawa2025,chentony2025,gross2025} and the Gaia-multi-peak selection \citep{manucci2022,manucci2023,ciurlo2023}, which are biased against obscured AGNs. Complementary to \citet{glikman2023}, spatially-resolved near- to mid-IR spectral energy distributions via JWST \citep{li2025} and machine learning algorithms designed to identify close ($<1''$) candidate AGN companions to known quasars in Euclid Q1 imaging \citep{ulivi2025} have now been used to identify further high redshift dual AGN candidates.}

\subsection{Outline for this Work}
In this work, we use the AllWISE point source catalog \citep{allwise_source_catalog_2013}{, the }WISE two-band color criterion from \citet{stern2012}{, and a visual morphological inspection of optical imaging }to develop a {new} sample of spatially-resolved mid-IR {dual AGN candidates in the local universe (sample median redshift $z\approx0.24$)}. Section~\ref{sec:selection} details the selection strategy for the mid-IR dual AGN candidate sample. {Details on} the data processing and basic analysis steps for our follow-up optical spectroscopic observations, along with the basics of the analysis of the mid-IR properties of these mid-IR pairs {are provided} in Section~\ref{sec:followup_obs} and \ref{sec:analysis}. Section~\ref{sec:results} {presents} this new sample of dual mid-IR AGNs and candidates, including basic details on the mid-IR color space, BPT properties, merger environments, and properties as a function of separation. We discuss in Section~\ref{sec:discussion} {this }new sample of mid-IR dual AGNs and candidates in the context of several other dual AGN samples from the literature. We provide our conclusions and follow-up observation plans in Section~\ref{sec:conclusion}. Additional information on the sample, such as rejected candidates, are provided in the Appendix. All magnitudes listed or referenced in this work are given in the Vega system, and throughout this work we assume the following cosmology: $\textrm{H}_0 = 70$\,km\,s$^{-1}$\,Mpc$^{-1}$, $\Omega_M=0.3$, and $\Omega_\Lambda=0.7$.

\section{Selection Methodology}
\label{sec:selection}

\subsection{\textit{WISE} Mid-IR Color Selection}
\label{sec:colorselection}
To assemble our sample of mid-IR dual AGN candidates, we began by drawing an all-sky list of mid-IR AGNs from the AllWISE Source Catalog \citep{cutri2014} that satisfy the \citet{stern2012} two-band \textit{WISE} color cut of $W1[3.4\rm{\mu m}]-W2[4.6\rm{\mu m}]\geq0.8$ down to a magnitude limit of $W2<15.05$ mag, a highly reliable selection method for bright mid-IR AGNs \citep[$\sim95$\% reliability,][]{stern2012}\footnote{Note, this color cut may not be as statistically complete for dual AGNs as the $W1-W2>0.5$ color cut proposed in \citet[][though this was proposed for dual AGNs specifically in late-stage mergers]{blecha2018}: in that work, dual AGNs were predicted to sustain colors of $W1-W2\geq0.5$ for longer periods of the merger cycle than more extreme colors like $W1-W2\geq0.8$.}. It is possible that we may have missed fainter mid-IR AGNs that would have been selected with {the 90\% reliability W1-W2} color criteria defined in \citet[][{see Equations 4 and 5 in Section~2 of that work}]{assef2018}{, where the two band $W1-W2$ color cut is a function of $W2$ magnitude}; in this work we only provide a secondary assessment of all sources with the \citet{assef2018} criteria. To reduce foreground Galactic contaminants, we limited the sample of mid-IR sources to Galactic latitudes $|b|>10$ deg. We conservatively required AGNs in the sample to be detected with a signal-to-noise (SNR) ratio $>10$ in the first two \textit{WISE} bands (\textit{W1} and \textit{W2})---corresponding to A-rating detections under the photometric quality flags contained in the \texttt{ph\_qual} field---and sources were further required to show detections in the $W3$ and $W4$ bands with SNRs $>7$ (high B-quality) and $>2$ (at least C-quality), respectively. All objects with potential photometric issues, such as halo contamination, contamination from diffraction spikes, ghost images, etc., were removed, leaving only a sample of AGN candidates with clean contamination and confusion flags (\texttt{cc\_flags}~=~``0000''). 

We matched this parent sample of mid-IR AGN {candidates} onto itself with a match tolerance of $60\arcsec$ using \textsc{topcat} \citep{2005ASPC..347...29T}, allowing all possible matches rather than simply the ``best matches''; we removed self-matches and limited the matched list to only pairs. {Groupings of three or more mid-IR sources -- referred to as ``N-tuple'' matches -- }were placed in a separate list for further, separate inspection; this list was also subject to the criteria outlined here. {Finally, we removed any pairs for which the WISE magnitudes were identical for AGN 1 and 2 (a particularly important criterion for matches with $5.5\arcsec<\theta<8\arcsec$), as these may be small-separation, same-Tile duplicates that were not properly removed from the AllWISE catalog}\footnote{{https://irsa.ipac.caltech.edu/data/WISE/docs\\/release/AllWISE/expsup/sec2\_2.html}}. 

A match tolerance of $60\arcsec$ was chosen to avoid selecting large numbers of interloping higher redshift pairs with incredibly large physical separations (by $z=0.1$, an angular separation of $\theta=60\arcsec$ already corresponds to $r_p\approx110$\,kpc) while also selecting pairs in the local universe ($z<0.1$) with separations between $\sim10-100$\,kpc. {Given the point spread function (PSF) of the AllWISE blurred imaging ($\sim8\arcsec$) as well as the native \textit{WISE} $W1$, $W2$, and $W3$ image resolution ($\sim6.1\arcsec$, $\sim6.4\arcsec$, $\sim6.5\arcsec$), mid-IR AGN pairs with separations $\theta>8''$ were considered firmly detected, spatially resolved pairs of AGN candidates. However, active source deblending\footnote{https://irsa.ipac.caltech.edu/data/WISE/docs/release/All-Sky/expsup/sec4\_4c.html\#wpro} within the WISE pipeline can yield mid-IR sources with separations $<6.1''$ and $<6.4''$ in the W1 and W2 bands within the AllWISE catalog. In an attempt to retain closely separated, potential mid-IR AGN pairs and push the WISE imaging angular resolution to its limits, we adopted a lower bound to our match tolerance of 5.5''. The mid-IR selection criteria and matching process described above yielded a total of 6,573 mid-IR AGN pair candidates, where 6,529 mid-IR AGN candidate pairs exhibited separations $\theta>8''$ and 44 exhibited separations $5.5\arcsec<\theta<8\arcsec$.} {Due to their closer separations, the latter pairs will} require additional scrutiny to ensure the presence of multiple AGNs, including supervised forced mid-IR photometry and source deblending as well as multiwavelength AGN diagnostics; unless otherwise stated in the text here, these additional tests are left to a future work. 

{As will be discussed in Section~\ref{sec:morphclass}, a critical component of the selection strategy in this work is the use of optical imaging to classify the host galaxies of these mid-IR AGN pair candidates; cases of mid-IR pairs hosted in galaxy mergers would represent promising dual AGN candidates.  To that end, this program required both large areal coverage and imaging of sufficient depth in the optical regime to identify morphological signs of galaxy interactions. Three major survey fields were considered: Pan-STARRS1 \citep[][$\sim30,000$ deg$^2$]{chambers2016}, the Sloan Digital Sky Survey \citep[][$\sim14,000$ sq. deg]{York2000,Abazajian2009,Aihara2011,almeida2023}, and the DESI Legacy Imaging Surveys \citep[][DESI LS, $\sim14,000$ deg$^2$ in the original survey, $\sim19,000$ deg$^2$ in data release 9 with at least one pass in all three filters in DR9]{dey2019}. The sample was ultimately limited to the footprint of DR9 of the DESI LS \citep[][]{dey2019} due to the large sky coverage and significant depth across the survey footprint. The DESI LS is in fact a product of three individual surveys\footnote{https://datalab.noirlab.edu/data/legacy-surveys}: the DECam Legacy Survey (DeCaLS), which provides $g$-, $r$-, $z$-band imaging across $\sim9500$ deg$^{2}$ to depths of $g=24.7$\,AB, $r=23.9$\,AB, $z=23.0$\,AB from the Dark Energy Camera on the Blanco 4\,m telescope; the Mayall z-band Legacy Survey (MzLS), which provides $z$-band imaging down to a depth of $z=23.0$\,AB from the MOSAIC instrument on the Mayall 4\,m telescope across $\sim5300$ deg$^{2}$; and the Beijing-Arizona Sky Survey (BASS), which provides $g$- and $r$-band imaging to depths of $g=24.4$\,AB and $r=23.9$\,AB using the 90Prime instrument at the Steward Observatory Bok telescope across $\sim5000$ deg$^{2}$. In particular, the DESI LS offered superior depth relative to both SDSS and Pan-STARRS1, which would be crucial for any morphological examination. This reduced the sample from 6,573 to 4,518 mid-IR pairs, where 4492 systems exhibited separations $\theta>8''$ and 26 exhibited separations $5.5\arcsec<\theta<8\arcsec$. In the following sections, we discuss the separation and velocity difference criteria for dual AGNs adopted in this work, the morphological classification of the sample, and the removal of contaminants within the sample.}

\subsection{$r_p$ and $|\Delta v|$ Criteria for Mid-IR Dual AGNs in this Work}
\label{sec:defining_duals}

{Many prior dual AGN searches have focused specifically upon the latest merger stages, where the galaxy nuclei have projected separations ($r_p$) of $r_p\lesssim10$ kpc \citep[e.g.,][]{liu2013,fu2012,fu2015a,satyapal2017,pfeifle2019a} or $\lesssim20-30$ kpc \citep[e.g.,][]{comerford2018,stemo2021,li2025}. This emphasis on later merger stages was typically motivated by specific selection strategies \citep[e.g., double-peaked emission lines in a single 3'' diameter spectroscopic fiber, e.g,][]{wang2009,liu2010a,smith2010} or closed-box simulations \citep[probing limited parameter spaces, e.g.,][]{vanwassenhove2012} that suggested more correlated AGN activity at smaller separations. These criteria, unfortunately, exclude the vast majority of the life cycle of a galaxy merger \citep[e.g.,][]{patton2024} and therefore potentially exclude a significant fraction of dual AGNs.} 

\citet{pfeifle2025} recently proposed physically motivated selection criteria for dual AGNs in terms of projected separation and velocity difference based on interacting pairs of galaxies in Illustris-TNG100 \citep{patton2024}: {projected} separations of $0.03 \,{\rm{kpc}} \lesssim r_p \leq 110\,{\rm{kpc}}$ with line-of-sight velocity differences $|\Delta v| \lesssim 600\,\rm{km}\,\rm{s}^{-1}$, where this velocity difference is designed for redshifts measured from narrow emission lines or absorption lines that trace the host systemic velocities. This criteria {is similar to the criteria adopted in \citet[][$r_p < 100\,{\rm{kpc}}$, $|\Delta v| < 600\,\rm{km}\,\rm{s}^{-1}$]{liu2011b} and \citet[][$r_p < 100\,{\rm{kpc}}$, $|\Delta v| < 300\,\rm{km}\,\rm{s}^{-1}$]{koss2012}. D}esigned to be 95\% complete for simulated pairs of interacting galaxies in Illustris-TNG100 \citep{patton2024,pfeifle2025}, {this} criteria will exclude some small fraction ($\sim5$\%) of  dual AGNs {in reality}. We allow for a slightly expanded form of this criteria: when redshifts are available, dual AGNs and candidates are required to show physical, projected separations $r_p\leq130$\,kpc and velocity differences $|\Delta v|<700\,\rm{km}\,\rm{s}^{-1}$, but with two important caveats: (1) if a pair of AGNs is found to be at a {projected} separation $110\,\rm{kpc}<r_p<130\,\rm{kpc}$, we only consider it to be a dual AGN or candidate if $|\Delta v|\ll600\,$km\,s$^{-1}$; conversely, (2) we only consider AGN pairs with velocity differences $600\,{\rm{km}}\,{\rm{s}}^{-1}<|\Delta v|<700\,\rm{km}\,\rm{s}^{-1}$ if the AGNs exhibit a pair separation $r_p\ll110$\,kpc. These requirements are informed by the distribution of interacting galaxy pairs in Illustris-TNG 100 \citep[][see Figure~2 in the latter work]{patton2024,pfeifle2025}, where there are virtually no interacting pairs simultaneously characterized by large projected separations ($r_p>110$\,kpc) and high line-of-sight velocity differences ($|\Delta v|>600$\,km\,s$^{-1}$). Furthermore, these expanded criteria and stipulations aim to avoid potentially throwing out a small fraction of galaxy mergers which may show slightly larger than expected separations or velocity differences. Note: we do not require archival redshifts for selection, and these criteria only become applicable when removing contaminants based on archival and follow-up spectra (Section~\ref{sec:contam_removal_phasei} and \ref{sec:contam_removal_phaseii}) and when performing follow-up analyses on the morphologically selected candidates (Section~\ref{sec:spec_analysis}).

\subsection{Removal of Contaminating Pairs Based on Redshift (Phase I)}
\label{sec:contam_removal_phasei}
To avoid biasing the sample toward optical spectroscopic selection, we did not require spectroscopic redshifts for any mid-IR pair during the initial sample selection. However, {in order to reduce the workload during the visual morphological assessment (see Section~\ref{sec:morphclass}),} we utilized archival {survey} redshift information to remove contaminating pairs of AGN candidates, such as foreground-background pairs and other non-mergers. We cross-matched the parent list of matched AGN pairs to the Sloan Digital Sky Survey (SDSS) data release 16 \citep[DR16,][]{ahumada2020} spectroscopic sample via a TAP query in \textsc{topcat} \citep{2005ASPC..347...29T} to search for available spectroscopic redshifts for each AGN in a pair. Rather than selecting only AGNs/pairs with archival spectroscopic redshifts in SDSS DR16, we used the criteria in Section~\ref{sec:defining_duals} to remove contaminant foreground-background pairs exhibiting discordant redshifts ($|\Delta v|>700$\,km\,s$^{-1}$, when both redshifts were available) or pairs with separations inconsistent with interacting galaxies ($>130$\,kpc), where these physical values were computed using the available {SDSS} spectroscopic redshifts. {1,150 mid-IR pairs were classified as contaminants and excluded from any further analysis (reducing the candidate sample to 3,368).} This process of removing interloping pairs based on archival spectroscopy was iterative; once this initial cull was complete, we proceeded {through} a complete optical visual morphological classification of the remaining sample (described below in Section~\ref{sec:morphclass}), and then returned to this process of interloping pair removal using new survey and follow-up data {that became available after our morphological assessment} (see Section~\ref{sec:contam_removal_phaseii} for further discussion). 

\begin{figure*}
    \centering
    \includegraphics[width=1.0\linewidth]{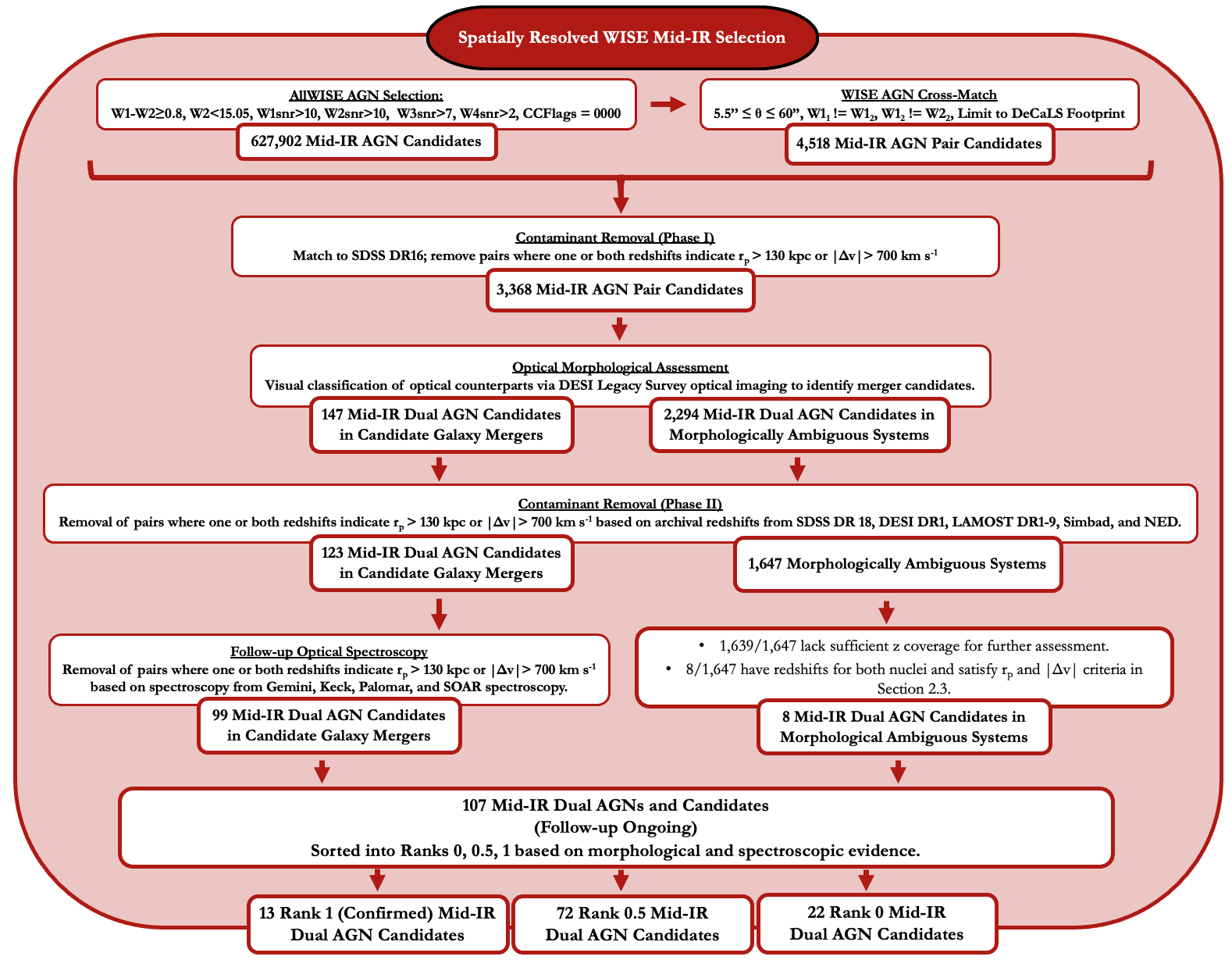}
    \caption{A flowchart detailing the selection methodology outlined in Sections~\ref{sec:colorselection}-\ref{sec:conf_ranks}, including the follow-up spectroscopy and analysis discussed in Section~\ref{sec:followup_obs}-\ref{sec:analysis}.}
    \label{fig:placeholder}
\end{figure*}

\subsection{Morphological Classification}
\label{sec:morphclass}
Rather than relying upon the availability of spectroscopic redshifts to select our mid-IR dual AGN candidates \citep[e.g.,][]{liu2011b,koss2012,fu2015a,satyapal2017,pfeifle2019a}, we rely upon morphological evidence of galaxy pairs and/or galaxy interactions to identify the bulk of our subsample of mid-IR dual AGNs/candidates. The early phase of our selection strategy is thus relatively blind with respect to optical spectroscopy (except when removing contaminants), which comes with the benefit that we can develop a relatively large sample of mid-IR dual AGN candidates; requiring archival spectroscopic redshifts would have severely limited the pool of dual AGN candidates identified hereafter (see Section~\ref{sec:whystartfromimaging}).

Two of the authors {visually} classified the {host morphologies of the} sample (3,368 mid-IR pairs) {by eye} using {DESI LS} \citep{dey2019} $grz$ optical imaging\footnote{{Drawn from the DESI LS Sky Viewer}: https://www.legacysurvey.org/viewer .}, {where the separate $g$, $r$, and $z$ images were inspected individually alongside a tricolor $grz$ image}. R.W.P.\ classified all mid-IR AGN pairs with separations $\theta<20\arcsec$ and a subset of pairs with $\theta>20\arcsec$, and M.M.\ classified all remaining pairs with separations $20\arcsec<\theta<60\arcsec$. Mergers were identified based on morphological evidence of interactions in the DESI LS imaging, including but not limited to: tidal tails or debris plumes, ripples, shells, double-nuclei, and warped or disrupted disks. We also considered close galaxy pairs to be merger candidates if they exhibited potential tidal features (asymmetries, fuzziness, extended low surface brightness features, etc.), particularly when the constituent galaxies exhibited similar angular sizes and optical colors. Mid-IR AGN pairs were classified into one of the following general categories: 

\begin{enumerate}
    \item ``m'': A clear galaxy merger or a galaxy merger candidate {with discernible or possible morphological indications of interactions}. 
    \item ``bf'': A foreground/background pair. {Here, }one mid-IR source {in a given pair} had a clear optical counterpart but the other mid-IR source had a very faint counterpart or no observable optical counterpart; it is possible that some of these could be minor mergers at higher redshift. 
    \item ``bb'': Background-background {pair. Here,} no optical counterparts were observed in the optical imaging; the counterparts were both presumed to reside at much higher redshifts and were therefore {considered} irrelevant. 
    \item ``a'': Ambiguous {pair}. Here, systems were {found to be } point-like and/or {lacked} discernible morphological indications {of} interactions, but could still reside at the same systemic redshift. 
\end{enumerate}
This visual morphological examination resulted in a total of 147 spatially-resolved mid-IR dual AGN candidates residing in galaxy mergers or candidate galaxy mergers. {A total of 2294 mid-IR pairs were classified as ``ambiguous'' pairs, 806 were classified as ``foreground-background'' pairs, and 121 were classified as ``background-background'' pairs; the latter two categories of objects were discarded, reducing the full sample to 2441 mid-IR pairs.} It is worth noting that {surface brightness decreases as $(1+z)^{-4}$ \citep{tolman1930,tolman1934}, and hence higher redshift mergers and/or low surface brightness tidal features are intrinsically more difficult to identify relative to local redshift galaxy mergers. O}ur reliance upon morphological evidence of tidal features {therefore} biases this sample toward lower redshifts and mergers exhibiting stronger tidal distortions. 

We also examined the optical imaging for all N-tuple AllWISE mid-IR AGN groupings (forming groups of three or more mid-IR AGN candidates) to see if any of these constituted multi-AGN candidates (ex: triple, quadruple AGNs, etc.). However, no convincing cases of spatially-resolved, N-tuple mid-IR AGN candidates were uncovered in the D{ESI Legacy Survey} optical imaging, nor were any further cases of dual AGN candidates identified. 

\subsection{Removal of Contaminating Pairs Based on Redshift (Phase II)}
\label{sec:contam_removal_phaseii} 
As mentioned in Section~\ref{sec:contam_removal_phasei}, the sample underwent additional rounds of contaminant removal after the morphological classification process outlined in Section~\ref{sec:morphclass}. Originally, we had limited our search for archival redshifts {to the SDSS DR16 \citep{ahumada2020} via a TAP query in Topcat \citep{2005ASPC..347...29T}; upon completion of the morphological classification described in Section~\ref{sec:morphclass}, we then retrieved any additional archival spectroscopic redshift measurements\footnote{{E.g., available from surveys such as the 6dF Galaxy Survey \citep[6dFGS,][]{jones2004}, or from individual literature articles.}} from }the SIMBAD astronomical database \citep{wenger2000} and the NASA Extragalactic Database \citep[NED,][]{helou1991} {for} the 147 systems classified as potential or clear mergers. {However, u}pon the recent release of the Dark Energy Survey Instrument \citep[DESI,][]{DESIcollab2024} spectroscopic data release 1 \citep[DR1,][]{DESICollab2025}, we revisited this approach and systematically searched for {additional} spectroscopic redshifts for the entire mid-IR pairs sample---not just those classified as mergers---with a match tolerance of $3\arcsec$ across the DESI DR1, SDSS DR18 \citep{almeida2023}, Large Sky Area Multi-Object Fiber Spectroscopic Telescope \citep[LAMOST,][]{cui2012} survey DR1-9 \citep{ai2016,jin2023}, as well as archival redshifts reported on SIMBAD and NED. {In particular, this search allowed us to retrieve new survey redshifts for the morphologically ambiguous pairs, some of which could exhibit consistent redshifts and represent mergers despite an apparent lack of tidal signatures.} 

Redshifts with warning flags (consistent with ``bad'' redshifts) {within the DESI, SDSS, and LAMOST surveys }were removed, and when a given survey provided multiple, consistent redshift entries for a single object, we averaged the redshift measurements. In a few cases where the redshifts for a given source within a single survey were inconsistent, we visually inspected the spectra and removed inaccurate redshifts. In cases where multiple surveys provided a redshift measurement for a single object, we ensured the redshift measurements were relatively consistent across the surveys, and we typically prioritized survey measurements in the following fashion based on spectral resolution and depth: DESI\,$>$\,SDSS\,$>$\,LAMOST (and these survey measurements almost always superseded measurements from SIMBAD and NED). Once these archival spectroscopic redshifts were gathered into a central table, we again removed all AGN pairs with physical separations $>130$\,kpc (when redshifts were available for one or both AGNs) and velocity differences $>700$\,km\,s$^{-1}$ (when redshifts were available for both AGNs); see Section~\ref{sec:defining_duals} for more details on these criteria. 

{Twenty-four contaminants were identified among the 147 morphologically selected dual AGN candidates: one candidate was found to be the well-known gravitational lens QSO B0957+5608B \citep{walsh1979} and rejected, while 23 merger candidates were found to be projected pairs and rejected. This reduced the morphologically selected merger sample to 123. Among the 2294 ambiguous mid-IR pairs, 10 were found to exhibit $r_p\leq130$\,kpc and $|\Delta v|\leq700$\,km\,s$^{-1}$, consistent }at face value{ with our selection criteria for merging/interacting galaxies}. {However, two of these (J124702.51+402410.3 and J165501.32+260517.4, at redshifts $z\approx0.72$ and $z\approx1.9$, respectively) were ultimately rejected due to their simultaneously large separations ($\sim79-97$\,kpc) and large velocity differences ($600<|\Delta v|<700$\,km\,s$^{-1}$), which are inconsistent with that expected of interacting galaxies; $<1$\% of interacting galaxies are expected to exhibit such combinations of $r_p$ and $|\Delta v|$ \citep[see Figure 2 in][{see also \citealp{patton2024}}]{pfeifle2025}.}\footnote{{A third high redshift AGN pair, J144003.15+390840.1 ($z\approx1.27$) exhibited an $r_p\approx94$\,kpc and a $|\Delta v|\approx230$\,km\,s$^{-1}$, and was assigned a confidence rank of +0.5; this system conforms to the dual AGN definitions outlined above and in \citet{pfeifle2025}. While no foreground lensing galaxy was observed in the DESI LS imaging, deeper imaging and higher quality spectra are likely required to completely rule out the lens scenario.}} {645 out of 2294 ambiguous pairs did not conform to the definitions} in Section~\ref{sec:defining_duals} and {were rejected, while the remaining 1639 ambiguous pairs still lack sufficient spectroscopic coverage for classification, but future SDSS and DESI data releases will likely aid in their classification. Similarly, deeper and higher resolution imaging would aid in the morphological assessment of these morphologically ambiguous pairs, since the host galaxies are insufficiently resolved in the survey imaging.}

{In total, 123 dual AGN candidates morphologically selected as being hosted in mergers remained. While the primary focus of this work is on the dual AGN candidates in morphologically selected merger candidates, we supplemented these candidates with the 8 morphologically ambiguous dual AGN candidates that conformed to the $r_p$ and $|\Delta v|$ criteria and caveats in Section~\ref{sec:defining_duals}, bringing the sample size to 131. To continue removing contaminants, we began a ground-based optical spectroscopic follow-up campaign to determine the spectroscopic redshifts and emission line properties for this sample of dual AGN candidates. We discuss a confidence rank system for this sample next in Section~\ref{sec:conf_ranks}, which details the conservative requirements in this work for a system to be classified as a bona fide dual AGN. This then leads the way for the details of our follow-up observations and analysis in Sections~\ref{sec:followup_obs}-\ref{sec:analysis}.}

\subsection{Confidence Ranks and Caveats for Mid-IR Dual AGNs in this Work}
\label{sec:conf_ranks}

We adopt the dual AGN confidence rank convention established in \citet{pfeifle2025}, which runs from $+1$ to $-1$ in half integer steps and denotes the evidence-based likelihood for the presence of a dual AGN in a given system. Initially, we assign each mid-IR dual AGN candidate selected during our morphological assessment in Sections~\ref{sec:morphclass} a confidence rank of ``Rank 0'' (baseline candidates{), and additional lines of evidence are needed to adjust the confidence rank up or down.}

{Spectroscopic observations are needed to ensure that a given pair of mid-IR AGNs reside at the same redshift and at separations consistent with galaxy mergers, even if circumstantial morphological evidence in favor of a merger is present (see Figure~\ref{fig:rank1_duals}). Thus, spectroscopic redshifts are one crucial piece of evidence required in the majority of cases to adjust the confidence rank of a given dual AGN candidate up or down. For example, mid-IR dual AGN candidates with redshifts for both nuclei and that satisfy the $r_p$ and $|\Delta v|$ criteria outlined in Section~\ref{sec:defining_duals} are elevated to Rank ``+0.5'' candidates. In a minority of cases, morphological evidence can be used to elevate the confidence rank of a given system; for systems lacking a redshift for one or both nuclei, but for which morphological evidence favors a merging or interacting system\footnote{For example, strong and clear tidal tails and where the system is isolated from other nearby objects are unlikely to be source confused, e.g., J042138.05-200312.6.}, we assign a confidence rank of ``+0.5''. Follow-up spectroscopic redshifts are still required for confirmation of such systems. Conversely, candidates that do not satisfy these $r_p$ and $|\Delta v|$ criteria are rejected and demoted to Rank ``$-1$''; such rejected candidates are not included in our analysis.} 

{To achieve Rank 1 (``confirmed''), a dual mid-IR AGN candidate must exhibit (a) consistent spectroscopic redshifts indicating that the hosts reside in the same system, and (b) multiwavelength evidence supporting the presence of two accreting SMBHs. This latter requirement is motivated by the potential for} (1) color contamination among the closest pairs, and (2) non-AGN contaminants driving the mid-IR colors. Recall from Section~\ref{sec:selection} that the WISE W1 and W2 PSFs are 6.1'' and 6.4''{, the AllWISE blurred imaging has a PSF of $\sim8''$}, and that we included pairs with separations down to 5.5'' that were actively deblended by the WISE photometric pipeline and showed distinct WISE magnitudes. {While the active deblending performed by the WISE pipeline suggests the presence of two closely separated ($<8''$) mid-IR sources in these cases, color contamination remains a possibility for these systems.} Unless otherwise stated in this work, systems with separations $5.5''\leq\theta\leq8''$ are considered Rank +0.5 candidate mid-IR dual AGNs; we leave to a future work the more rigorous analyses --- including careful and supervised deblending, spectroscopic analysis, and/or spectral energy decompositions --- that are needed to confirm or reject these systems.

As we will discuss in Section~\ref{sec:results}, the median redshift for the sample is $z\approx0.24$, and as such there is the possibility that some fraction of these mid-IR dual AGN candidates -- even those found to be at consistent redshifts and residing in truly interacting systems -- do not actually host accreting SMBHs but are instead driven purely by star formation. \citet{assef2018} and \citet{barrows2023} noted that for systems at higher redshifts ($z>0.1$), dust attenuation of bluer stellar emission and the presence of prominent PAH features could conspire to allow purely star forming galaxies to mimic the mid-IR colors of AGNs and be accidentally included when selecting AGN pairs. In fact, even local redshift ($z<0.1$) starburst galaxies have also been shown to mimic AGN colors \citep{asmus2020}. 

Given the potential for ambiguity in differentiating between AGNs and purely star forming systems based on mid-IR color alone, in order to elevate a given dual AGN candidate to ``Rank 1'' in this work we conservatively require {both mid-IR AGN candidates to show optical spectroscopic evidence of accreting SMBHs either via }Seyfert-like {narrow emission line ratios on the Baldwin, Phillips, Terlevich} \citep[BPT,][]{baldwin1981,kewley2001,kauffmann2003,kewley2006} {diagram }or the presence of broad optical emission lines. Mid-IR dual AGN candidates showing Composite-, LINER, and [HII]-like line ratios are held at confidence ranks of ``+0.5'' and require further analysis. {While this is a conservative approach, i}n future works various multiwavelength information -- not just optical {diagnostics} -- will be used to support the confirmation of mid-IR dual AGNs in this sample \citep[see][for the multiwavelength -- mid-IR, optical, and radio --  confirmation of a triple AGN in this sample]{schwartzman2025}. In such works, we will investigate the potential presence of non-AGN contaminants in the sample and adjust any confidence ranks accordingly. We describe the identification of spectroscopically confirmed mid-IR dual AGNs and rejection of contaminants in Section~\ref{sec:results}, which incorporates spectroscopic redshift and BPT information obtained through our ongoing optical long-slit spectroscopic programs (see Section~\ref{sec:followup_obs}).

\section{Follow-Up Optical Spectroscopic Observations and Reduction}
\label{sec:followup_obs}

{For mid-IR dual AGN candidate systems lacking sufficient spectroscopic redshift and emission line flux information for one or both sources, we have pursued follow-up optical spectroscopic observations. These previous and ongoing observing programs target not only the individual mid-IR AGN candidates, but also additional nearby candidate companion galaxies that could be associated with the mid-IR dual AGN candidates. } 

\subsection{Gemini Observations and Processing}
\label{sec:gemini_obs}
We obtained follow-up long slit spectra for {48} mid-IR dual AGN candidate {systems} ({114} total {sources} observed, {including any nearby candidate galaxy companions}) through a Gemini North GMOS 2025A Band 3 program (GN-2025A-Q-306), GMOS 2025B {and 2026A} poor weather programs on Gemini North and South (GN-2025B-Q-404, GS-2025B-Q-407{, GN-2026A-2026A-405, GS-2026A-Q-407), and a 2026B Gemini South Band 3 program (GS-2026B-Q-305)}. For all observations, we used a 1\farcs{}0 wide slit and, when possible, the slit was aligned along the two nuclei and/or companion nuclei. For systems where one nucleus had a previously determined spectroscopic redshift, we used the R400 grating (simultaneous wavelength coverage: 416\,nm). For systems with no prior spectroscopic redshift measurements, we instead used the R150 grating (simultaneous wavelength coverage: 1071\,nm). The grating central wavelength for each observation was typically set to the blaze wavelength, but it was adjusted accordingly when required to ensure coverage of relevant emission lines such as [O III] and H$\alpha$. {A blocking filter (GG455 or OG515, depending on the central wavelength) was used} with each grating to avoid second order contamination. Wavelength and Y-direction dithers were used to fill in the spectral and spatial coverage gaps (caused by the CCD chip gaps and the GMOS slit bridges, respectively)\footnote{https://www.gemini.edu/instrumentation/gmos/observation-preparation} as well as account for cosmic rays and bad pixels. Flats and arcs were obtained with each set of observations, and one spectrophotometric standard star was observed per GMOS configuration per semester. The Gemini long slit spectra were reduced using the Data Reduction for Astronomy from Gemini Observatory North and South (\textsc{DRAGONS}) v4 package  \citep{labrie2023}. For this work we followed the standard processing procedures outlined in the online documentation.

\subsection{Keck and Polomar Observations and Processing}
\label{sec:keck_obs}

Long-slit optical spectra were acquired for seven systems (14 total {sources}): two systems were observed with the DEep Imaging Multi-Object Spectrograph \citep[DEIMOS;][]{faber2003} on the Keck-II Telescope at Mauna Kea Observatory and the remaining five systems were observed with the Double Spectrograph \citep[DBSP;][]{oke1982} on the Hale 200-inch telescope at Palomar Observatory. 

The DEIMOS observations were obtained on UT 2024 February 17, which was a challenging night that suffered from extremely high winds (50-70 mph) that caused telescope closures and poor, variable seeing.  We observed two systems for 300~s each with the 600ZD grating ($\lambda_{\rm blaze} = 7500$~\AA) and through a 1\farcs5 wide slit aligned with the two nuclei.  We used the Python Spectroscopic Data Reduction Pipeline \citep[PypeIt,][]{prochaska2020} to reduce the data, using calibration flats and arcs obtained the same night.  Given the poor conditions, an archival sensitivity function was used to produce relative flux calibration of the data.

The DBSP observations were spread across three observing runs between August and October 2024.  One source was observed on UT 2024 August 11 for 300~s on a relatively clear, but non-photometric night with 1\farcs3 - 1\farcs4 seeing.  Three sources were observed on the two-night run of UT 2024 September 7-8, with integrations that ranged from 600~s to 1200~s.  Both nights suffered from clouds and the seeing for these nights ranged from 1\farcs5 to 2\farcs0.  The final target was observed for 600~s on UT 2024 October 2, which was a photometric night with 1\farcs2 - 1\farcs8 seeing.  All DBSP observations used the D55 dichroic, the 1\farcs5 wide slit, the 600 $\ell\, {\rm mm}^{-1}$ grating on the blue arm ($\lambda_{\rm blaze} = 4000$~\AA), and the 316 $\ell\, {\rm mm}^{-1}$ grating on the red arm ($\lambda_{\rm blaze} = 7500$~\AA).  We processed the data following standard procedures with IRAF using calibration flats, arcs, and standard stars observed during the same observing runs.  Given the variable conditions for some of these nights, the flux calibration is only relative.

\subsection{SOAR Observations and Processing}
\label{sec:soar_obs}

We obtained additional follow-up long slit spectra of 20 systems (55 total {sources}, including additional companions) using the Goodman High Throughout Spectrograph \citep[HTS,][]{clemens2004} on the SOAR Telescope (program ID 2025B-117238). For these observations, we used a 1\farcs0  wide long-slit with the 400M2 grating, the red camera, and the GG-455 blocking filter to avoid second order contamination. Flats and biases were obtained at the start of each night, arcs were obtained with each set of observations, and a spectrophotometric standard star was observed once per observing run (four nights/standards total).

The SOAR spectroscopic observations were processed using the Goodman HTS Pipeline\footnote{https://soardocs.readthedocs.io/projects/goodman-pipeline/en/latest/}, most often using the automated pipeline commands \textsc{redccd} and \textsc{redspec}. In a handful of cases, the automated detection routine did not identify and extract the spectra of our optical sources; for these sources, we manually extracted the 1D calibrated spectra from the 2D images using customized forms of the Goodman pipeline modular processing routines.

\section{Analysis} 
\label{sec:analysis}

\subsection{Optical Spectroscopic Analysis}
\label{sec:spec_analysis}
The 1D Gemini, Keck, Palomar, and SOAR spectra were fit using the Bayesian AGN Decomposition Analysis for Sloan Digital Sky Survey (SDSS) Spectra (\textsc{BADASS}) software package \citep{sexton2021}. We also refit all archival SDSS and DESI spectra with BADASS. {{BADASS is an open-source\footnote{https://github.com/remingtonsexton/BADASS3} spectral fitting code that derives parameter and uncertainty estimations using the Bayesian affine-invariant Markov-Chain Monte Carlo sampler \textsc{emcee} \citep{foreman-mackey2013} and assesses parameter convergence using autocorrelation analysis \citep{sexton2021}. BADASS can fit spectra from virtually any facility, and it allows for a significant degree of customization: components can be toggled on or off, and the code can fit for custom lists of individual narrow, broad, and absorption line features, the stellar line-of-sight velocity distribution, narrow and broad FeII emission, the AGN power law continuum, outflow emission components, and UV iron emission.}} During the fitting process, we typically started with the default \textsc{BADASS} settings before freeing parameters for individual components \citep[for a brief description of default line parameter settings, see Section 2.1.2 in][]{sexton2021}, and we focused on (and fit separately) two narrow regions of the spectra: (1) the H$\beta$ and [O III]$\lambda$5007 emission lines, and (2) the H$\alpha$, [N II]$\lambda$6858, [S II]$\lambda\lambda$6717,6733, and [O I]$\lambda$6302 emission lines. For the preliminary optical spectroscopic analysis in this work, we use only a simple AGN power law (and, when necessary, an additional polynomial to handle spectral curvature) to model the continuum and simple Gaussian components to model the narrow emission lines; when necessary, we included Gaussian components to model broad line emission or outflows. Furthermore, we limited our BPT analysis to archival survey observations and follow-up observations obtained prior to the 2025B semester, though we present all spectroscopic redshifts obtained up through 2026B (including ongoing programs). A more thorough spectroscopic analysis will be presented in a future work focusing on the spectroscopic properties of the complete sample. 

After fitting the spectroscopic emission lines, we computed the BPT \citep[][]{baldwin1981} emission line ratios to judge the dominant optical photoionizing source. {BPT diagrams rely upon four spectroscopic emission line ratios ([OIII]/H$\beta$, [NII]/H$\alpha$, [SII]/H$\alpha$, [OI]/H$\alpha$) and are commonly used to classify emission line galaxies into four categories: Seyferts, Low-Ionization Nuclear Emission-line Regions (LINERs), AGN-HII ``Composite'' nuclei, and star forming HII regions \citep{baldwin1981,veilleux1987,kewley2001,kauffmann2003,kewley2006,schawinski2007}. Seyfert galaxies are commonly associated with accreting SMBHs and can be divided into Seyfert 1s and Seyfert 2s depending upon the presence of absence of broad spectroscopic emission lines \citep{khachikian1974,antonucci1993,urry1995}. Forming a separate branch on the BPT diagram \citep{kewley2006}, LINER optical spectra are dominated by low ionization species \citep{heckman1980,ho1997} and a variety a ionization sources have been proposed for LINERs, including: accreting SMBHs \citep[e.g.,][]{ho1997}, shocks \citep[e.g.,][]{heckman1980,allen2008,rich2014,rich2015}, and stellar photoionization \citep[e.g.,][]{taniguchi2000,singh2013}. Composite galaxies exhibit line ratios that are intermediate between the classes of Seyferts, LINERs, and HII regions \citep{kewley2006}, lying between the star forming locus and Seyfert and LINER galaxy branches on the BPT diagram.}

In order to classify nuclei on the BPT diagram, we require $\geq3\sigma$ detections for the relevant emission lines \citep{kewley2001,kewley2006}; in this work, we also rely upon upper and lower limits for classification. For cases where H$\beta$ was not detected but [OIII] was detected at $\geq3\sigma$, we computed the lower limit on the flux ratio; only Seyferts and one LINER in the sample exhibit a lower limit on the log([OIII]/H$\beta$) ratio. We computed the relevant upper limit on the flux ratio when H$\beta$ was detected at $\geq3\sigma$ but [OIII] was not detected; only [HII] regions and Composites in the sample exhibit an upper limit on the log([OIII]/H$\beta$) ratio. We rely upon these optical classifications and the presence of broad lines as indicators of AGNs. 12 systems exhibited pairs of Seyfert-Seyfert nuclei, one exhibited a Seyfert nucleus +  broad line AGN pair; all 12 were elevated to Rank 1 mid-IR dual AGNs in this work. One additional system did not have sufficient spectral coverage for BPT classification, but was previously confirmed as a bona fide triple AGN \citep[J121857.42+103551.2,][]{schwartzman2025} and was therefore raised to a Rank 1 dual AGN in this work. 

{Our follow-up spectroscopic observations also allowed us to identify an additional 25 foreground-background contaminants among the remaining morphologically selected merger candidates; these 25 systems were rejected, thereby reducing the merger candidate sample to 98, and reducing the full dual AGN candidate sample (including morphologically ambiguous systems) to 107. At this point, 118/147 morphologically selected merger candidates have been examined via archival or follow-up observations: 48/147 have been identified as contaminants, and 70/147 systems have nuclei that have been spectroscopically confirmed to reside at consistent redshifts (and are thus confirmed mergers). 29 out of 147 systems still await redshift measurements. Relative to the 118 merger-selected systems with archival or follow-up observations, our morphological selection strategy so far boasts a success rate of $\approx59\%$ (70/118) for identifying mid-IR dual AGN candidates in interacting galaxies.}

\begin{figure*}
    \centering
    \includegraphics[width=0.8\linewidth]{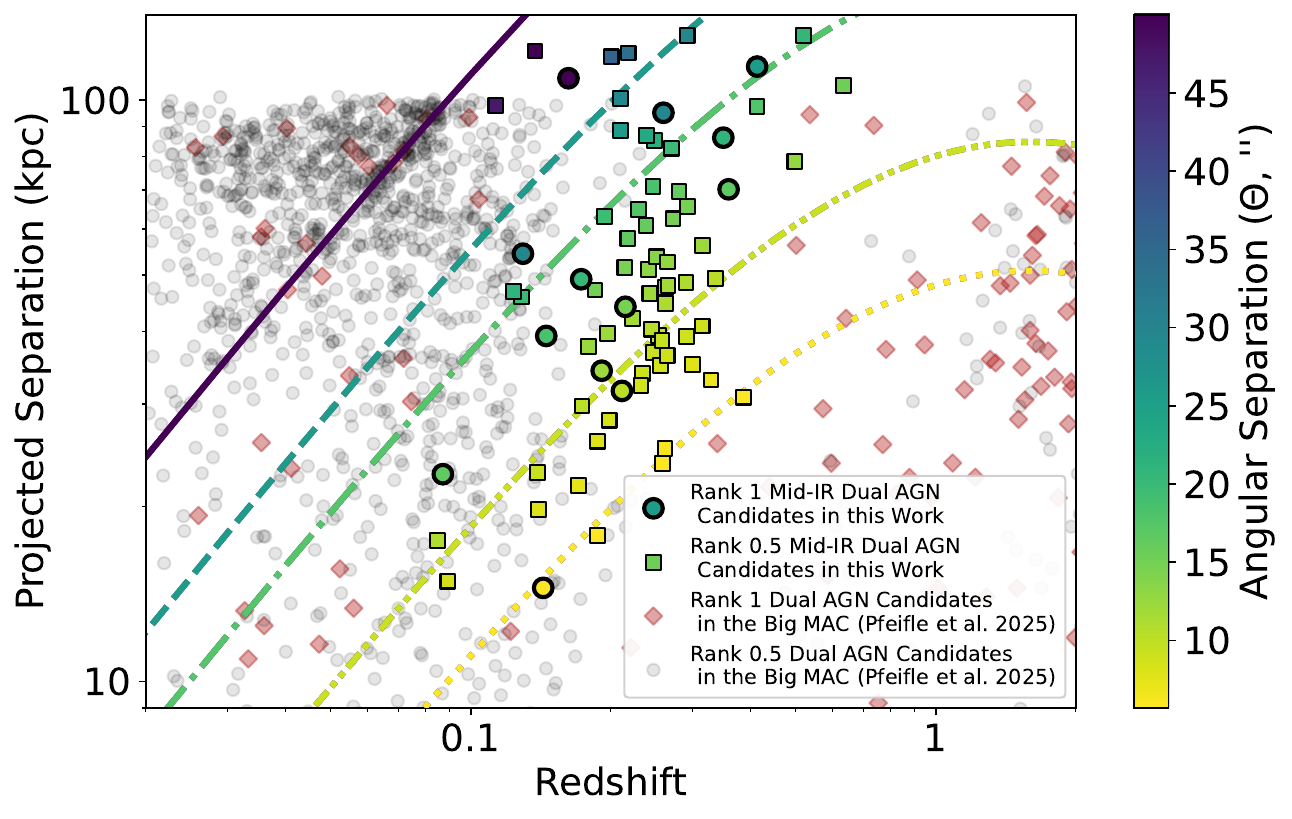}
    \caption{The redshift distribution plotted against the projected separation (kpc) distribution for the sample of Rank 1 and 0.5 spatially-resolved mid-IR dual AGNs in this sample. The data are color coded according to the angular separation of the mid-IR AGN pairs, with the color map provided on the auxiliary color bar; Rank 1 (confirmed) candidates are represented by circles with dark outlines, while Rank 0.5 candidates (with redshifts for both nuclei) are squares. Solid, dashed, dash-dotted, dash-dot-dotted, and dotted curves indicate angular separations of 60\arcsec, 30\arcsec, 20\arcsec, 15\arcsec, and 10\arcsec, respectively, and share the same color map as the dual AGN data points. For comparison, the projected, physical separations and redshifts of dual AGN candidates identified in the {observational }literature {between 1970 and 2020 independent of selection strategy} are also plotted \citep[drawn from the Big Multi-AGN Catalog, or Big MAC, Ranks $>0$;][]{pfeifle2025}{; Rank 1 (or ``confirmed'') dual AGNs from the literature are plotted as red diamonds, while Rank 0.5 candidates are plotted as light gray circles.} {Many of the confirmed and candidate dual AGNs drawn from the Big MAC originally come from the optical spectroscopic pairs sample in \citet{liu2011b}, the hard X-ray selected dual AGNs in \citet{koss2012}, and the high redshift dual quasars identified in \citet{hennawi2006,hennawi2010} and \citet{eftekharzadeh2017}. This sample of mid-IR dual AGN candidates occupies a slightly higher redshift space than most dual AGN candidates {identified (up to 2020)} in the local universe.}}

    \label{fig:rp_vs_z}
\end{figure*}

\begin{figure}
    \centering
    \includegraphics[width=1.0\linewidth]{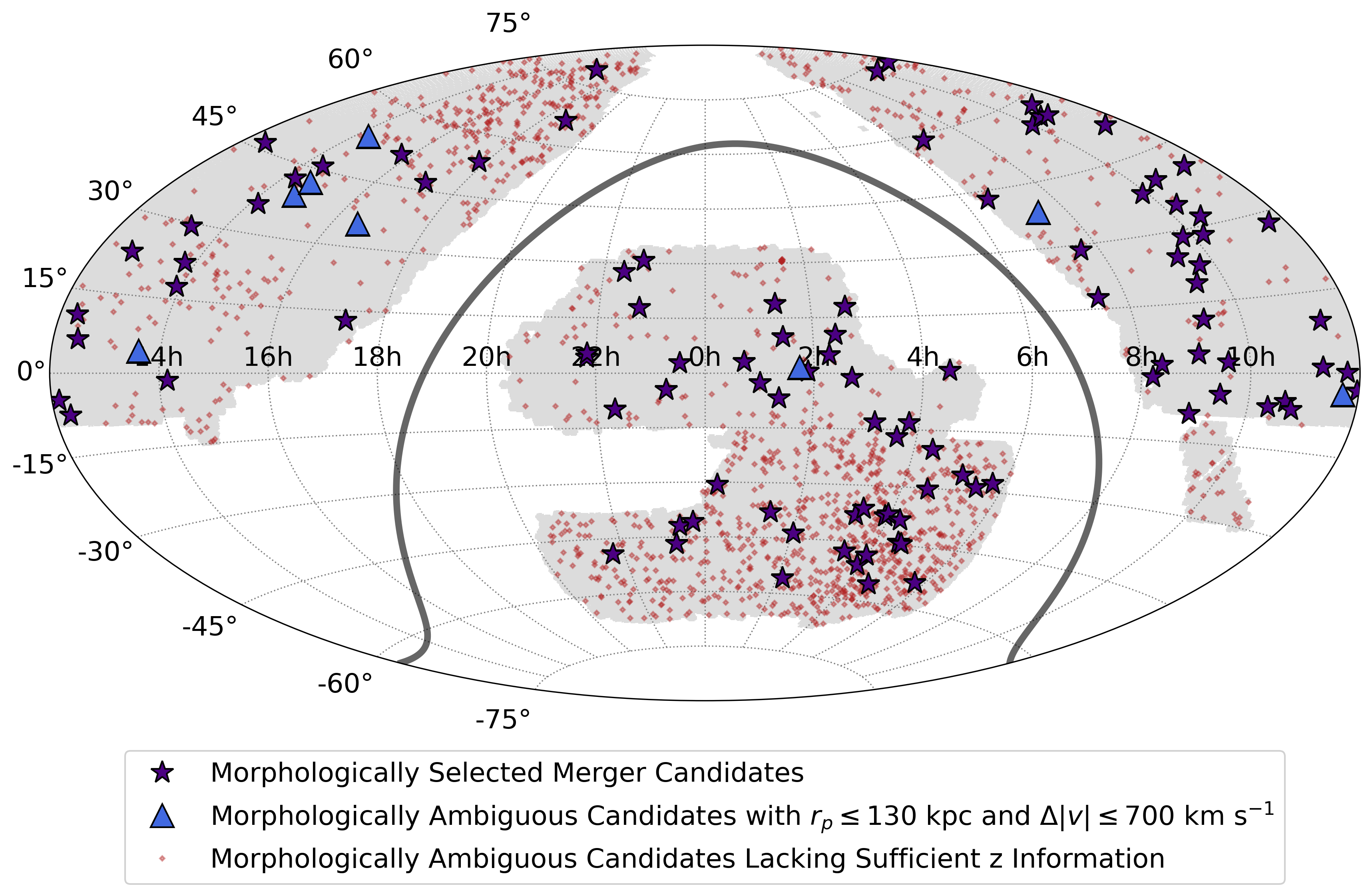}
    \caption{{The sky distribution of mid-IR AGN candidate pairs selected in Section~\ref{sec:selection}, after removal of contaminants identified in Sections~\ref{sec:selection} and \ref{sec:analysis}. The gray line indicates the Galactic plane, and the gray shaded regions denote the footprint of the DESI LS \citep{dey2019}. Morphologically selected merger candidates are denoted by indigo stars, morphologically ambiguous pairs that satisfy the $r_p$ and $|\Delta v|$ criteria in Section~\ref{sec:defining_duals} are denoted by blue triangles. Morphologically ambiguous mid-IR AGN pair candidates that lack sufficient redshift coverage to reject or confirm as mergers are indicated by small red circles.}}
    \label{fig:aitoff_project}
\end{figure}

\begin{figure}
    \centering
    \includegraphics[width=1.0\linewidth]{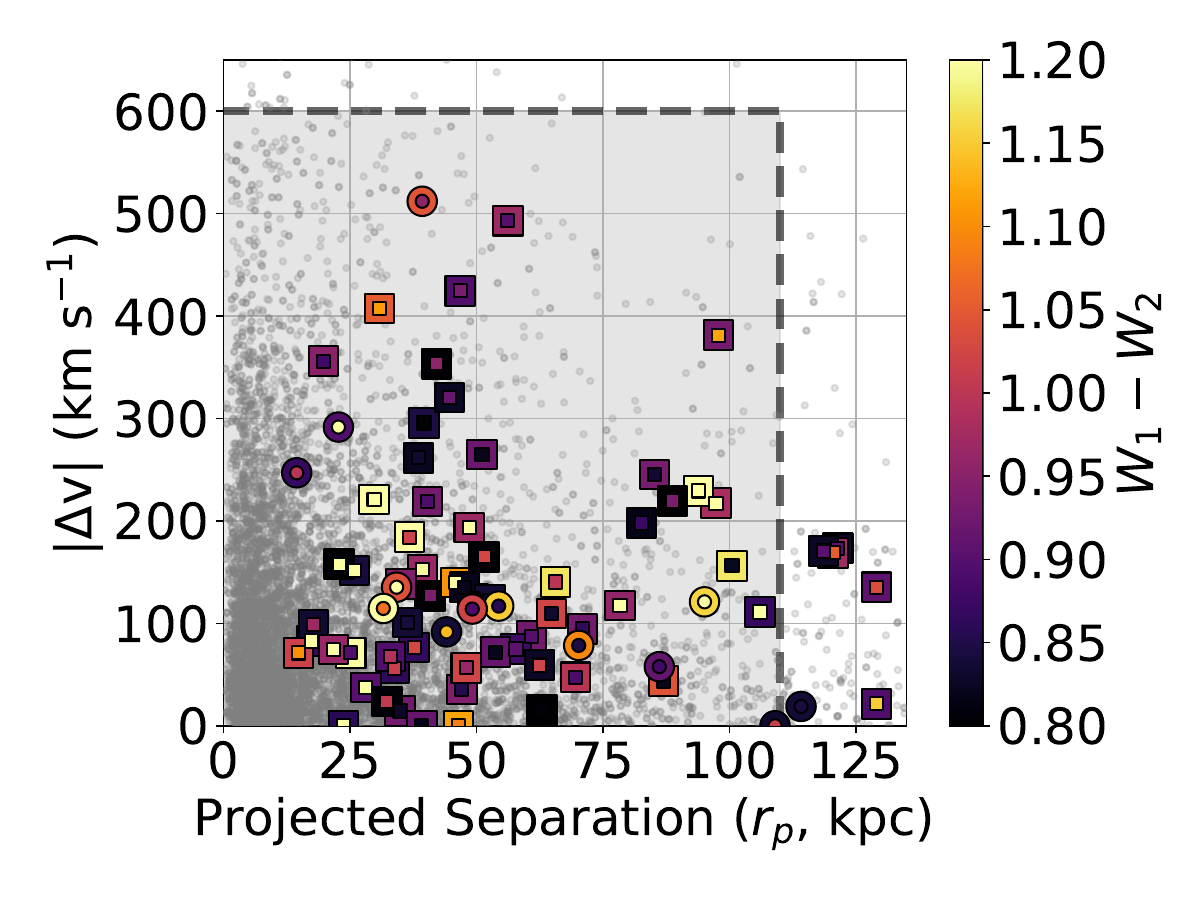}\\
    \includegraphics[width=1.0\linewidth]{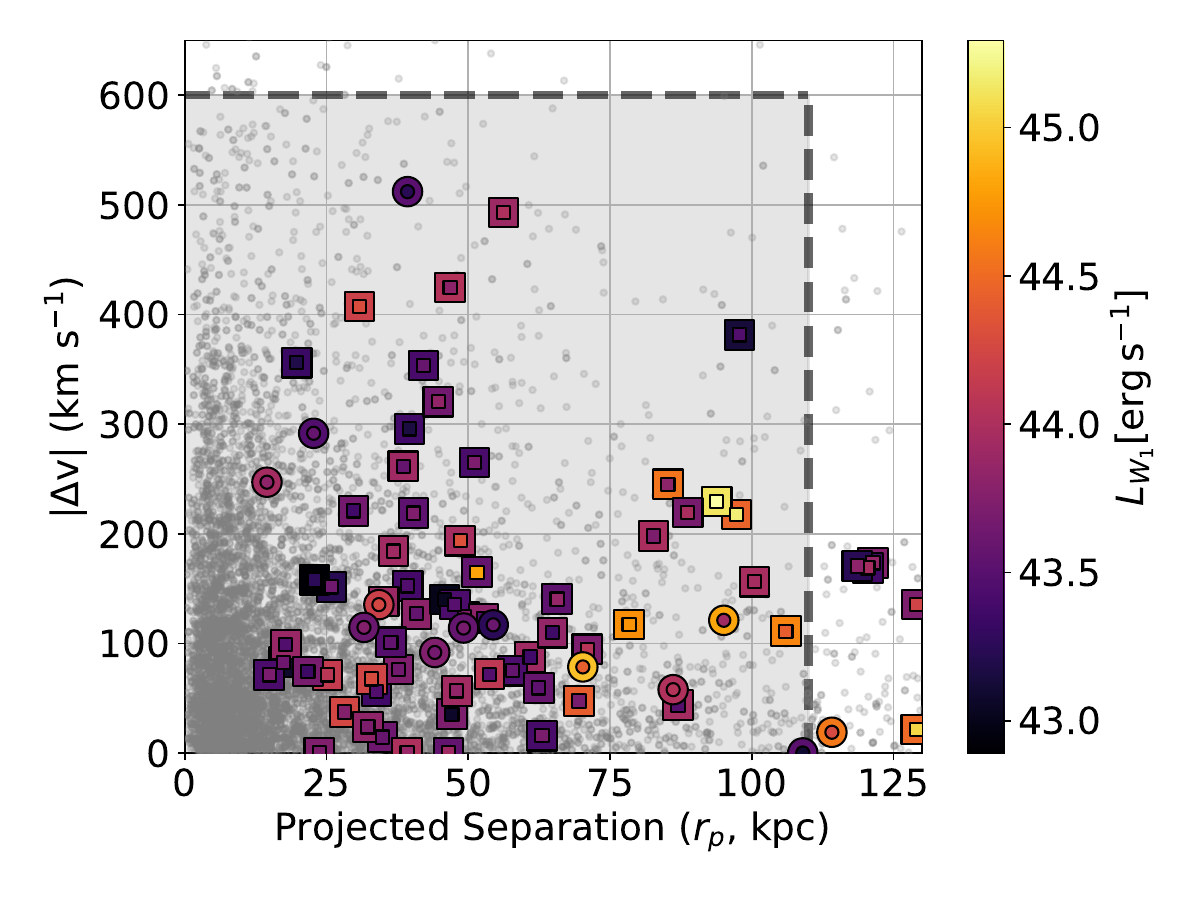}
    \caption{Projected, physical separation (kpc) and velocity difference (km s$^{-1}$) distribution for Rank 1 and 0.5 spatially-resolved mid-IR dual AGNs, color coded by (top) WISE $W1-W2$ color and (bottom) WISE W1 luminosity. In each panel, Rank 1 mid-IR dual AGN pairs are represented by two concentric circles: the inner circle represents AGN 1, and the outer circle represents AGN 2, and the colors correspond to the scaling given on the auxiliary color bars. Similarly, Rank 0.5 candidates (where both nuclei have redshifts) are marked as concentric squares. The gray dashed lines and shaded gray box represent the recommended dual AGN criteria from \citet{pfeifle2025}, while the light gray points represent 2D projected separations and velocity differences of interacting galaxy pairs drawn from Illustris-TNG100 that have undergone a close pass $\leq10$\,kpc \citep[at an arbitrary simulation snapshot, see Section 2 of][see also \citealp{patton2024}]{pfeifle2025}.}

    \label{fig:rp_vs_dV}
\end{figure}

\section{Results} 
\label{sec:results}

\subsection{A Sample of Spatially-Resolved Mid-IR Dual AGN Candidates}
\label{sec:general_sample}

Our spatially resolved mid-IR dual AGN candidate sample currently consists of {107} systems {(after removing 48 contaminants identified so far)} within the DESI LS footprint{, 99 of which were optically classified as galaxy merger candidates, while the remaining 8 were classified as morphologically ambiguous.} {Based on morphological and spectroscopic evidence, this sample has been sorted into 13 Rank 1 (``confirmed'') dual AGNs, 72 Rank 0.5 dual AGN candidates, and 22 Rank 0 dual AGN candidates.} 

As described in Section~\ref{sec:selection}, these {107} AGN pairs were selected using the two-band, {magnitude limited} color criterion $W1-W2\geq0.8$ \citep[$W2<15.05$,][]{stern2012} and they also satisfy the two-band, magnitude-dependent, 90\% reliability criteria from \citet{assef2018}. In {78/107} mid-IR dual AGN candidates {(which includes all 8 morphologically ambiguous systems)}, the constituent AGN candidates in a given pair have been confirmed to reside at the same redshift ($|\Delta v|\leq600$\,km\,s$^{-1}$) with separations $r_p<130\,$kpc and are therefore considered interacting and/or merging galaxy systems. {Considering as a lower limit the 75 systems confirmed to reside in mergers at $z\lesssim0.5$, the sky density of mid-IR dual AGN candidates would be $>4\times10^{-3}$ deg$^{-2}$ in the redshift range $0.07<z<0.5$.}

Thirteen of 78 of these systems with {distinct} nuclear redshift measurements have been identified as Seyfert-Seyfert pairs, exhibited broad lines indicative of AGNs, or were previously confirmed as dual AGNs (see Sections~\ref{sec:bpt_analysis} and \ref{sec:bigmac}), and these are therefore classified as Rank 1 (confirmed) dual AGNs; 65/78 of these systems have redshift measurements for both nuclei but are classified as Rank 0.5 candidates because they require further scrutiny to rule out starburst contaminants. For the 78 Rank 1 and Rank 0.5 candidates that have redshifts measured for both nuclei, we plot the redshift versus projected separation distribution in Figure~\ref{fig:rp_vs_z}, where the mid-IR dual AGN candidates are color coded by angular separation. {For a comparison to the larger literature, we overlay on Figure~\ref{fig:rp_vs_z} in red and gray data points the distribution of Rank 1 and 0.5 dual AGN candidates, respectively, from the Big Multi-AGN Catalog \citep[Big MAC,][]{pfeifle2025}, which is a literature complete catalog of dual AGN, binary AGN, recoiling AGN, and N-tuple AGN candidates identified between 1970 and 2020.}   

\begin{figure*}[p]
\centering
    \subfloat{\includegraphics[width=0.2\linewidth]{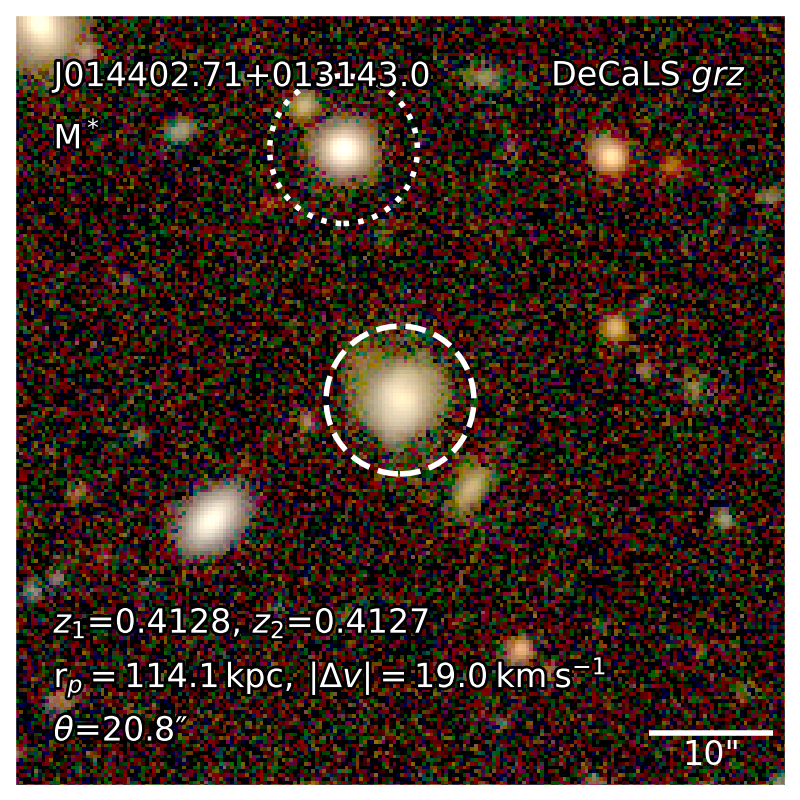}}
    \subfloat{\includegraphics[width=0.2\linewidth]{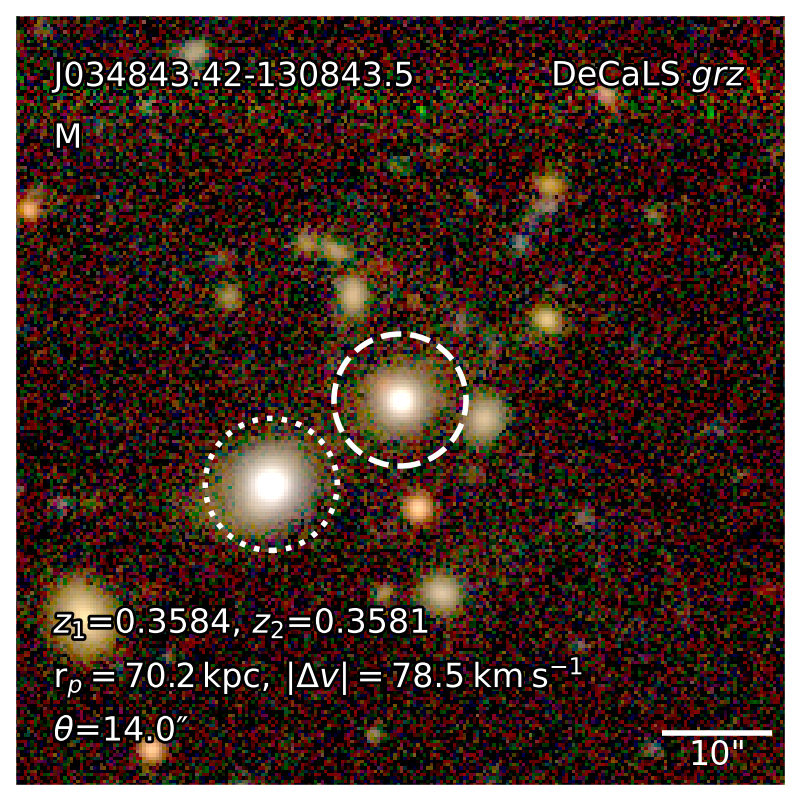}}
    \subfloat{\includegraphics[width=0.2\linewidth]{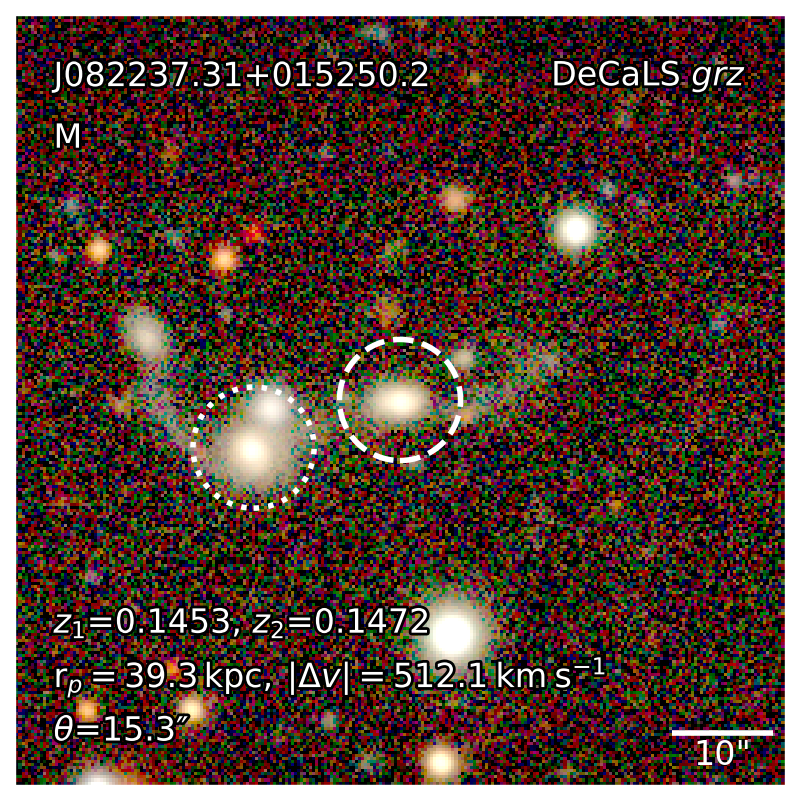}}
    \subfloat{\includegraphics[width=0.2\linewidth]{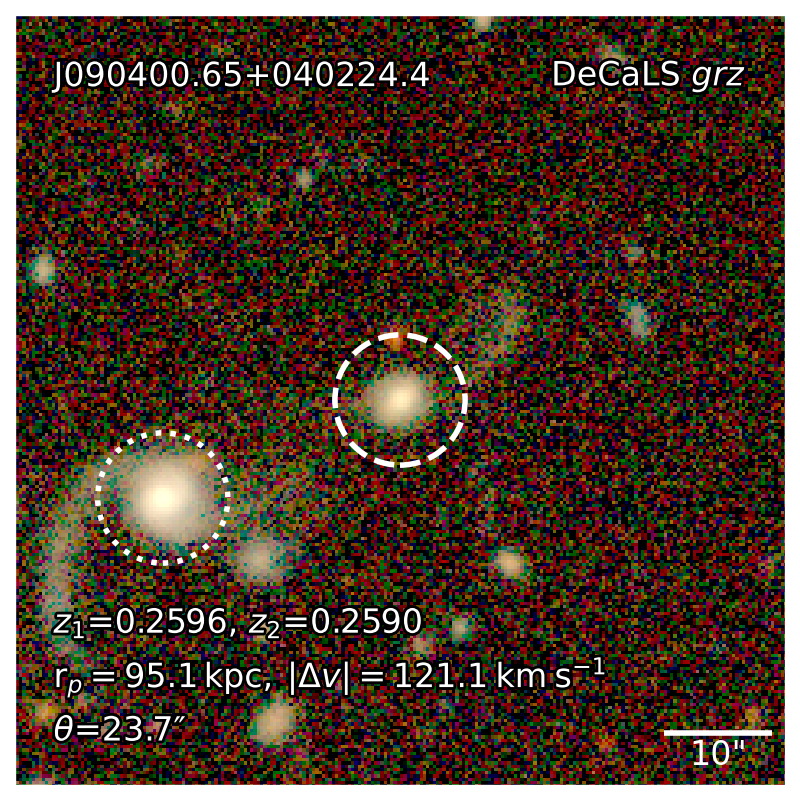}} 
    \subfloat{\includegraphics[width=0.2\linewidth]{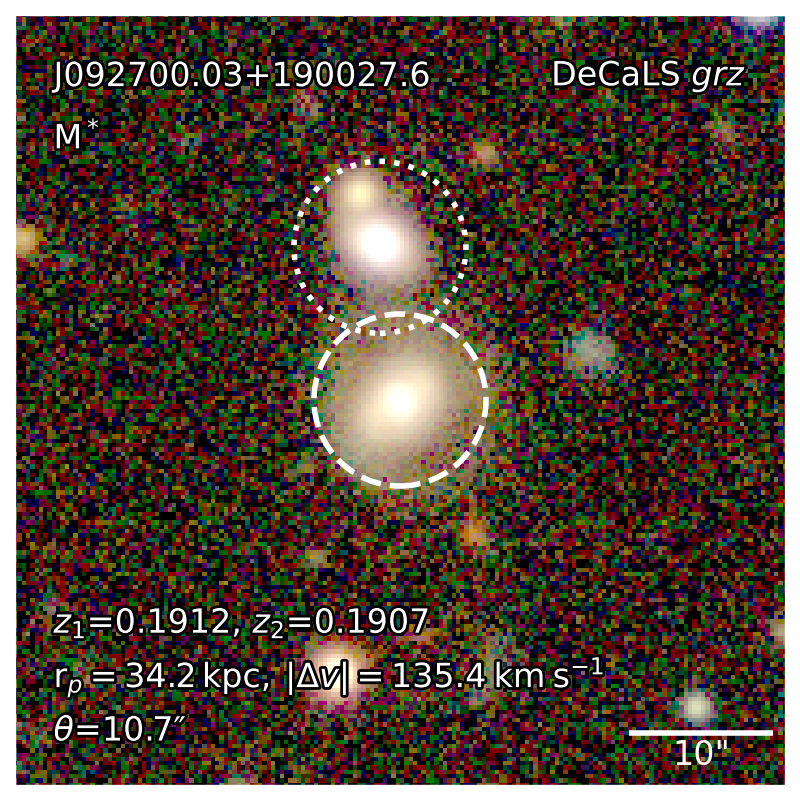}}\\
    \vspace{-4.5mm}
    \subfloat{\includegraphics[width=0.2\linewidth]{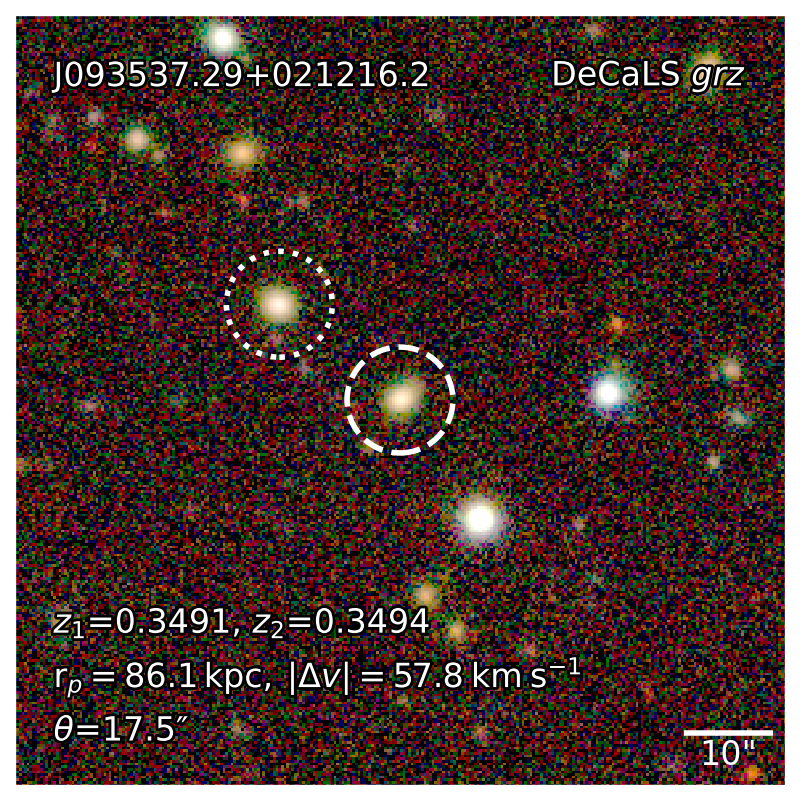}}
    \subfloat{\includegraphics[width=0.2\linewidth]{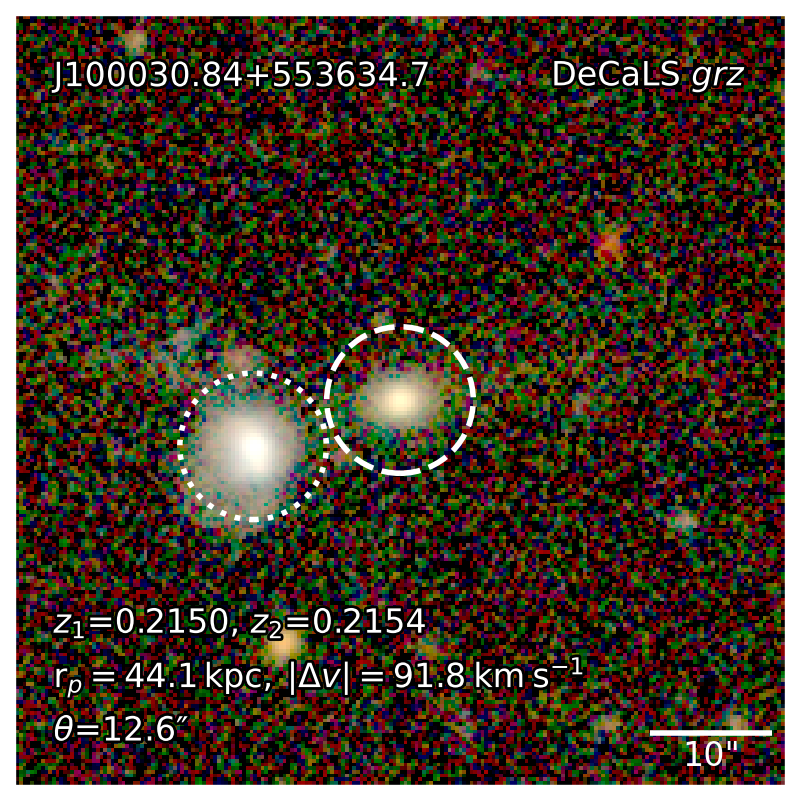}}
    \subfloat{\includegraphics[width=0.2\linewidth]{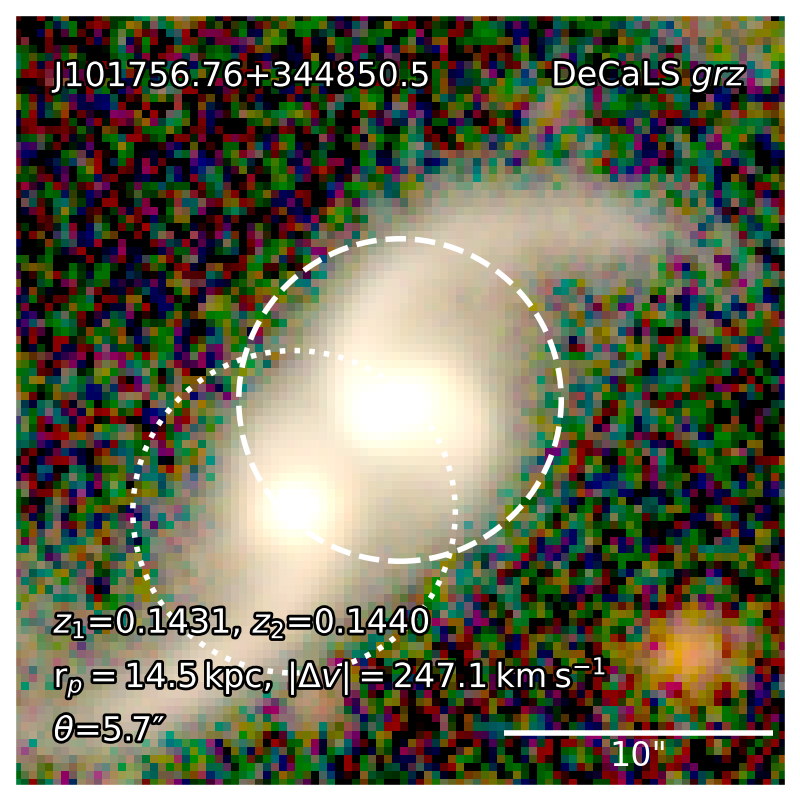}}
    \subfloat{\includegraphics[width=0.2\linewidth]{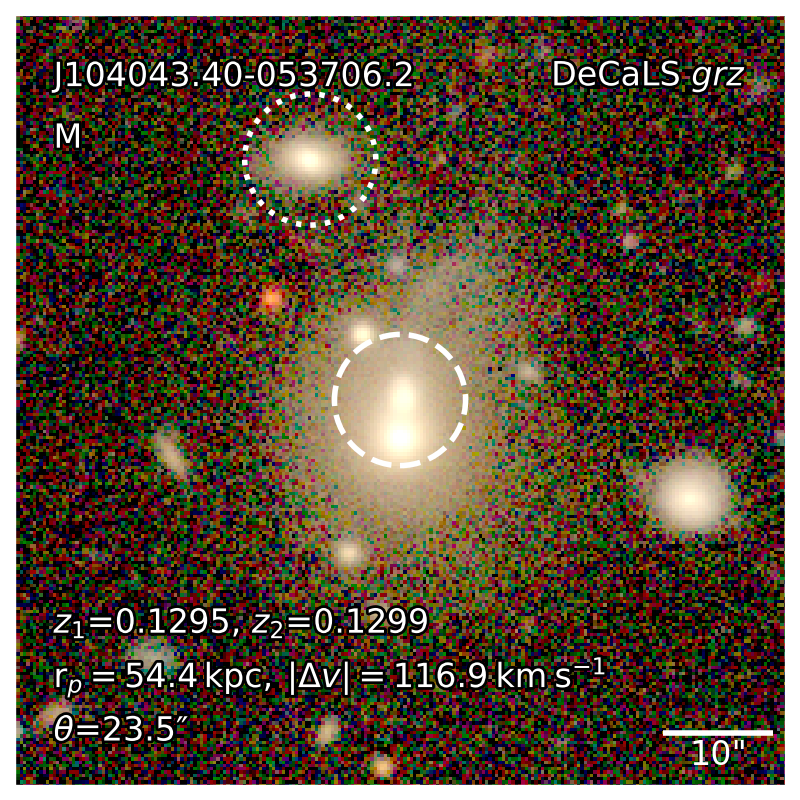}}
    \subfloat{\includegraphics[width=0.2\linewidth]{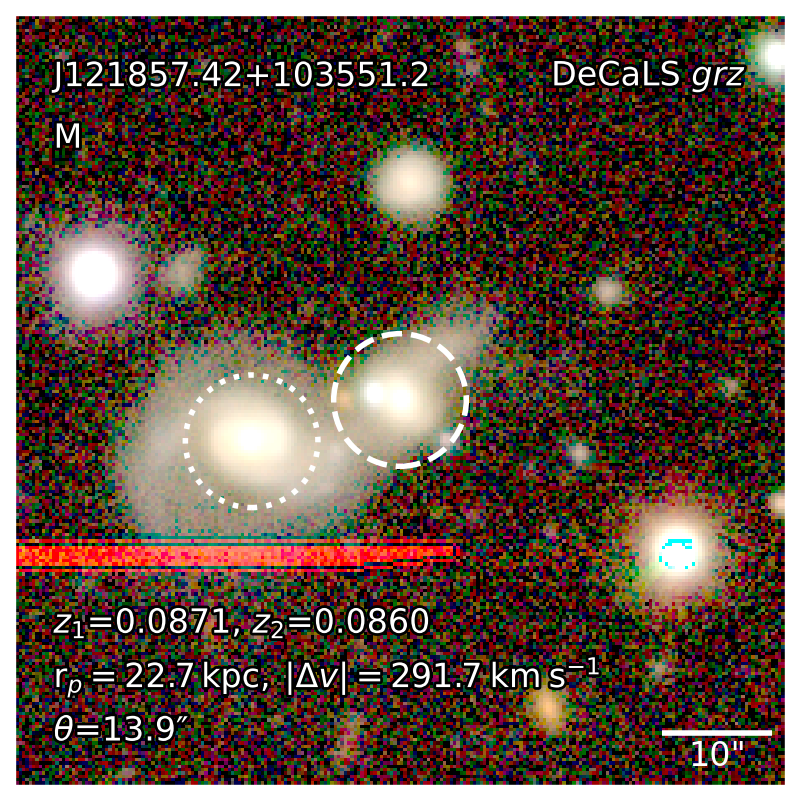}}\\
    \vspace{-4.5mm}
    \subfloat{\includegraphics[width=0.2\linewidth]{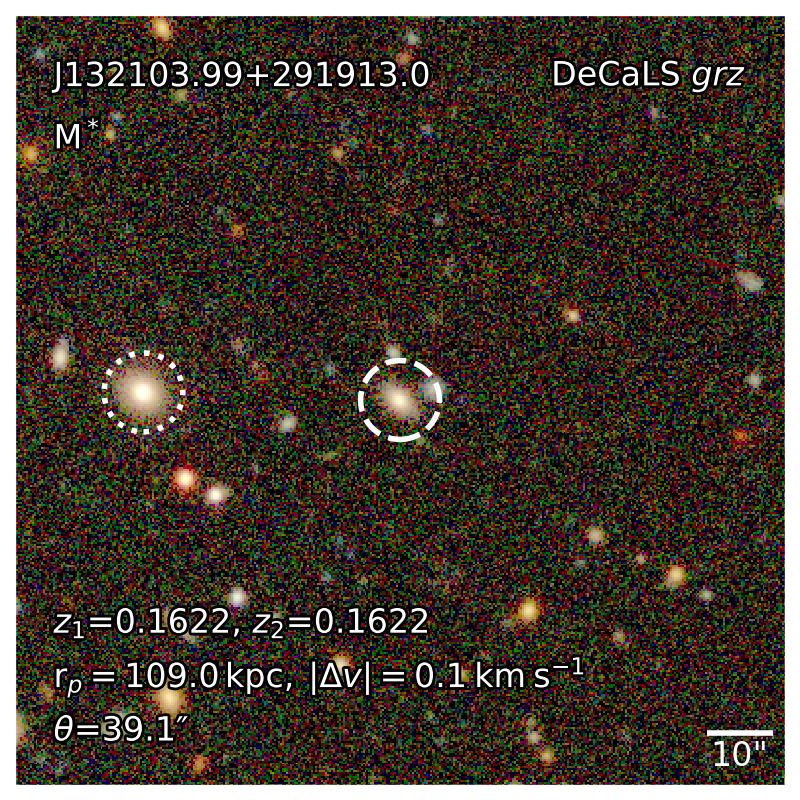}}
    \subfloat{\includegraphics[width=0.2\linewidth]{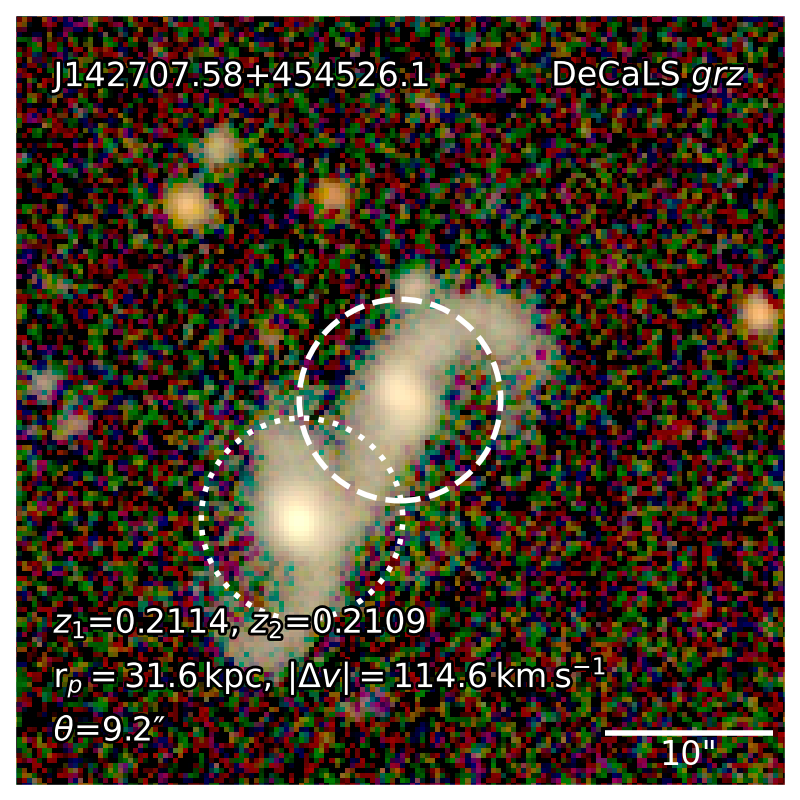}}
    \subfloat{\includegraphics[width=0.2\linewidth]{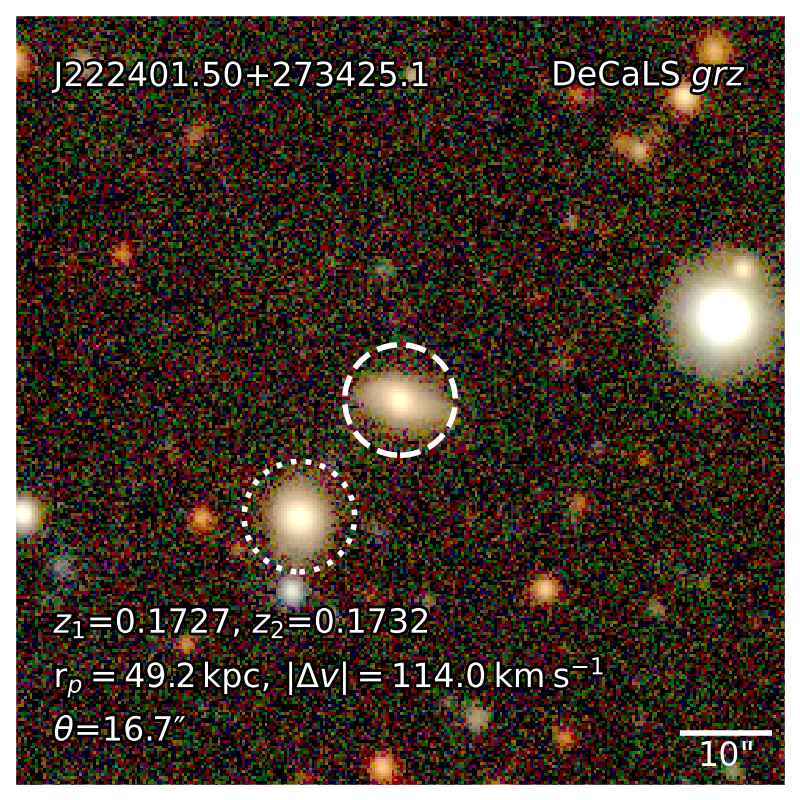}}\\
\caption{Rank 1 mid-IR dual AGNs. For each panel, the WISE designation for ``AGN 1'' is given in the top left corner, and a 10'' scale bar is given in the bottom right corner. The redshifts of the two AGNs ($z_1$ and $z_2$), projected separation ($\rm{r}_p$), absolute velocity difference ($|\Delta v|)$, and angular separation ($\theta)$ of the two mid-IR AGNs are given in the bottom left corner of each panel. White, 6'' diameter dashed and dotted circles denote the WISE positions AGN 1 and AGN 2, respectively, for a given pair. Panels marked with an ``M'' below the AGN 1 designation are multi-mergers, where at least one additional companion was spectroscopically confirmed to reside within 130\,kpc and $\leq700$\,km\,s$^{-1}$; ``M$^*$'' markers indicate multi-merger candidates, where additional candidate companions are seen in the imaging, but spectroscopic redshift verification is still needed. {For reference, J014402.71+013143.0, was originally marked as morphologically ambiguous.} {J121857.42+103551.2 was confirmed in \citet{schwartzman2025} as a triple AGN.}}
\label{fig:rank1_duals}
\end{figure*}

The rank 1 and 0.5 sample currently exhibits a median redshift $\tilde{z}=0.242^{+0.068}_{-0.077}$, where the upper and lower bounds provide the 84$^{\rm{th}}$ and 16$^{\rm{th}}$ percentile values of the redshift distribution; these candidates occupy a gap in redshift space relative to known dual AGN candidates from the literature (up to 2020). Rank 1 and 0.5 mid-IR dual AGN candidates are found across a large variety of physical, 2D separations, from $\sim 14.5-129$\,kpc (median separation: $47.8^{+46.9}_{-17.7}$ kpc). Rank 1 and rank 0.5 mid-IR dual AGN candidates are listed in Tables~\ref{tab:rank1} and \ref{tab:rank0.5}. 7 additional Rank 0.5 dual AGN candidates (lacking redshifts for both nuclei) are listed in Table~\ref{tab:rank0.5} and 22 Rank 0 dual AGN candidates are listed in Table~\ref{tab:rank0}; follow-up spectra from ongoing and future spectroscopic campaigns will be used to confirm or reject all remaining Rank +0.5 and 0 candidates. DESI LS $grz$ tricolor images of Rank 1, Rank 0.5, and Rank 0 candidates are shown in Figures~\ref{fig:rank1_duals}, \ref{fig:rank0.5_duals}, and \ref{fig:rank0_duals}, respectively.

The sample distribution of $|\Delta v|$ and $r_p$ is shown in Figure~\ref{fig:rp_vs_dV}, where each Rank 1 and 0.5 system is represented by a set of concentric circles or squares, respectively (again, including only systems that have redshift measurements for both nuclei). {As expected through the use of morphological selection to identify mergers and merger candidates, the sample appears to mimic the $r_p$ and $|\Delta v|$ distributions of simulated interacting galaxies within Illustris-TNG100 \citep[][displayed as gray dots in Figure~\ref{fig:rp_vs_dV}]{patton2024}.} The vast majority ({71/78}) of our mid-IR dual AGN candidates with redshifts for both nuclei conform to the dual AGN definitions outlined in \citet{pfeifle2025} and do not require the expanded {$r_p$} and {$|\Delta v|$} criteria described in Section~\ref{sec:defining_duals}. Among the six Rank 1 and 0.5 mid-IR dual AGNs in this sample with separations $110\,\rm{kpc}<r_p<130\,\rm{kpc}$, all have velocity differences $|\Delta v|\leq250$\,km\,s$^{-1}$ (in other words, $\ll600$\,km\,s$^{-1}$) and are therefore included in the final sample; these systems make up {$7.1^{+2.7}_{-2.2}\%$ of the sample of 107, and $9.7^{+3.6}_{-3.0}\%$ of Rank 1 and 0.5 systems with complete redshifts}\footnote{Using the \textsc{scipy} function \textsc{special.betaincinv} to calculate the percentile values, where the upper and lower bounds are the 84$^{\rm{th}}$ and 16$^{\rm{th}}$ percentile values.} \citep[{consistent at the 2-3$\sigma$ level with} predictions from galaxy pairs in Illustris-TNG100,][see also Figure~2 in \citealp{pfeifle2025}]{patton2024} and could represent early-stage or long-lived mergers and interactions. Two of these are confirmed to reside in multi-mergers or groups and an additional two potentially reside in multi-mergers or groups, which may be conducive to producing larger separation dual AGNs; encounters in group environments can result in complex orbital dynamics and disrupt subsequent pericentric passages \citep[see Figure~10 in][]{patton2024}.

\begin{table*}[p]
\begin{center}
\caption{Rank 1 Confirmed Spatially-Resolved Mid-IR Dual AGNs}
\label{tab:rank1}
\begin{tabular}{ccccccccccc}
\hline
\hline
\noalign{\smallskip}
\noalign{\smallskip}
$\rm{designation}_1$ &  $z_1$ & $z_1$ source & W1-W2$_1$ &  $\rm{designation}_2$ &  $z_2$ & $z_2$ source & W1-W2$_2$ & $\theta_{\rm{MIR}}$ &  $r_{p\,{\rm{(MIR)}}}$ &  $|\Delta v|$ \\
 &  &  &  &  &  &  &  & (\arcsec{}) & (kpc) & km s$^{-1}$\\
(1) & (2) & (3) & (4) & (5) & (6) & (7) & (8) & (9) & (10) & (11)  \\
\noalign{\smallskip}
\noalign{\smallskip}
\hline
\noalign{\smallskip}

J014402.71+013143.0 & 0.413 &   D & 0.84 & J014403.02+013203.4 & 0.413 &   D & 0.84 & 20.8 & 114.1 &   19.0 \\
J034843.42-130843.5 & 0.358 &   D & 0.85 & J034844.22-130851.2 & 0.358 &   D & 1.09 & 14.0 &  70.2 &   78.5 \\
J082237.31+015250.2$^{\dagger}$ & 0.145 & G & 0.96 & J082238.28+015245.5$^{\dagger}$ & 0.147 & G & 1.04 & 15.3 &  39.3 & 512.1 \\
J090400.65+040224.4 & 0.260 &   D & 1.33 & J090402.11+040215.4 & 0.259 &   D & 1.16 & 23.7 &  95.1 &  121.1 \\
J092700.03+190027.6 & 0.191 & G & 1.51 & J092700.13+190038.3 & 0.191 & G & 1.04 & 10.7 &  34.2 &  135.4 \\
J093537.29+021216.2 & 0.349 &   D & 0.88 & J093538.20+021227.0 & 0.349 &   D & 0.91 & 17.5 &  86.1 &  57.8 \\
J100030.84+553634.7$^{\dagger}$ & 0.215 &   S & 1.14 & J100032.26+553630.9$^{\dagger}$ & 0.215 &   D & 0.84 & 12.6 &  44.1 &  91.8 \\
J101756.76+344850.5 & 0.143 &  G & 1.00 & J101757.08+344846.4 & 0.144 &   S & 0.87 &  5.7 &  14.5 & 247.1 \\
J104043.40-053706.2 & 0.129 &   D & 0.86 & J104043.95-053644.2 & 0.130 & G & 1.15 & 23.5 &  54.4 & 116.9 \\
J121857.42+103551.2 & 0.087 &   K & 1.34 & J121858.33+103547.5 & 0.086 &   K & 0.89 & 13.9 &  22.7 &  291.7 \\
J132103.99+291913.0$^{\dagger}$ & 0.162 & G & 1.01 & J132106.98+291914.2$^{\dagger}$ & 0.162 & G & 0.83 & 39.1 & 109.0 &   0.1 \\
J142707.58+454526.1 & 0.211 &   D & 1.07 & J142708.14+454519.0 & 0.211 & G & 1.26 &  9.2 &  31.6 &  114.6 \\
J222401.50+273425.1 & 0.173 &   D & 0.89 & J222402.32+273412.5 & 0.173 &   D & 1.02 & 16.7 &  49.2 & 114.0 \\
\noalign{\smallskip}
\hline
\end{tabular}
\end{center}
\tablecomments{Mid-IR dual AGNs spectroscopically confirmed to be at the same redshift and confirmed as Rank 1 based on Seyfert-Seyfert emission line ratios, broad lines, or other evidence in prior works. Col 1-4: \textit{WISE} designation, spectroscopic redshift, redshift source, and \textit{WISE} $W1-W2$ color for AGN 1. Col 5-8:  \textit{WISE} designation, spectroscopic redshift, redshift source, and \textit{WISE} $W1-W2$ color for AGN 2. For redshift sources, D=DESI, K=Keck, G=Gemini, S=SDSS. Col 9-10: angular (\arcsec{}) and projected, physical separation (kpc) of the two \textit{WISE} AGNs. Col 11: velocity difference between the two AGNs/galaxies. {J121857.42+103551.2 was confirmed in \citet{schwartzman2025} as a triple AGN. $^{\dagger}$: Systems identified as spatially resolved mid-IR dual AGN candidates in \citet{barrows2023}.}
}
\end{table*}

\begin{figure*}
\centering
    \subfloat{\includegraphics[width=0.2\linewidth]{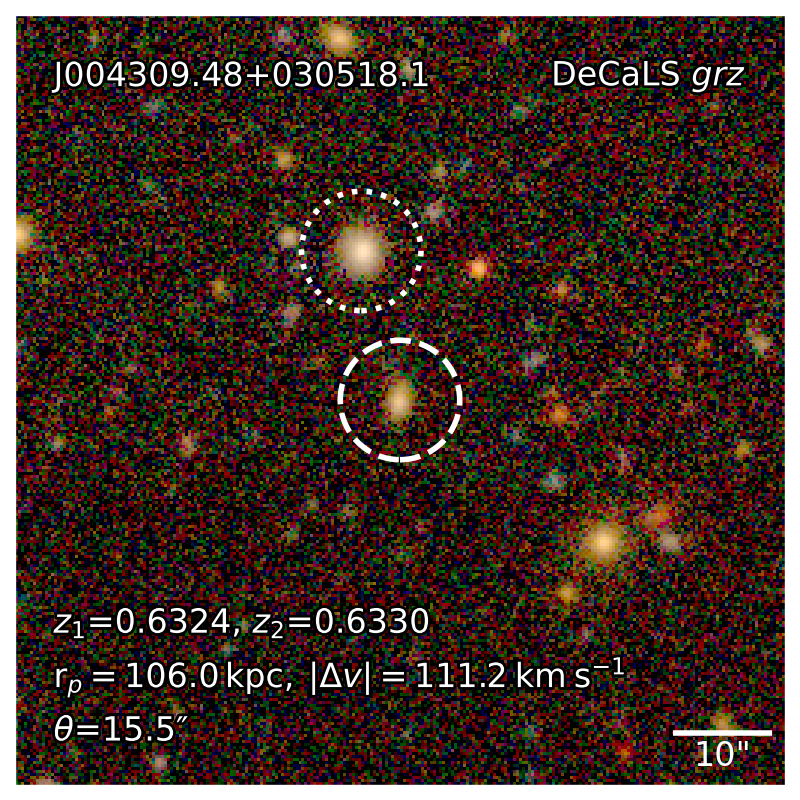}}
    \subfloat{\includegraphics[width=0.2\linewidth]{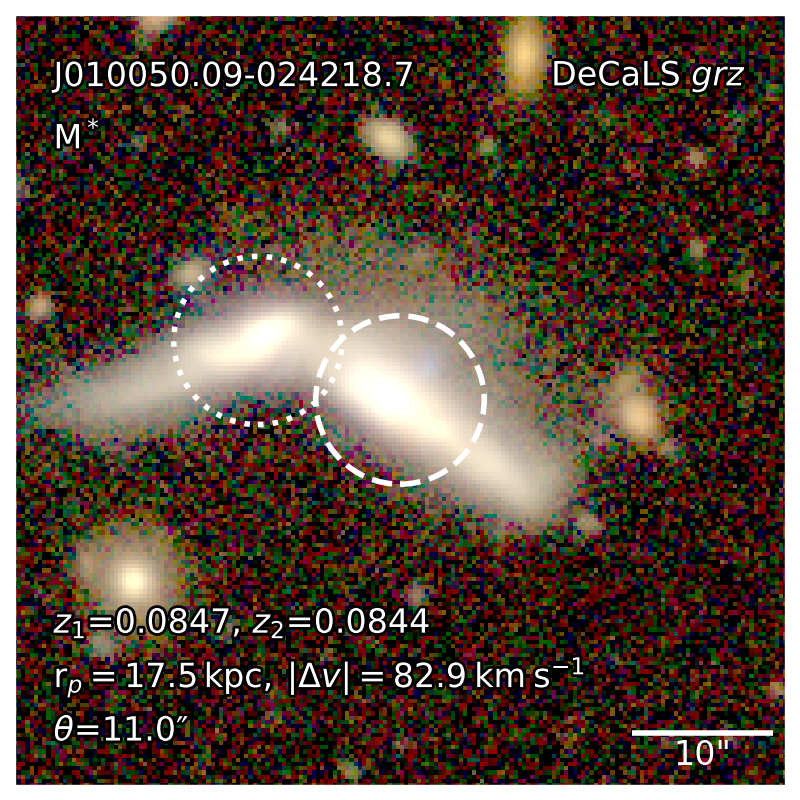}}
    \subfloat{\includegraphics[width=0.2\linewidth]{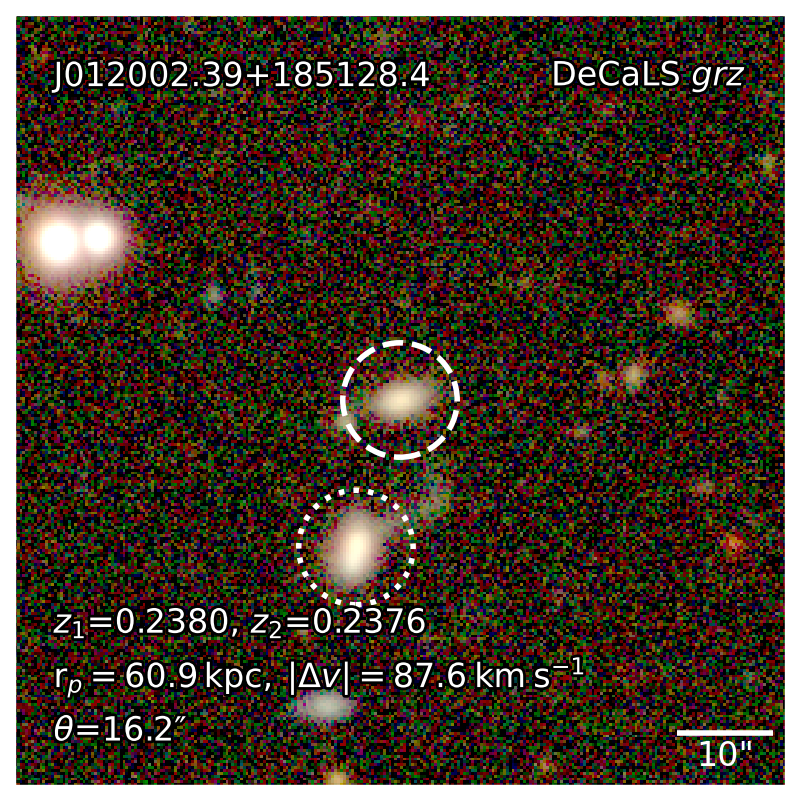}}
    \subfloat{\includegraphics[width=0.2\linewidth]{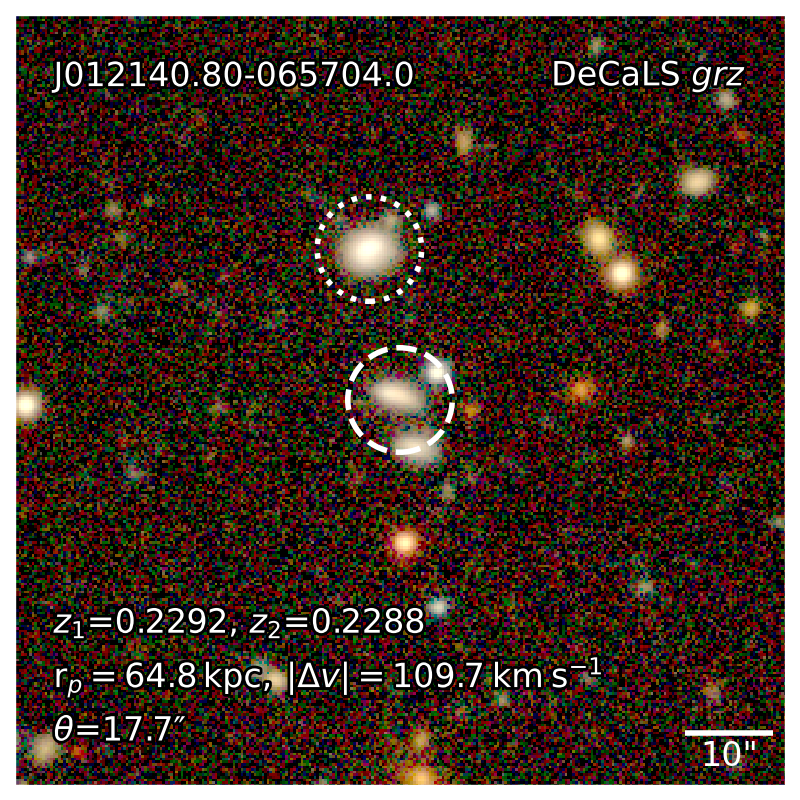}}
    \subfloat{\includegraphics[width=0.2\linewidth]{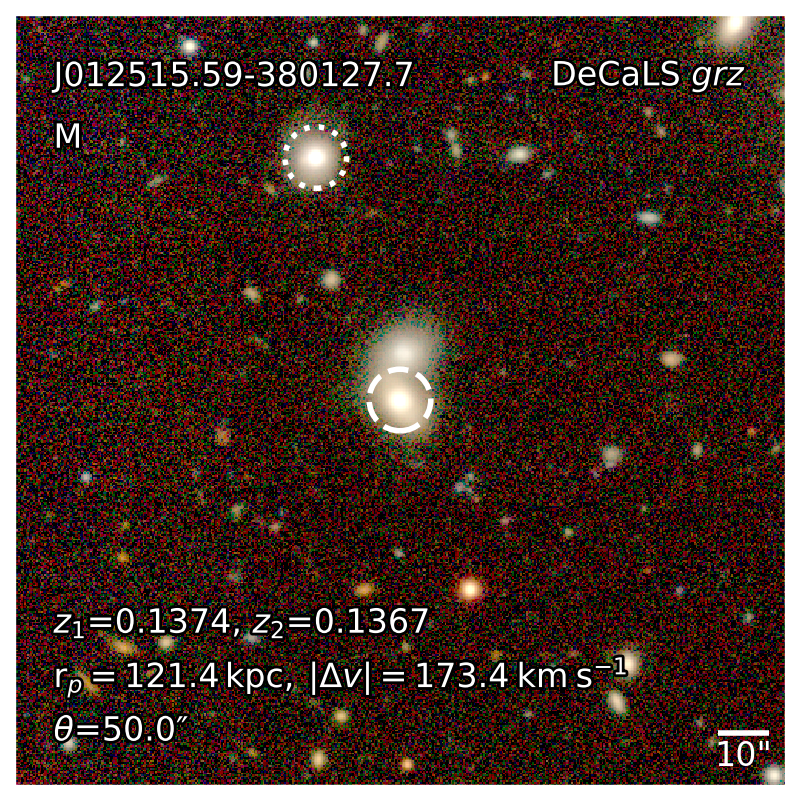}}\\
    \vspace{-4.5mm}
    \subfloat{\includegraphics[width=0.2\linewidth]{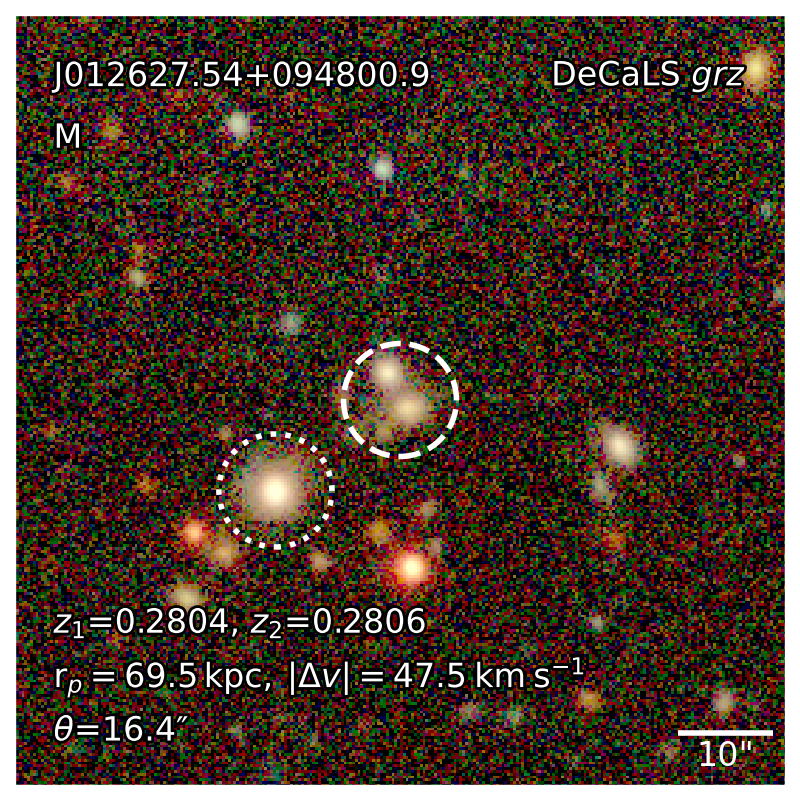}}
    \subfloat{\includegraphics[width=0.2\linewidth]{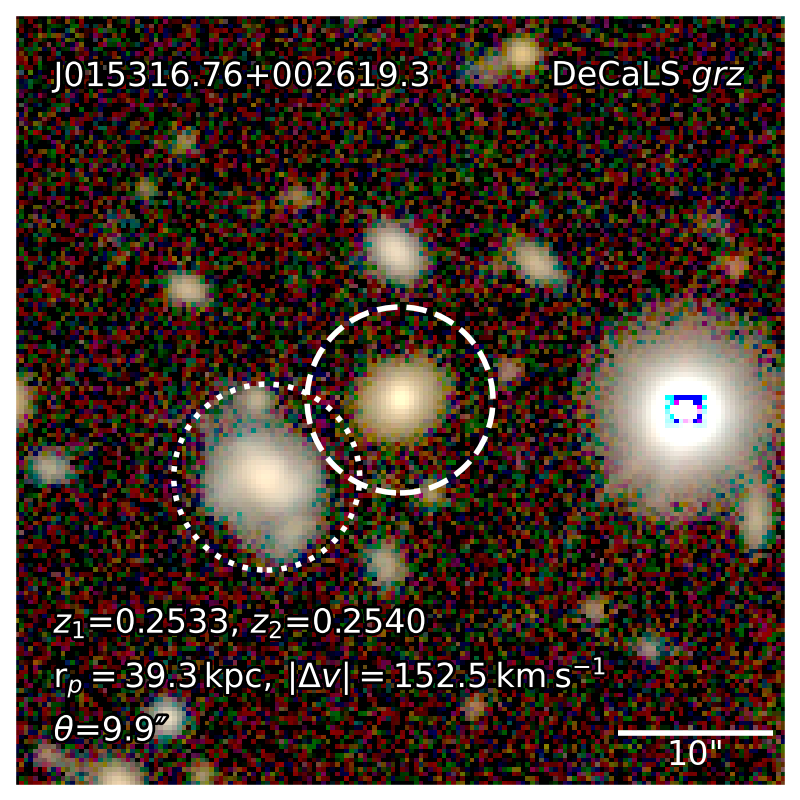}}
    \subfloat{\includegraphics[width=0.2\linewidth]{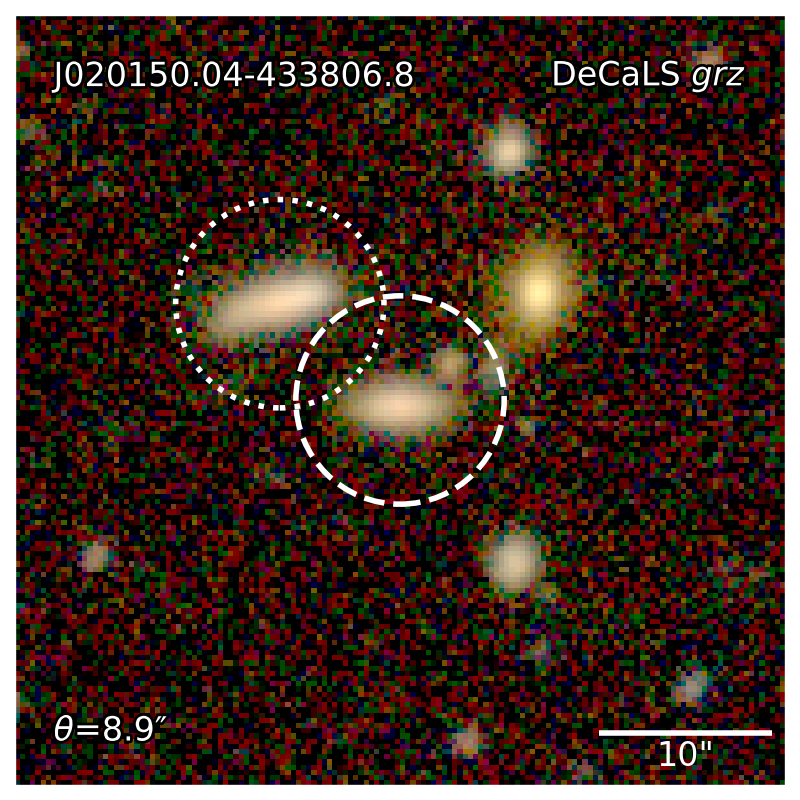}}
    \subfloat{\includegraphics[width=0.2\linewidth]{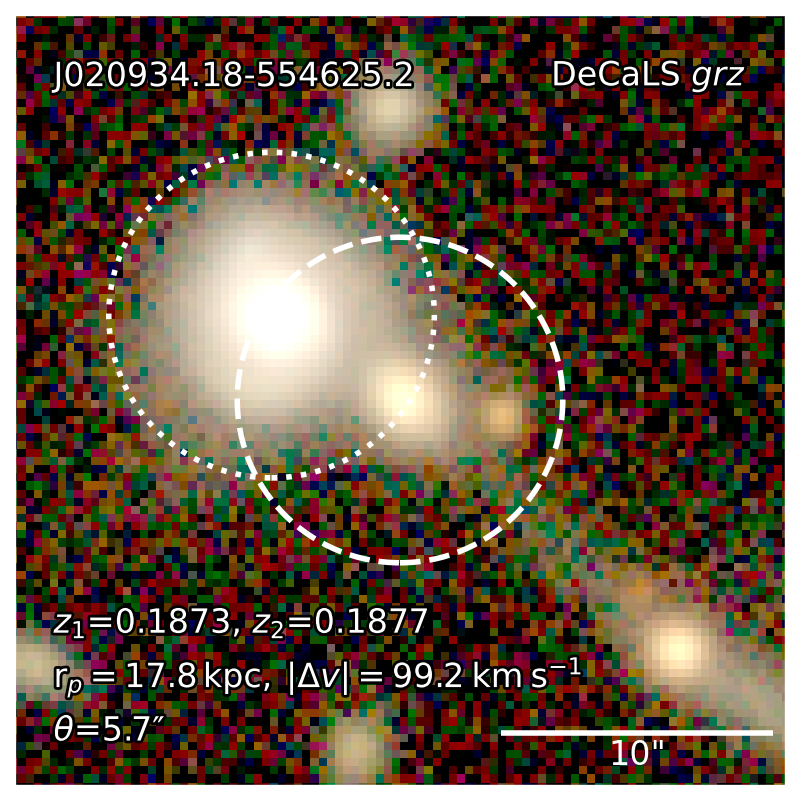}}
    \subfloat{\includegraphics[width=0.2\linewidth]{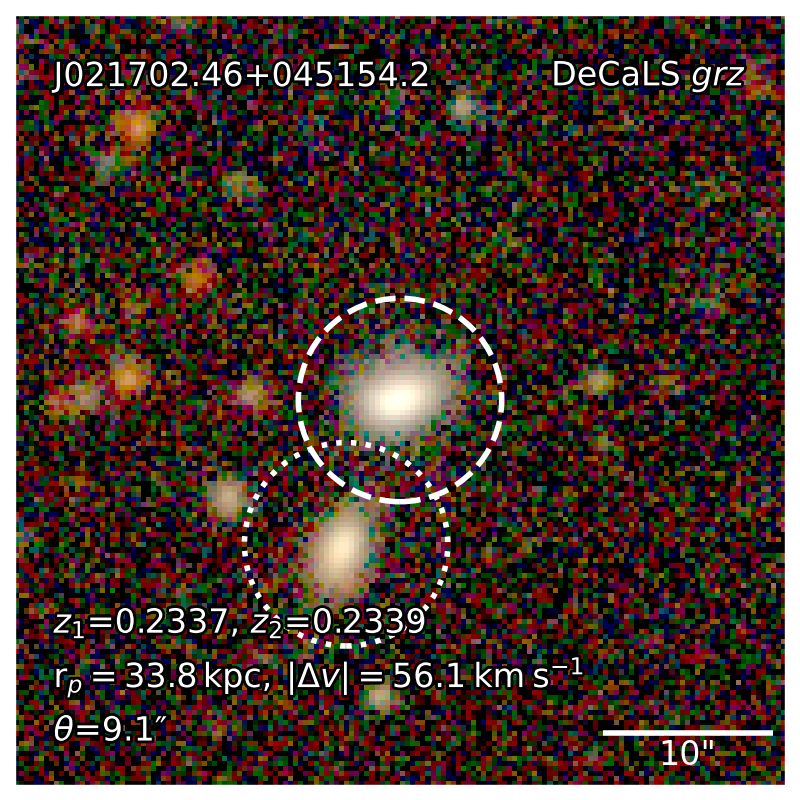}}\\
    \vspace{-4.5mm}
    \subfloat{\includegraphics[width=0.2\linewidth]{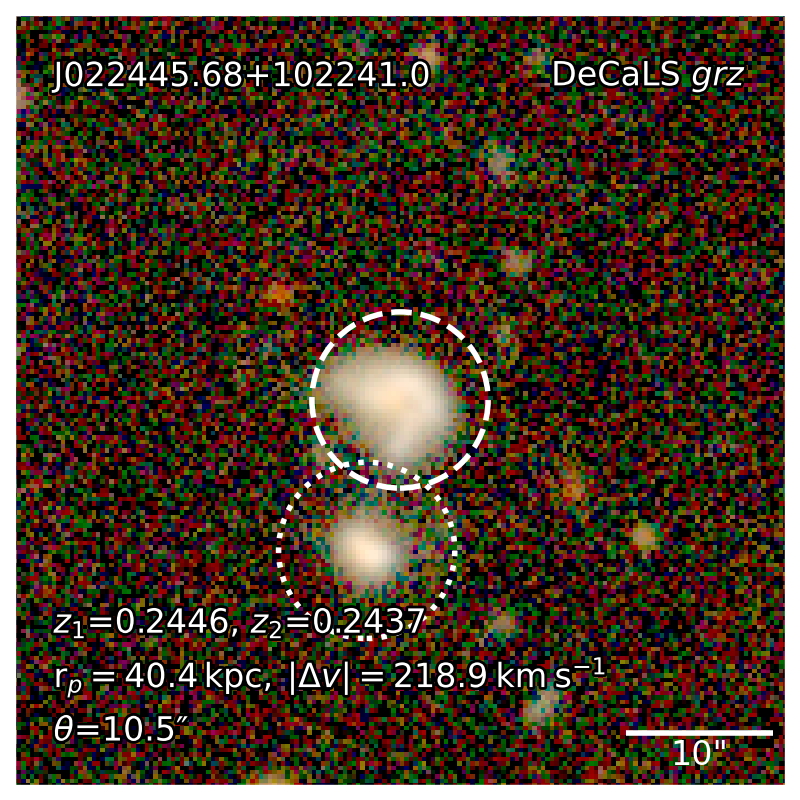}}
    \subfloat{\includegraphics[width=0.2\linewidth]{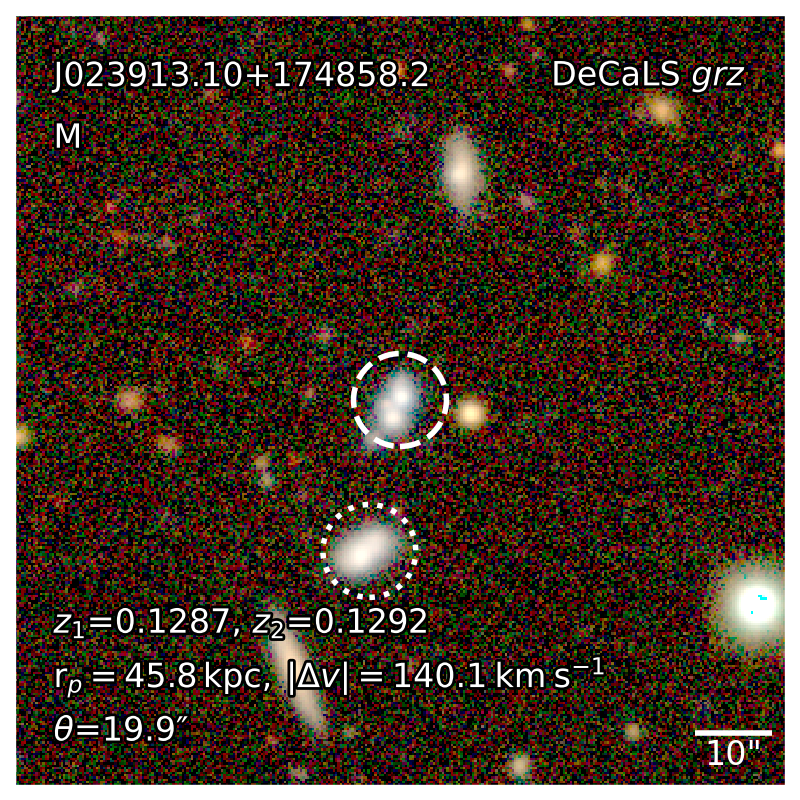}}
    \subfloat{\includegraphics[width=0.2\linewidth]{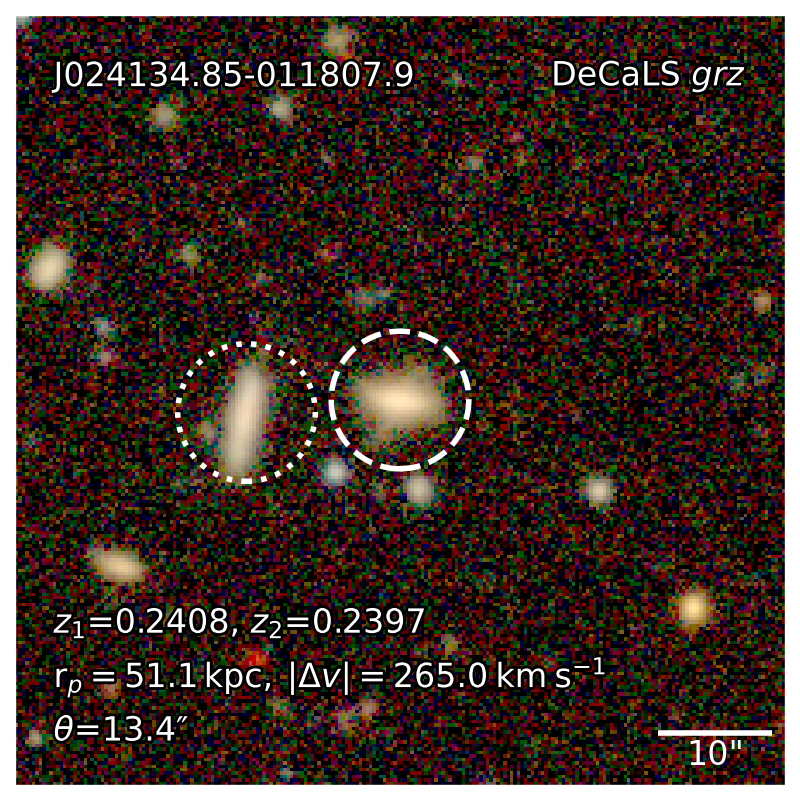}}
    \subfloat{\includegraphics[width=0.2\linewidth]{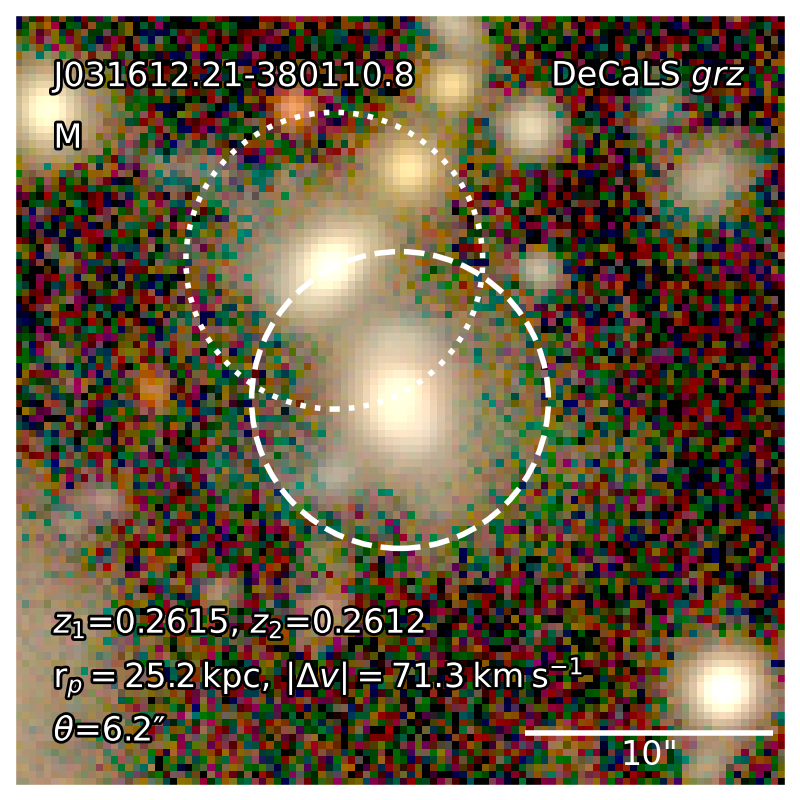}}
    \subfloat{\includegraphics[width=0.2\linewidth]{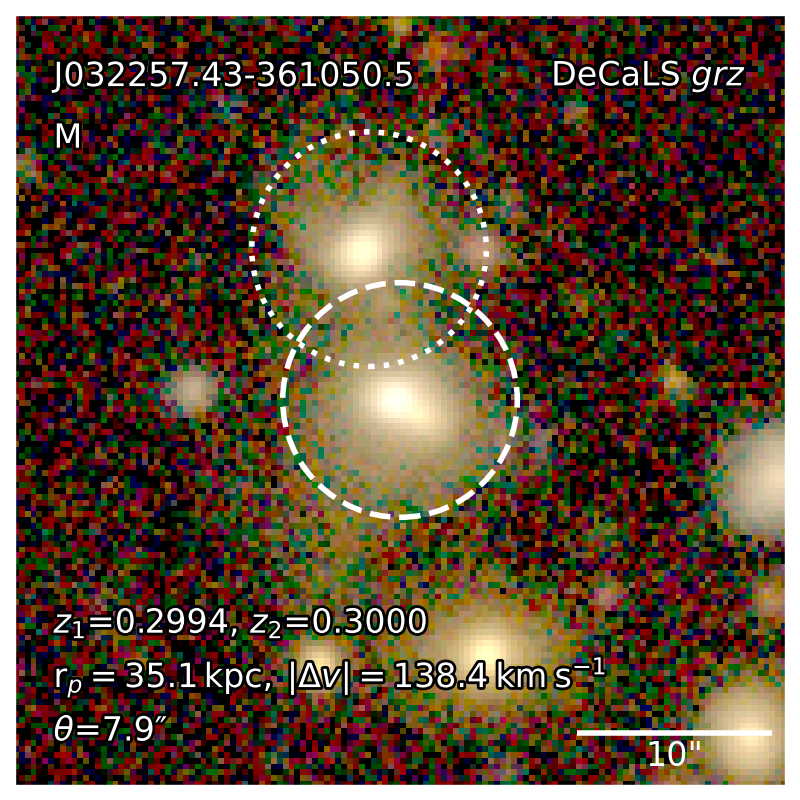}}\\
    \vspace{-4mm}
    \subfloat{\includegraphics[width=0.2\linewidth]{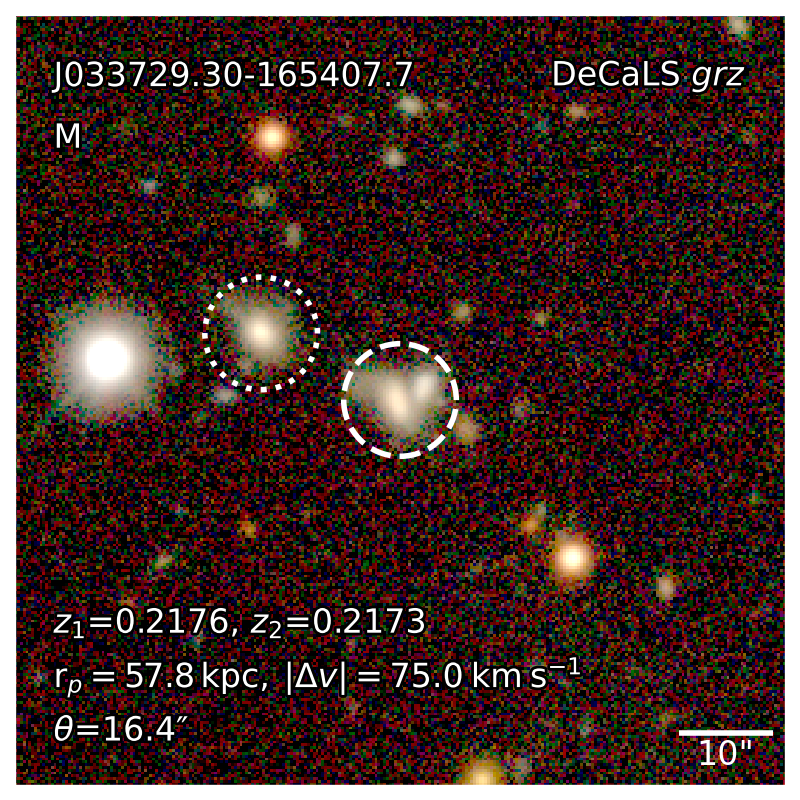}}
    \subfloat{\includegraphics[width=0.2\linewidth]{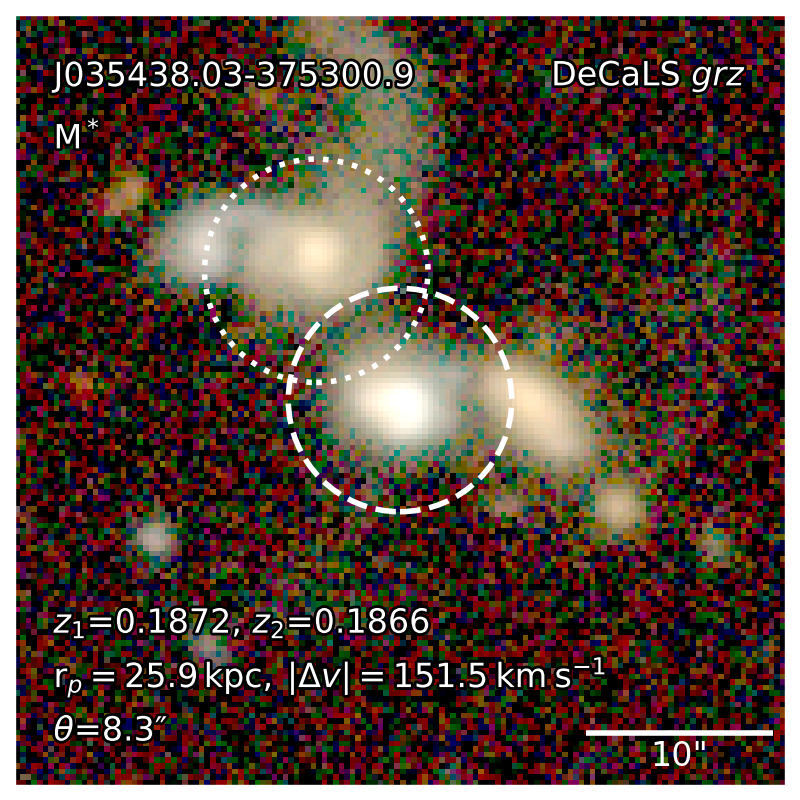}}  
    \subfloat{\includegraphics[width=0.2\linewidth]{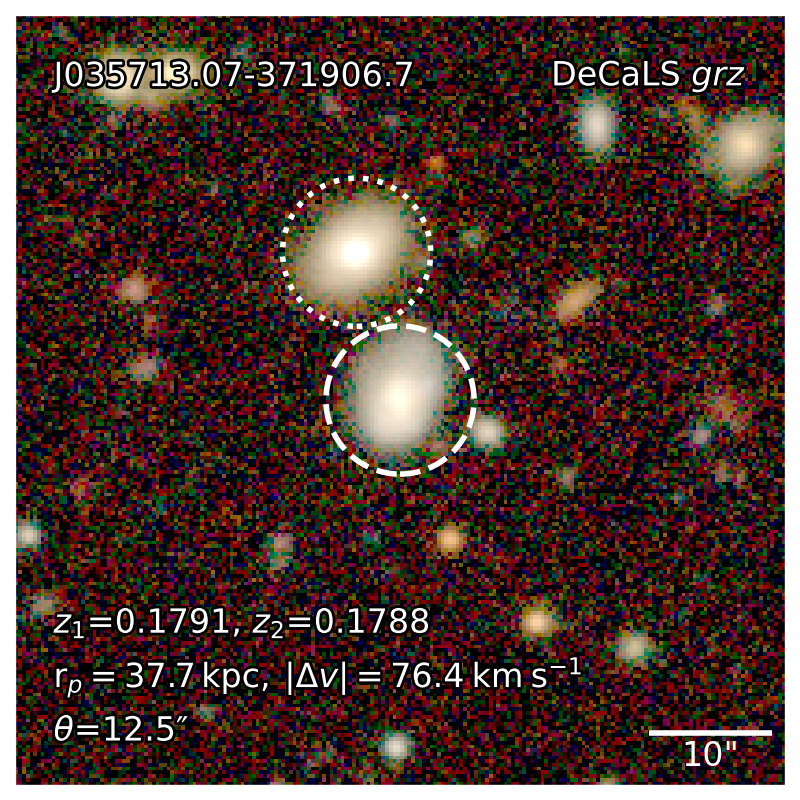}}  
    \subfloat{\includegraphics[width=0.2\linewidth]{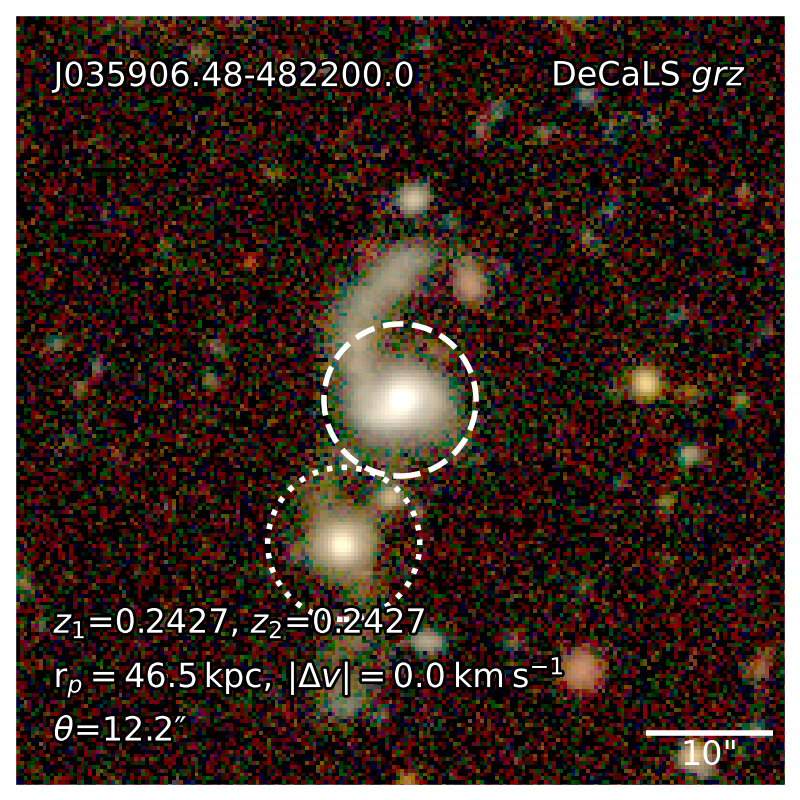}}    
    \subfloat{\includegraphics[width=0.2\linewidth]{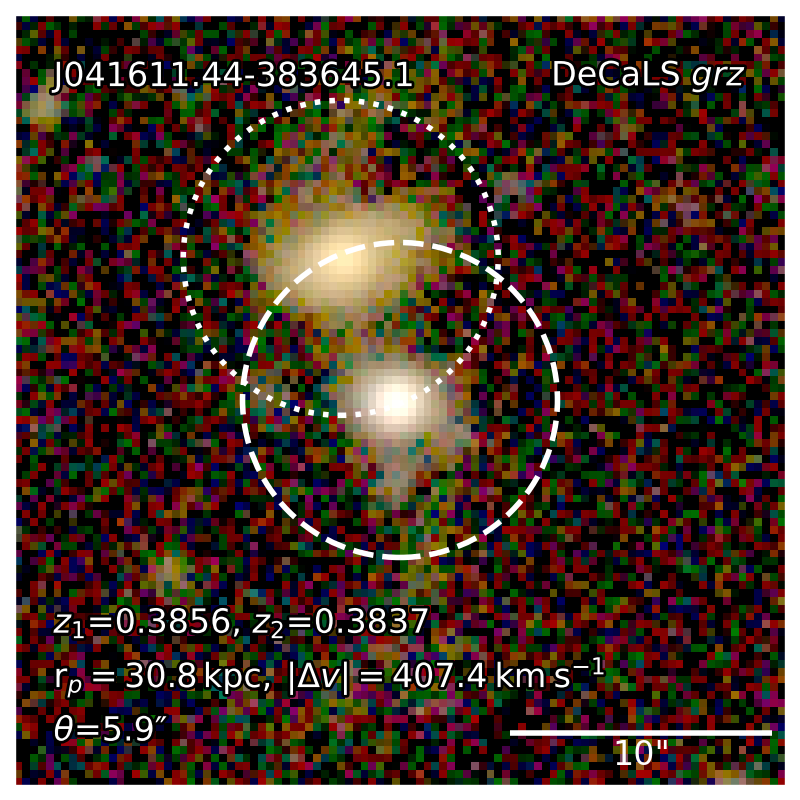}}\\   
    \vspace{-4.5mm}
    \subfloat{\includegraphics[width=0.2\linewidth]{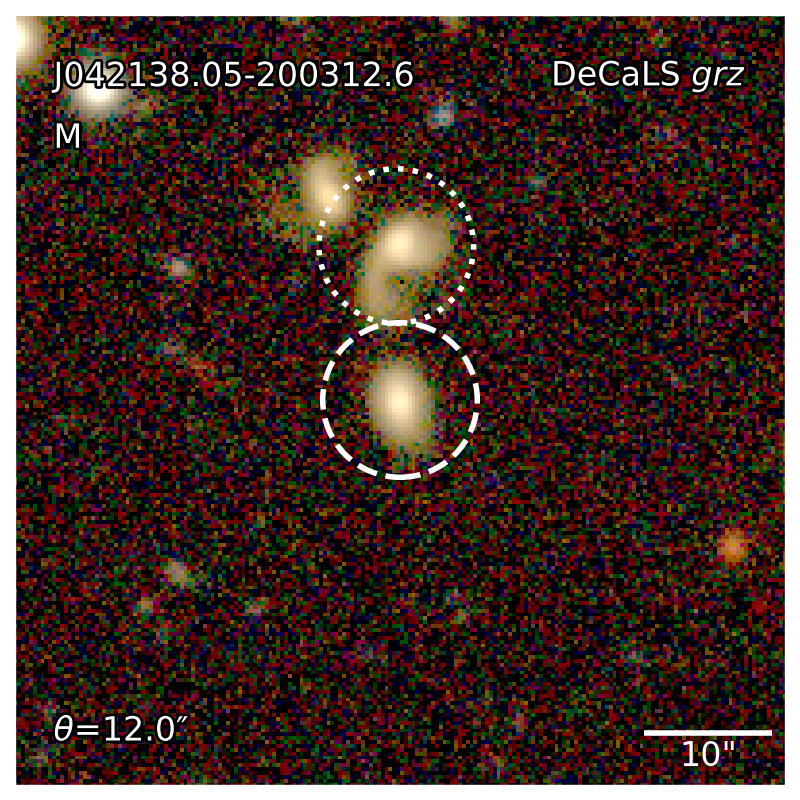}}
    \subfloat{\includegraphics[width=0.2\linewidth]{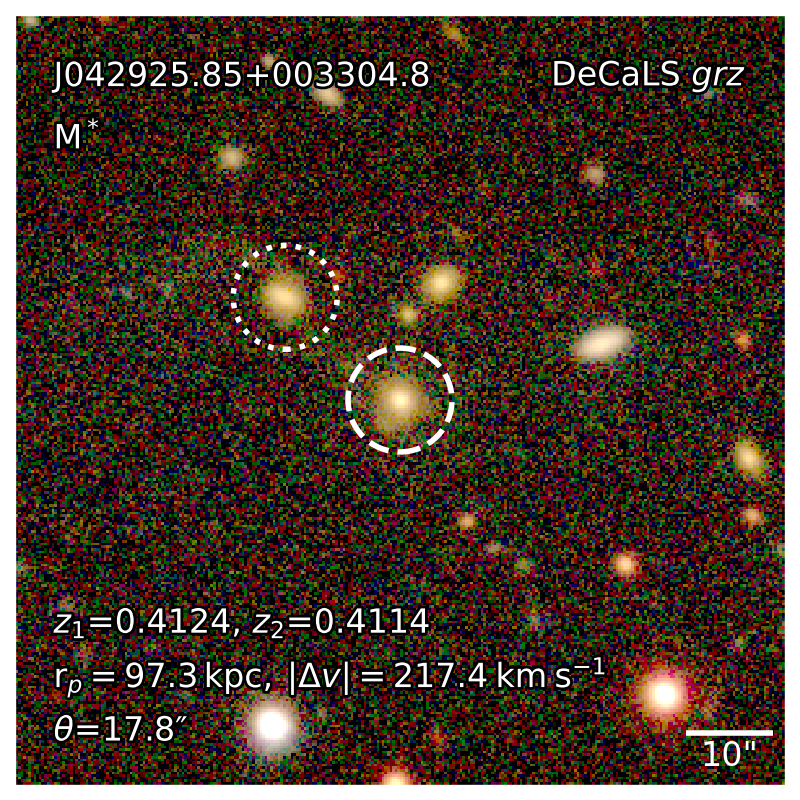}}
    \subfloat{\includegraphics[width=0.2\linewidth]{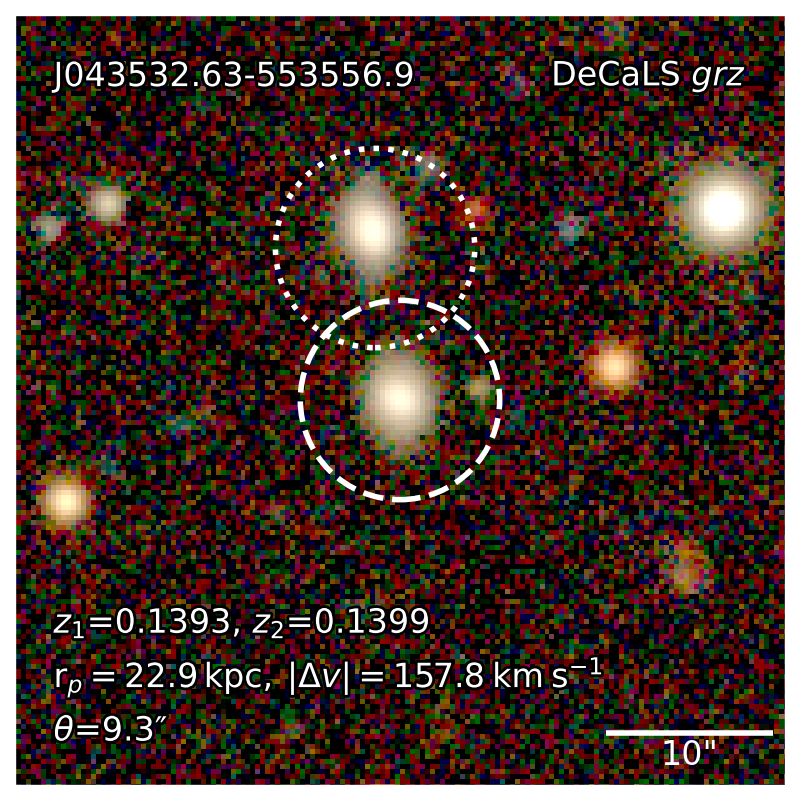}}
    \subfloat{\includegraphics[width=0.2\linewidth]{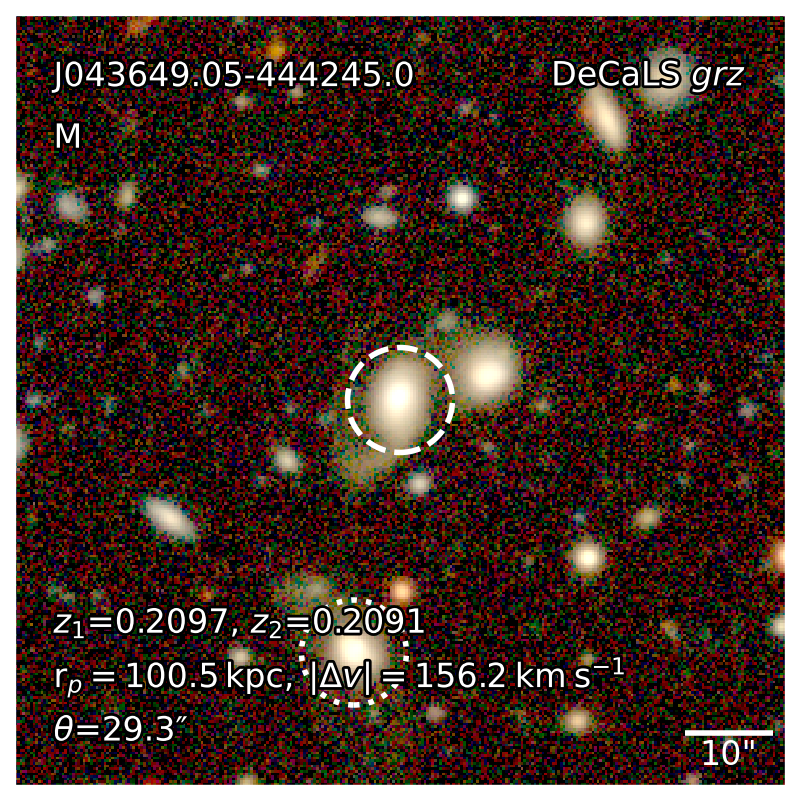}}
    \subfloat{\includegraphics[width=0.2\linewidth]{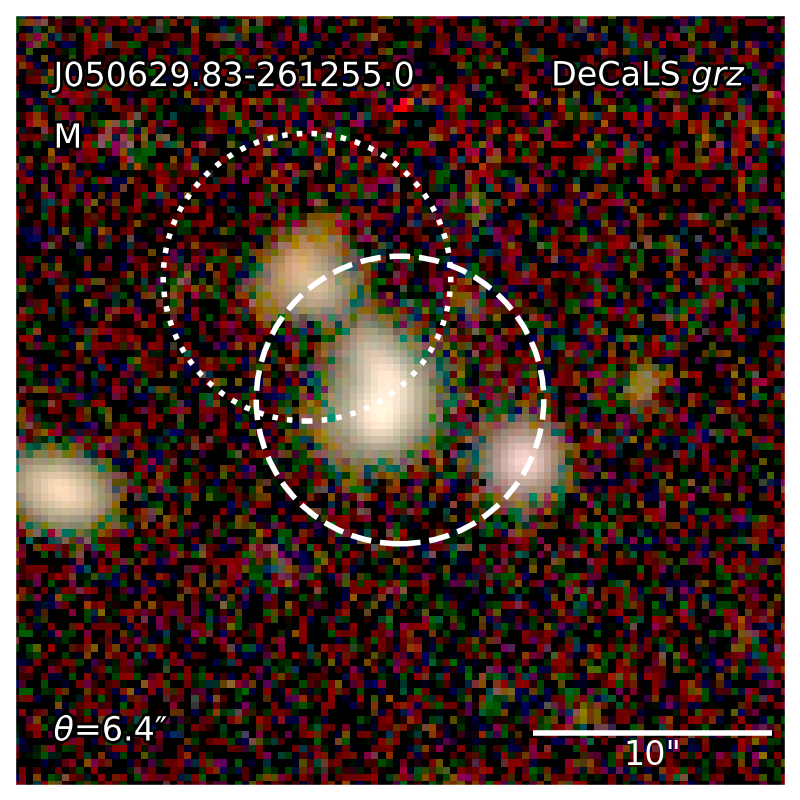}}\\
    \vspace{-4.5mm}
    \subfloat{\includegraphics[width=0.2\linewidth]{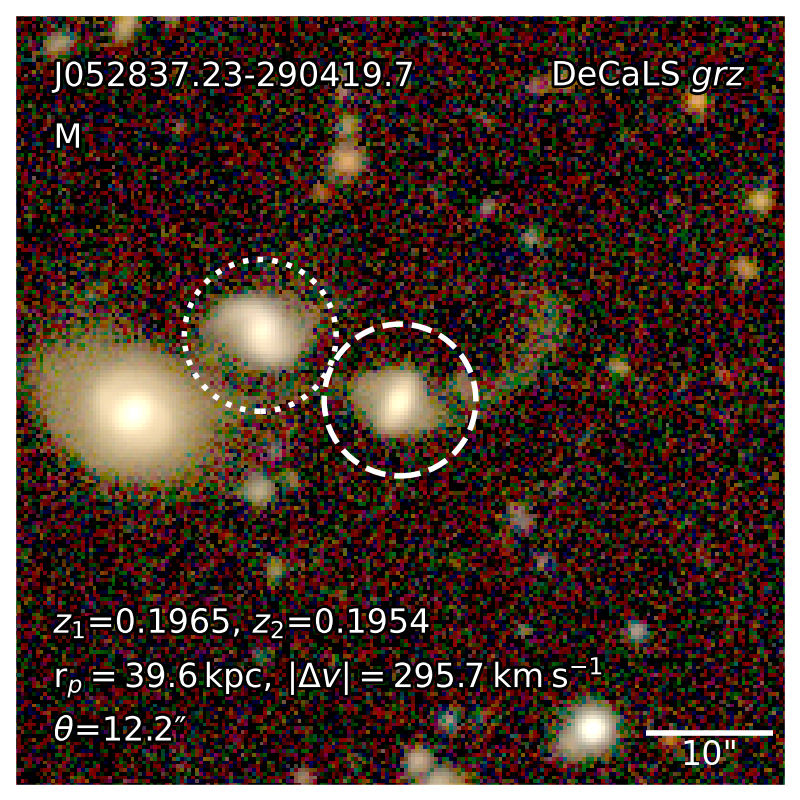}}
    \subfloat{\includegraphics[width=0.2\linewidth]{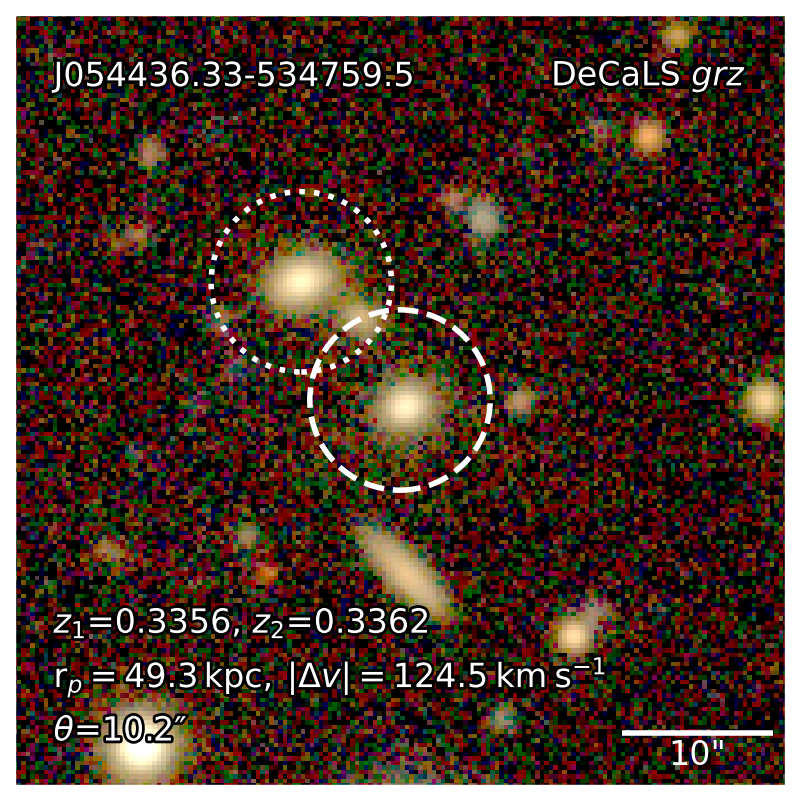}}
    \subfloat{\includegraphics[width=0.2\linewidth]{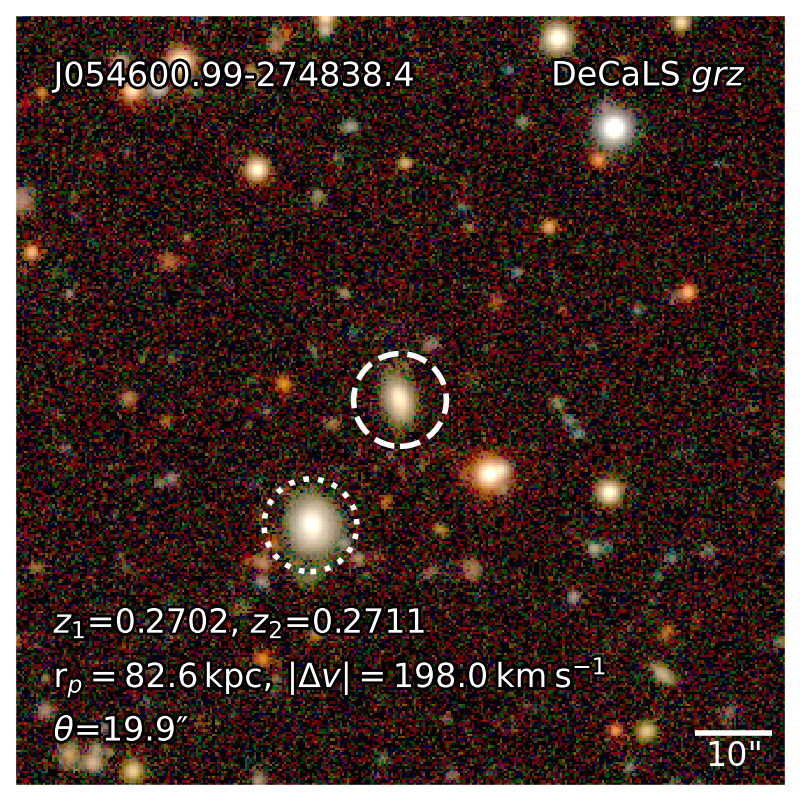}}
    \subfloat{\includegraphics[width=0.2\linewidth]{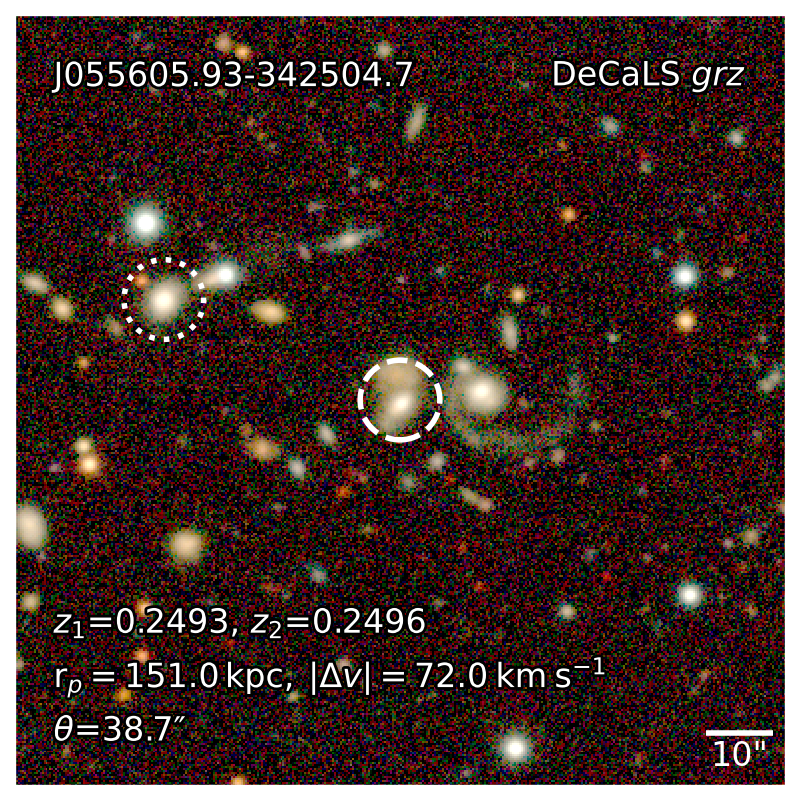}}
    \subfloat{\includegraphics[width=0.2\linewidth]{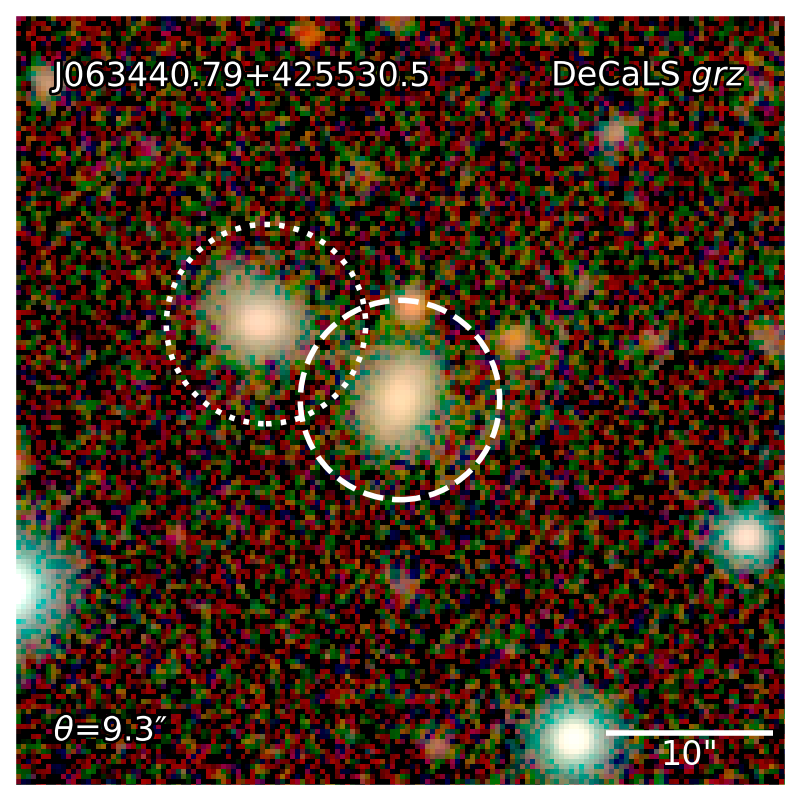}}\\
    \caption{Rank 0.5 Mid-IR Dual AGN Candidates. Each panel is constructed in an identical fashion to those in Figure~\ref{fig:rank1_duals}. {For reference, J072320.17+382254.4, J114228.86-035311.6, J133603.27+042131.7, J141821.52+532022.0, J144003.15+390840.1, J144012.58+421330.4, and J163215.62+352453.5 were originally classified as morphologically ambiguous (see Section~\ref{sec:morphclass}).}}
    \label{fig:rank0.5_duals}
\end{figure*}

\begin{figure*}[p]
\ContinuedFloat
\centering
    \subfloat{\includegraphics[width=0.2\linewidth]{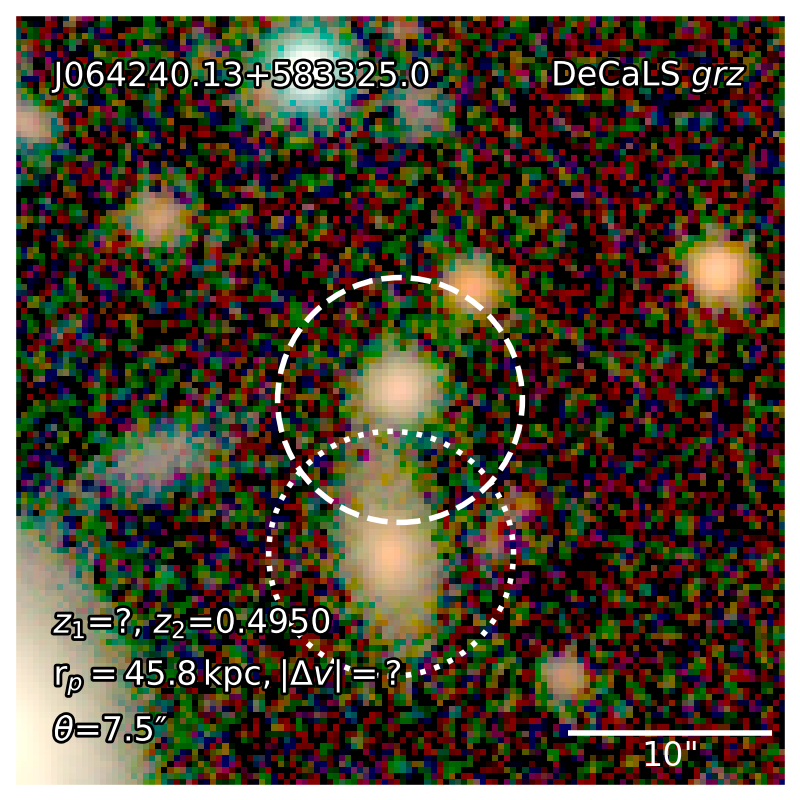}}
    \subfloat{\includegraphics[width=0.2\linewidth]{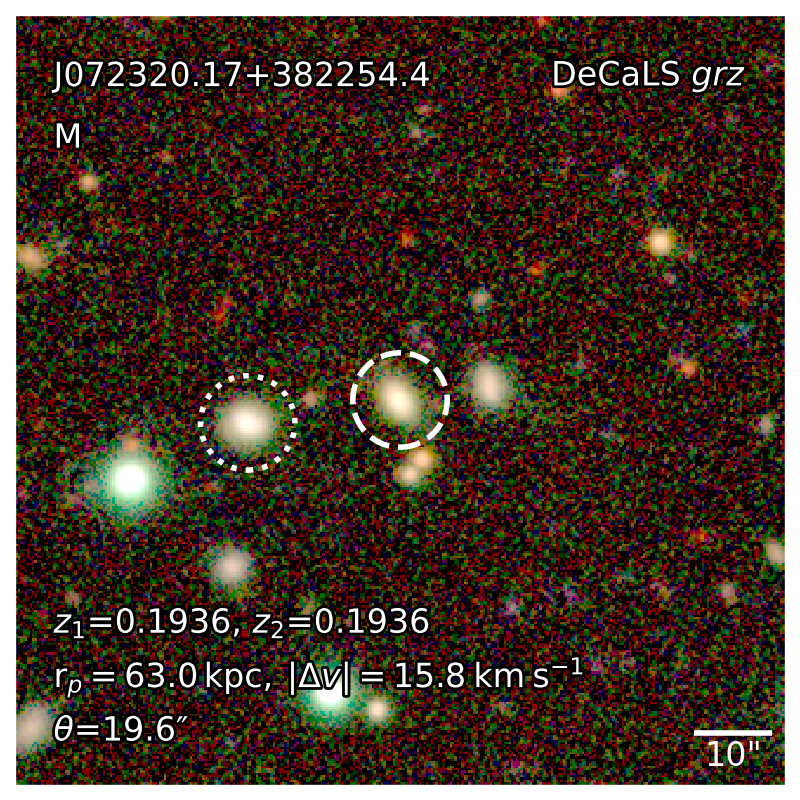}}
    \subfloat{\includegraphics[width=0.2\linewidth]{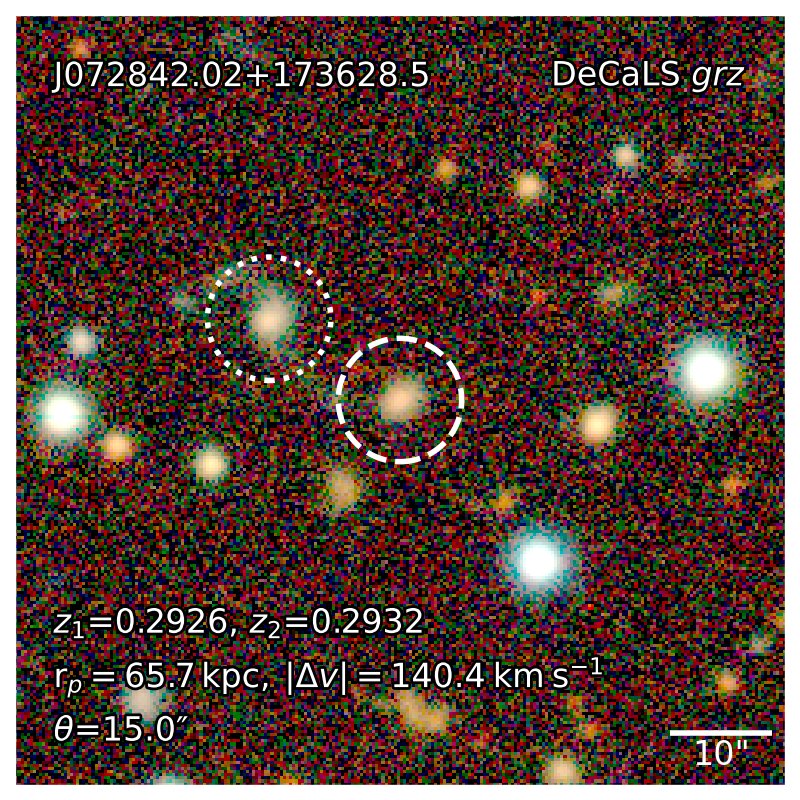}}
    \subfloat{\includegraphics[width=0.2\linewidth]{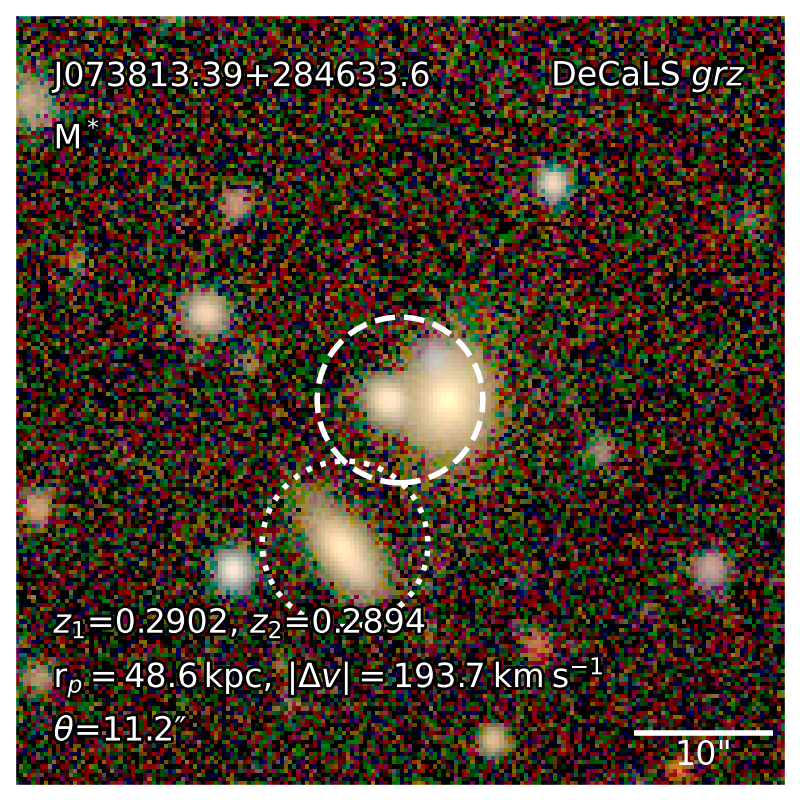}}
    \subfloat{\includegraphics[width=0.2\linewidth]{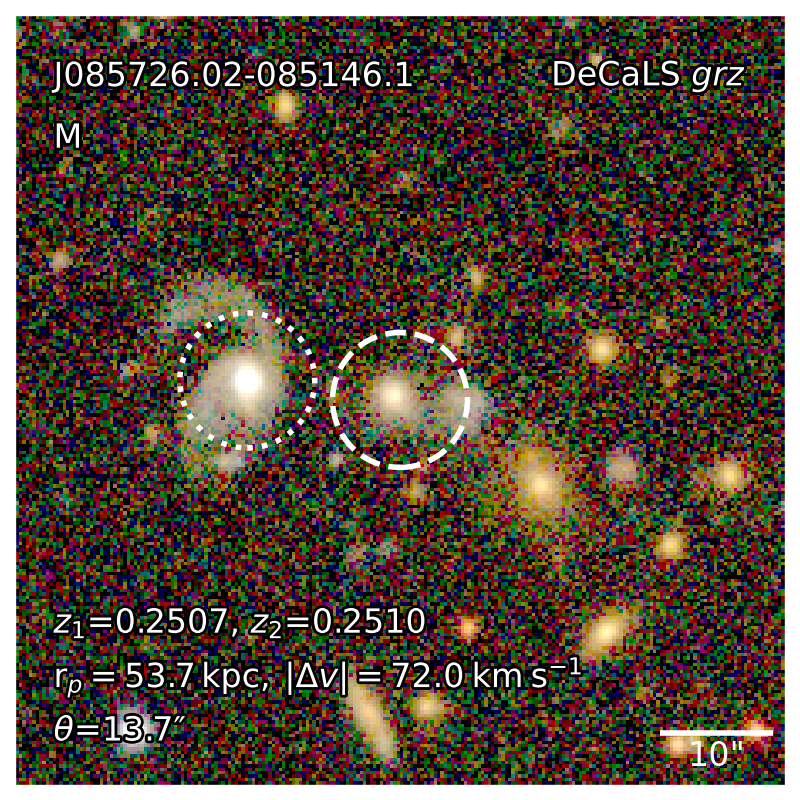}}\\
    \vspace{-4.5mm}
    \subfloat{\includegraphics[width=0.2\linewidth]{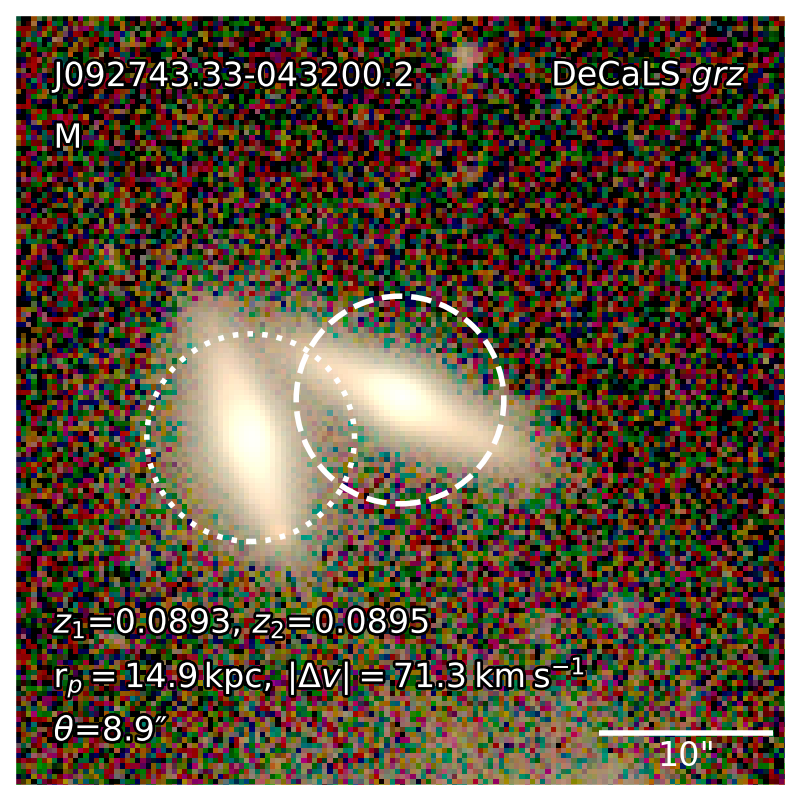}}
    \subfloat{\includegraphics[width=0.2\linewidth]{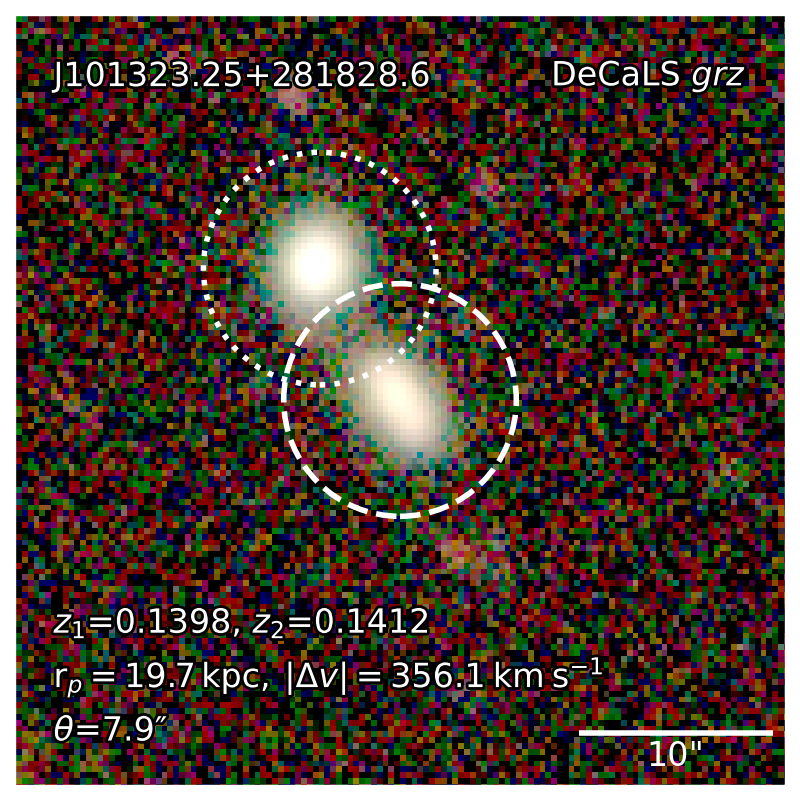}}
    \subfloat{\includegraphics[width=0.2\linewidth]{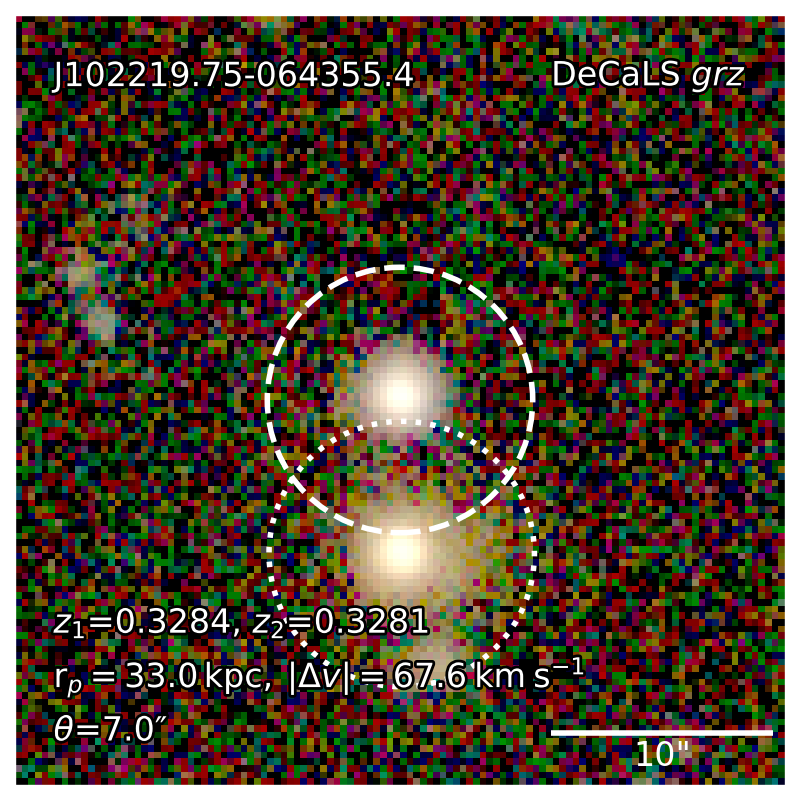}}
    \subfloat{\includegraphics[width=0.2\linewidth]{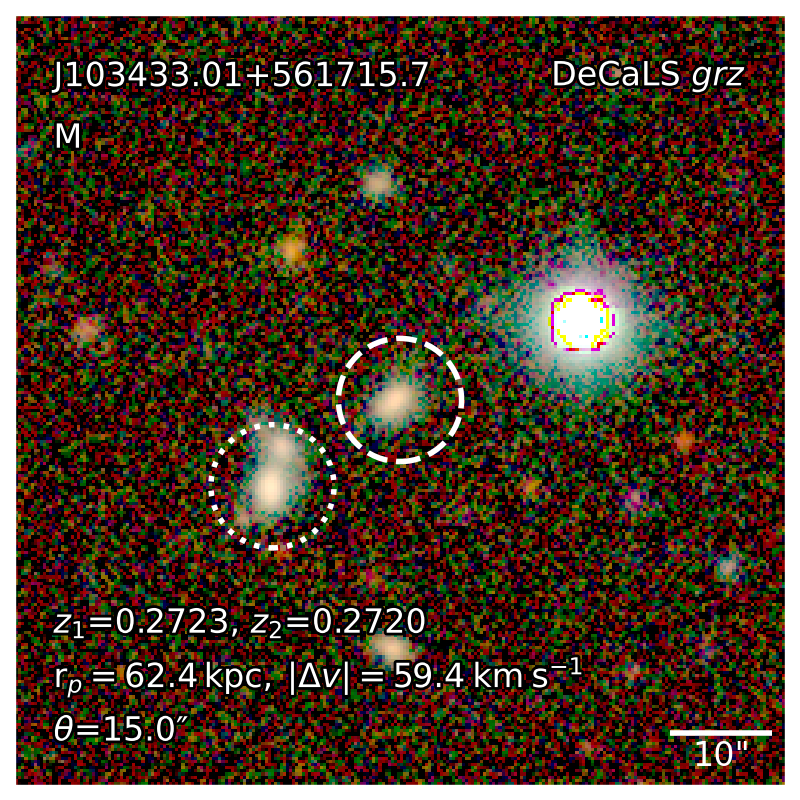}}
    \subfloat{\includegraphics[width=0.2\linewidth]{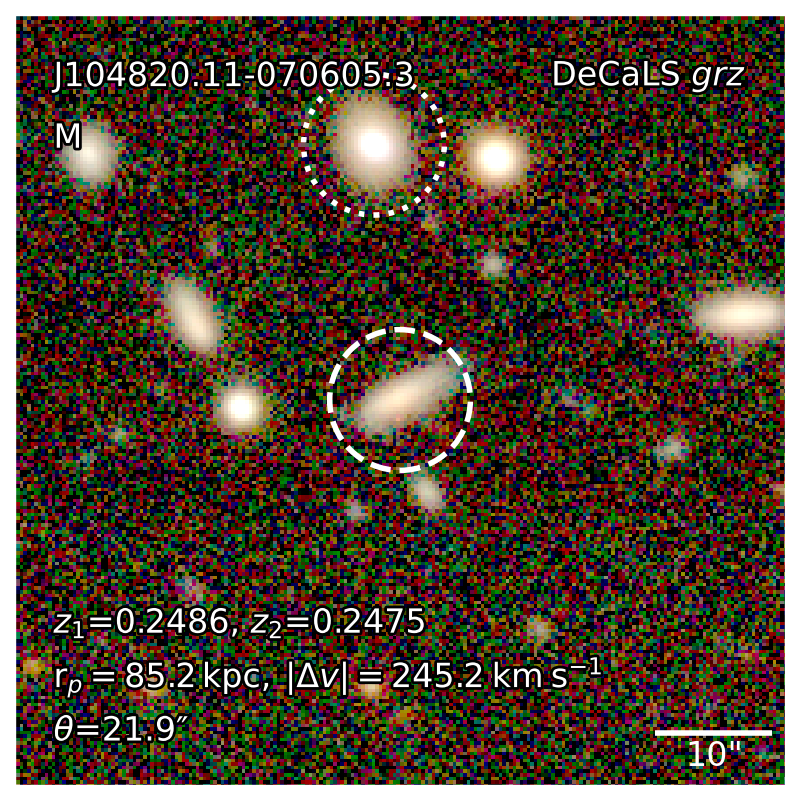}}\\
    \vspace{-4.5mm}
    \subfloat{\includegraphics[width=0.2\linewidth]{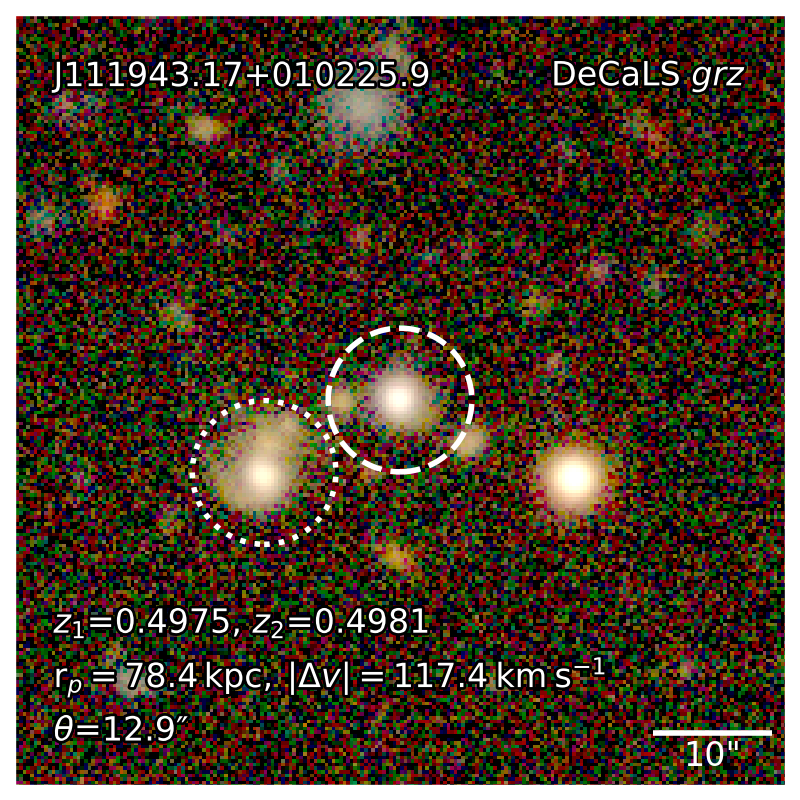}}
    \subfloat{\includegraphics[width=0.2\linewidth]{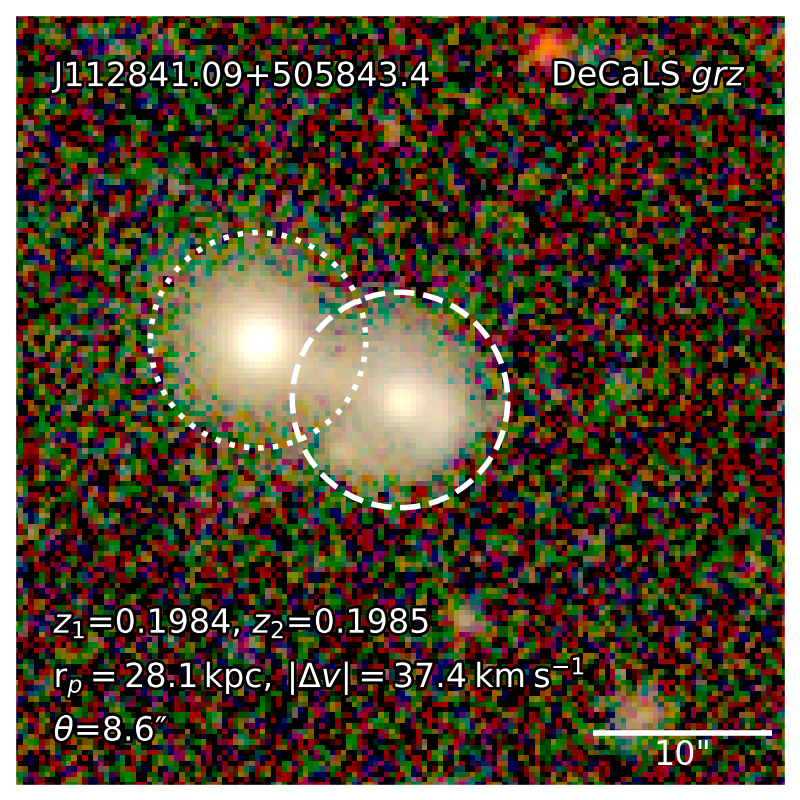}}
    \subfloat{\includegraphics[width=0.2\linewidth]{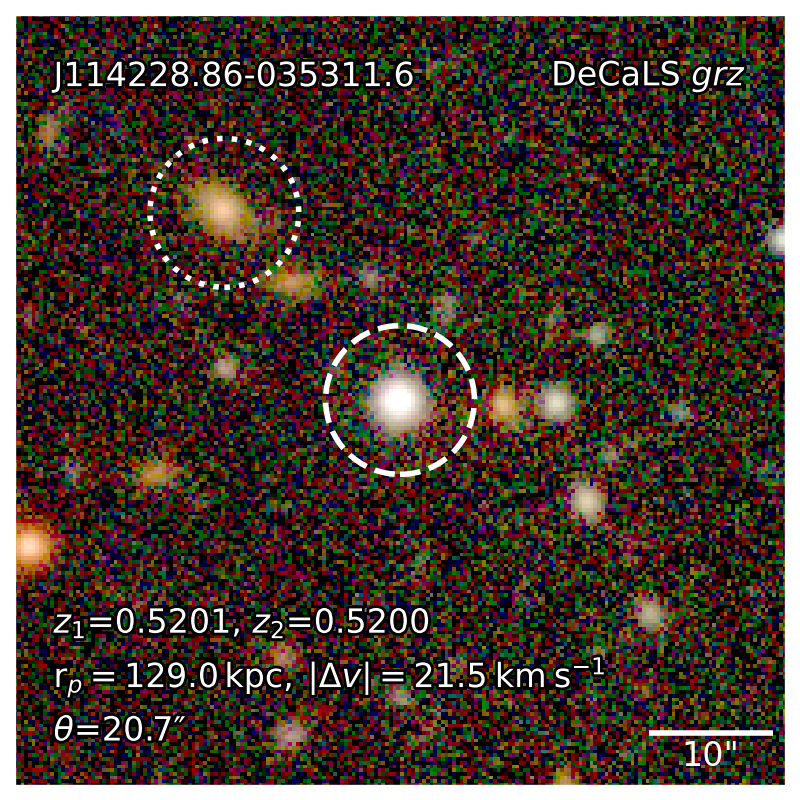}}
    \subfloat{\includegraphics[width=0.2\linewidth]{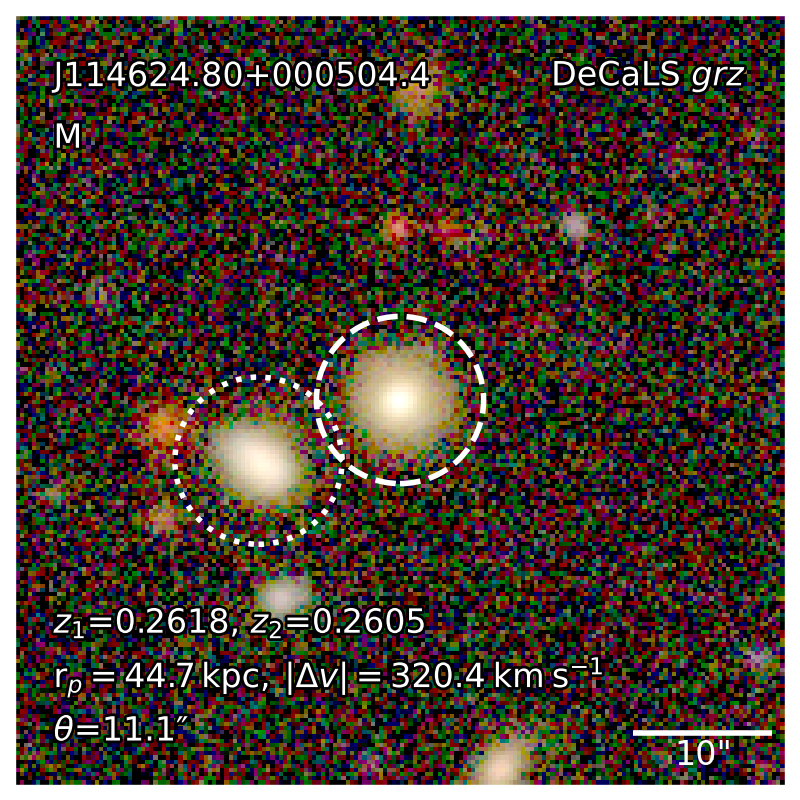}}
    \subfloat{\includegraphics[width=0.2\linewidth]{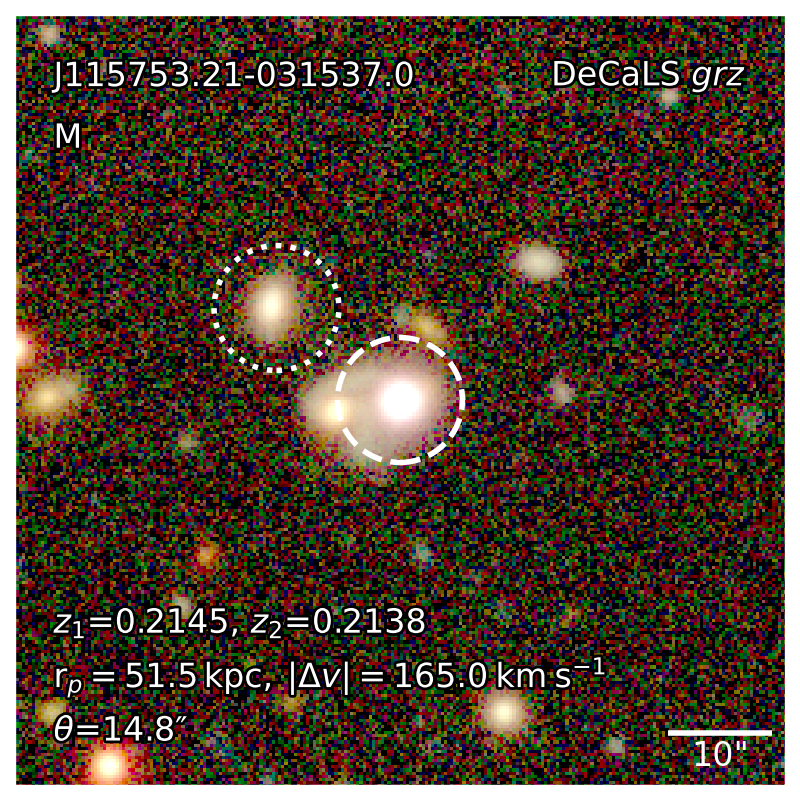}}\\
    \vspace{-4.5mm}
    \subfloat{\includegraphics[width=0.2\linewidth]{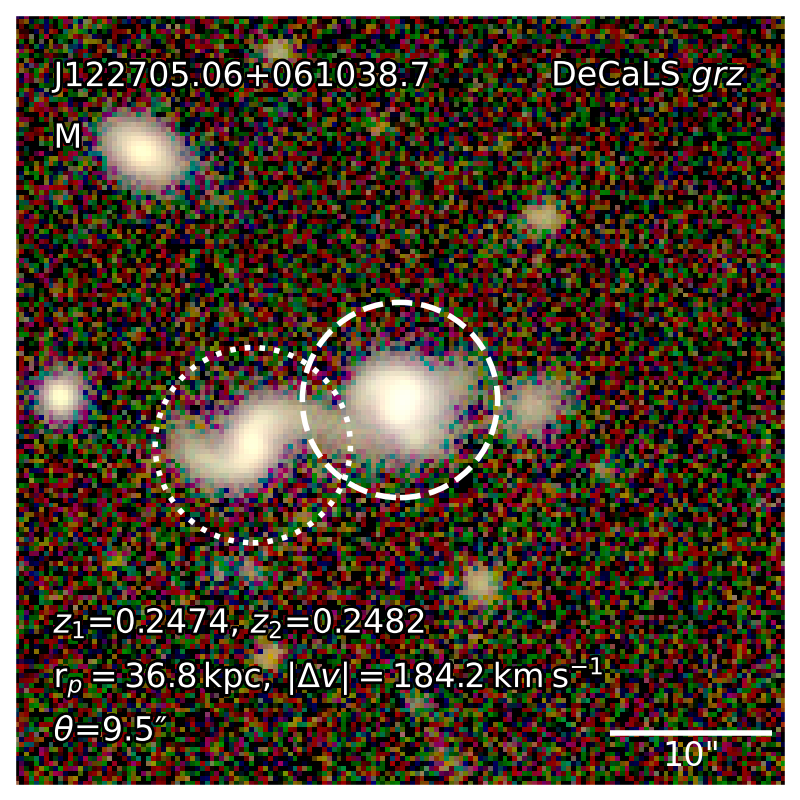}}
    \subfloat{\includegraphics[width=0.2\linewidth]{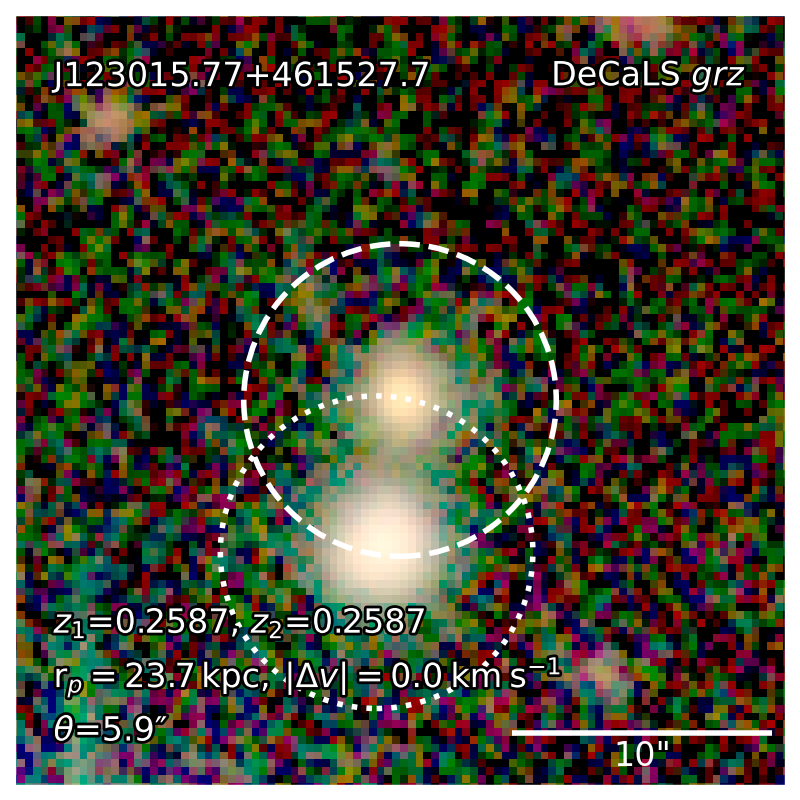}}
    \subfloat{\includegraphics[width=0.2\linewidth]{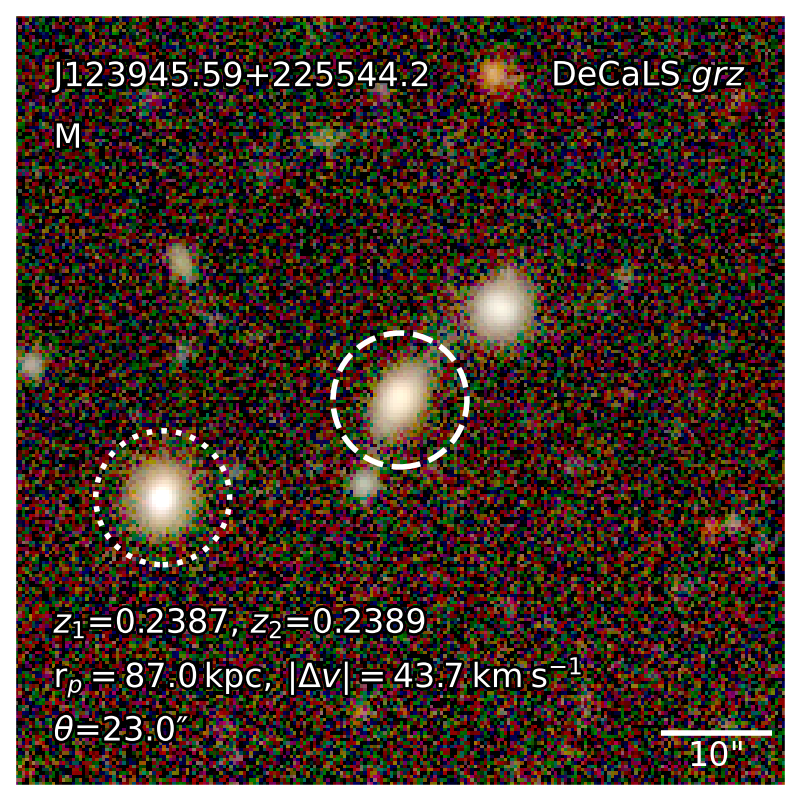}}
    \subfloat{\includegraphics[width=0.2\linewidth]{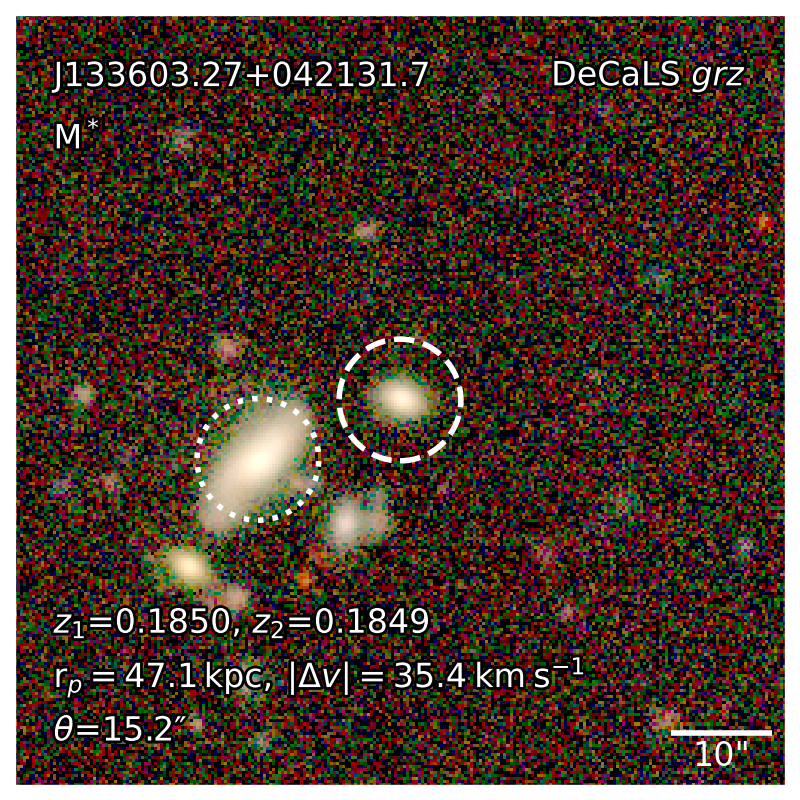}}
    \subfloat{\includegraphics[width=0.2\linewidth]{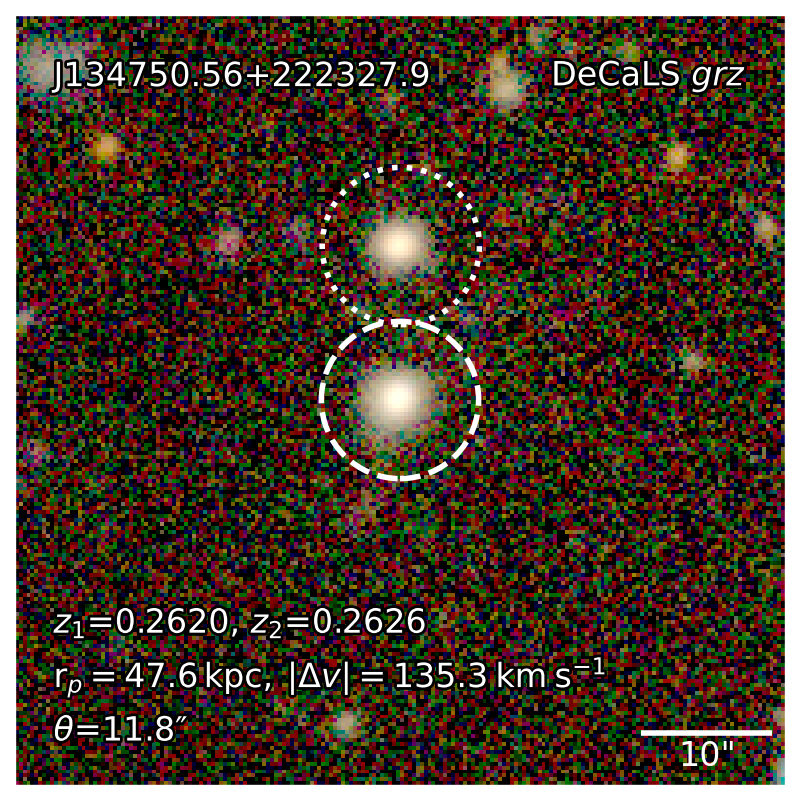}}\\
    \vspace{-4.5mm}
    \subfloat{\includegraphics[width=0.2\linewidth]{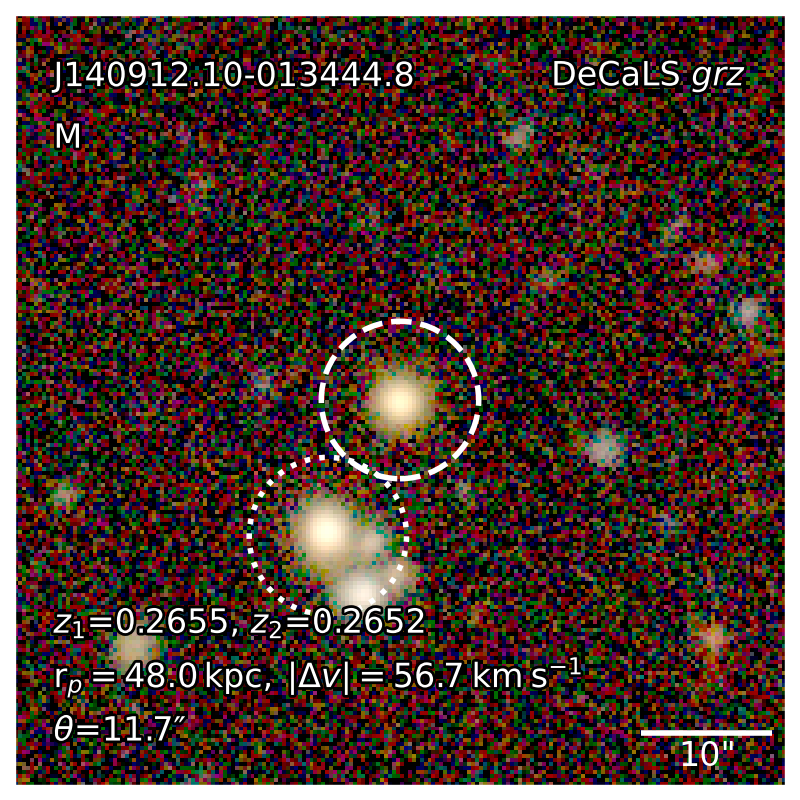}}
    \subfloat{\includegraphics[width=0.2\linewidth]{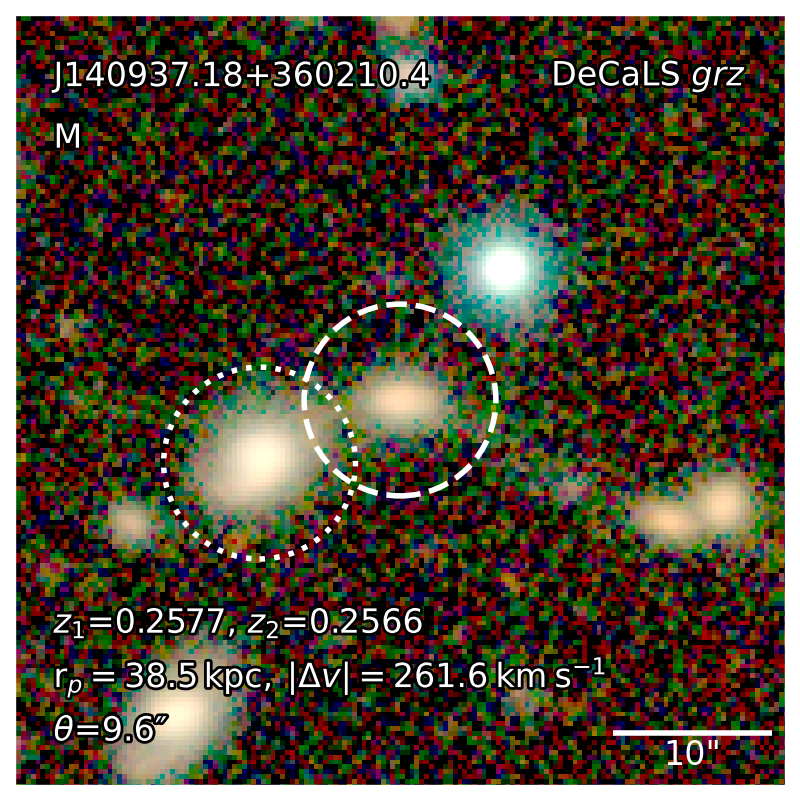}}
    \subfloat{\includegraphics[width=0.2\linewidth]{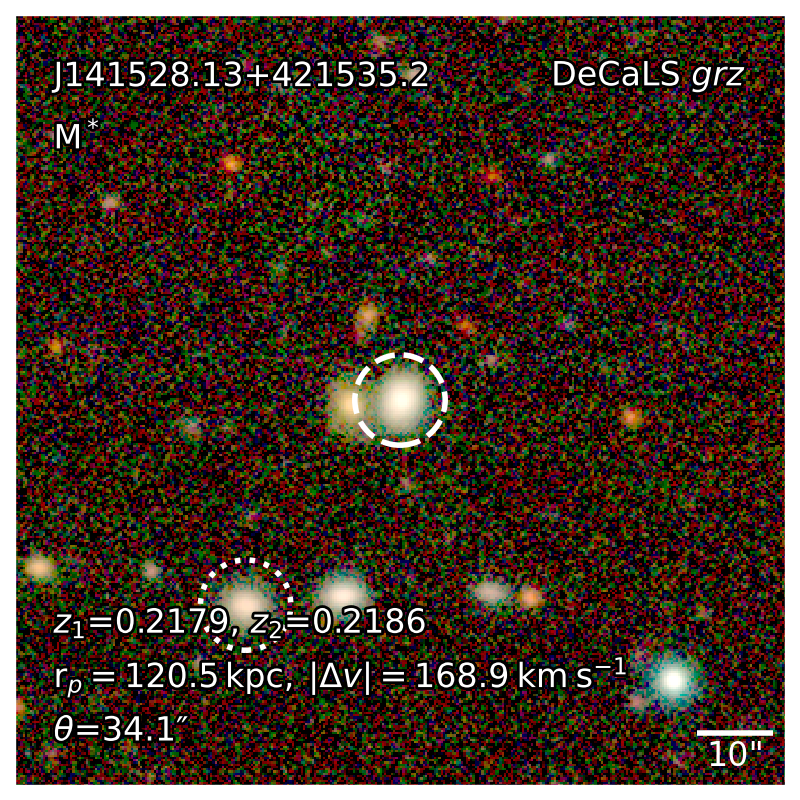}}
    \subfloat{\includegraphics[width=0.2\linewidth]{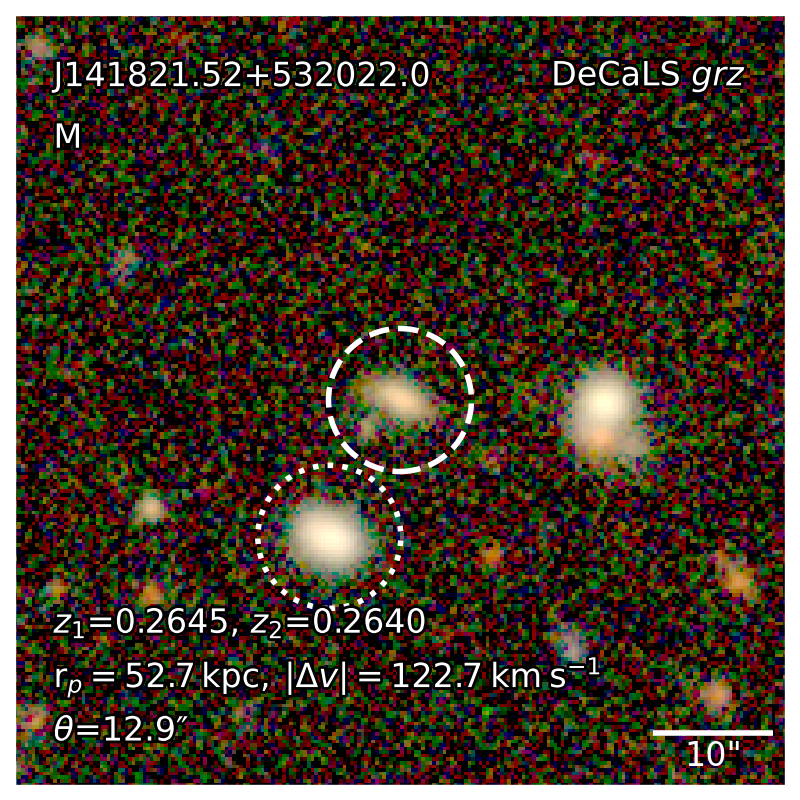}}
    \subfloat{\includegraphics[width=0.2\linewidth]{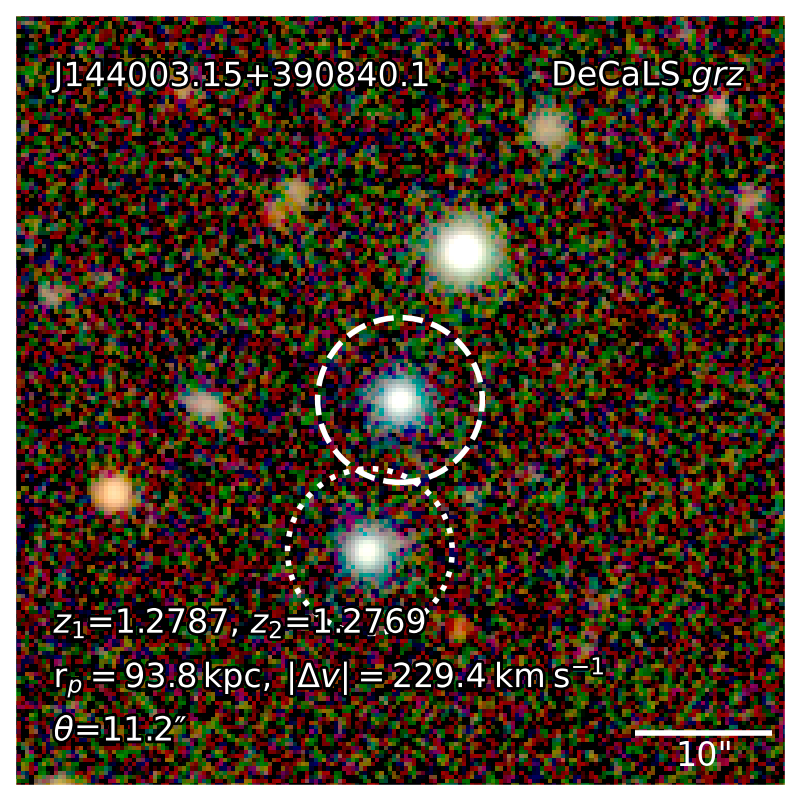}}\\
    \vspace{-4mm}
    \subfloat{\includegraphics[width=0.2\linewidth]{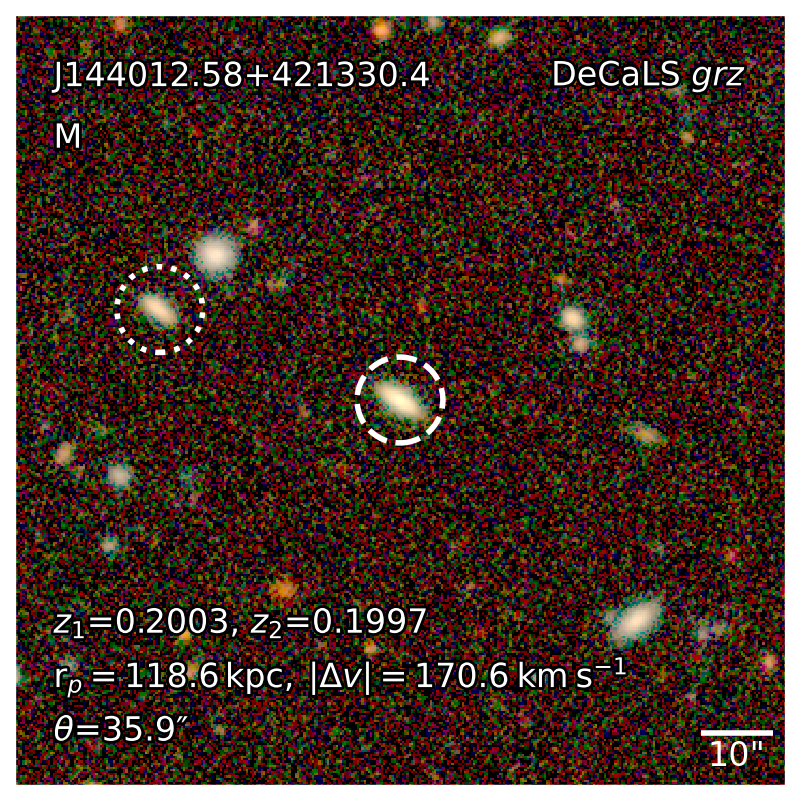}}
    \subfloat{\includegraphics[width=0.2\linewidth]{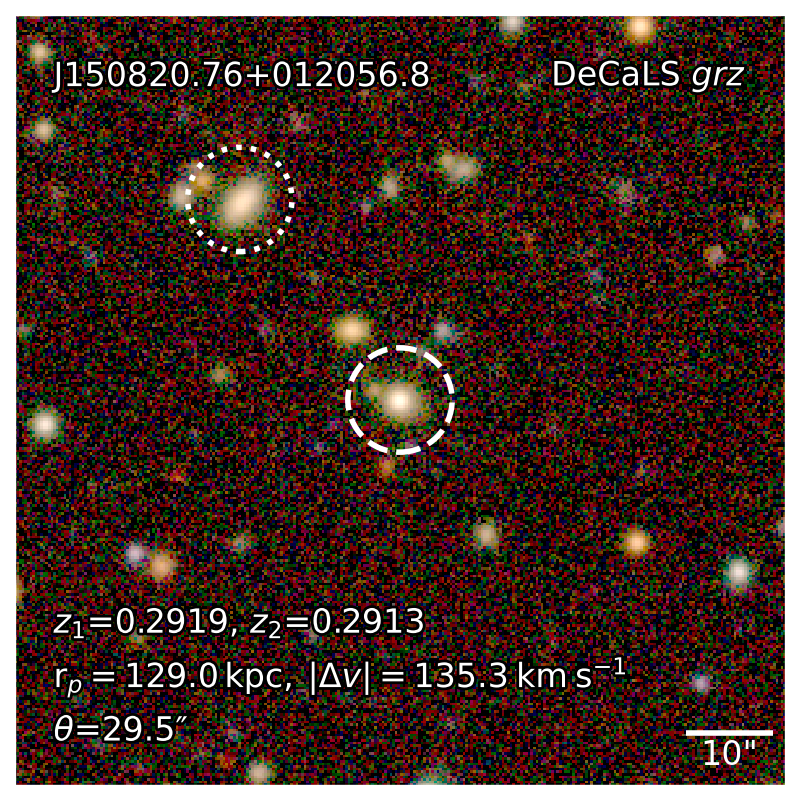}}
    \subfloat{\includegraphics[width=0.2\linewidth]{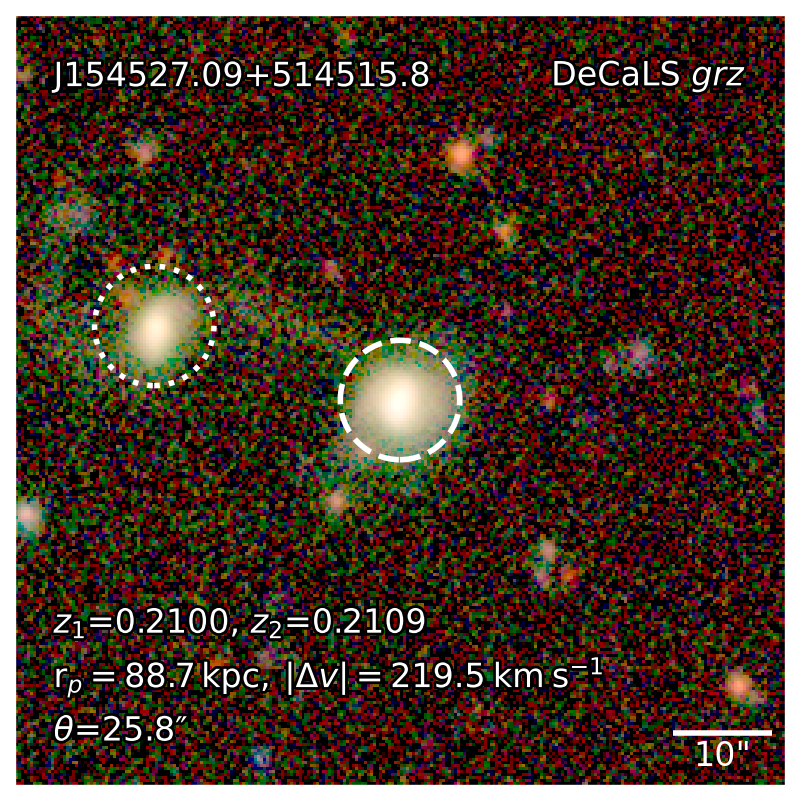}}
    \subfloat{\includegraphics[width=0.2\linewidth]{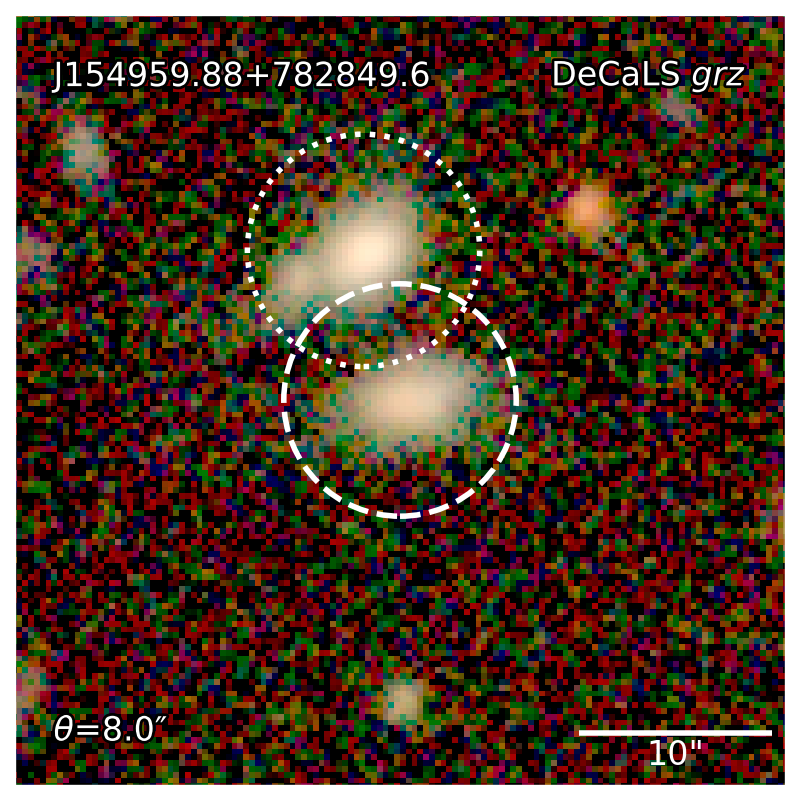}}
    \subfloat{\includegraphics[width=0.2\linewidth]{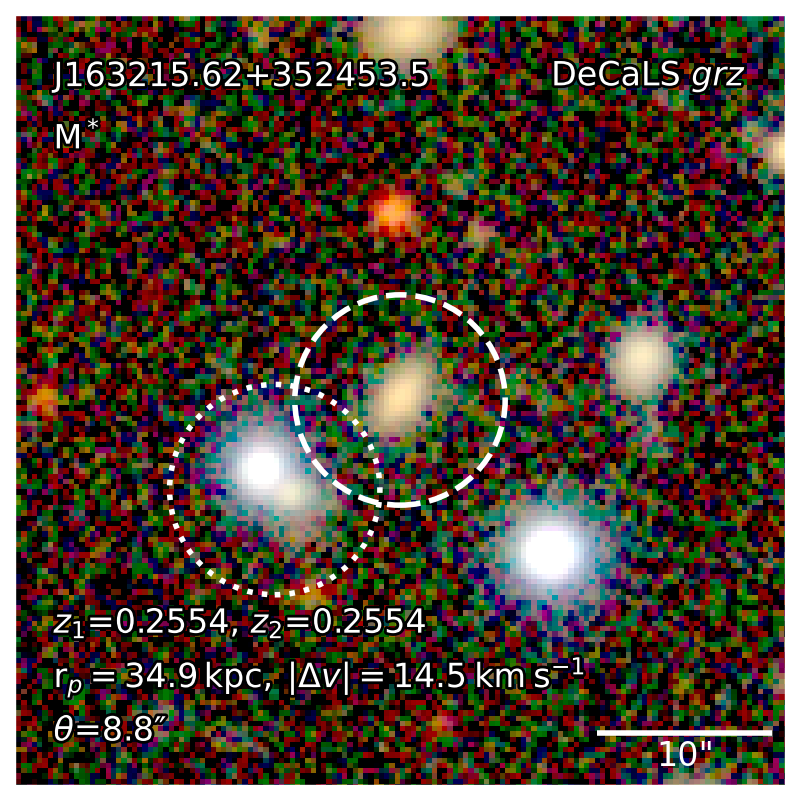}}\\
    \caption{Continued}
\end{figure*}

\begin{figure*}
\ContinuedFloat
\centering
    \subfloat{\includegraphics[width=0.2\linewidth]{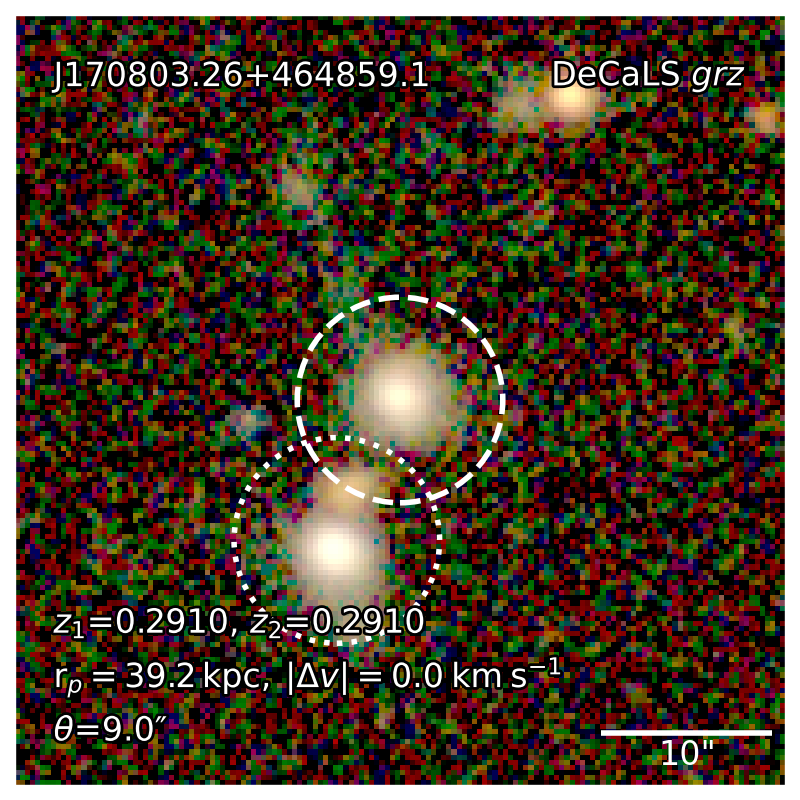}}
    \subfloat{\includegraphics[width=0.2\linewidth]{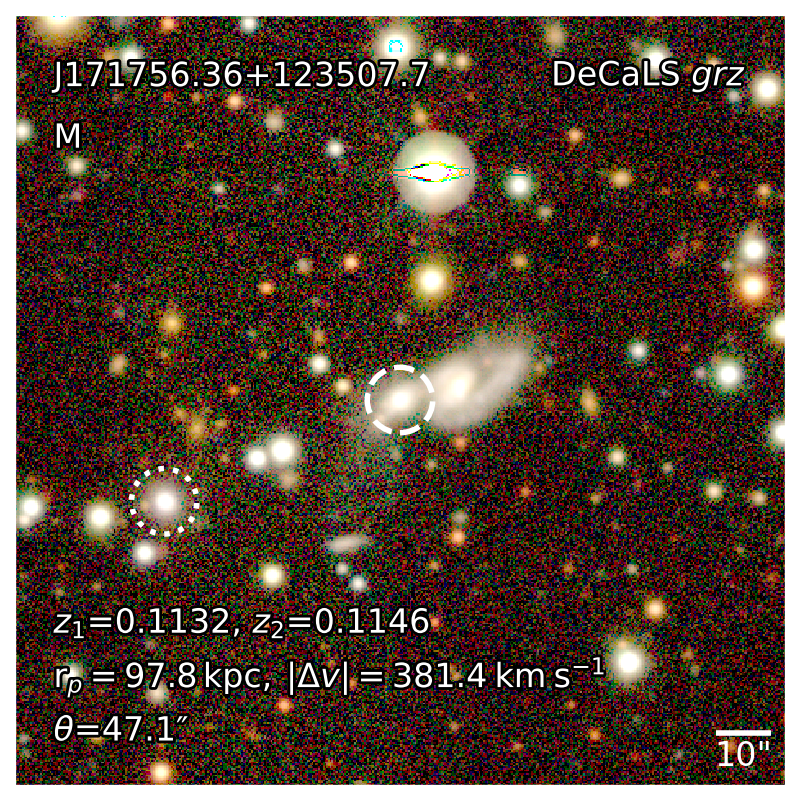}}
    \subfloat{\includegraphics[width=0.2\linewidth]{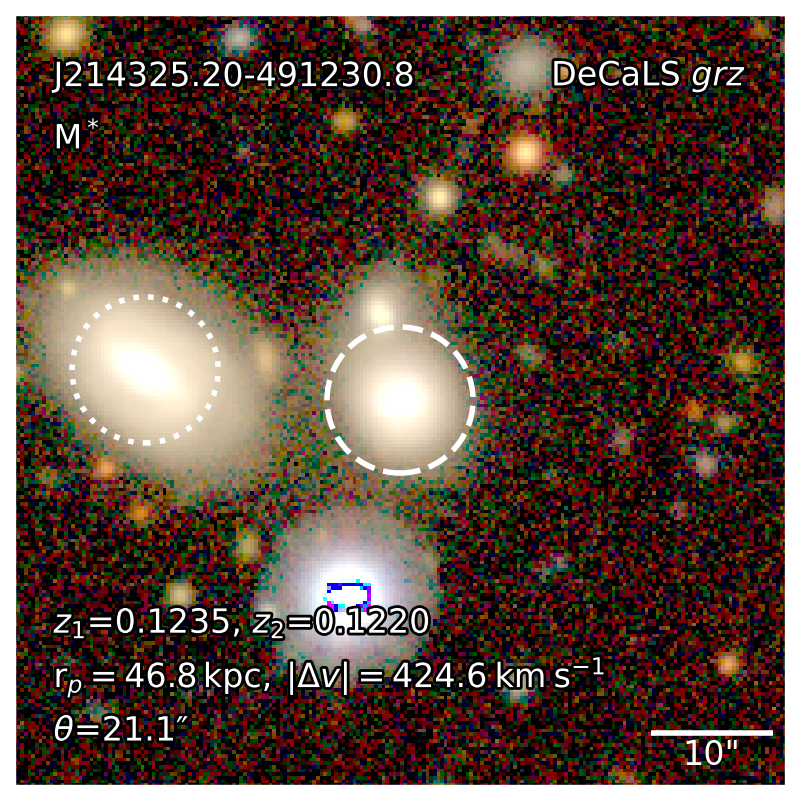}}
    \subfloat{\includegraphics[width=0.2\linewidth]{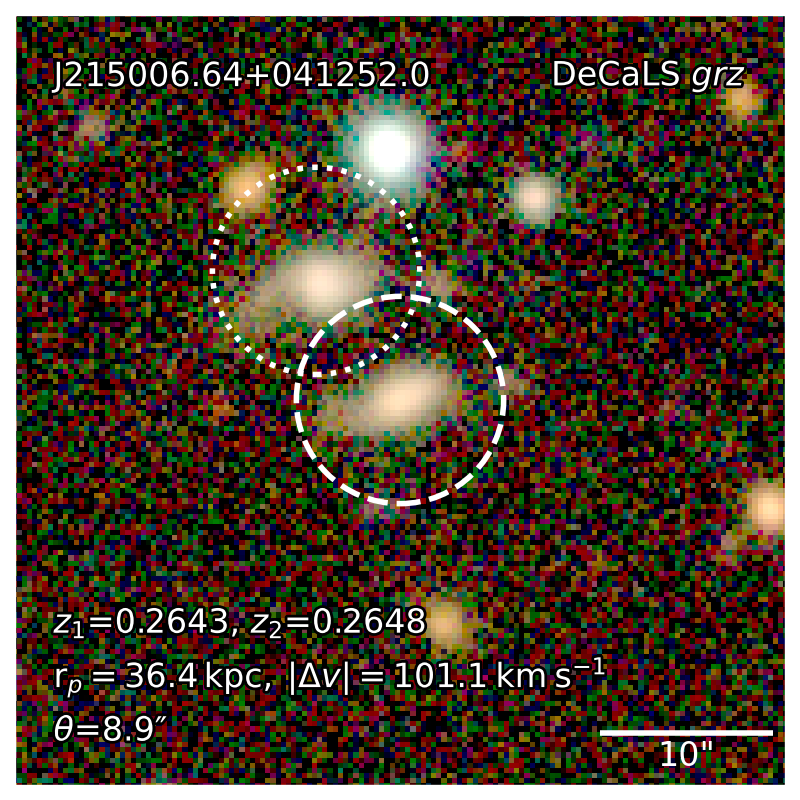}}
    \subfloat{\includegraphics[width=0.2\linewidth]{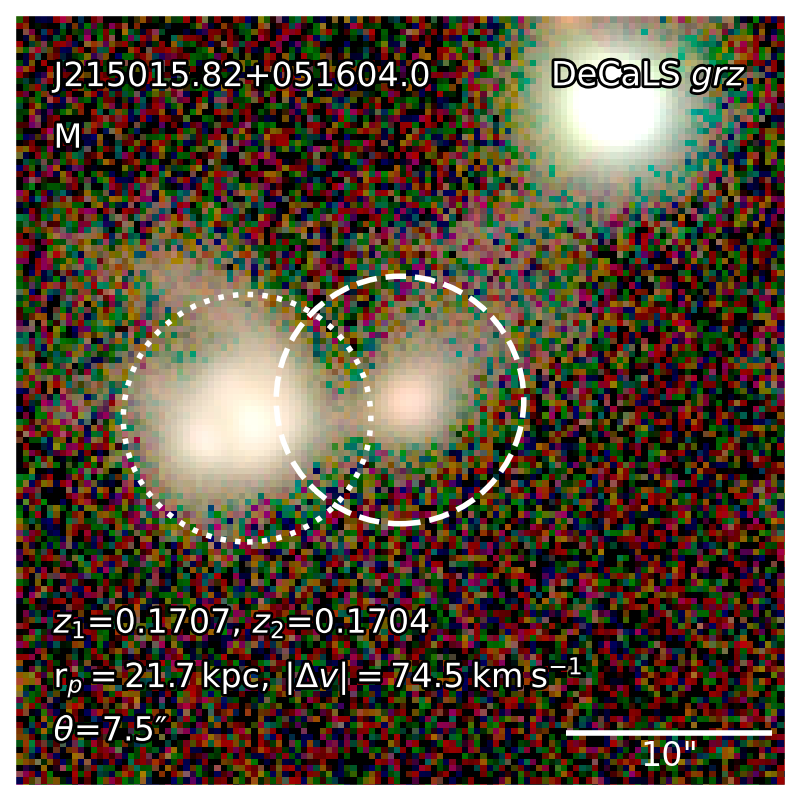}}\\
    \vspace{-4mm}
    \subfloat{\includegraphics[width=0.2\linewidth]{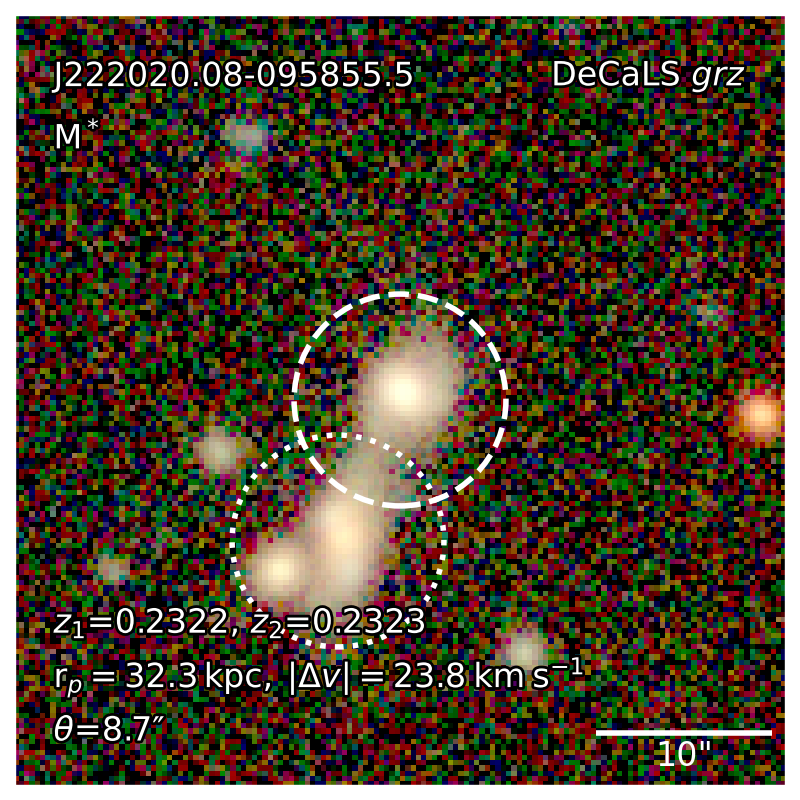}}
    \subfloat{\includegraphics[width=0.2\linewidth]{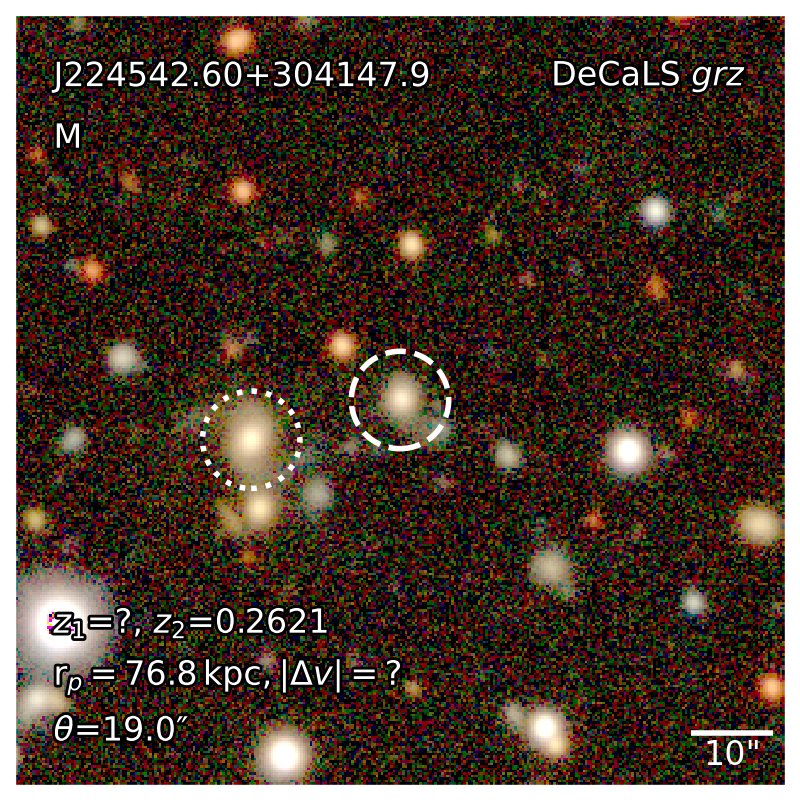}}
    \subfloat{\includegraphics[width=0.2\linewidth]{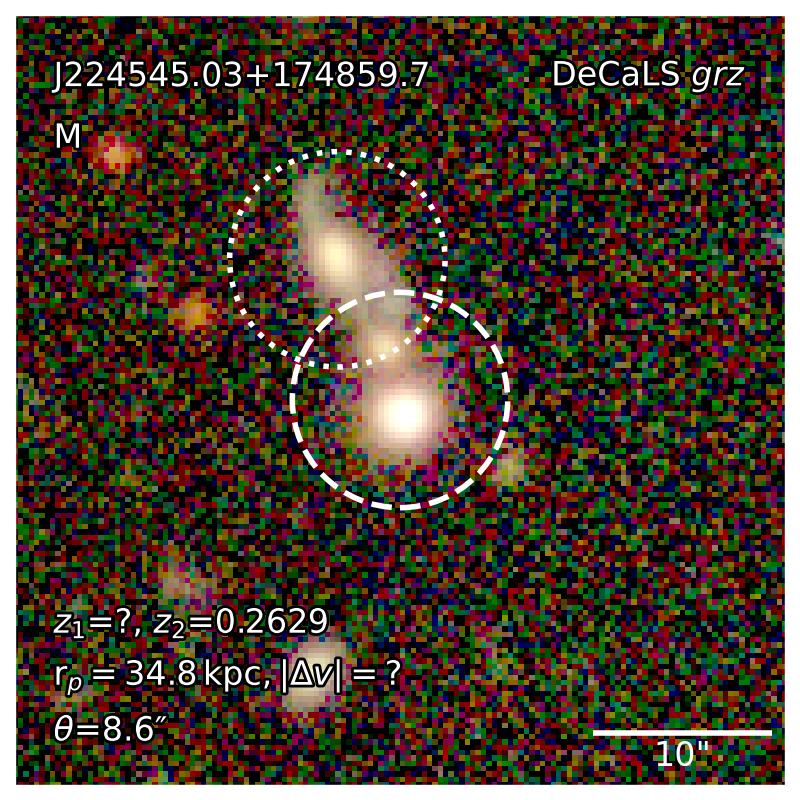}}
    \subfloat{\includegraphics[width=0.2\linewidth]{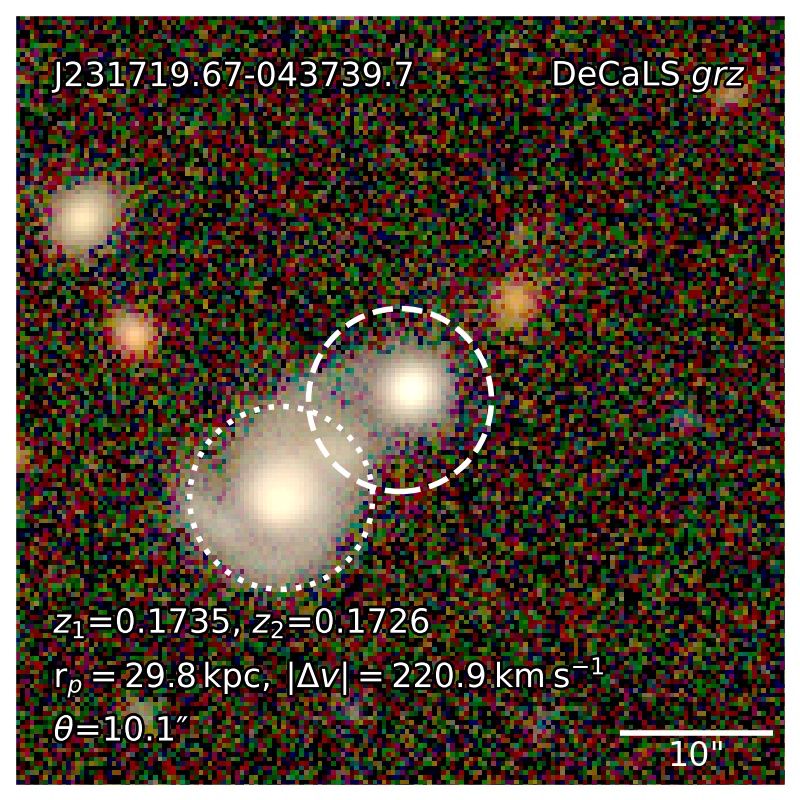}}
    \subfloat{\includegraphics[width=0.2\linewidth]{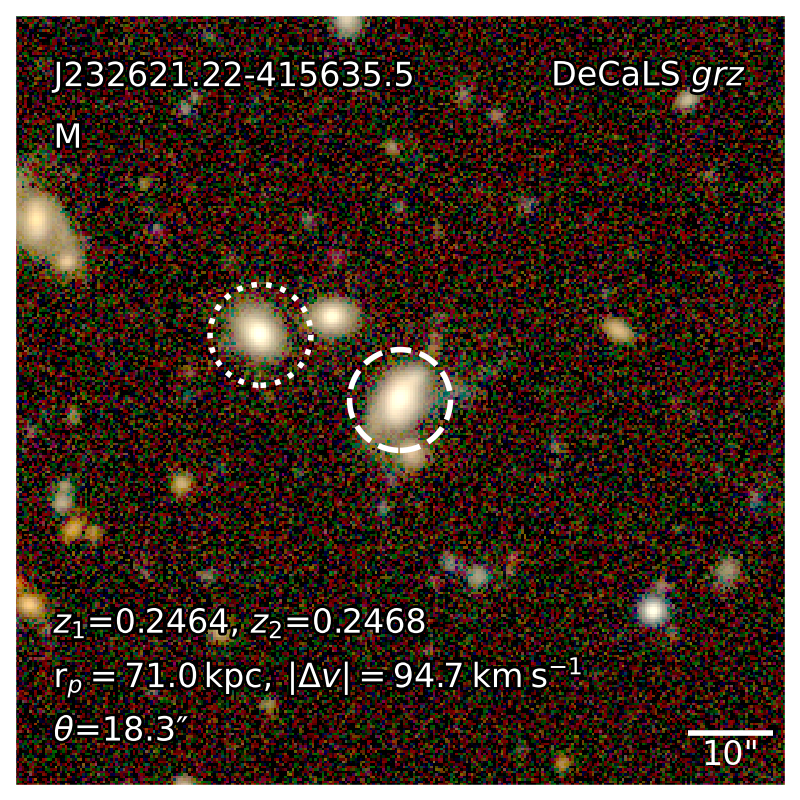}}\\
    \vspace{-4mm}
    \subfloat{\includegraphics[width=0.2\linewidth]{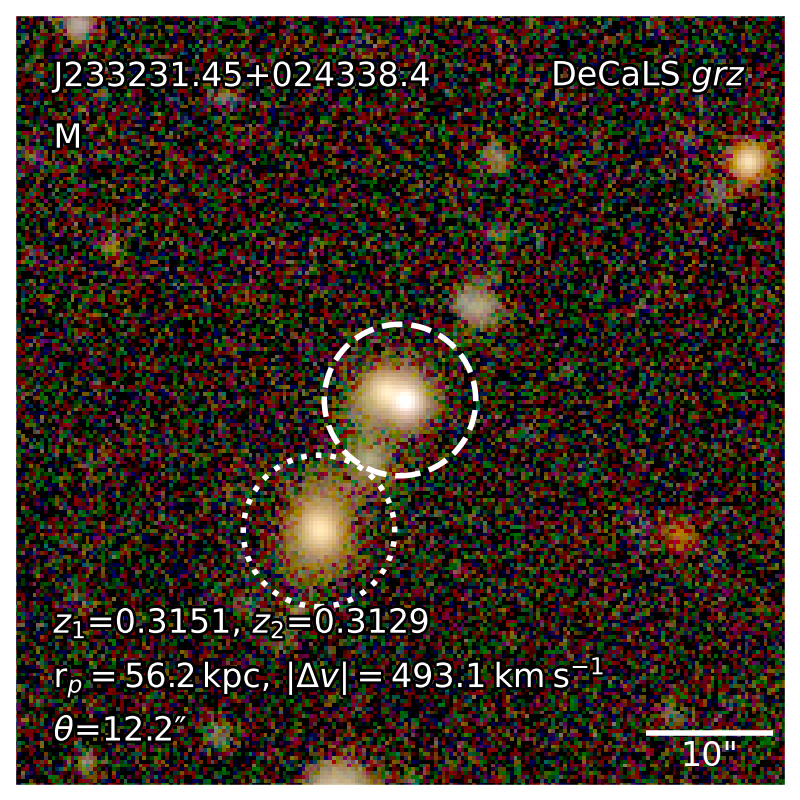}}
    \subfloat{\includegraphics[width=0.2\linewidth]{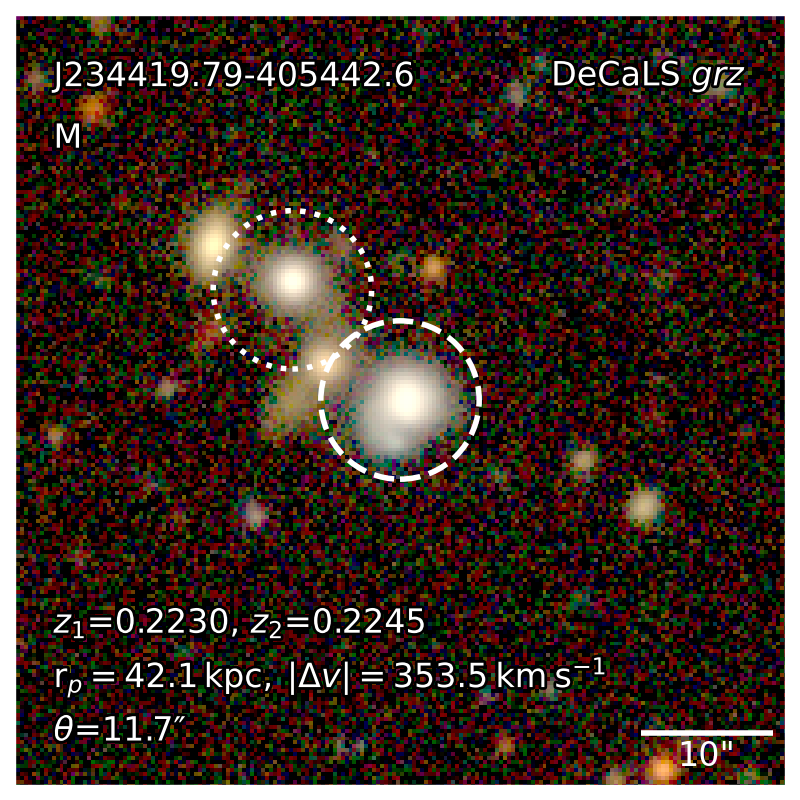}}\\    
    \caption{Continued}
\end{figure*}

\begin{table*}
\begin{center}
\caption{Rank 0.5 Spatially-Resolved Mid-IR Dual AGN Candidates}
\label{tab:rank0.5}
\begin{tabular}{ccccccccccc}
\hline
\hline
\noalign{\smallskip}
\noalign{\smallskip}
$\rm{designation}_1$ &  $z_1$ & $z_1$ source & W1-W2$_1$ &  $\rm{designation}_2$ &  $z_2$ & $z_2$ source & W1-W2$_2$ & $\theta_{\rm{MIR}}$ &  $r_{p\,{\rm{(MIR)}}}$ &  $|\Delta v|$ \\
 &  &  &  &  &  &  &  & ('') & (kpc) & km s$^{-1}$\\
(1) & (2) & (3) & (4) & (5) & (6) & (7) & (8) & (9) & (10) & (11)  \\
\noalign{\smallskip}
\noalign{\smallskip}
\hline
\noalign{\smallskip}
J010050.09-024218.7 & 0.085 &   SR & 1.71 & J010050.77-024214.5 & 0.084 &   SR & 0.87 & 11.0 &  17.5 &   82.9 \\
J012002.39+185128.4 & 0.238 &   D & 0.90 & J012002.71+185112.9 & 0.238 &   D & 0.93 & 16.2 &  60.9 &   87.6 \\
J012140.80-065704.0 & 0.229 & G & 0.83 & J012141.03-065646.7 & 0.229 & G & 1.02 & 17.7 &  64.8 &  109.7 \\
J012515.59-380127.7 & 0.137 &   SR & 0.90 & J012516.98-380040.4 & 0.137 &   SR & 0.83 & 50.0 & 121.4 &  173.4 \\
J012627.54+094800.9 & 0.280 &   D & 0.89 & J012628.43+094751.3 & 0.281 &   D & 1.00 & 16.4 &  69.5 &  47.5 \\
J015316.76+002619.3 & 0.253 &   D & 1.20 & J015317.34+002614.3 & 0.254 &   D & 0.96 &  9.9 &  39.3 & 152.5 \\
J020150.04-433806.8 & \dots &    \dots & 0.94 & J020150.68-433801.3 &   \dots &    \dots & 0.80 &  8.9 & \dots &    \dots \\
J020934.18-554625.2 & 0.187 &   SR & 0.97 & J020934.74-554622.0 & 0.188 &   SR & 0.83 &  5.7 &  17.8 &  99.2 \\
J021702.46+045154.2 & 0.234 & G & 1.02 & J021702.68+045145.7 & 0.234 & G & 0.86 &  9.1 &  33.8 &  56.1 \\
J022445.68+102241.0 & 0.245 & G & 0.89 & J022445.83+102230.8 & 0.244 & G & 0.93 & 10.5 &  40.4 &  218.9 \\
J023913.10+174858.2 & 0.129 & G & 1.26 & J023913.38+174838.7 & 0.129 & G & 1.11 & 19.9 &  45.8 & 140.1 \\
J024134.85-011807.9 & 0.241 &   SR & 0.82 & J024135.74-011809.0 & 0.240 &   SR & 0.92 & 13.4 &  51.1 &  265.0 \\
J031612.21-380110.8 & 0.261 &   SR & 0.89 & J031612.43-380105.2 & 0.261 &   SR & 1.36 &  6.2 &  25.2 &   71.3 \\
J032257.43-361050.5 & 0.299 &   SR & 0.86 & J032257.56-361042.8 & 0.300 &   SR & 0.94 &  7.9 &  35.1 & 138.4 \\
J033729.30-165407.7 & 0.218 &   SR & 0.90 & J033730.33-165400.6 & 0.217 &   SR & 0.88 & 16.4 &  57.8 &   75.0 \\
J035438.03-375300.9 & 0.187 &   SR & 1.78 & J035438.41-375254.0 & 0.187 &   SR & 0.84 &  8.3 &  25.9 &  151.5 \\
J035713.07-371906.7 & 0.179 &   SR & 1.03 & J035713.36-371854.8 & 0.179 &   SR & 0.87 & 12.5 &  37.7 &   76.4 \\
J035906.48-482200.0 & 0.243 & G & 1.09 & J035906.92-482211.3 & 0.243 & G & 1.12 & 12.2 &  46.5 &    0.0 \\
J041611.44-383645.1 & 0.386 &   SR & 1.11 & J041611.63-383639.6 & 0.384 &   SR & 1.05 &  5.9 &  30.8 &  407.4 \\
J042138.05-200312.6$^{\dagger}$ & 0.239 &    \dots & 0.82 & J042138.07-200300.7$^{\dagger}$ &   \dots &    \dots & 0.83 & 12.0 & \dots &    \dots \\
J042925.85+003304.8 & 0.412 &   D & 1.42 & J042926.73+003316.6 & 0.411 &   D & 0.98 & 17.8 &  97.3 &  217.4 \\
J043532.63-553556.9 & 0.139 &   SR & 1.94 & J043532.81-553547.8 & 0.140 &   SR & 0.80 &  9.3 &  22.9 & 157.8 \\
J043649.05-444245.0 & 0.210 &   SR & 0.82 & J043649.54-444313.9 & 0.209 &   SR & 1.17 & 29.3 & 100.5 &  156.2 \\
J050629.83-261255.0 & 0.239 &    \dots & 0.81 & J050630.12-261249.9 &   \dots &    \dots & 1.08 &  6.4 & \dots &    \dots \\
J052837.23-290419.7 & 0.197 &   SR & 0.80 & J052838.07-290414.6 & 0.195 &   SR & 0.85 & 12.2 &  39.6 &  295.7 \\
J054436.33-534759.5 & 0.336 &   SR & 0.80 & J054437.07-534751.6 & 0.336 &   SR & 1.22 & 10.2 &  49.3 & 124.5 \\
J054600.99-274838.4 & 0.270 &   SR & 0.87 & J054601.87-274854.7 & 0.271 &   SR & 0.82 & 19.9 &  82.6 & 198.0 \\
J055605.93-342504.7 & 0.249 &                SR & 0.89 & J055608.80-342449.5 & 0.250 &     SR & 0.90 & 38.7 & 151.0 &    72.0 \\
J063440.79+425530.5 & \dots &    \dots & 1.01 & J063441.53+425535.0 &   \dots &    \dots & 1.07 &  9.3 & \dots &    \dots \\
J064240.13+583325.0 &   \dots &    \dots & 1.87 & J064240.18+583317.5 & 0.495 & G & 1.22 &  7.5 &  45.8 &    \dots \\
J072320.17+382254.4$^{\dagger}$ & 0.194 &   D & 0.81 & J072321.82+382251.5$^{\dagger}$ & 0.194 &   D & 0.80 & 19.6 &  63.0 &   15.8 \\
J072842.02+173628.5 & 0.293 & G & 1.00 & J072842.91+173636.4 & 0.293 & G & 1.17 & 15.0 &  65.7 & 140.4 \\
J073813.39+284633.6 & 0.290 &   D & 1.31 & J073813.69+284623.2 & 0.289 &   D & 0.97 & 11.2 &  48.6 &  193.7 \\
J085726.02-085146.1 & 0.251 &   SR & 0.82 & J085726.94-085144.4 & 0.251 &   SR & 0.92 & 13.7 &  53.7 &  72.0 \\
J092743.33-043200.2 & 0.089 & G & 1.10 & J092743.91-043202.4 & 0.090 & G & 1.02 &  8.9 &  14.9 &  71.3 \\
J101323.25+281828.6 & 0.140 & G & 0.88 & J101323.57+281835.4 & 0.141 & G & 0.95 &  7.9 &  19.7 & 356.1 \\
J102219.75-064355.4 & 0.328 &   SR & 0.99 & J102219.75-064402.3 & 0.328 &   SR & 0.89 &  7.0 &  33.0 &   67.6 \\
J103433.01+561715.7$^{\dagger}$ & 0.272 &   D & 1.02 & J103434.51+561707.3$^{\dagger}$ & 0.272 & G & 0.83 & 15.0 &  62.4 &   59.4 \\
J104820.11-070605.3 & 0.249 &   D & 0.84 & J104820.26-070543.5 & 0.248 &   D & 0.93 & 21.9 &  85.2 &  245.2 \\
J111943.17+010225.9 & 0.498 &   S & 1.28 & J111943.93+010219.9 & 0.498 &   D & 0.96 & 12.9 &  78.4 & 117.4 \\
J112841.09+505843.4 & 0.198 &   D & 1.20 & J112841.93+505846.8 & 0.199 &   D & 0.90 &  8.6 &  28.1 &  37.4 \\
J114228.86-035311.6 & 0.520 &   D & 1.15 & J114229.80-035256.5 & 0.520 &   D & 0.89 & 20.7 & 129.0 &   21.5 \\
J114624.80+000504.4 & 0.262 &   GA & 0.92 & J114625.47+000500.0 & 0.260 &   D & 0.82 & 11.1 &  44.7 &  320.4 \\
\dots & \dots & \dots & \dots & \dots & \dots & \dots & \dots & \dots & \dots & \dots\\
\noalign{\smallskip}
\hline
\end{tabular}
\end{center}
\tablecomments{Rank 0.5 spatially resolved mid-IR dual AGN candidates, considered to be ``high confidence'' dual AGN candidates based on spectroscopic redshift information and/or morphological properties. This table follows an identical column structure to that given in Table~\ref{tab:rank1}. The full version of this table is available in electronic form online. {$^{\dagger}$: Systems identified as spatially resolved mid-IR dual AGN candidates in \citet{barrows2023}.}
}
\end{table*}

\subsection{Identified Contaminants}
\label{sec:identifiedcontaminants}
A total of 48/147 mid-IR dual AGN candidates morphologically selected {as galaxy mergers} turned out to be non-dual AGN contaminants, which we detail here. Forty-five out of 48 {of these contaminants} were in fact projected foreground-background pairs and/or had separations inconsistent with mergers based on archival and follow-up spectroscopy (Section~\ref{sec:contam_removal_phaseii} and \ref{sec:followup_obs}). {One system,} J100120.68+555355.4, {was} originally classified as a merger candidate {but} turned out to be the well known gravitational lens QSO B0957+5608B at $z=1.41$ \citep[][]{walsh1979} and was therefore rejected. One candidate (J223559.62-512011.6) was conservatively rejected from the sample after SOAR follow-up spectroscopy revealed that a close ($<5''$) companion to one of the nuclei was a bright foreground star; the optical spectroscopic signal from the star overwhelmed any spectroscopic emission from the background AGN. {The remaining contaminant (J035448.34-505623.6) was spectroscopically identified as a star-quasar pair separated by 22\farcs1 in a follow-up Gemini observation. }For completeness, we also include in Table~\ref{tab:midIR_rejects} and Appendix Figure~\ref{fig:rejects} all rejected mid-IR dual AGN candidates.

\subsection{Morphological Examination}
\subsubsection{The Success Rate of Morphological Selection}

The selection strategy outlined in Section~\ref{sec:selection} prioritized using optical imaging to identify galaxy mergers and merger candidates hosting mid-IR AGN candidate  pairs, while optical spectroscopy was used to remove contaminants. This morphological selection originally yielded a subsample of 147 mid-IR dual AGN candidates {in galaxy mergers or merger candidates}. Archival and follow-up optical spectra for 118/147 of these systems revealed 48 contaminants (foreground-background pairs, pairs with separations $>130$\,kpc, lenses, etc.), yielding a tentative failure rate of $\approx40.6\%$. Complete spectroscopic coverage of the remainder of the sample is necessary to determine the true success/failure rate of this morphology-prioritized selection strategy, but these results tentatively suggest that morphological selection can be used with a success rate of $\approx58\%$ in future dual AGN studies. Note, as this selection strategy relies upon the observation of tidal features, this strategy is biased against merger mass ratios, host morphologies, and orbital configurations that are less conducive to tidal disturbances and the production of tidal tails \citep[e.g.,][]{mihos1998,dubinski1996,dimatteo2007,barnes2016,saleh2024}. {The observation of tidal features is also a function of the restframe wavelength of the imaging filters as well as a function of the DESI LS imaging depth \citep{dey2019}\footnote{https://datalab.noirlab.edu/data/legacy-surveys\#ls-dr9}, which also biases the sample toward lower redshifts.} Similarly, the failure rate of this selection strategy is likely inflated due to the fact that we erred on the side of inclusion rather than exclusion during the morphological assessment, often including mid-IR pairs even when the hosts displayed only subtle asymmetries or very faint structures in the imaging that could either be potential low surface brightness tidal structures or simply noise (thereby including more potential contaminants in the pursuit of higher completeness). Had we limited our search only to those with the most dramatic and obvious tidal features and excluded all other merger candidates, the success rate of our selection strategy would have been higher (at the expense of lower completeness).

\subsubsection{The Fraction of Multi-Mergers}
\label{sec:frac_multi}

After identifying our subsample of mid-IR dual AGNs and candidates, we then re-examined the DESI LS optical imaging using a larger field of view to search for additional candidate galaxy companions associated with the mid-IR dual AGN candidates. As in our initial imaging selection (Section~\ref{sec:morphclass}), we searched for tidal features indicative of interactions, and we supplemented this morphological examination with spectroscopic information from the SDSS DR18 \citep{almeida2023}, LAMOST DR1-9 \citep{cui2012}, and DESI DR1 \citep{DESICollab2025} spectroscopic releases{, archival redshifts provided in the Simbad and NED archives, as well as with our own follow-up observations}. Among the 78 Rank 1 and Rank 0.5 dual AGN candidates with spectroscopic redshifts for both nuclei, 31 represent multi-mergers with at least one additional galaxy companion that has been spectroscopically confirmed to exhibit $r_p<130$\,kpc and $|\Delta v|<600-700$\,km\,s$^{-1}$ relative to either of the mid-IR AGNs in a given pair. This yields a multi-merger fraction of $39.7\%$ for the Rank 1 and 0.5 dual AGN candidates; this represents a lower limit due to the lack of complete spectroscopic coverage for all candidate companions (an additional 12, or 15.7\%, Rank 1 and 0.5 dual AGNs have a candidate companion galaxy not yet spectroscopically confirmed). Note, among only the 13 Rank 1 dual AGNs, 4/13 are associated with at least one additional spectroscopically confirmed companion (or a $\approx31\%$ multi-merger fraction among confirmed dual AGNs), and an additional 3/13 have candidate companion galaxies {(which require spectroscopic redshift verification)}. These fractions will also evolve as more Rank 0.5 and Rank 0 dual AGN candidates are spectroscopically confirmed; among the 29 Rank 0.5 and 0 dual AGN candidates that lack spectroscopic redshifts for one or both nuclei, 20 are potentially multi-mergers involving three or more galaxies. Complete spectroscopic coverage of the remainder of the sample is necessary to determine the final multi-merger fraction among spatially-resolved mid-IR dual AGNs.

\subsubsection{J055605.93-342504.7: A Triple Mid-IR AGN Candidate}
\label{sec:tenttriple}

During the morphological examination discussed in Section~\ref{sec:selection} and in Section~\ref{sec:frac_multi}, we often examined both the DESI LS optical and WISE mid-IR imaging to search around the mid-IR dual AGN candidates for any additional, nearby mid-IR AGN candidates. One system stood out from the rest of the sample: J055605.93-342504.7 and J055608.80-342449.5 (hereafter J0556-3425 S and NE) represent two mid-IR AGN candidates separated by $\sim$38.7''{; $grz$ imaging of the host system is shown in the sixth row of Figure~\ref{fig:rank0.5_duals}}. Follow-up SOAR optical long-slit spectroscopy showed that the AGNs reside at $z=0.249$ and $0.250$ and therefore exhibit a projected, physical separation of $\sim151$\,kpc and velocity difference $|\Delta v|=176.9$\,km\,s$^{-1}$; abiding strictly by the criteria outlined in Section~\ref{sec:selection}, these two mid-IR AGNs do not constitute a dual AGN, but they are nonetheless located physically close to one another. A third mid-IR source (J055604.95-342503.6, hereafter J0556-3425 W) resides in a companion galaxy 12.4'' ($\sim$48\,kpc) westward relative to J0556-3425 S, and strong tidal features and a $|\Delta v|=72.0$ km s$^{-1}$ suggest that the two galaxies are interacting. J0556-3425 W exhibits a WISE two band color of $W1-W2=0.722$; while this color does not satisfy the 95\% reliability criterion from \citet[][]{stern2012}, it does satisfy both the 90\% completeness and 90\% reliability two band magnitude-dependent color criterion from \citet[][]{assef2018}. Though our selection methodology described in Section~\ref{sec:selection} for the overarching mid-IR dual AGN sample relied upon the \citet{stern2012} color cut, we can nevertheless conclude that a mid-IR AGN candidate resides in J0556-3425 W, making J0556-3425 S and J0556-3425 W a dual AGN candidate. Tidal signatures are seen between J0556-3425 W and S; no tidal bridge or other structure connecting J0556-3425 NE and W can be seen in the imaging. J0556-3425 NE appears slightly asymmetrical with faint, potential tidal shell features. While J0556-3425 NE resides $\sim$150\,kpc away from J0556-3425 S (and therefore does not satisfy the criteria in Section~\ref{sec:selection}), multi-mergers can lead to complex orbital configurations and it remains possible that J0556-3425 NE is interacting with J0556-3425 S and W (or has in the past). In light of this and the consistency of the redshifts, we consider J0556-3425 NE, S, and W a candidate triple mid-IR AGN; this system warrants deep, follow-up optical, near-IR, and HI imaging to search for evidence of tidal structures connecting the three nuclei. If confirmed, J0556-3425 would represent the first mid-IR selected triple AGN; for now, we classify this system as a Rank 0.5 dual and triple AGN candidate.

\subsection{\textit{WISE} Colors and Luminosities}
\label{sec:wise_analysis}

\begin{figure}
    \centering
    \includegraphics[width=1.0\linewidth]{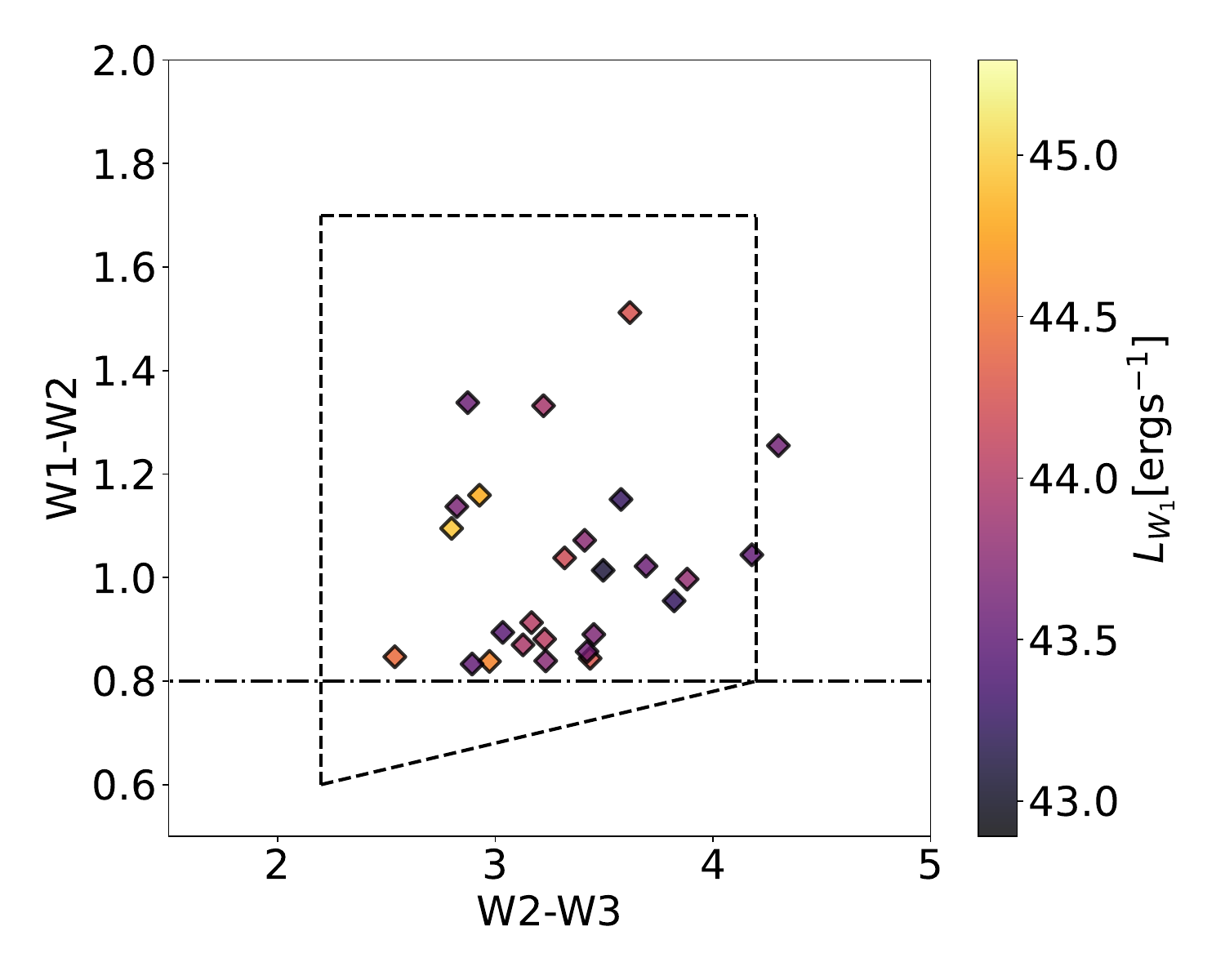}\\
    \includegraphics[width=1.0\linewidth]{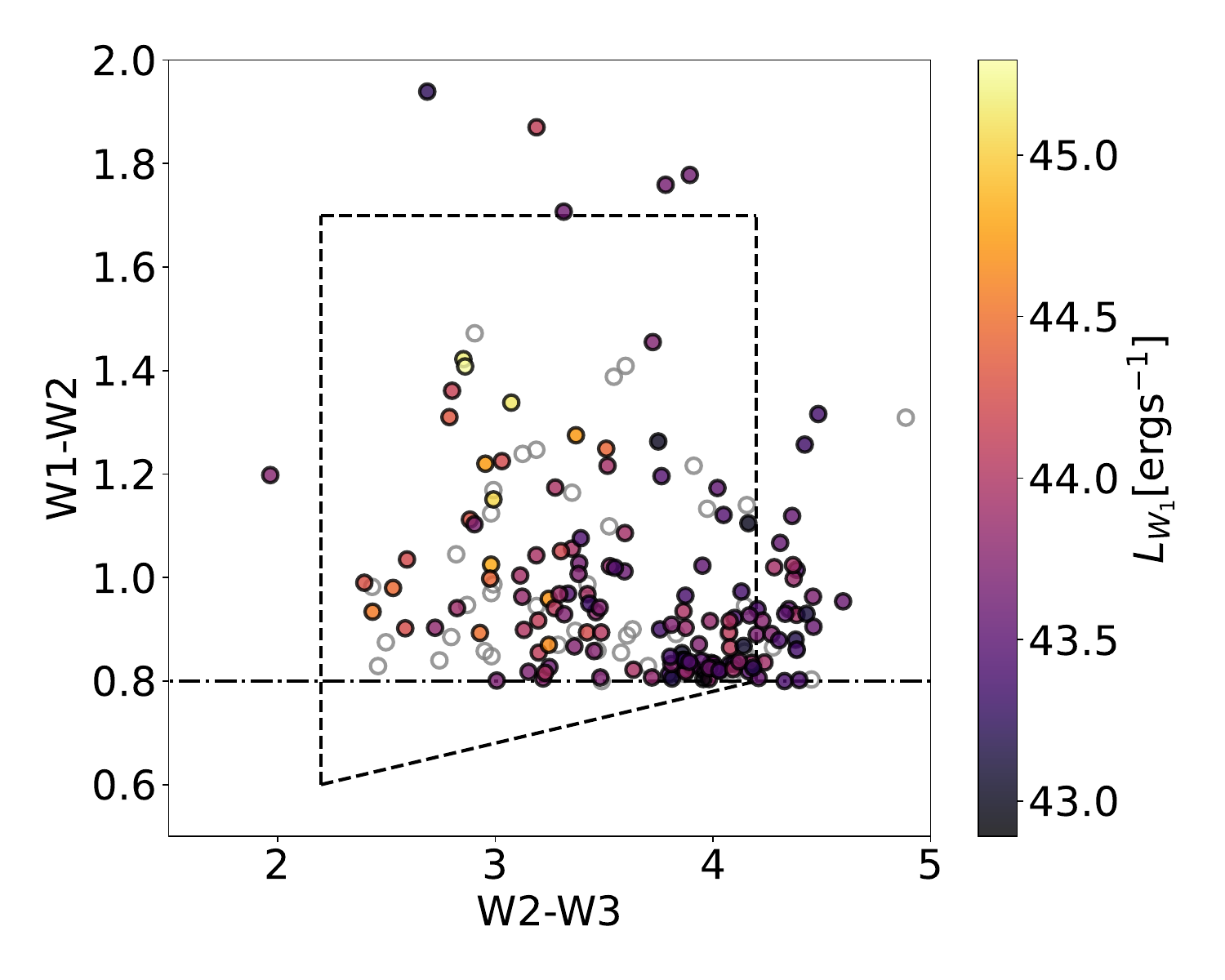}
    \caption{\textit{WISE} mid-IR color space for (top, diamond markers) Rank 1 and (bottom) Rank +0.5 and 0 mid-IR dual candidates in this sample. $W2-W3$ color is given on the x-axis and $W1-W2$ color is given on the y-axis. The data are color coded according to WISE W1 luminosity when redshifts are available, where the scale is given on the auxiliary axes; when redshifts are not available, the nuclei are marked as open-faced, grey circles. The dashed-dotted horizontal line represents the WISE AGN color cut $W1-W2\geq0.8$ from \citet{stern2012}; the dashed polygon represents the WISE AGN color wedge defined by \citet{jarrett2011}. The majority of sources in this sample exhibit WISE colors consistent with the color wedge from \citet{jarrett2011}.}
    \label{fig:midIR_colors}
\end{figure}

The top (bottom) panel of Figure~\ref{fig:midIR_colors} shows the distribution of WISE mid-IR colors for the Rank 1 (Rank +0.5 and 0) spatially-resolved mid-IR dual AGNs and candidates in our sample, where the data are color-coded according to WISE W1 luminosity. As discussed in Section~\ref{sec:analysis} and \ref{sec:results}, these mid-IR dual AGNs and candidates were selected using the two-band color cut $W1-W2\geq0.8$ \citep[$W2<15.05$,][]{stern2012}, and all of these mid-IR dual candidates also satisfy the magnitude dependent 90\% reliability mid-IR AGN criteria laid out in \citet{assef2018}. 83.2\% of the total nuclei also satisfy the more stringent three-band color wedge defined by \citet{jarrett2011}. In 12/13 (or 92.3\%) of Rank 1 mid-IR dual AGNs, 45/71 (or 63.4\%) of Rank 0.5 candidates, and 23/27 (or 85.2\%) of Rank 0 candidates, {the} constituent AGN candidates satisfy the three-band color wedge from \citet{jarrett2011}. Rank 1 mid-IR dual AGNs in this sample exhibit median nuclear W1 and W2 luminosities of $5.65^{+21.77}_{-3.59}\times10^{43}$ erg s$^{-1}$ and $6.44^{+21.99}_{-3.59}\times10^{43}$ erg s$^{-1}$, respectively (where the upper and lower bounds are the $84^{\rm{th}}$ and $16^{\rm{th}}$ percentiles, respectively). Among Rank 0.5 candidates where both nuclei have measured spectroscopic redshifts, these W1 and W2 median nuclear luminosities are $5.32^{+11.06}_{-2.79}\times10^{43}$ erg s$^{-1}$ and $5.24^{+11.05}_{-2.79}\times10^{43}$ erg s$^{-1}$, respectively.

Figure~\ref{fig:midIR_colors} tentatively suggests that more luminous Rank 1 and 0.5 AGN candidates exhibit bluer $W2-W3$ and redder $W1-W2$ colors. Motivated by this, we searched for any potential relationship between the WISE $W1$ and $W2$ luminosities and (1) the WISE $W1-W2$ colors and (2) the WISE $W2-W3$ colors. We used the SciPy function \texttt{pearsonr} \citep{virtanen2020} to calculate the Pearson correlation coefficient (which varies from +1 to -1 and measures the linear relationship between two datasets) and the $p$-value (which indicates the probability that datasets produced from an uncorrelated system could produce a correlation coefficient at least as extreme as the one computed with the given data).\footnote{https://docs.scipy.org/doc/scipy/reference/generated/scipy.\\stats.pearsonr.html} During this analysis, we included only Rank 0.5 and 1 dual AGN candidates for which spectroscopic redshifts were available for both nuclei, and we tested for correlations between the WISE colors and WISE luminosities under two distinct cases:
\begin{itemize}

    \item Case 1: $L_{W_{\alpha}}$ vs. $W1-W2$ or $W2-W3$, where the luminosities and colors of the nuclei were treated separately.
    \item Case 2: {total} $L_{W_{\alpha}}$ vs. {total} $W1-W2$ or {total} $W2-W3$, {treating the system as a whole}. {Here,} the luminosities {were summed, and the combined WISE colors were derived by summing the WISE fluxes, converting the total fluxes back to magnitudes, and then calculating the total magnitude-based color}
\end{itemize}
where $W_{\alpha}$ is a placeholder for the individual WISE bands ($W1$ and $W2$). 

For $L_{W1}$ vs. $W1-W2$, we found correlation coefficients and p-values of 0.15 ($6.4\times10^{-2}$) and 0.06 ($5.8\times10^{-1}$), respectively, for Cases 1 and 2. For $L_{W2}$ vs. $W1-W2$ we found higher correlation coefficients and smaller p-values: 0.33 ($3.9\times10^{-5}$) and 0.24 ($3.3\times10^{-2}$), respectively, for Cases 1 and 2, suggesting more of the variation in $W1-W2$ can be attributed to $L_{W2}$ than $L_{W1}$. The derived coefficients suggest that, for a given redshift requirement and treatment of the nuclei (individual or separate), $\gtrsim6-11\%$ of the variation in $W1-W2$ may be attributable to $L_{W2}${; the low p-values found for $W1-W2$ color vs. WISE luminosity casts doubt on a correlation between these two parameters}.
{For $L_{W2}$ vs. $W2-W3$, we found correlation coefficients and p-values of -0.57 ($p=1.4\times10^{-14}$) and -0.60 ($p=7.9\times10^{-9}$), respectively, for Cases 1 and 2.} For $L_{W2}$ vs. $W2-W3$, we found correlation coefficients and p-values of -0.55 ($p=1.5\times10^{-13}$) and -0.58 ($p=5.1\times10^{-8}$), respectively, for Cases 1 and 2. These coefficients suggest that $\gtrsim30-36\%$ of the variation in $W2-W3$ {may} be attributed to $L_{W1}$ and $L_{W2}$. 

\subsection{WISE AGN Luminosities as a Function of Separation}
\label{sec:wise_asf_sep}

Tentative inverse correlations have been reported in the literature between the [OIII] luminosity and separation \citep[e.g.,][]{liu2012,hou2020}, $2-10$\,keV X-ray luminosity and separation \citep[e.g.,][]{koss2012,hou2020}, $N_{\rm{H}}$ and separation \citep[for dual AGNs and mergers in general, e.g.,][]{ricci2017mnras,guainazzi2021,derosa2023}, and $E(B-V)$ with separation \citep[][]{barrows2023}. Motivated by these reported anti-correlations, we searched for any potential relationship between the projected, physical separation of the nuclei and (1) the WISE $W1-W2$ colors and (2) the $W1$, $W2$, and $W3$ luminosities of the AGNs. As in Section~\ref{sec:wise_analysis}, we used the \textsc{scipy} package \textsc{pearsonr} \citep{virtanen2020} to calculate the Pearson correlation coefficient, and we included only systems where each nucleus had a measured spectroscopic redshift. We tested for correlations between the mid-IR luminosities and separation under two distinct cases, limited to Rank 0.5 and 1 dual AGN candidates where both nuclei were required to have spectroscopic redshifts:
\begin{itemize}

    \item Case 1: $L_{W_{\alpha}}$ or $W1-W2$ vs. $r_p$, where the luminosities or colors of the nuclei were treated separately.
    \item Case 2: {total} $L_{W_{\alpha}}$ or {total} $W1-W2$ vs. $r_p$. {Here,} the luminosities {were summed, and the combined WISE colors were derived by summing the WISE fluxes, converting the total fluxes back to magnitudes, and then calculating the total magnitude-based color}.
\end{itemize}
where $W_{\alpha}$ is a placeholder for the individual WISE bands ($W1$, $W2$, and $W3$). 

For Case 1 and WISE $L_{W1}$, $L_{W2}$, and $L_{W3}$, we found correlation coefficients and p-values of: 0.33 ($p=3.1\times10^{-5}$), 0.28 ($p=4.5\times10^{-4}$), and 0.26 ($p=9.5\times10^{-4}$), respectively. For Case 2 and WISE $L_{W1}$, $L_{W2}$, and $L_{W3}$, we found correlation coefficients and p-values of: 0.39 ($p=4.7\times10^{-4}$), 0.33 ($p=3.4\times10^{-3}$), and 0.30 ($p=8.0\times10^{-3}$), respectively. These Pearson correlation coefficients suggest that, for a given WISE band and under specific spectroscopic redshift requirements, $\gtrsim7-15\%$ of the variation observed in the mid-IR luminosities {may} be attributed to the projected pair separation. For Cases 1 and 2 for WISE $W1-W2$ color, we found correlation coefficients of $-0.20$ and $-0.28$ and p-values of $1.3\times10^{-2}$ and $1.6\times10^{-2}$; the low p-values here cast doubt on a link between pair separation and WISE color at this point.

\begin{figure*}
    \centering
    \includegraphics[width=0.33\linewidth]{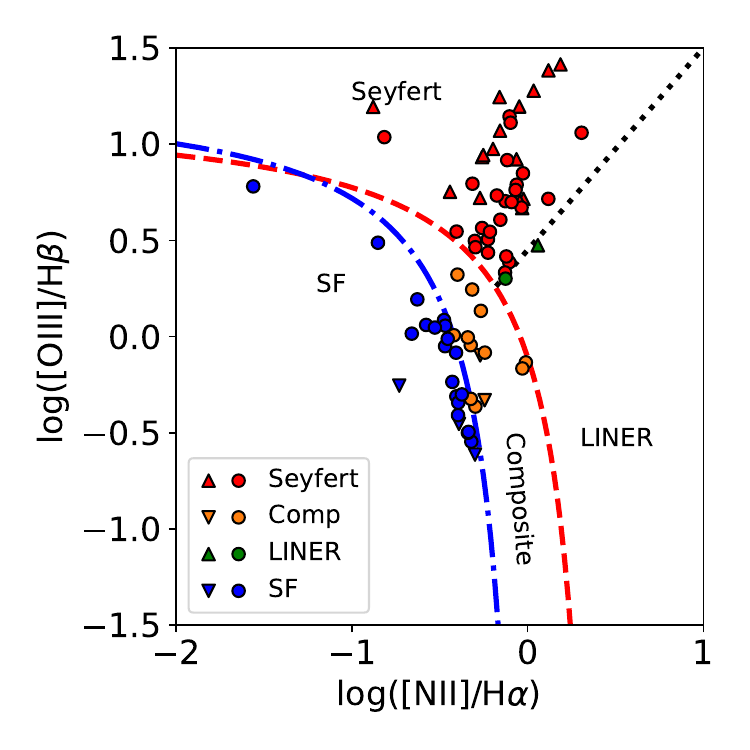}\hspace{-0.2mm}
    \includegraphics[width=0.33\linewidth]{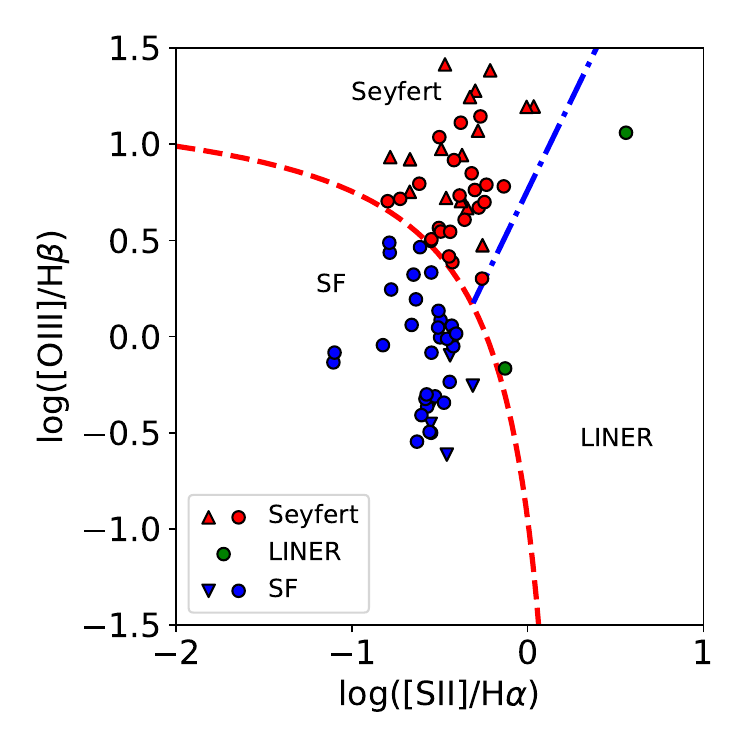}\hspace{-0.2mm}
    \includegraphics[width=0.33\linewidth]{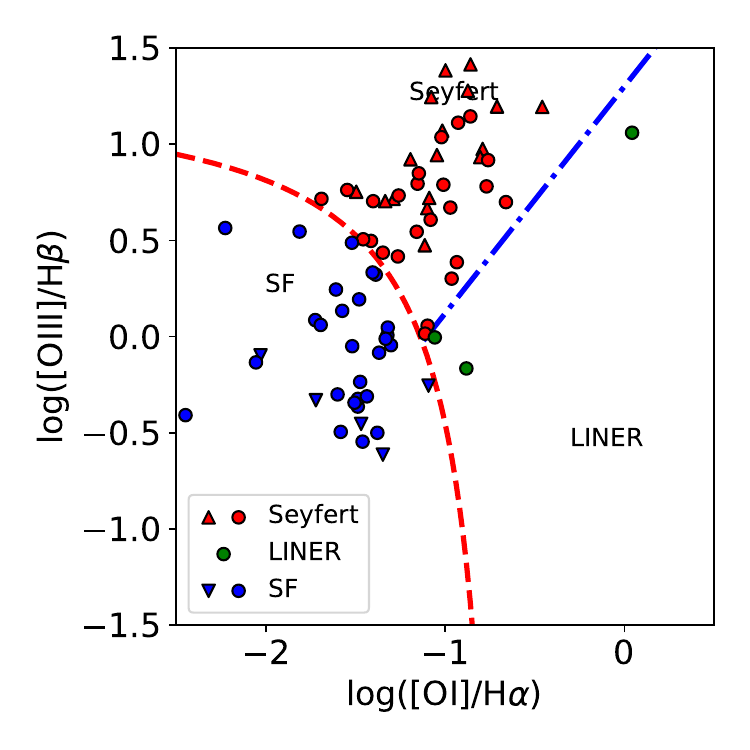}\hspace{-0.2mm}\\
    \caption{BPT \citep{baldwin1981,kewley2001,kauffmann2003,kewley2006} spectroscopic emission line ratio diagnostic plots for mid-IR dual AGNs with sufficient spectroscopic coverage. Left panel: log([N\,II]$\lambda6584$/H$\alpha$) diagnostic plot; the dashed red curve represents the theoretical starburst limit from \citet{kewley2001}, the dash-dotted blue curve is an empirically derived demarcation separating star forming nuclei from Composite (Seyfert-HII) nuclei \citep{kauffmann2003}, and the dotted black line \citep{schawinski2007} separates Seyferts from LINERs. Middle panel: log([S\,II]$\lambda6717,\lambda6731$/H$\alpha$) diagnostic plot; the red dashed and blue dash-dotted lines -- adopted from \citet{kewley2006} -- differentiate between Seyferts, LINERs, and star forming nuclei. Right panel: log([O\,I]$\lambda6300$/H$\alpha$) diagnostic plot; the red dashed and blue dash-dotted lines differentiate between Seyferts, LINERs, and star forming nuclei \citep{kewley2006}. }

    \label{fig:BPTs}
\end{figure*}

\subsection{BPT Optical Classes}
\label{sec:bpt_analysis}

We used the spectroscopic emission line fluxes obtained from archival observations (Sections~\ref{sec:contam_removal_phaseii}) and our follow-up observations (Section~\ref{sec:followup_obs}) to compute, when possible, the BPT \citep[][]{baldwin1981} emission line ratios and determine the optical BPT classes of these mid-IR dual AGNs and candidates. {We limit our BPT analysis to Rank 1 and 0.5 dual AGN candidates only.} A total of {81} nuclei across the {Rank 1 and 0.5} sample had readily accessible spectra with the relevant wavelength coverage sufficient for optical classification based on the [NII]/H$\alpha$ BPT diagram \citep{baldwin1981,kewley2001,kauffmann2003,kewley2006}. The BPT line ratios for {these} candidates, when available, are plotted in Figure~\ref{fig:BPTs}, where the left panel depicts the [NII]/H$\alpha$ diagram, the middle panel depicts the [SII]/H$\alpha$ diagram, and the right panel depicts the [OI]/H$\alpha$ diagram. Each panel also includes the classical demarcations separating Seyferts, LINERs, Composite nuclei, and HII regions \citep[][]{baldwin1981,kewley2001,kauffmann2003,kewley2006,schawinski2007}. While in this work we use Seyfert-like line ratios and the presence of broad optical lines to confirm the presence of bona fide accreting SMBHs in these mid-IR-selected AGNs (and, hence, elevate dual AGN candidates to confidence Rank 1), it is important to note that LINER-, Composite-, or HII-like line ratios do not rule out the presence of accreting SMBHs. While some of these non-Seyfert nuclei may be starbursts mimicking the mid-IR colors of AGNs \citep[see Section~\ref{sec:conf_ranks}; see also][]{barrows2021} and Composites and LINERs can be driven by {sources other than} AGN \citep[e.g.,][]{allen2008,cidfernandes2010,cidfernandes2011,singh2013,rich2015}, heavily dust enshrouded AGNs have been shown to be optically elusive and do not always show BPT evidence of AGNs at optical wavelengths \citep{komossa2003,satyapal2017,koss2018,pfeifle2019a}. This conservative BPT requirement will be revisited in a future work.

\begin{figure}
    \centering
    \includegraphics[width=1.0\linewidth]{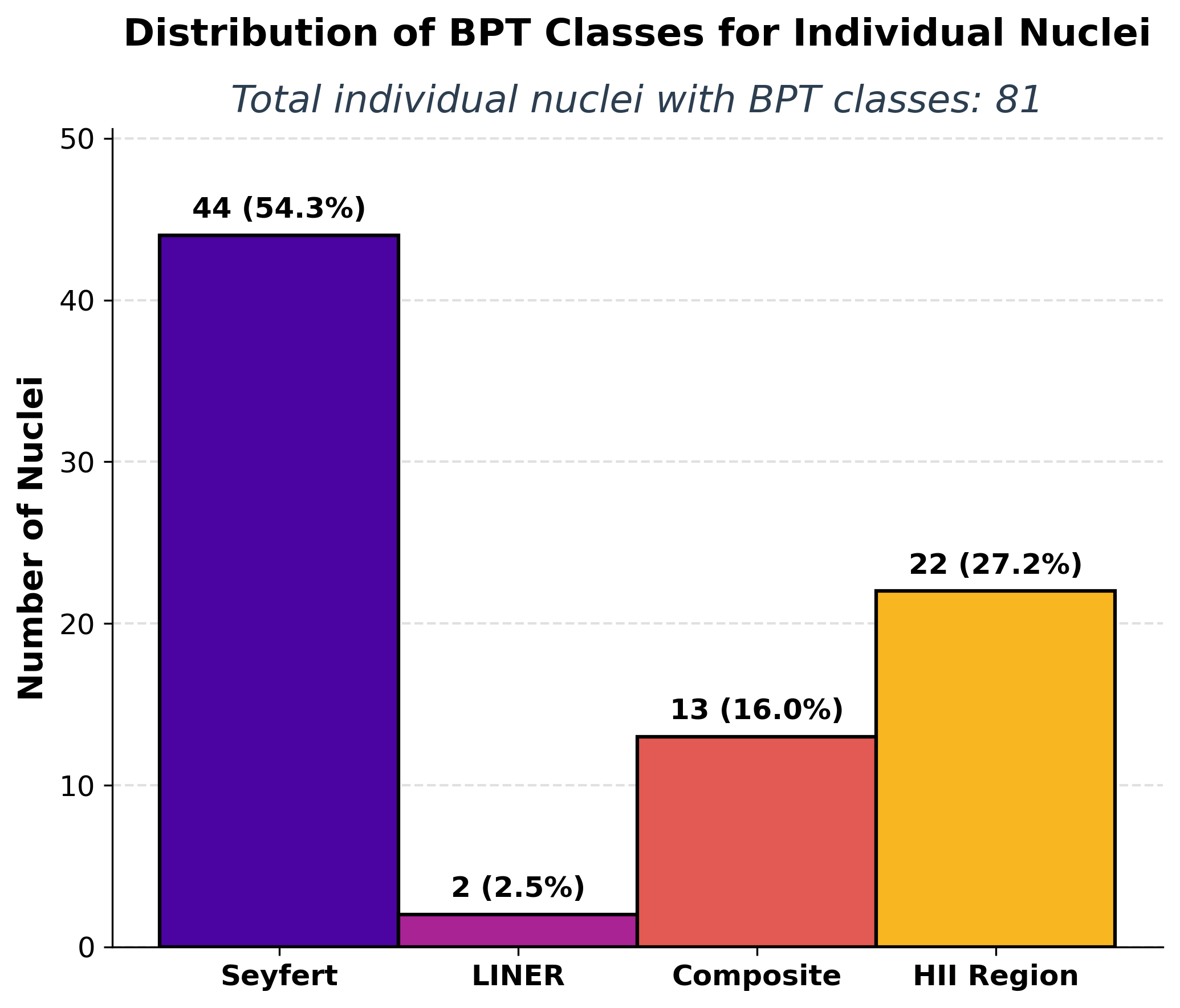}\hspace{-0.2mm}\\
    \includegraphics[width=1.0\linewidth]{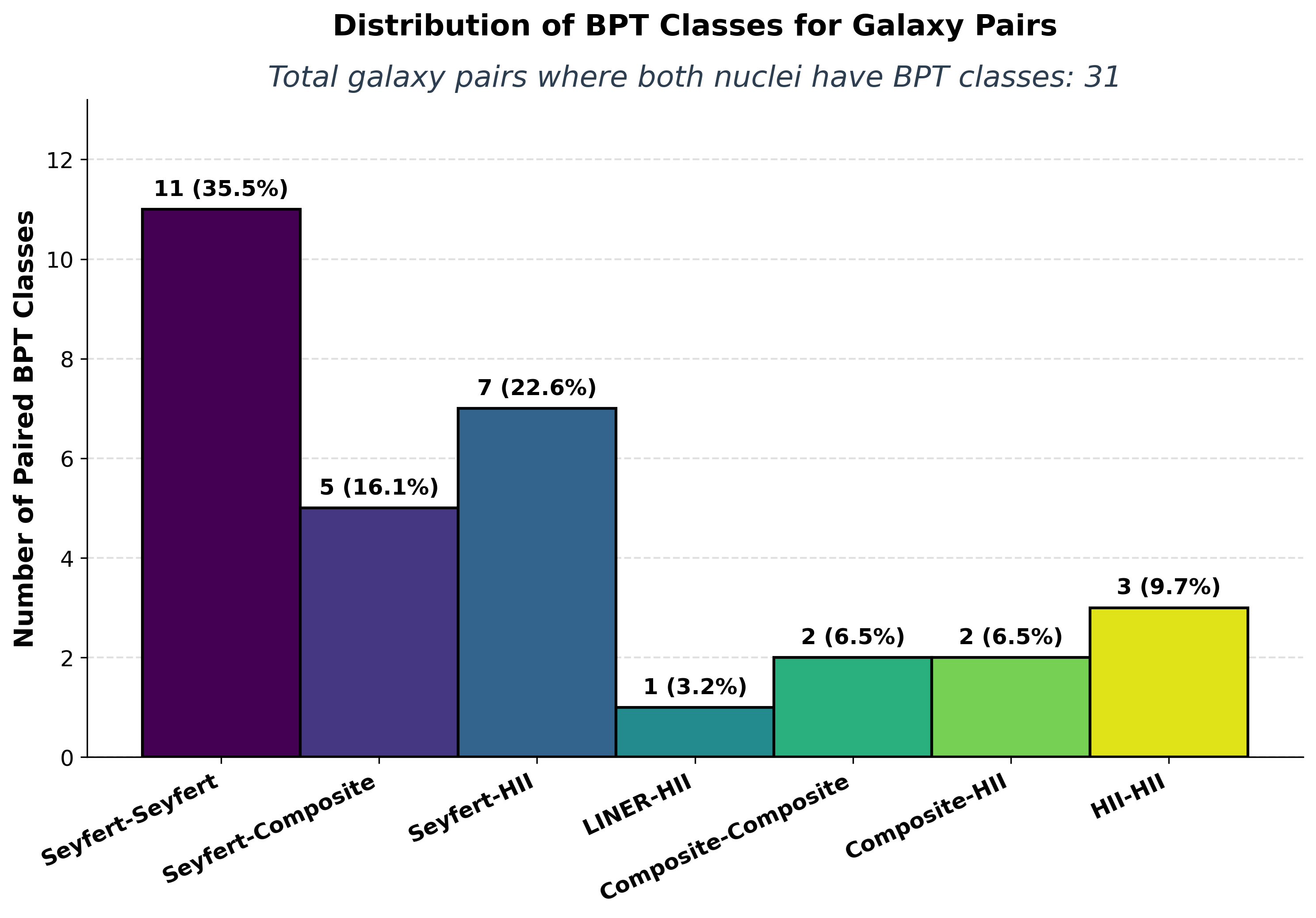}\hspace{-0.2mm}\\
    \caption{{(Top) The breakdown of optical BPT classes for individual nuclei for Rank 1 and 0.5 candidates. (Bottom) The breakdown of different BPT pair combinations, based upon the 31 Rank 1 and 0.5 candidates where both nuclei could be classified on the BPT diagram.}}

    \label{fig:bpt_fractions}
\end{figure}

{The BPT diagrams displayed in} Figure~\ref{fig:BPTs} demonstrate that this sample of mid-IR dual AGNs and candidates exhibits a clear diversity in optical type. {The top panel in Figure~\ref{fig:bpt_fractions} provides the statistical breakdown of BPT classes for the 81} {individual} rank 1 or rank 0.5 nuclei with BPT classifications from the [NII]{/H$\alpha$} diagram. 44 (54.3\%) are consistent with Seyferts, 2 (2.5\%) with LINERs, 13 (16.0\%) with Composites, and 22 (27.2\%) with star forming regions. {These individual nuclei with BPT classes are spread across 50 rank 0.5 and rank 1 dual AGN candidates.} {The lower panel of Figure~\ref{fig:bpt_fractions} displays the statistical breakdown of BPT pair combinations for 31 systems where both nuclei in a given system were classified on the BPT diagram.} In terms of BPT pairs, 11 (35.5\%) represent Seyfert-Seyferts, 5 (16.1\%) represent Seyfert-Composites, 7 (22.6\%) represent Seyfert-HII regions, 1 (3.2\%) represents a LINER-HII, 2 (6.5\%) represent Composite-Composites, 2 (6.5\%) represent Composite-HII regions, and 3 (9.7\%) represent HII-HII regions. No Seyfert-LINER, LINER-Composite, or LINER-LINER pairs have so far been identified.

\begin{figure}
    \centering
    \includegraphics[width=0.9\linewidth]{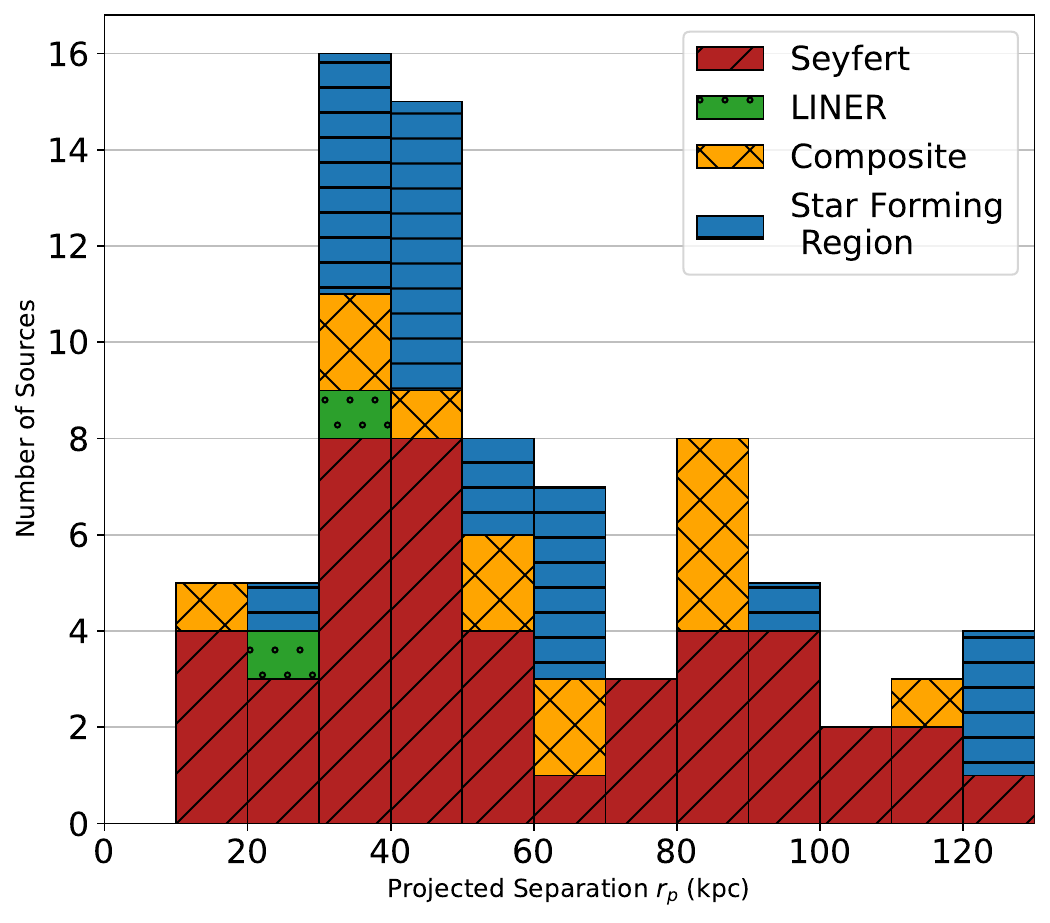}
    \caption{The distribution of projected separations (kpc) for Rank 1 and 0.5 mid-IR dual AGN as a function of BPT optical class. The projected separation bins are given on the x-axis, and the number of sources per bin is given on the y-axis. Seyferts, LINERs, composites, and {HII regions} are represented by red (diagonal hatched), green (dot hatched), orange (cross hatched), and blue (horizontal hatched) columns{, respectively, on this stacked histogram}.}
    \label{fig:sep_BPT_histogram}
\end{figure}

\subsection{BPT Optical Classes as a Function of Separation and WISE Properties}
\label{sec:bpt_analysis_sep_wise}

The {stacked} histogram in Figure~\ref{fig:sep_BPT_histogram} illustrates the relationship between the [NII] BPT optical class and projected separation for the Rank 1 and 0.5 mid-IR dual AGN candidates; red, green, orange, and blue hatched columns denote Seyferts, LINERs, Composites, and HII regions, respectively. While we do not have a complete census for the optical classes of the nuclei/mergers in this sample at this point, the current results for Rank 1 and 0.5 candidates suggest that Seyfert nuclei have a median separation $49.2^{+46.2}_{-17.8}$\,kpc, where the upper and lower bounds represent the $84^{\rm{th}}$ and $16^{\rm{th}}$ percentiles of the distribution and reflect the large standard deviation in separations. Similarly, Rank 1 and 0.5 Composite and HII nuclei show median separations of $62.4^{+24.7}_{-25.9}$\,kpc and $48.3^{+37.0}_{-10.8}$\,kpc, respectively. Here, we have excluded LINERs, as they make up a statistically insignificant fraction of the sample. These median separations may suggest that Seyfert and HII nuclei exhibit similar projected separations while Composite nuclei preferentially exhibit higher projected separations relative to Seyferts and HII regions, though the small number of sources and the large spread in the distributions of separations across all three BPT optical types prevents a more definitive result for now. 
\begin{figure}
    \centering
    \includegraphics[width=1.0\linewidth]{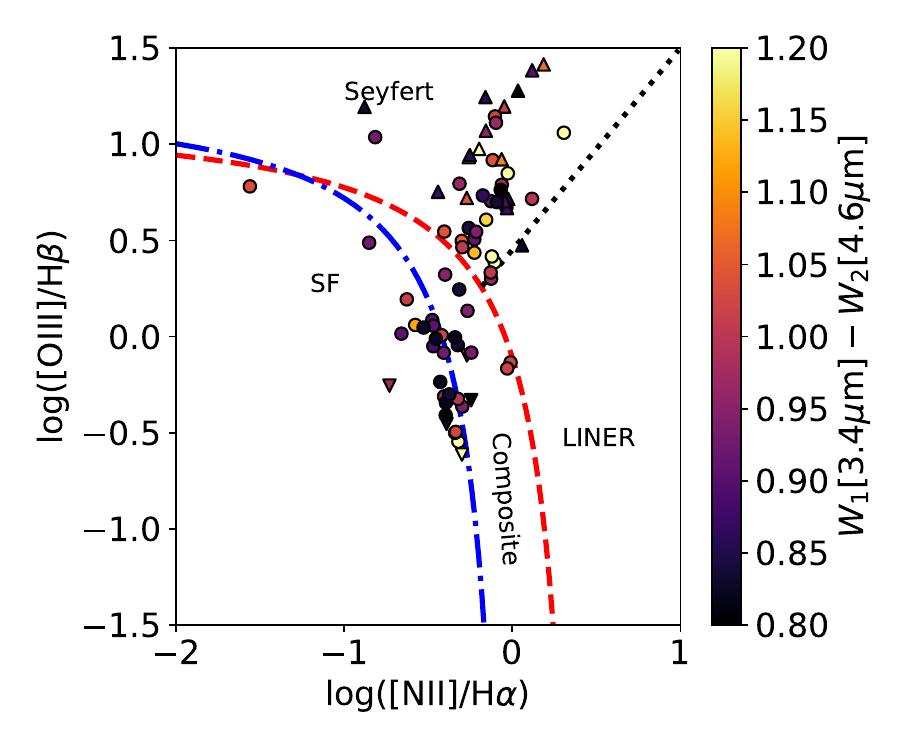}\\    \includegraphics[width=1.0\linewidth]{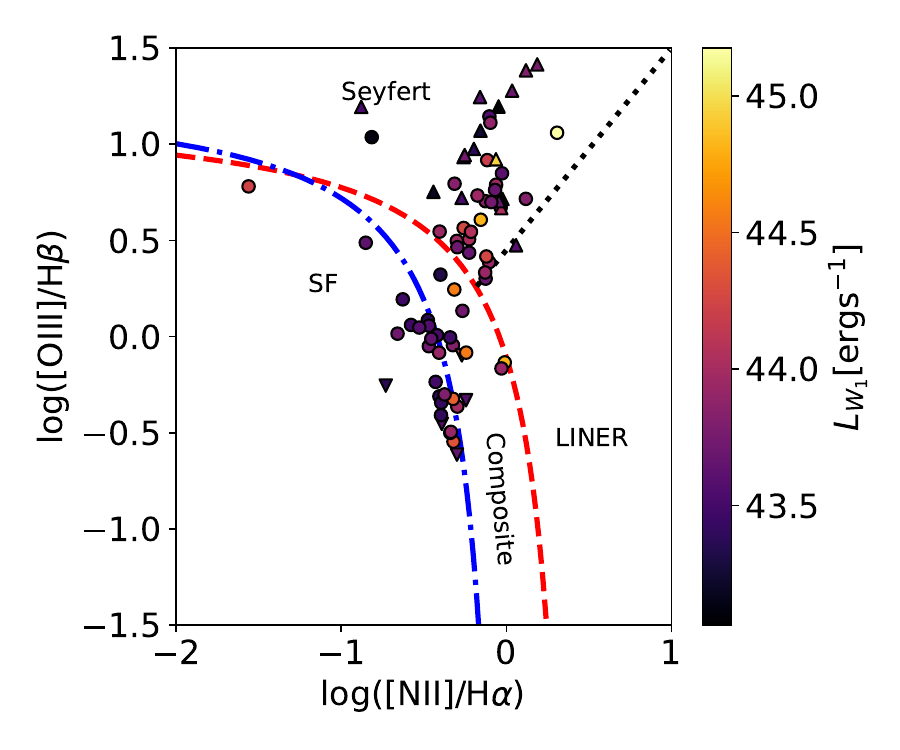}\\
    \caption{BPT \citep{baldwin1981} spectroscopic emission line ratio diagnostic plots for mid-IR dual AGN candidates with sufficient spectroscopic coverage, color coded by (top) $W1-W2$ color and (bottom) WISE W1 luminosity. The emission line ratios and red dashed, blue dash-dotted, and black dotted demarcations are identical to those depicted in the left panel of Figure~\ref{fig:BPTs}. }
    \label{fig:BPT_WISEprops}
\end{figure}

In the top panel of Figure~\ref{fig:BPT_WISEprops}, we show the [NII] BPT diagram and the emission line ratios of the Rank 1 and 0.5 dual mid-IR AGN candidates color coded by WISE $W1-W2$ color. As a visual aid, the maximum value on the auxiliary axis was set to $W1-W2=1.2$. Among Rank 1 and 0.5 candidates, Seyfert, Composite, and HII nuclei display median $W1-W2$ colors of $0.965^{+0.173}_{-0.112}$, $0.942^{+0.078}_{-0.108}$, and $0.918^{+0.172}_{-0.093}$, respectively, where again the upper and lower bounds represent the $84^{\rm{th}}$ and $16^{\rm{th}}$ percentiles of the distributions (and we have again ignored LINERs due to their statistical insignificance). Based on these median values, WISE color appears to tentatively increase when moving from HII nuclei, to Composites, to Seyferts, but the small sample sizes and large spreads in these distributions again prevent definitive conclusions for now.  

In Section~\ref{sec:wise_asf_sep}, we identified potential correlations between WISE luminosity and separation, with $L_{W1}$ versus separation yielding the largest Pearson-r correlation coefficient. Here, we check for any potential relationship between WISE $L_{W1}$ and BPT optical type. The bottom panel of Figure~\ref{fig:BPT_WISEprops} again depicts the [NII] BPT diagram and the emission line ratios of Rank 1 and 0.5 dual mid-IR AGN candidates, but the data are color coded according to WISE W1 luminosity rather than $W1-W2$ color; this same plot color coded by W2 luminosity (not shown) is qualitatively similar. Among Rank 1 and 0.5 candidates, Seyfert, Composite, and HII nuclei exhibit median W1 luminosities of $6.10^{+9.60}_{-2.98}\times10^{43}$\,erg\,s$^{-1}$, $8.39^{+27.86}_{-5.08}\times10^{43}$\,erg\,s$^{-1}$, and $4.10^{+4.34}_{-1.51}\times10^{43}$\,erg\,s$^{-1}$, respectively, where the upper and lower bounds again represent the $84^{\rm{th}}$ and $16^{\rm{th}}$ percentiles of the distributions. When considering the large spread in these distributions, these median values do not indicate a clear relationship between WISE W1 luminosity and optical type, though the $L_{W1}$ distributions for Seyferts and Composites do extend to higher luminosities than HII regions. We found a similar result when examining -- for a given AGN pair {(with BPT measurements for one or both nuclei)} -- whether the more luminous or less luminous of the nuclei prefer a particular optical type (e.g., when probing if more luminous nuclei tend to host more Seyferts than fainter nuclei across the pairs). Among the Rank 1 and 0.5 systems, we found 23 Seyferts, 1 LINER, 8 Composites, and 14 HII regions among the more luminous AGNs in these pairs; this result is very similar to that found for less luminous AGNs in these pairs: 21 Seyferts, 1 LINER, 5 Composites, and 8 HII regions.

\section{Discussion} 
\label{sec:discussion}

\subsection{The Importance of Prioritizing Candidate Selection First and Spectroscopy Second}
\label{sec:whystartfromimaging}

A significant number of prior works in the field of dual AGNs relied fundamentally upon the availability of optical spectroscopic observations either for optical BPT emission line ratio selection \citep[e.g.,][]{wang2009,liu2010b,smith2010,liu2011b} or more generally for spectroscopic redshift requirements \citep[e.g.,][]{koss2012,fu2015a,fu2015b,satyapal2017,pfeifle2019a}. Fiber collision limits on ground-based fiber spectroscopic surveys -- 55''-60'' for SDSS, $\sim89''$ for DESI -- {result in spectroscopic incompleteness for closely separated sources;}\footnote{With the exception being systems re-observed within a single data release or across different data releases.} {while it has been argued that this incompleteness can be corrected for in large-scale, statistical studies of AGNs in galaxy mergers \citep[e.g.,][]{ellison2008,patton2008,ellison2011} -- and specifically dual AGNs \citep{liu2011b} -- it has nonetheless limited many prior works searching for dual AGNs and focusing on smaller sample sizes \citep[e.g.,][]{fu2015a,satyapal2017,pfeifle2019a,gross2019}}. We can illustrate the severely limiting nature of requiring survey spectroscopic redshifts for initial selection by comparing {our sample }against the DESI DR1 spectroscopic release \citep{DESICollab2025}.

{Here, we limit} the mid-IR dual AGN candidate sample to the {78 mid-IR AGN candidates confirmed to reside in galaxy mergers (where spectroscopic redshifts have demonstrated that the nuclei exhibit pair separations $r_p\leq130$\,kpc and velocity differences $|\Delta v|\leq700$\,km\,s$^{-1}$)\footnote{We also include J055605.93-342504.7, as per Section~\ref{sec:tenttriple}.}, and we focus specifically on the 70/78 systems that were morphologically classified as mergers (we ignore the morphologically ambiguous pairs for this comparison). Of these 70 systems selected via their morphology and shown to host nuclei at the same redshift}, {34} systems ($\approx48.5\%$) have spectra available through the DESI DR1 for at least one nucleus, {and o}nly {15/70} ($\approx{21.4}\%$) systems have DESI DR1 spectra for both AGN candidates. Requiring spectroscopic redshifts for initial selection would ultimately have {removed 55/70 bona fide galaxy mergers hosting mid-IR dual AGN candidates (a reduction of} $\approx{78.5}\%$){; among the 13 Rank 1 (confirmed) dual AGNs identified in this sample so far, only four have DESI spectra for both nuclei, meaning that 9/13 (or $\approx$69\%) of the confirmed dual AGNs in this sample would have been missed in a spectroscopy-prioritized search.} This simple exercise demonstrates that few mid-IR dual AGN candidates would be uncovered when requiring spectroscopic redshifts for initial selection, and underscores the potential dangers of restricting dual AGN searches to only systems with archival spectra for one or both nuclei \citep[{such as in }][{etc.}]{koss2012,fu2015a,satyapal2017,pfeifle2019a}. Moreover, this exercise emphasizes the necessity of appropriately prioritizing (when possible) other wavebands and selection techniques prior to requiring spectroscopic information\footnote{For studies not relying specifically upon spectroscopic pair selection.}.

\subsection{Overlap with Prior Samples of Dual AGNs}
\label{sec:overlap}

\subsubsection{Overlap with \citet{barrows2023}}
\label{sec:disc_overlapwbarrows}

\citet{barrows2023} was the first to assemble a sample of spatially resolved mid-IR dual AGN candidates. These probabilistic pairs (with separations $\leq100$ kpc) were drawn from the catalog of WISE selected mid-IR AGNs presented in \citet{barrows2021}; these mid-IR AGNs candidates reside within the SDSS footprint, satisfy the {WISE} two-band, magnitude dependent {mid-IR} criteria established in \citet[][{Equations 1-4 in Section 2 of that work}]{assef2018}{, and rely upon probabilistic distributions of photometric redshifts for their pairing}. \citet{barrows2023} used spectral energy decompositions to derive photometric redshifts, AGN obscuration and bolometric luminosities, and host galaxy properties like stellar mass and star formation rate. Notably, \citet{barrows2023} found an increase in the mid-IR AGN pair fraction with increasing host galaxy specific star formation rate (suggesting a corelation between star formation and AGN triggering), an increase in AGN triggering with decreasing separation \citep[similar to findings in prior works, e.g.,][]{satyapal2014,weston2017}, and that --- among dual AGNs --- the AGNs in the less-massive hosts may be biased toward higher Eddington ratios relative to the AGNs hosted in the massive companions. 

To identify any overlap between the sample of probabilistic pairs presented in  \citet{barrows2023} and the present work, we matched the sample of 107 dual AGN candidates (Ranks 1, 0.5, and 0) to the \citet{barrows2023} sample with a match tolerance of 0.5''. A total of {13} overlapping systems were identified{, and we have marked these in Tables~\ref{tab:rank1},~\ref{tab:rank0.5}, and~\ref{tab:rank0} using $\dagger$ symbols}. This modest overlap of the two samples is natural given the partial overlap of the SDSS and DESI survey footprints as well as the fact that all AGNs in the current work satisfy the 90\% reliability color criteria from \citet{assef2018}, despite the differences in selection strategies \citep[SED fitting and probabilistic redshift selection in][and morphological selection here]{barrows2023}. The overlap between the two samples is in fact promising, as it suggests that we are finding similar populations of mid-IR AGNs within the SDSS footprint and that our DESI LS morphological approach for identifying mid-IR dual AGN candidates complements the work presented in \citet{barrows2023}. {For completeness, we also matched our Rank -1 candidates (contaminants) against the sample from \citet{barrows2023} and identified two overlapping systems, which may suggest additional foreground-background pairs could reside within the \citet{barrows2023} sample; we have marked these two systems with $\dagger$ symbols in Table~\ref{tab:midIR_rejects}.} We leave a more rigorous comparison against the probablistic pairs from \citet{barrows2023} to a future work where we will incorporate complete spectroscopic results and spectral energy decompositions and study our sample as a function of mass separation, obscuration, etc.

\subsubsection{Comparison against the Big Multi-AGN Catalog (Big MAC)}
\label{sec:bigmac}

To check for other overlap with previous samples of dual, binary, and recoiling AGN candidates (up to 2020), we matched each nucleus in this mid-IR dual AGN sample against the Big Multi-AGN Catalog \citep[Big MAC,][]{pfeifle2025} using a match tolerance of 10''. Only two systems overlap with the Big MAC: J121857.42+103551.2 and J165501.32+260517.4. 

J121857.42+103551.2 was identified as a dual AGN candidate by \citet{liu2011b} based on the detection of a narrow line Seyfert 2 AGN in J121857.42+103551.2 (selected in this work as a mid-IR AGN), while the second putative AGN they identified (displaying Composite BPT emission line ratios) was in a third galaxy located $\approx 60$'' to the southeast of J121857.42+103551.2 rather than the secondary mid-IR AGN identified in J121858.33+103547.5 here. Intriguingly, this southeastern companion galaxy hosted a bright radio source in VLA All Sky Survey \citep[VLASS,][]{lacy2020} S-band radio imaging; in concert with the composite BPT line ratios of this third nucleus and the presence of a dual mid-IR AGN nearby, this evidence suggested that this triple galaxy system hosted a triple AGN. \citet{schwartzman2025} recently confirmed this system as a triple radio AGN using follow-up Very Large Array (VLA) S, X, and Ku band imaging, where all three AGNs {were found to be} radio emitting and classified as radio quiet, the most luminous of which resides in the southeastern companion galaxy. This is the third triple AGN to be confirmed in the local universe \citep[see][]{turner2001,koss2012,pfeifle2019b,liu2019}, and the first triple radio AGN reported in the literature. 

{J165501.32+260517.4 was one of two morphologically ambiguous pairs that, as discussed in Section~\ref{sec:contam_removal_phaseii}, was rejected due to it exhibiting simultaneously large $r_p$ and $|\Delta v|$ values; here we describe this system in more detail with evidence from the literature in support of our Rank -1 classification. } J165501.32+260517.4 was selected as a lens candidate by 
\citet{inada2010} during a systematic search for lensed quasars in the SDSS. However, \citet{inada2010} rejected J165501.32+260517.4 as a lens candidate based on differences in the quasar SEDs and the slightly different redshifts derived from the SDSS fiber spectra; the sizable velocity difference \citep[$|\Delta v|\approx 1139$\,km\,s$^{-1}$, based on the redshifts reported in Table 3 of][]{inada2010} also led to its rejection as a ``binary quasar''. As discussed in Section~\ref{sec:identifiedcontaminants}, J165501.32+260517.4 exhibits a velocity difference of $|\Delta v|\approx 631$\,km\,s$^{-1}$ based on more recent measurements from SDSS and DESI, and the system exhibits a projected separation of 78.6\,kpc. Considering that interacting galaxies with projected separations in the range $75-85$ kpc are instead expected to exhibit median velocity differences $\approx56^{+112}_{-40}$\,km\,s$^{-1}$ \citep[][where the upper and lower bounds represent the 84$^{\rm{th}}$ and 16$^{\rm{th}}$ percentiles]{patton2024,pfeifle2025}, the spectroscopic evidence favors the scenario where these quasars are {clustered} {\citep[e.g.,][]{hennawi2006,hennawi2010}} but {likely }not hosted in an interacting {or merging} system, { in agreement with \citet{inada2010} and }as outlined above in Section~\ref{sec:identifiedcontaminants} {where we rejected it}. 

As noted in Section~\ref{sec:general_sample} and shown in Figure~\ref{fig:rp_vs_z}, {the} sample of mid-IR dual AGN {candidates identified within the present work} occupies a slightly higher redshift space than the majority of local universe dual AGN candidates identified up to 2020 \citep[Ranks 0.5 and 1 from][]{pfeifle2025}. This {is likely }a systematic effect driven by our selection strategy: at $z=0.04$, dual AGNs with separations $\gtrsim47.5$ kpc would be excluded by our angular separation cut off of 60'', and bright spatially resolved mid-IR dual AGNs with clean quality and contamination flags in WISE W1 and W2 but poor quality and contamination flags in W3 and W4 would have been excluded. We will investigate the presence of any missing, low redshift spatially resolved mid-IR dual AGNs in a future work.

\subsubsection{Overlap with \citet{zhang2021a}}
\label{sec:overlap_zhang}
J101756.76+344850.5 was recently included in the radio preselected dual AGN candidate sample of \citet{zhang2021a}, {where selection of candidates required: (1) two optical nuclei separated by $<8''$, (2) at least one optical nucleus to have an SDSS fiber spectrum in DR12/DR14 \citep{alam2015,abolfathi2018} and be identified as an AGN at $z<0.25$ \citep{paris2017,paris2018}, and (3) at least one radio detection in the Faint Images of the Radio Sky at Twenty Centimeters
(FIRST) survey \citep{becker1995,helfand2015}}. Using spatially-resolved, follow-up long slit optical spectroscopy, \citet{zhang2021a} demonstrated that J1017+3448 hosts an optical dual AGN system, wherein the constituents are a Type 1 broadline AGN and a Seyfert 2. In this work, we identified J1017+3448 as a dual mid-IR AGN candidate based on the presence of two WISE sources in the AllWISE catalog exhibiting mid-IR colors consistent with AGNs \citep{stern2012}. {$grz$ imaging of the host system for J1017+3448 is shown in the third row of Figure~\ref{fig:rank1_duals}}. The constituent AGNs are separated by only $\sim5.7''$, which is below the nominal WISE $W1$ and $W2$ resolutions and suggests that the mid-IR colors are contaminated. Nevertheless, the mid-IR emission in this system is visually extended in the WISE imaging -- overlapping the positions of both optical nuclei -- and these WISE sources were actively deblended by the WISE processing pipeline into two distinct sources (which enabled their inclusion in this sample). unWISE forced photometry \citep{lang2016} drawn from unblurred (native resolution) WISE imaging still yields two mid-IR AGNs with $W1-W2=0.912$ and $W1-W2=0.917$ for the NW and SE sources, respectively (versus $W1-W2=0.997$ and $W1-W2=0.870$ in AllWISE). Given the results from \citet[][a Type 1 AGN + Seyfert 2 pair, with which our own spectroscopic results agree]{zhang2021a} and the AllWISE and unWISE partially deblended photometry, we reconfirm this system as a dual optical AGN and we consider it a strong mid-IR dual AGN candidate and elevate it Rank 1. We leave a more thorough deblending and estimation of the distinct WISE colors of these nuclei for a future work, but the current evidence points to the presence of two mid-IR AGNs in this system.

\subsection{Comparison to Optical Spectroscopic Pairs from \citet{liu2011b}}

Given that contaminant removal and final verification for the present sample of mid-IR dual AGNs and candidates relied upon optical spectroscopy, it is natural to compare this sample against one where selection depended solely upon the availability of optical spectroscopic observations. \citet{liu2011b} conducted a broad search for optical spectroscopic AGN pairs within the SDSS DR7, requiring nuclear pair separations $\leq100$\,kpc, line-of-sight velocity differences $|\Delta v|\leq600$\,km\,s$^{-1}$, and optical BPT emission line ratios \citep{baldwin1981} consistent with either Seyferts, LINERs, or Composites, or emission line signatures consistent with narrow line quasars, broad line AGNs, and broad line quasars. In total, \citet{liu2011b} identified 1244 dual AGN candidates, 39 triple AGN candidates, two quadruple AGN candidates, and one quintuple AGN candidate. 

We first examined the fraction of mid-IR AGNs within the \citet{liu2011b} sample and checked for any potential overlap between the \citet{liu2011b} sample and the mid-IR dual AGN sample here. We limited the \citet{liu2011b} sample to pairs\footnote{Using the \citet{stern2012} cut, there is only a single mid-IR AGN among the optical triplets, and no mid-IR AGNs among the optical quadruplets or quintuplet. Using the \citet{assef2018} criteria, there are only three mid-IR AGNs among the optical triplets; no mid-IR duals were found among the triplets.} with angular separations $\geq5.5''$ (as adopted in Section~\ref{sec:selection}), which limits the 
\citet{liu2011b} sample to 1149 pairs. Across the 1149 pairs with $\theta>5.5''$, 42 pairs host at least one mid-IR AGN satisfying the {two band, magnitude limited color cut from} \citet{stern2012} and 68 pairs host at least one mid-IR AGN satisfying the {two band, magnitude dependent, 90\% reliability color cut from} \citet[][{see Equation 4 in Section 2 of that work}]{assef2018} . There are no spatially resolved mid-IR dual AGNs satisfying the \citet{stern2012} color cut within the optical pairs, but there are three pairs hosting spatially resolved mid-IR dual AGNs wherein each nucleus satisfies the 90\% reliability color criterion from \citet{assef2018} as well as the more stringent three-band color criteria defined by \citet{jarrett2011}: (1) J094554.41+423840.0 and J094554.49+423818.7 ($r_p=30.2$\,kpc), (2) J103853.29+392151.2 and J103855.95+392157.6 ($r_p=33.6$\,kpc), and (3) J131700.82+004505.7 and J131701.42+004509.7 ($r_p=24.1$\,kpc). These systems result in a spatially-resolved mid-IR dual AGN fraction of $\approx0.3\%$  within the $\theta>5.5''$ pairs sample from \citet{liu2011b}. In the optical, two of these dual AGNs host a broad line AGN and narrow line Seyfert AGN while the third dual AGN is a LINER-LINER pair.

While we find only three mid-IR dual AGNs coincident among the pairs sample from \citet[][where the mid-IR AGNs were coincident with the optical nuclei studied]{liu2011b}, at least one more mid-IR dual AGN lurks within the sample: J1218+1035, which is in fact simultaneously a dual mid-IR AGN and a triple radio AGN recently confirmed in \citet{schwartzman2025}, as discussed above in Section~\ref{sec:bigmac}. It is therefore possible that additional triple AGNs reside within the \citet{liu2011b} sample (besides the optically-selected candidates within that work), but detecting these elusive systems likely requires a more intensive multiwavelength approach. Any additional nuclei -- residing in the same system as a given optical pair -- lacking a fiber spectrum would not be recovered using optical fiber selection alone, and identification of these kinds of systems therefore demands follow-up optical observations and AGN assessments at non-optical wavelengths \citep[e.g.,][]{schwartzman2025}.

We can also make a preliminary comparison between the optical spectroscopic properties of this mid-IR dual AGN sample and the optical pairs from \citet{liu2011b}. As we demonstrated in Section~\ref{sec:bpt_analysis}, the nuclei among our Rank 1 and 0.5 mid-IR dual AGNs (where both nuclei have measured redshifts) exhibit a variety of optical BPT classes: 54.3\% show emission line ratios consistent with Seyferts, 16.0\% consistent with Composites, and 27.2\% consistent with HII regions, while very few (2.5\%) exhibit line ratios consistent with LINERs. This contrasts starkly with the $\theta>5.5''$ pairs from \citet{liu2011b}, where 9.1\% are Seyferts, 52.3\% are Composites, and 35.3\% are LINERs (and 3.2\% are broad line AGNs). A similar difference between the samples is found when examining the BPT pair combinations: among the spatially-resolved mid-IR dual AGNs with BPT classifications, Seyfert-Seyferts and Seyfert-HII nuclei represent the two dominant populations so far (35.5\% and 22.6\%, respectively); among the $\theta>5.5''$ optical pairs, Composite-Composite and LINER-Composite systems are the dominant populations (28.4\% and 34.6\%, respectively), with Seyfert-Seyferts representing only 0.70\% of the optical pair sample. An even larger disparity between the two samples is found when excluding star forming nuclei \citep[like in][]{liu2011b} from the mid-IR dual AGN sample: Seyfert-Seyferts and Seyfert-Composites then become the dominant populations (61.1\% and 27.8\%, respectively), while Composite-Composites make up $11.1\%$. Not only would all of our dual AGN candidates (and the individual nuclei) exhibiting HII-like line ratios be missed in an optical spectroscopic search like that in \citet{liu2011b}, our sample exhibits substantially lower fractions of LINERs and Composite nuclei and larger fractions of Seyfert nuclei. This could suggest that, by selecting in the mid-IR, we are simultaneously selecting intrinsically more powerful AGNs (potentially resulting in a higher fraction of Seyferts) as well as intrinsically more heavily obscured, optically elusive AGNs \citep[which could potentially explain a portion of the HII nuclei in the optical, where the host galaxy emission is dominating the spectroscopic emission lines, see also][]{satyapal2017,koss2018,pfeifle2019a}. The lower fraction of LINERs in our sample may also suggest that we are selecting systems less influenced by shocks \citep[e.g.,][]{allen2008}. A more thorough comparison  between the sample from \citet{liu2011b} and the mid-IR dual AGNs in this work will be presented in a follow-up study focusing on the complete optical spectroscopic coverage and properties of our sample.

\section{Conclusion} 
\label{sec:conclusion}

Dual AGNs are predicted to represent important sites of vigorous, obscured, merger-triggered SMBH growth across cosmic time \citep{blecha2018,capelo2017,derosa2019,chen2023,foord2024}. The properties and pairing rates of dual AGNs provide key expectations for the properties of gravitationally-bound and inspiralling binary SMBHs and, by extension, for SMBH binary-driven gravitational wave emission \citep{amaro-seoane2023,agazie2023b}. A variety of multiwavelength selection strategies and analyses are required for a statistically significant, holistic understanding of the dual AGN population \citep{derosa2019}. In this work, we have identified a sample of spatially-resolved mid-IR dual AGN candidates, where mid-IR AGN pairs were identified via the WISE two band mid-IR color criteria defined by \citet[][$W1-W2 \geq 0.8$, $W2<15$ mag]{stern2012}, and a visual morphological assessment of optical DESI LS imaging was used to search for signatures of galaxy mergers.  Here we summarize the mid-IR and {preliminary} optical spectroscopic results of our ongoing dual AGN investigation: 

\begin{itemize}
    \item We initially identified a sample of 147 spatially-resolved mid-IR dual AGN candidates {in galaxy merger candidates }using our mid-IR color and optical morphological selection strategy. An additional 8 mid-IR dual AGN candidates, originally classified as morphologically ambiguous, were {found to have consistent redshifts and separations consistent with interacting galaxies} based on archival spectroscopic measurements, bringing the sample size to 155. Spectroscopic redshift measurements have so far confirmed that 78/155 mid-IR dual AGN candidates {are hosted} in interacting systems where the constituent nuclei reside at the same redshift ($|\Delta|\leq700$ km s$^{-1}$) and have {projected} separations $r_p\leq130$ kpc. 13 mid-IR dual AGN candidates have been confirmed (``Rank 1'') as bona fide dual AGNs. {48/155 candidates have so far been identified as non-dual AGN contaminants.}  29 candidates still require optical spectroscopy for redshift verification. 

    \item {Our selection approach, which prioritized morphological identification of dual mid-IR AGNs in galaxy merger candidates and used spectroscopy only to verify redshifts and remove contaminants, boasts a success rate of $\sim59\%$ ($\sim40.6\%$ failure rate) in identifying dual mid-IR AGN candidates in interacting galaxies. Had DESI DR1 spectroscopic observations of both nuclei in each system been required for initial selection, 55/70 ($\approx78.5$\%) of the morphologically identified -- and redshift confirmed -- mergers would have been missed; 9/13, or $\approx69\%$, of confirmed (Rank 1) dual AGNs in this sample would also have been missed by extension.} 
    \item Among the 78 Rank 1 and 0.5 mid-IR dual AGN candidates where each nucleus has a measured spectroscopic redshift, 31 (or $39.7\%$) have at least one additional companion within $r_p\leq130$\,kpc and $|\Delta v|\leq700$ km s$^{-1}$ based on archival or follow-up optical spectroscopic observations. {31\% of Rank 1 (confirmed only) systems have at least one companion.} These systems are therefore considered multi-mergers. 

    \item Preliminary results based on the available data show that these mid-IR dual AGN candidates exhibit a wide variety of BPT optical types. Among the 78 Rank 1 and 0.5 mid-IR dual AGN candidates in this sample, optical spectra for 81/156 nuclei were available and sufficient to classify the BPT types. Among these 81 nuclei, 54.3\% are optically classified as Seyferts, 2.5\% as LINERs, 16.0\% as composite nuclei, and 27.2\% as NII nuclei based on the [N II] BPT diagram. Further analyses are needed to confirm {any} narrow line, non-Seyfert mid-IR AGN candidates as bona fide AGNs; nonetheless, if a significant portion of non-Seyferts here are indeed driven by AGNs, then optically-elusive AGNs represent a substantial  portion of this population. 
    \item {Among the 31 Rank 1 and 0.5 dual AGN candidates where BPT line ratios could be computed for both nuclei, Seyfert-Seyferts and Seyfert-HII nuclei BPT pairings make up 35.5\% and 22.6\%}, respectively; the dominance of these two BPT-pairings may suggest preferential selection of simultaneously powerful AGNs as well as heavily dust reddened, optically elusive AGNs where the hosts dominate the optical emission. 

    \item Approximately $30-36\%$ of the variation in the nuclear WISE $W2-W3$ colors {may} be attributed to the $W1$ and $W2$ luminosities, while only $6-11\%$ of the variation in the $W1-W2$ colors {may be} attributed to the $W2$ luminosity, based on the derived Pearson $r$ correlation coefficients. Similarly, $7-15\%$ of the variation in $W1$, $W2$, and $W3$ luminosity {may} be attributed to the projected separations. These percentages depend upon the specific WISE band, redshift requirements, and treatment of the nuclei (individual or combined).
\end{itemize}

This is the first search for dual AGNs to rely upon a selection strategy involving both spatially resolved mid-IR color selection and morphological evidence of interactions. {Despite its angular resolution often being considered a weakness,} we have {paired the mid-IR imaging of }WISE with the depth and coverage of DESI LS to specifically target dual mid-IR AGNs in early-to-intermediate merger stages. In doing so, this program focuses on dual AGN population(s) that are often underrepresented in the literature. Though this is not the first study to focus on non-optical dual AGN selection strategies \citep[e.g.,][]{koss2012,imanishi2014,fu2015a,satyapal2017,pfeifle2019a,kim2019,glikman2023,barrows2023} nor the first to utilize mid-IR colors to identify dual AGN candidates \citep{satyapal2017,ellison2017,pfeifle2019a,glikman2023,barrows2023}, the diverse spectroscopic properties and environments identified in this sample thus far highlight the strength and importance of multiwavelength strategies in the hunt for populations of dual AGNs across the merger sequence.

Exhaustive multiwavelength follow-up observations are underway to confirm and characterize this sample across the radio, optical, and X-ray regimes. For example, in the first work of this series, \citet{schwartzman2025} confirmed the first triple radio AGN using follow-up VLA observations of one of the mid-IR dual AGNs identified in this work (J1218+1035), and in this work we used mid-IR colors in concert with BPT emission line ratios and broad emission lines to confirm 12 dual mid-IR AGNs (not including J1218+1035). In a series of follow-up works, we {intend to study} (1) the radio properties of this sample using VLASS epoch 1-3 imaging, (2) the complete optical spectroscopic {properties} for the sample using archival and follow-up observations from ongoing ground-based programs, and (3) the X-ray properties of a small subsample of these dual AGNs using Chandra and XMM-Newton.

\begin{acknowledgements}
R. W. P. gratefully acknowledges support through an appointment to the NASA Postdoctoral Program at Goddard Space Flight Center, administered by ORAU through a contract with NASA. We thank Riley Decolibus, Sujeoung Lee, and Maria Tsedrik for assisting with the September 2024 Palomar observations. We thank Scott Barrows for sharing his catalog of probabilistic WISE AGN pairs with us to check for any potential overlap in sources with the sample in this work.

This publication makes use of data products from the Wide-field Infrared Survey Explorer, which is a joint project of the University of California, Los Angeles, and the Jet Propulsion Laboratory/California Institute of Technology, funded by the National Aeronautics and Space Administration. 

This publication was supported by observations obtained at the international Gemini Observatory (Program ID GN-2025A-Q-306, GN-2025B-Q-405, GS-2025A-Q-407), a program of NSF NOIRLab, which is managed by the Association of Universities for Research in Astronomy (AURA) under a cooperative agreement with the U.S. National Science Foundation on behalf of the Gemini Observatory partnership: the U.S. National Science Foundation (United States), National Research Council (Canada), Agencia Nacional de Investigación y Desarrollo (Chile), Ministerio de Ciencia, Tecnología e Innovación (Argentina), Ministério da Ciência, Tecnologia, Inovações e Comunicações (Brazil), and Korea Astronomy and Space Science Institute (Republic of Korea). 

This publication was supported by observations obtained at Keck Observatory, which is a private 501(c)3 non-profit organization operated as a scientific partnership among the California Institute of Technology, the University of California, and the National Aeronautics and Space Administration. The Observatory was made possible by the generous financial support of the W. M. Keck Foundation. The authors wish to recognize and acknowledge the very significant cultural role and reverence that the summit of Maunakea has always had within the Native Hawaiian community. We are most fortunate to have the opportunity to conduct observations from this mountain.

We acknowledge that many discoveries and analyses of confirmed and candidate multi-AGNs have relied upon telescopes and facilities built upon lands considered sacred or culturally important to historically marginalized, underrepresented, and disregarded communities; we thank these communities for the privilege of observing on grounds considered sacred, and we urge the scientific community to not only acknowledge these communities in future works, but to specifically bear in mind and engage with these communities when designing and planning future observatories and facilities. 

\end{acknowledgements}


\appendix 

\section{Rank 0 Mid-IR Dual AGN Candidates}
\label{app:rank0_duals}

\begin{figure*}[p]
\centering
    \subfloat{\includegraphics[width=0.2\linewidth]{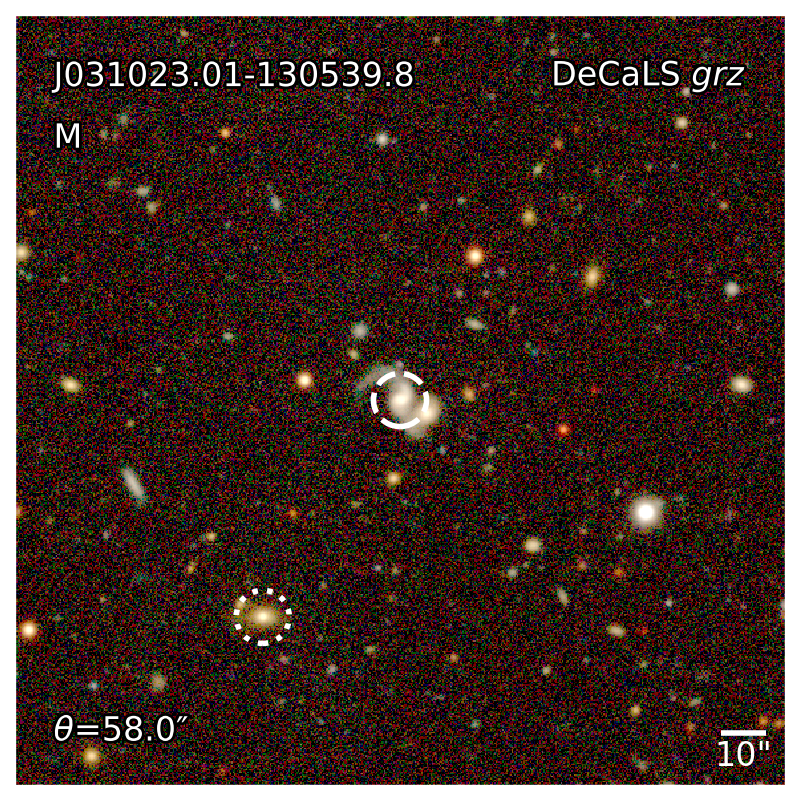}}
    \subfloat{\includegraphics[width=0.2\linewidth]{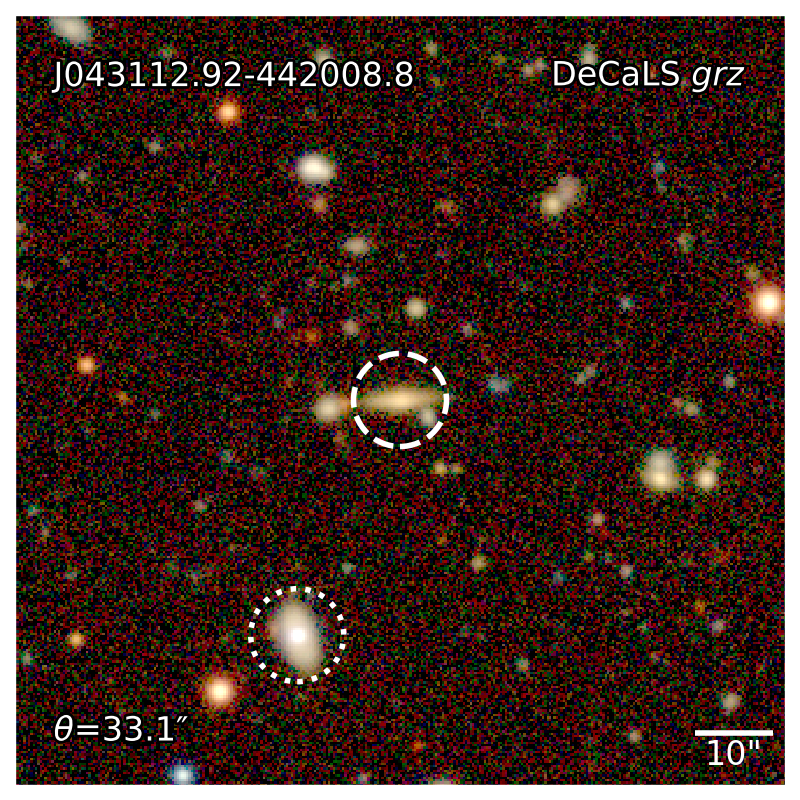}}
    \subfloat{\includegraphics[width=0.2\linewidth]{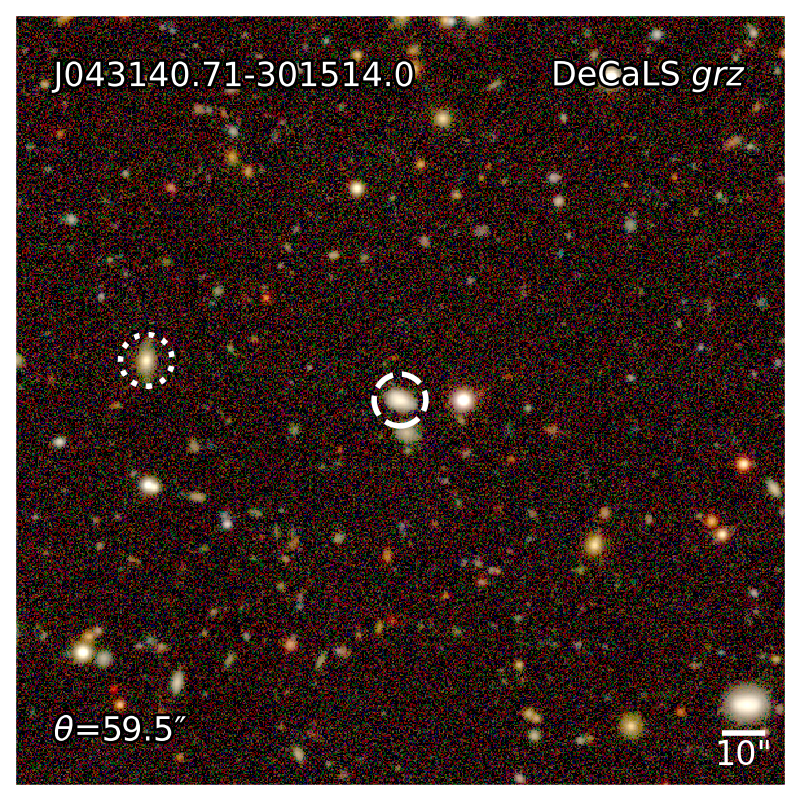}}
    \subfloat{\includegraphics[width=0.2\linewidth]{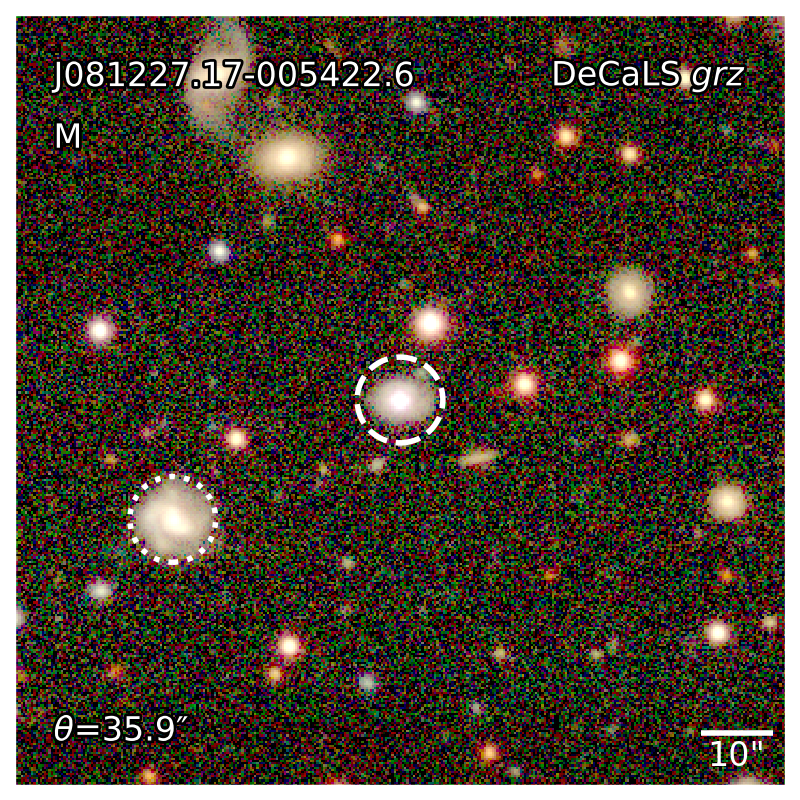}}
    \subfloat{\includegraphics[width=0.2\linewidth]{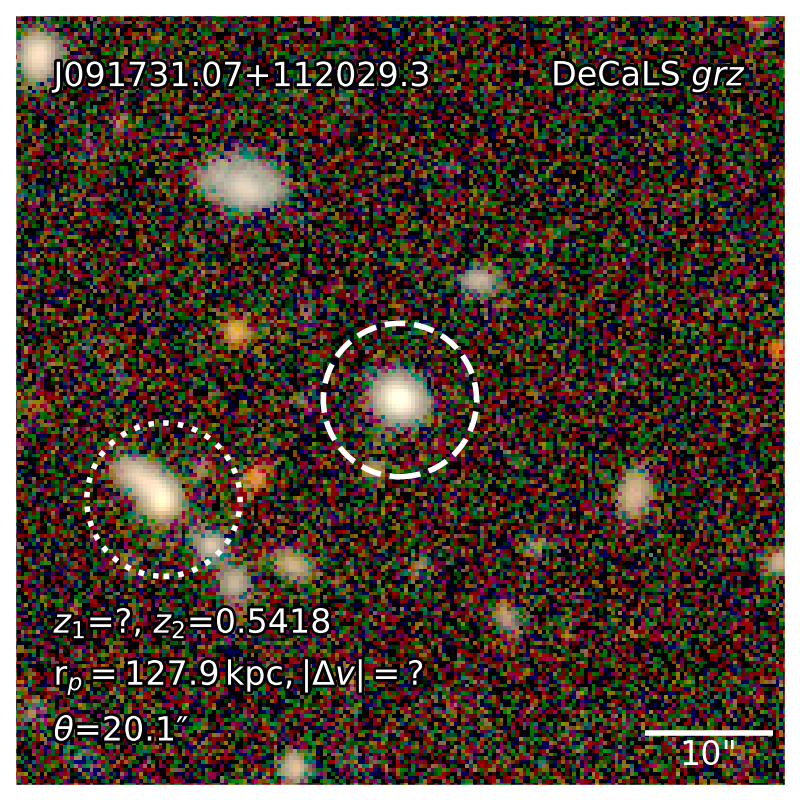}}\\
    \vspace{-4.5mm}
    \subfloat{\includegraphics[width=0.2\linewidth]{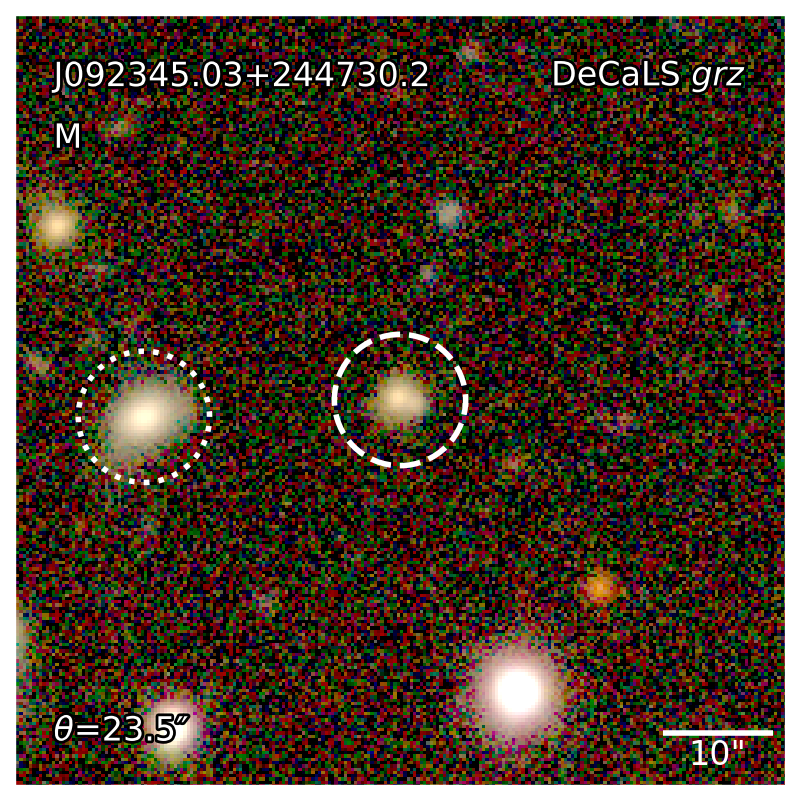}}
    \subfloat{\includegraphics[width=0.2\linewidth]{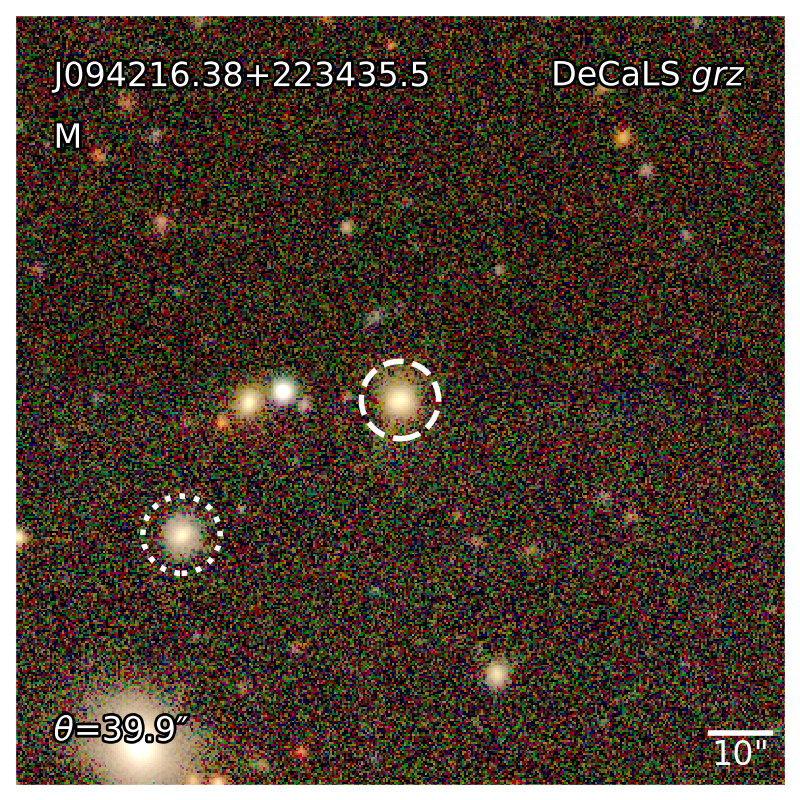}}
    \subfloat{\includegraphics[width=0.2\linewidth]{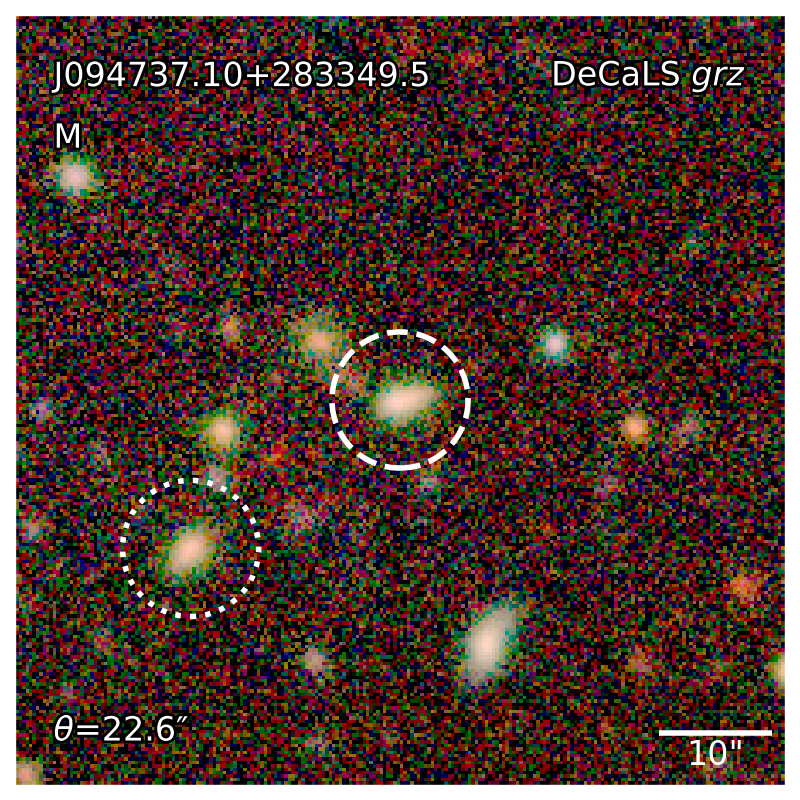}}
    \subfloat{\includegraphics[width=0.2\linewidth]{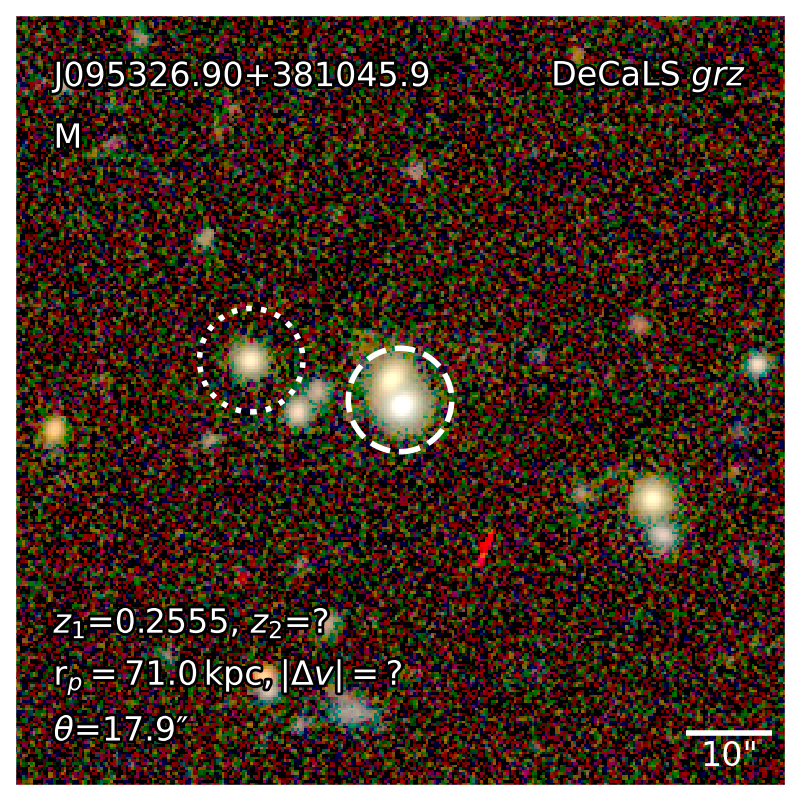}}
    \subfloat{\includegraphics[width=0.2\linewidth]{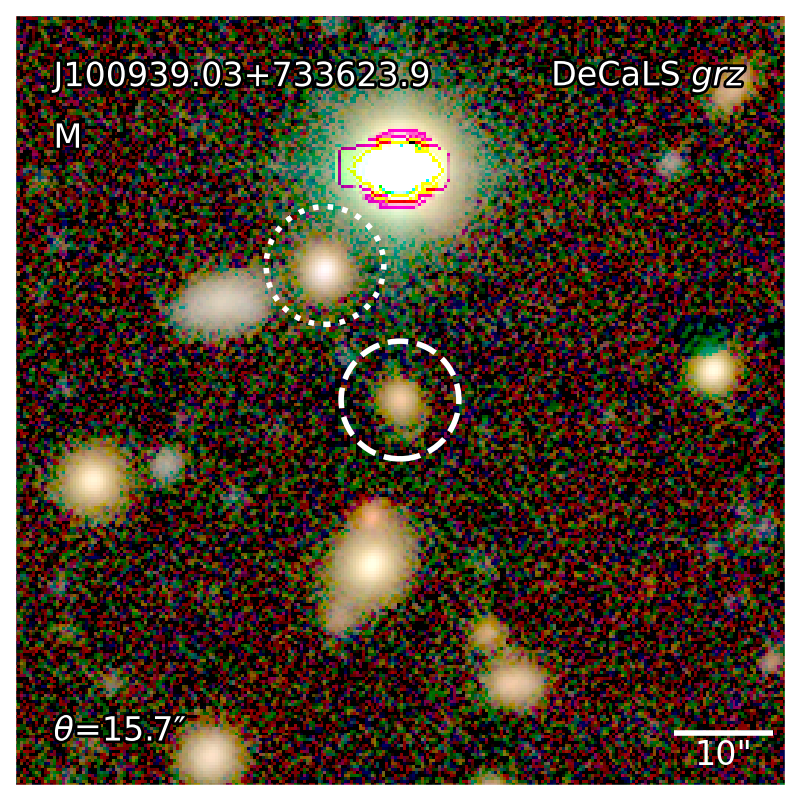}}\\
    \vspace{-4.5mm}
    \subfloat{\includegraphics[width=0.2\linewidth]{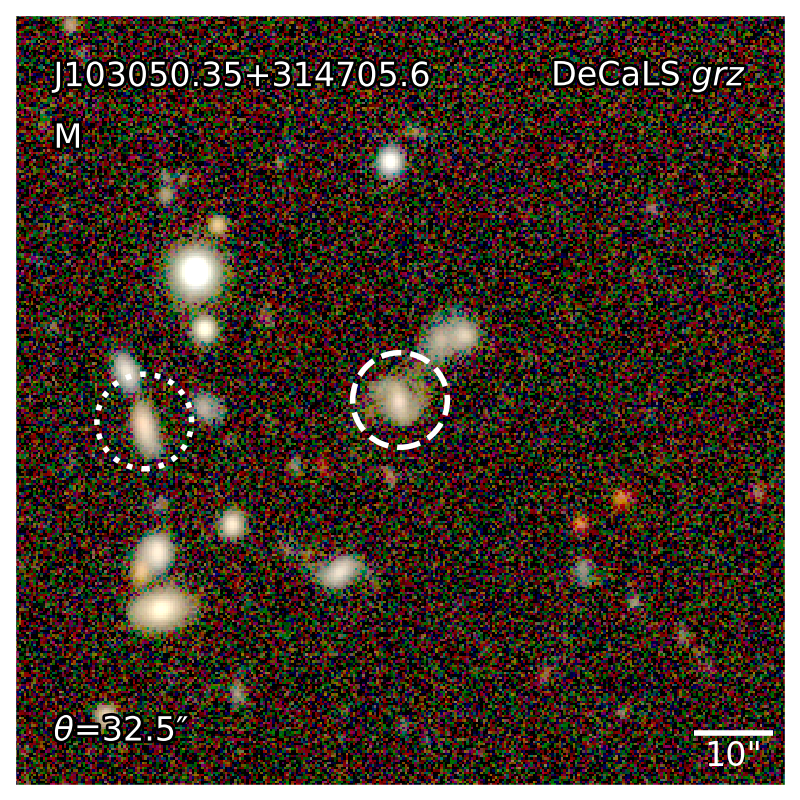}}
    \subfloat{\includegraphics[width=0.2\linewidth]{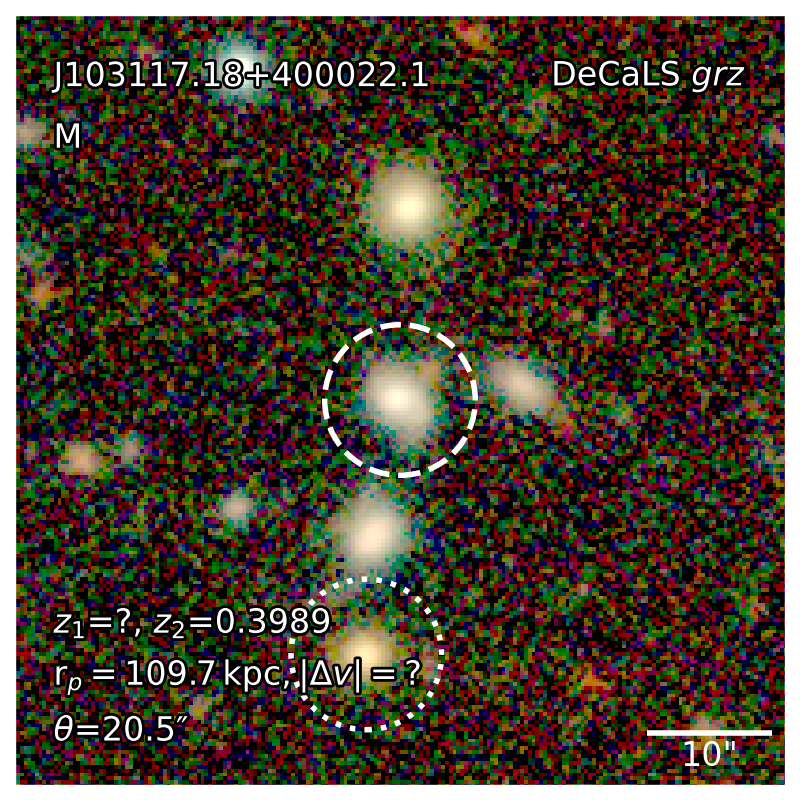}}
    \subfloat{\includegraphics[width=0.2\linewidth]{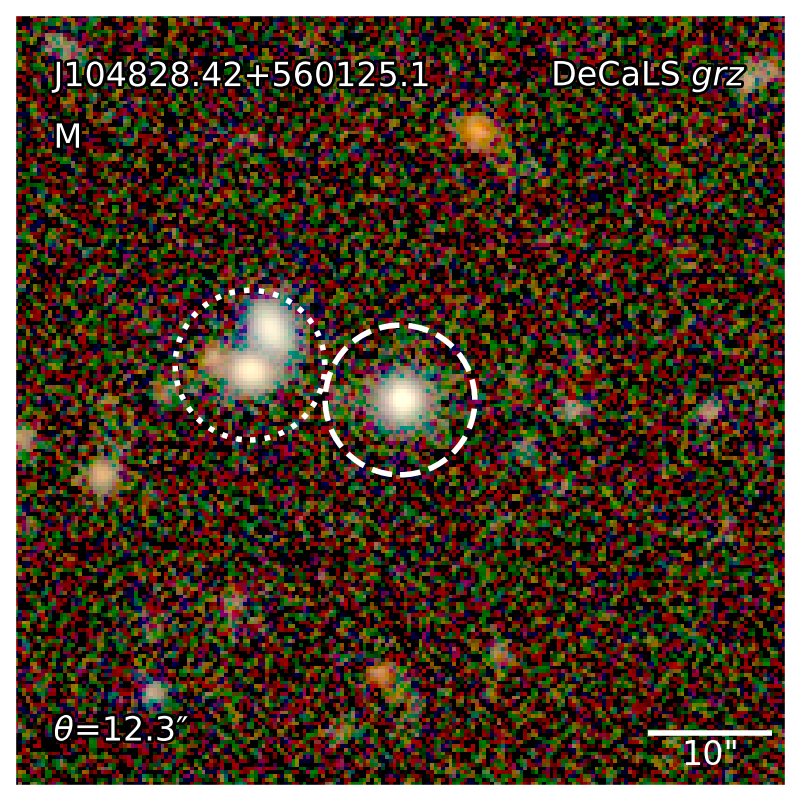}}
    \subfloat{\includegraphics[width=0.2\linewidth]{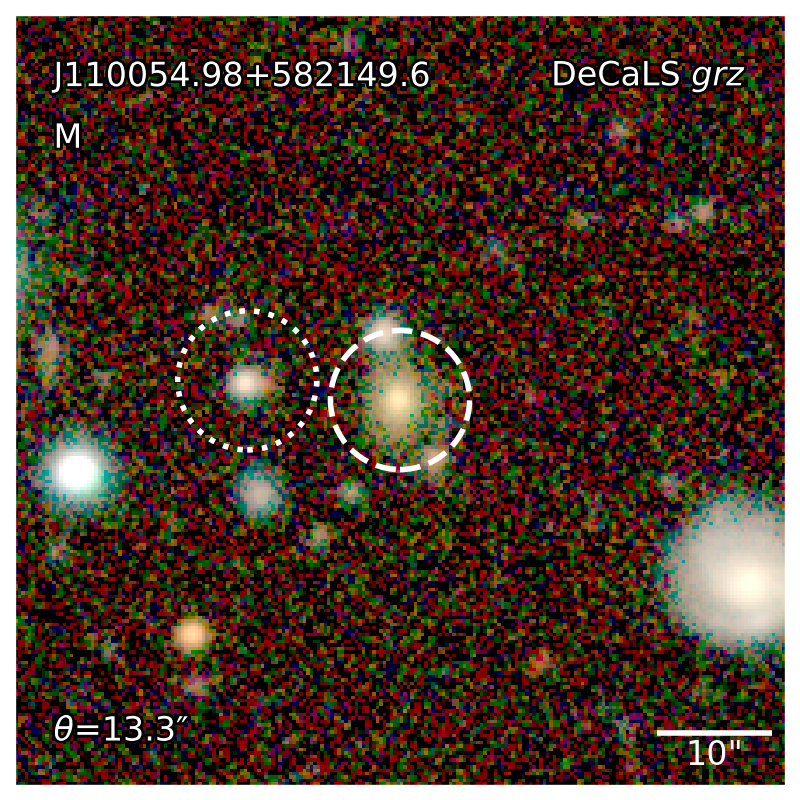}}    \subfloat{\includegraphics[width=0.2\linewidth]{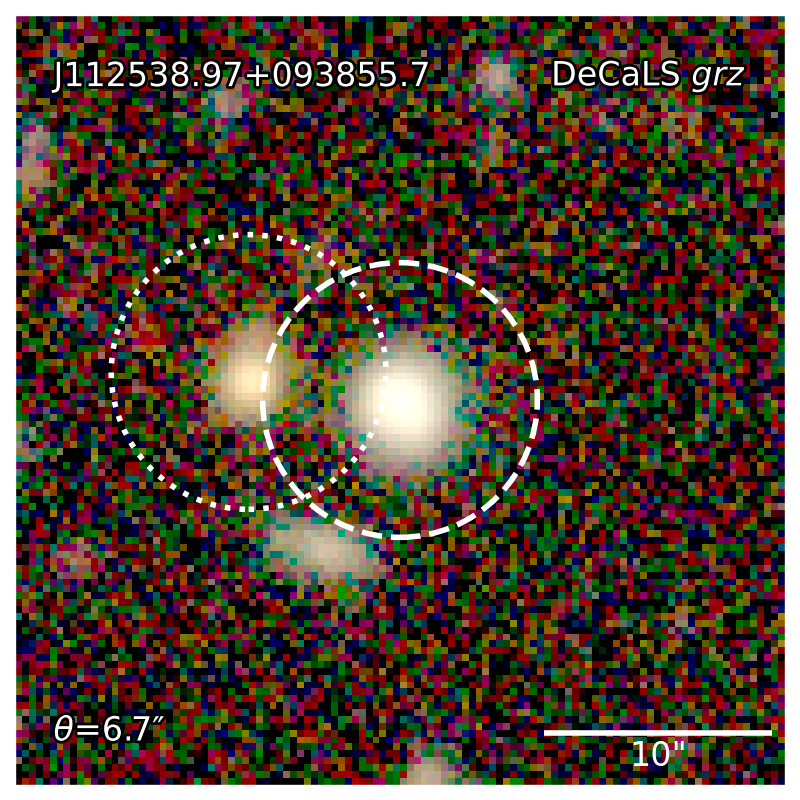}}\\
    \vspace{-4.5mm}
    \subfloat{\includegraphics[width=0.2\linewidth]{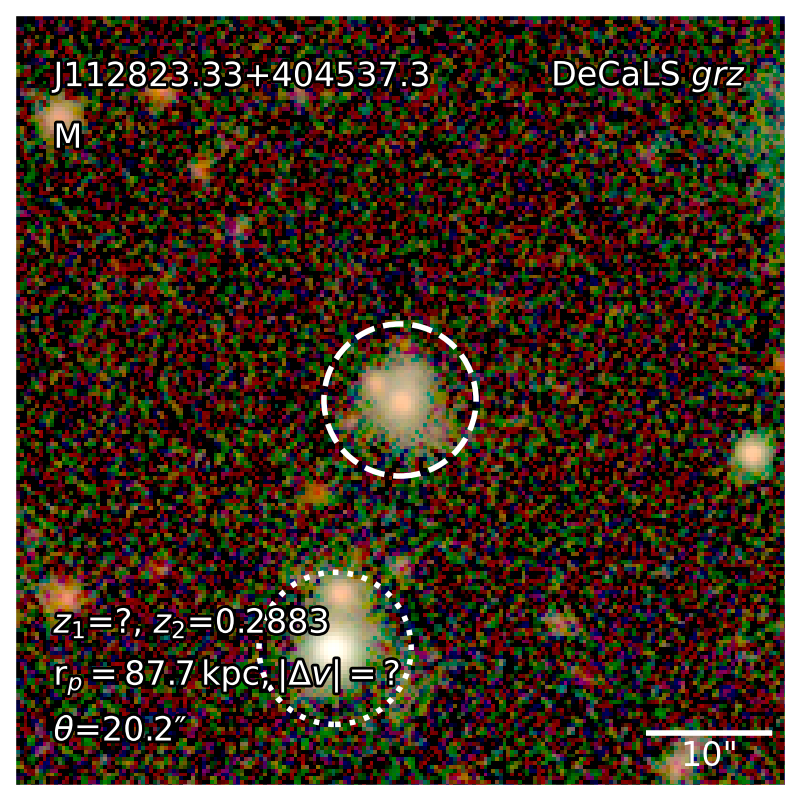}}
    \subfloat{\includegraphics[width=0.2\linewidth]{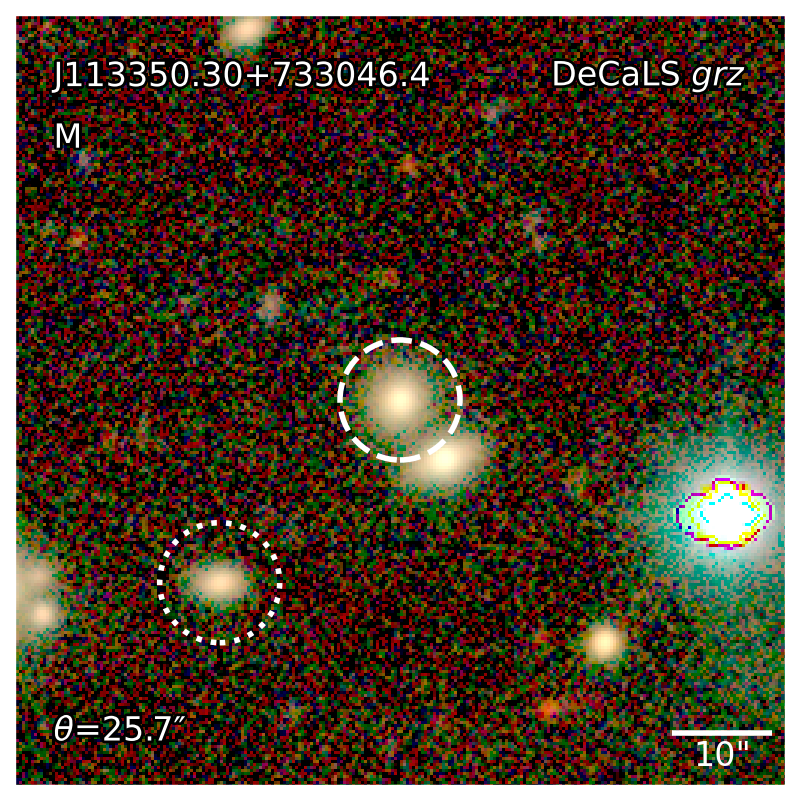}}
    \subfloat{\includegraphics[width=0.2\linewidth]{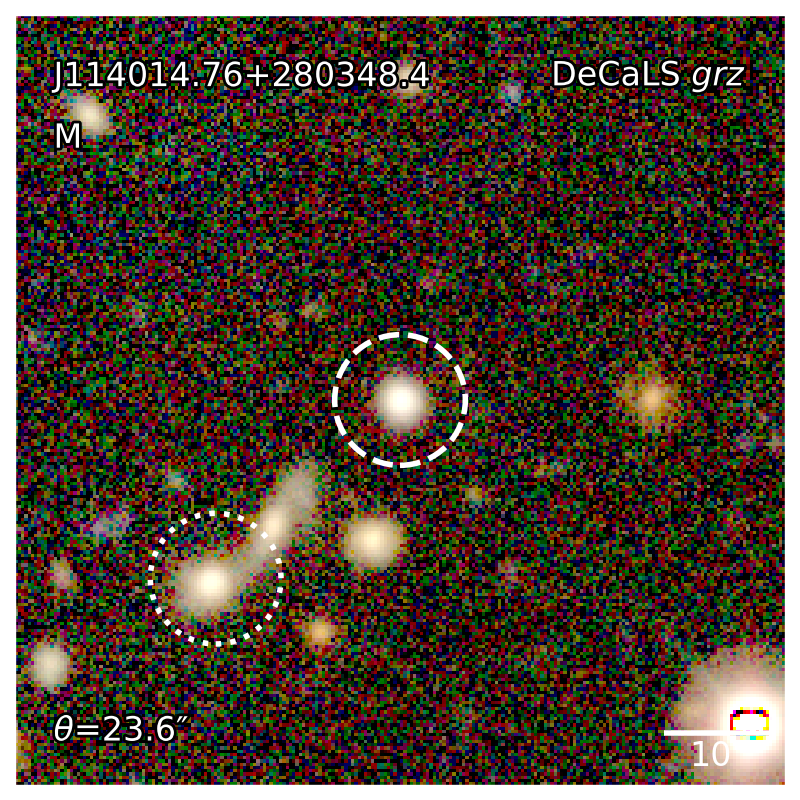}}
    \subfloat{\includegraphics[width=0.2\linewidth]{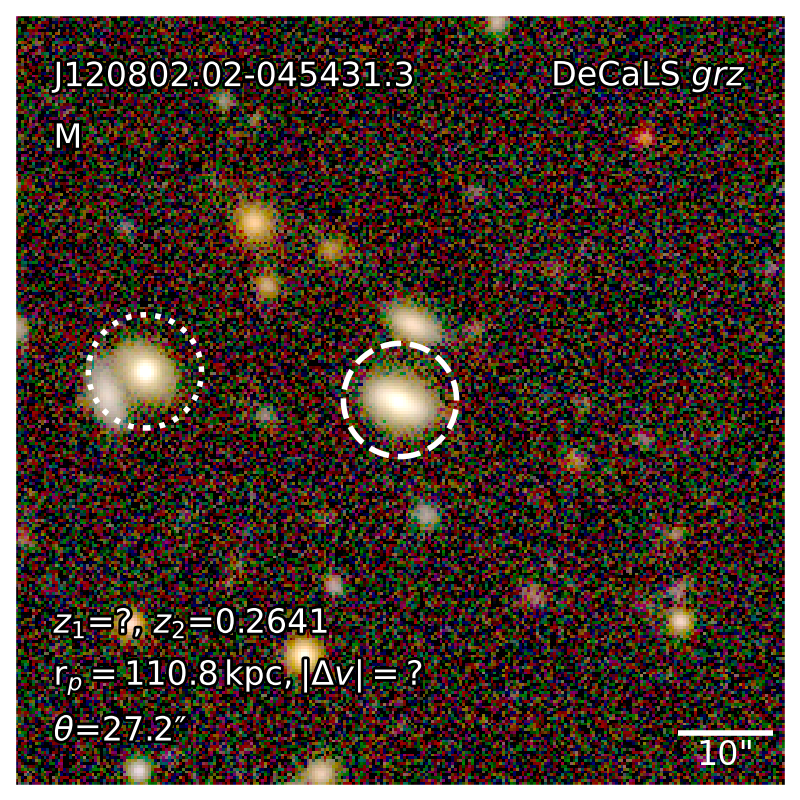}}
    \subfloat{\includegraphics[width=0.2\linewidth]{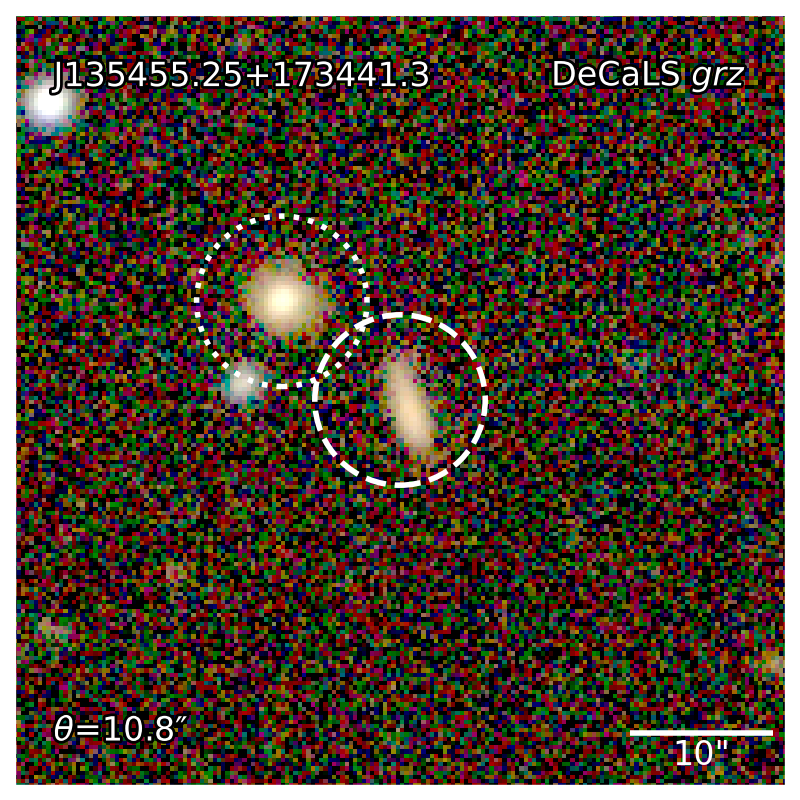}}\\
    \vspace{-4mm}
    \subfloat{\includegraphics[width=0.2\linewidth]{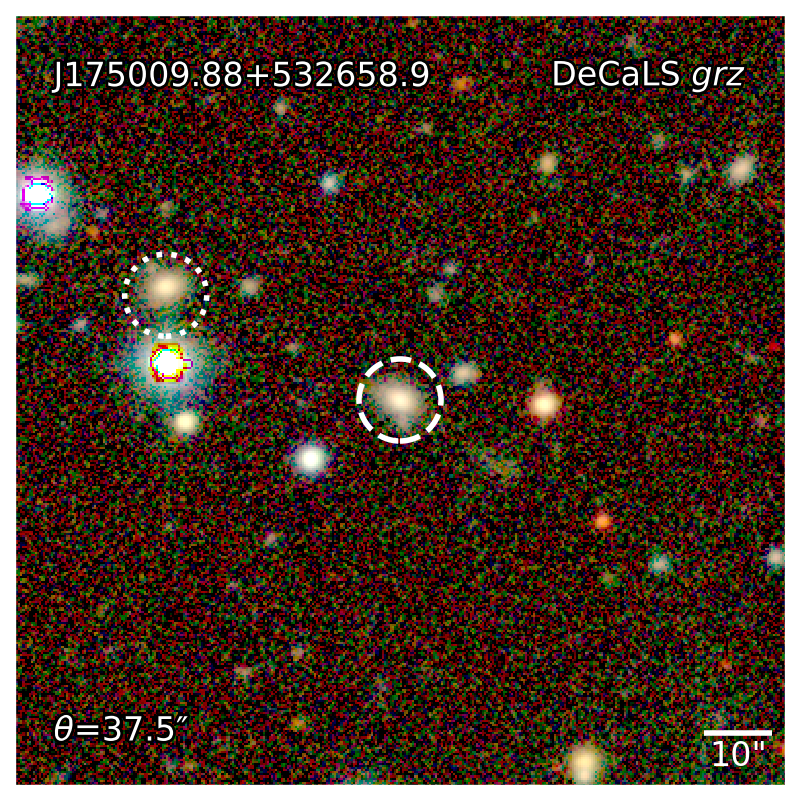}}
    \subfloat{\includegraphics[width=0.2\linewidth]{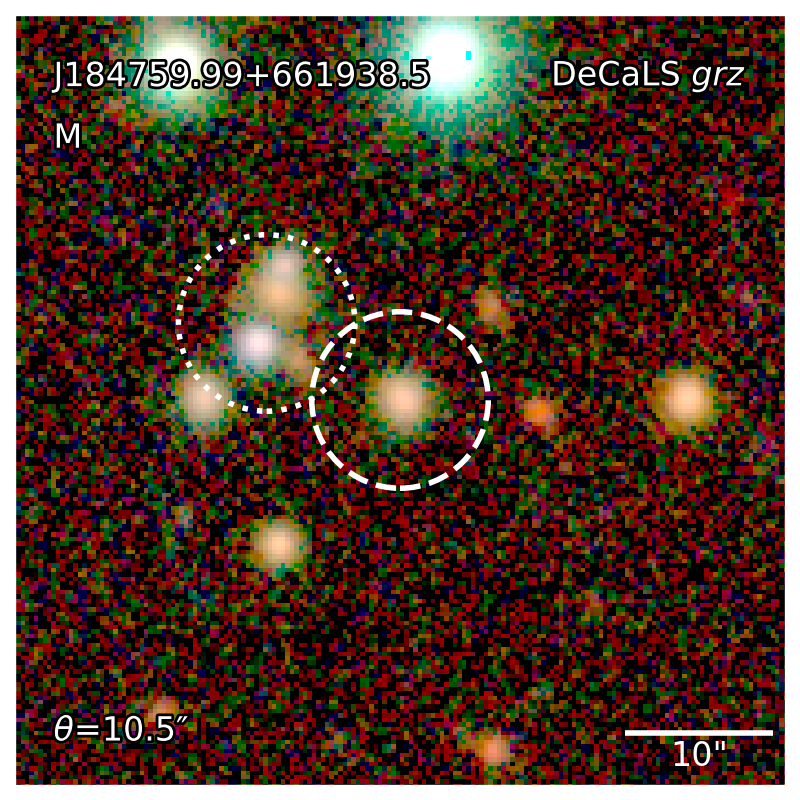}}
    \caption{Rank 0 mid-IR dual AGN candidates. Each panel is constructed in an identical fashion to those in Figures~\ref{fig:rank1_duals} and \ref{fig:rank0.5_duals}.}
    \label{fig:rank0_duals}
\end{figure*}

In Figure~\ref{fig:rank0_duals}, we show the remaining Rank 0 mid-IR dual AGN candidates in our sample. These candidates still require spectroscopic follow-up in order to ensure the nuclei in a given pair reside at the same redshift, with separations and velocity differences consistent with interacting galaxy pairs.

\begin{table*}
\begin{center}
\caption{Rank 0 Spatially-Resolved Mid-IR Dual AGN Candidates }
\label{tab:rank0}
\begin{tabular}{ccccccccccc}
\hline
\hline
\noalign{\smallskip}
\noalign{\smallskip}
$\rm{designation}_1$ &  $z_1$ & $z_1$ source & W1-W2$_1$ &  $\rm{designation}_2$ &  $z_2$ & $z_2$ source & W1-W2$_2$ & $\theta_{\rm{MIR}}$ &  $r_{p\,{\rm{(MIR)}}}$ &  $|\Delta v|$ \\
 &  &  &  &  &  &  &  & ('') & (kpc) & km s$^{-1}$\\
(1) & (2) & (3) & (4) & (5) & (6) & (7) & (8) & (9) & (10) & (11)  \\
\noalign{\smallskip}
\noalign{\smallskip}
\hline
\noalign{\smallskip}
J031023.01-130539.8 & \dots &  \dots & 1.31 & J031025.13-130628.8 &   \dots &  \dots & 1.22 & 58.0 & \dots & \dots \\
J043112.92-442008.8 & \dots &  \dots & 0.88 & J043114.16-442039.1 &   \dots &  \dots & 0.86 & 33.1 & \dots & \dots \\
J043140.71-301514.0 & \dots &  \dots & 0.83 & J043145.24-301504.8 &   \dots &  \dots & 1.41 & 59.5 & \dots & \dots \\
J081227.17-005422.6 & \dots &  \dots & 1.16 & J081229.29-005439.3 &   \dots &  \dots & 0.84 & 35.9 & \dots & \dots \\
J091731.07+112029.3 & \dots &  \dots & 0.87 & J091732.33+112021.5 & 0.542 & D & 1.24 & 20.1 & 127.9 & \dots \\
J092345.03+244730.2 & \dots &  \dots & 1.13 & J092346.75+244728.6 &   \dots &  \dots & 0.95 & 23.5 & \dots & \dots \\
J094216.38+223435.5 & \dots &  \dots & 0.83 & J094218.83+223414.6 &   \dots &  \dots & 1.10 & 39.9 & \dots & \dots \\
J094737.10+283349.5 & \dots &  \dots & 0.86 & J094738.50+283336.5 &   \dots &  \dots & 0.89 & 22.6 & \dots & \dots \\
J095326.90+381045.9 & 0.256 & S & 0.85 & J095328.36+381050.5 &   \dots &  \dots & 0.89 & 17.9 &  71.0 & \dots \\
J100939.03+733623.9 & \dots &  \dots & 0.99 & J100940.84+733637.6 &   \dots &  \dots & 0.97 & 15.7 & \dots & \dots \\
J103050.35+314705.6 & \dots &  \dots & 1.14 & J103052.89+314702.9 &   \dots &  \dots & 0.89 & 32.5 & \dots & \dots \\
J103117.18+400022.1 & \dots &  \dots & 0.86 & J103117.41+400001.8 & 0.399 & D & 1.12 & 20.5 & 109.7 & \dots \\
J104828.42+560125.1$^{\dagger}$ & \dots &  \dots & 0.86 & J104829.86+560127.9$^{\dagger}$ &   \dots &  \dots & 0.82 & 12.3 & \dots & \dots \\
J110054.98+582149.6 & \dots &  \dots & 0.90 & J110056.66+582151.3 &   \dots &  \dots & 1.25 & 13.3 & \dots & \dots \\
J112538.97+093855.7 & \dots &  \dots & 0.95 & J112539.41+093856.9 &   \dots &  \dots & 0.94 &  6.7 & \dots & \dots \\
J112823.33+404537.3 & \dots &  \dots & 0.81 & J112823.78+404517.7 & 0.288 & S & 1.04 & 20.2 &  87.7 & \dots \\
J113350.30+733046.4 & \dots &  \dots & 0.98 & J113354.54+733028.1 &   \dots &  \dots & 1.17 & 25.7 & \dots & \dots \\
J114014.76+280348.4$^{\dagger}$ & \dots &  \dots & 0.85 & J114016.04+280331.9$^{\dagger}$ &   \dots &  \dots & 0.87 & 23.6 & \dots & \dots \\
J120802.02-045431.3 & \dots &  \dots & 1.39 & J120803.83-045428.2 & 0.264 & D & 0.84 & 27.2 & 110.8 & \dots \\
J135455.25+173441.3 & \dots &  \dots & 0.81 & J135455.83+173448.3 &   \dots &  \dots & 1.47 & 10.8 & \dots & \dots \\
J175009.88+532658.9 & \dots &  \dots & 0.80 & J175013.71+532714.3 &   \dots &  \dots & 0.80 & 37.5 & \dots & \dots \\
J184759.99+661938.5 & \dots &  \dots & 0.84 & J184801.50+661943.8 &   \dots &  \dots & 0.90 & 10.5 & \dots & \dots \\

\noalign{\smallskip}
\hline
\end{tabular}
\end{center}
\tablecomments{Rank 0 spatially-resolved mid-IR dual AGN candidates. This table follows an identical column structure to that given in Tables~\ref{tab:rank1} and \ref{tab:rank0.5}. {$^{\dagger}$: Systems identified as spatially resolved mid-IR dual AGN candidates in \citet{barrows2023}.}
}
\end{table*}

\section{Rejected Mid-IR Dual AGN Candidates}
\label{app:rejected_candidates}

In Figure~\ref{fig:rejects} we show the DESI LS tricolor images of Rank -1 (rejected) dual mid-IR AGN candidates described in Section~\ref{sec:identifiedcontaminants}. Table~\ref{tab:midIR_rejects} lists the details for these rejected candidates, including names, redshifts, separations, and velocity differences.

\begin{table*}
\begin{center}
\caption{Spectroscopically Rejected Mid-IR Dual AGN Candidates}
\label{tab:midIR_rejects}
\begin{tabular}{ccccccccccc}
\hline
\hline
\noalign{\smallskip}
\noalign{\smallskip}
$\rm{designation}_1$ &  $z_1$ & $z_1$ source & W1-W2$_1$ &  $\rm{designation}_2$ &  $z_2$ & $z_2$ source & W1-W2$_2$ & $\theta_{\rm{MIR}}$ &  $r_{p\,{\rm{(MIR)}}}$ &  $|\Delta v|$ \\
 &  &  &  &  &  &  &  & ('') & (kpc) & km s$^{-1}$\\
(1) & (2) & (3) & (4) & (5) & (6) & (7) & (8) & (9) & (10) & (11)  \\
\noalign{\smallskip}
\noalign{\smallskip}
\hline

J000046.83-111548.1 & 0.171 &                SR & 0.83 & J000050.33-111549.7 & 0.259 &     SR & 0.83 & 51.5 & 206.7 & 21674.5 \\
J000729.21-324331.8 &   \dots &                 \dots & 0.82 & J000732.88-324315.4 & 0.286 & NED & 1.08 & 49.1 & 211.3 &      \dots \\
J002551.33+041731.2 & 0.344 &                D & 0.90 & J002554.16+041801.2 &   \dots &      \dots & 0.83 & 51.8 & 253.0 &      \dots \\
J010125.63+144726.9 & 1.356 &                S & 1.55 & J010125.82+144736.5 & 0.291 &   G & 0.95 &  9.9 &  83.7 & 160904.3 \\
J011902.13-630627.8 & 0.078 &                SR & 0.93 & J011904.35-630607.5 & 0.235 &     SR & 0.83 & 25.3 &  94.4 & 40342.6 \\
J012040.06+023436.3 & 0.312 &              G & 1.02 & J012040.94+023434.8 & 0.326 &   G & 1.13 & 13.2 &  62.3 &  3095.7 \\
J012148.49-404856.9 & 0.422 &                SR & 1.41 & J012149.56-404858.6 & 0.214 &     SR & 1.19 & 12.3 &  68.0 &  46817.6 \\
J012719.43-352145.7 & 0.171 &              G & 0.93 & J012719.50-352123.2 & 0.279 &   G & 0.83 & 22.5 &  95.2 & 26448.3 \\
J020303.14-074152.0 & 0.067 & P16 & 0.97 & J020303.92-074149.7 & 0.210 &     GA & 0.84 & 11.9 &  40.9 & 37417.7 \\
J021109.70-151248.1 & 0.340 &              G & 0.94 & J021109.87-151328.8 & 0.181 &   G & 0.90 & 40.8 & 197.6 &  37440.5 \\
J023041.83-573056.5 & 0.374 &              G & 0.86 & J023044.39-573101.6 & 0.266 &   G & 1.32 & 21.3 & 109.5 &  24326.2 \\
J023437.17-210251.6 & 0.210 &              G & 0.86 & J023438.53-210339.1 & 0.306 &   G & 1.12 & 51.3 & 231.5 & 22649.2 \\
J032347.41-474433.0 & 0.239 &  \dots & 0.99 & J032352.23-474505.1 &   \dots &  \dots & 0.97 & 58.2 & \dots & \dots \\
J035448.34-505623.6 & 0.239 &  \dots & 0.83 & J035450.39-505612.9 &   \dots &  \dots & 1.28 & 22.1 & \dots & \dots \\
J052938.08-335921.1 & 0.274 &                SR & 1.10 & J052940.52-335918.2 & 0.285 &     SR & 0.92 & 30.4 & 130.8 &  2637.0 \\
J061044.30-433751.8 & 0.427 &                SR & 1.14 & J061046.17-433748.4 & 0.182 &     SR & 0.90 & 20.6 & 115.4 &  55702.9 \\
J073639.88+254222.1 & 0.172 &                D & 0.81 & J073642.26+254246.8 & 0.383 &     D & 0.80 & 40.5 & 211.7 & 48879.7 \\
J081117.66+072027.3 & 0.238 &                D & 0.99 & J081120.00+072033.2 & 0.287 &     D & 1.04 & 35.3 & 152.5 & 11528.6 \\
J090123.60-011806.2 & 0.247 &                D & 1.46 & J090124.37-011715.6 & 0.178 &     D & 0.89 & 51.8 & 200.8 &  17037.5 \\
J091706.37+744200.8 & 0.299 &                D & 0.88 & J091718.86+744208.8 &   \dots &      \dots & 0.85 & 50.1 & 222.5 &      \dots \\
J092751.41+072940.9 &   \dots &                 \dots & 0.95 & J092752.20+073014.8 & 0.255 &     D & 1.07 & 35.9 & 142.5 &      \dots \\
J094128.78+033431.7 & 0.290 &                D & 1.01 & J094131.08+033426.3 &   \dots &      \dots & 0.86 & 34.9 & 151.8 &      \dots \\
J101847.67+135710.3 & 0.198 &              G & 0.94 & J101850.23+135706.5 & 0.209 &   G & 0.84 & 37.5 & 127.8 &  2762.8 \\
J103723.97+290221.2 &   \dots &                 \dots & 0.94 & J103725.36+290153.0 & 0.444 &     D & 0.91 & 33.5 & 191.4 &      \dots \\
J110314.64+392352.0$^{\dagger}$ & 0.154 &                S & 0.97 & J110317.56+392340.8$^{\dagger}$ & 0.219 &     D & 1.08 & 35.6 & 126.1 & 16384.1 \\
J111232.99+290532.4 &   \dots &                 \dots & 0.84 & J111235.61+290600.7 & 0.323 &     D & 1.20 & 44.5 & 208.4 &      \dots \\
J112321.75+005734.7 & 0.101 &              2d & 0.85 & J112323.91+005730.3 & 0.201 &     D & 1.20 & 32.7 & 108.2 & 25849.3 \\
J112925.88+154628.1 & 0.324 &              G & 0.85 & J112926.88+154626.0 & 0.200 &   G & 1.12 & 14.7 &  68.9 &  29308.0 \\
J121334.66+325615.9 & 0.222 &                S & 0.87 & J121336.02+325608.0 & 0.169 &     D & 1.19 & 18.8 &  67.2 &  13275.9 \\
J124605.35+071128.2 & 2.085 & M12 & 1.27 & J124608.75+071156.5 &   \dots &      \dots & 0.87 & 57.9 & 482.1 &      \dots \\
J124702.51+402410.3$^{\dagger}$ & 0.724 &                D & 0.97 & J124703.60+402415.1$^{\dagger}$ & 0.720 &     S & 0.98 & 13.3 &  96.6 &    659.5 \\
J131823.22+061256.4 & 0.442 &                S & 1.07 & J131824.38+061229.4 &   \dots &      \dots & 0.84 & 32.1 & 182.9 &      \dots \\
J142421.44-050102.1 & 0.234 &                D & 0.82 & J142423.92-050111.2 & 0.236 &     D & 0.82 & 38.1 & 142.9 &   459.6 \\
J145303.50+015544.4 & 0.220 &                GA & 0.82 & J145305.80+015501.3 & 0.098 &     S & 1.45 & 55.2 & 195.8 &  31209.8 \\
J160735.51+064706.4 & 0.330 &                D & 0.82 & J160735.66+064716.6 & 0.193 &     D & 1.32 & 10.4 &  49.6 &  32284.6 \\
J161208.42+255505.4 & 0.253 &                D & 0.96 & J161211.15+255508.3 &   \dots &      \dots & 1.05 & 37.0 & 146.0 &      \dots \\
J165501.32+260517.4 & 1.904 &                S & 1.15 & J165502.01+260516.2 & 1.910 &     D & 1.36 &  9.3 &  78.6 &   630.8 \\
J171730.88+195927.3 & 0.175 &                D & 0.95 & J171733.19+195907.0 & 0.230 &     D & 0.83 & 38.4 & 141.1 & 13728.3 \\
\dots & \dots &       \dots        & \dots & \dots & \dots &     \dots & \dots & \dots &  \dots &  \dots\\
\noalign{\smallskip}
\noalign{\smallskip}
\hline
\end{tabular}
\end{center}
\tablecomments{Spectroscopically rejected (Rank -1) dual AGN candidates. The layout of this table is identical to that in Table~\ref{tab:rank1}. For redshift sources, G=Gemini, GA=GAMA, K=Keck, SR=SOAR, S=SDSS, D=DESI, 2d=2dFGRS, P16=\citet{pierre2016}, M12=\citet{maddox2012}. {$^{\dagger}$: Systems identified as spatially resolved mid-IR dual AGN candidates in \citet{barrows2023}, but found to be projected pairs spectroscopically in this work. The full version of this table is available in electronic form online.}
}
\end{table*}

\begin{figure*}[p]
\centering
    \subfloat{\includegraphics[width=0.2\linewidth]{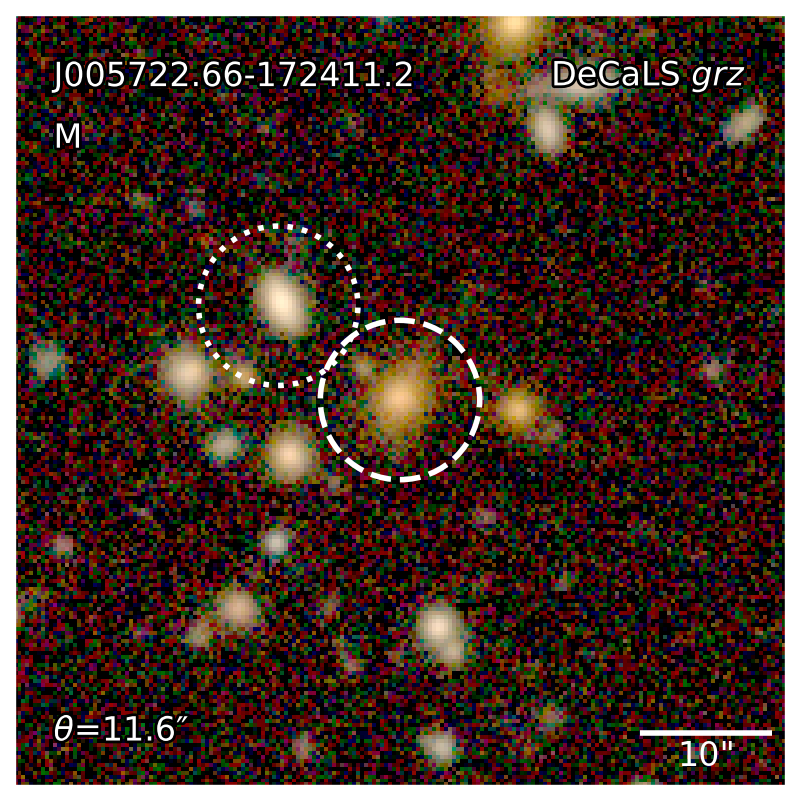}}
    \subfloat{\includegraphics[width=0.2\linewidth]{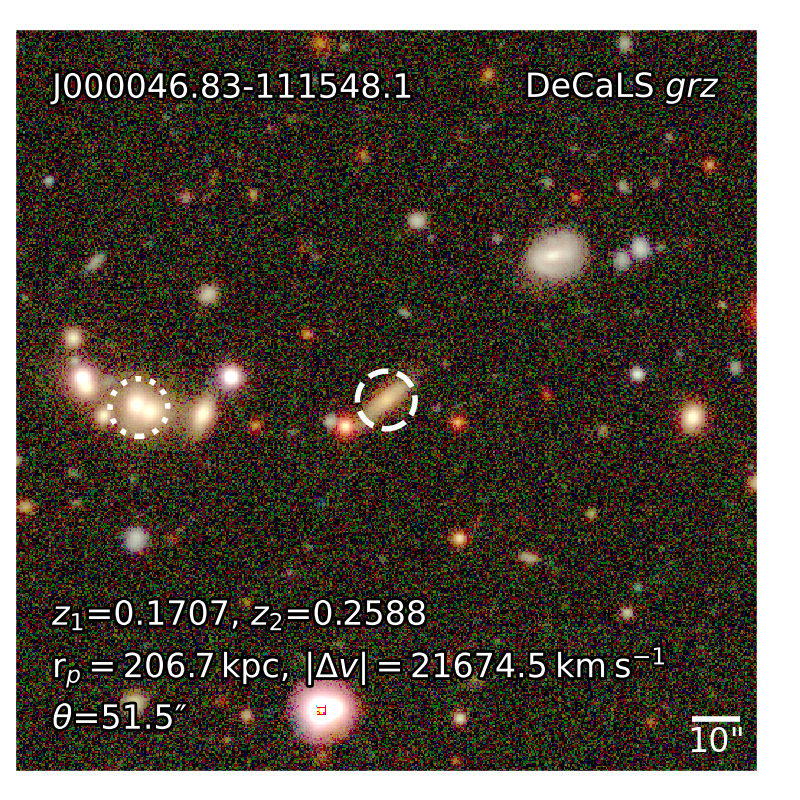}}
    \subfloat{\includegraphics[width=0.2\linewidth]{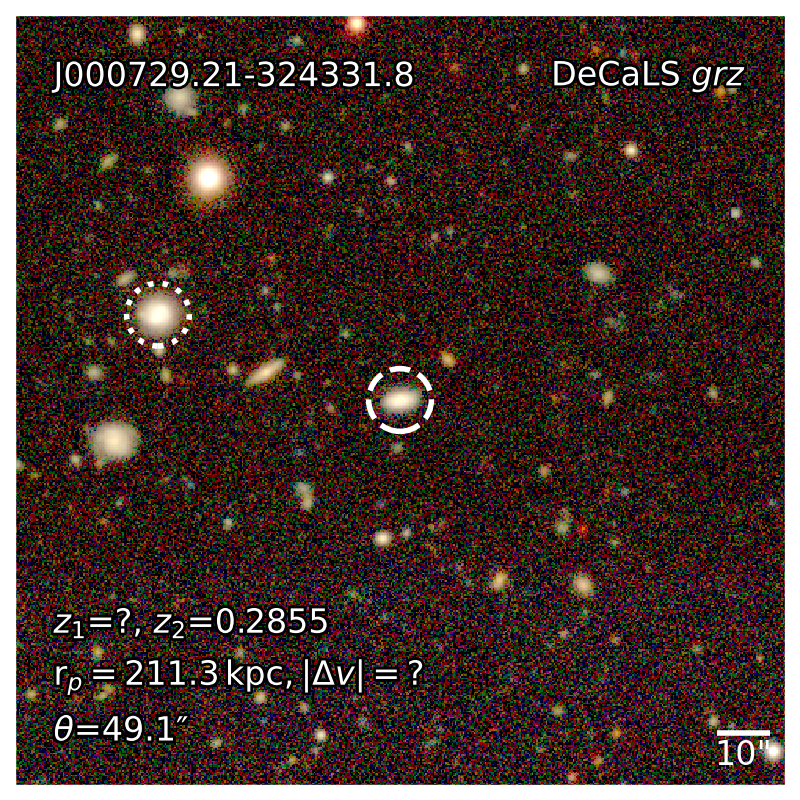}}
    \subfloat{\includegraphics[width=0.2\linewidth]{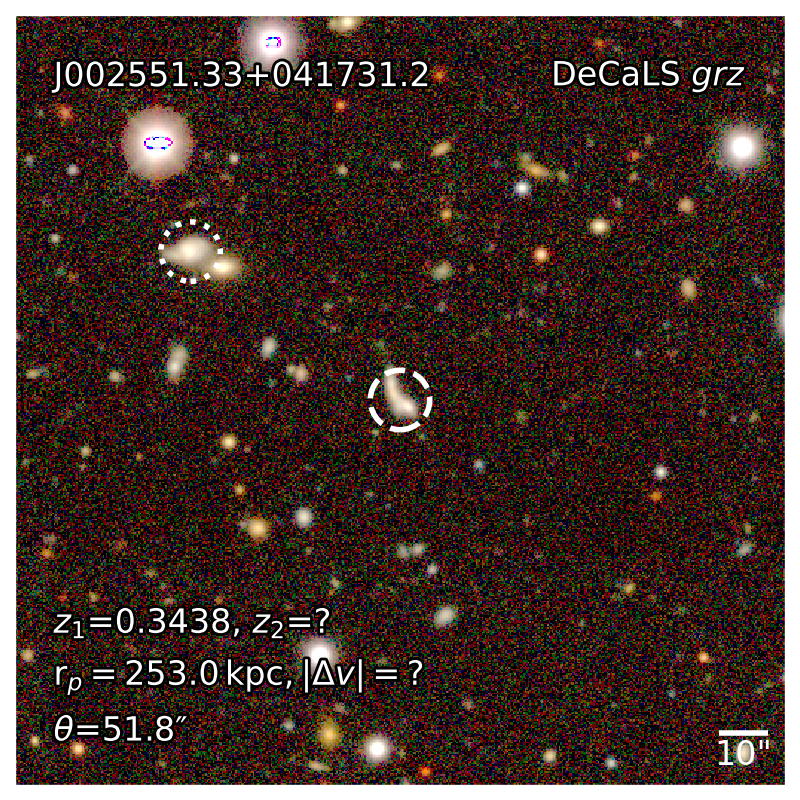}}
    \subfloat{\includegraphics[width=0.2\linewidth]{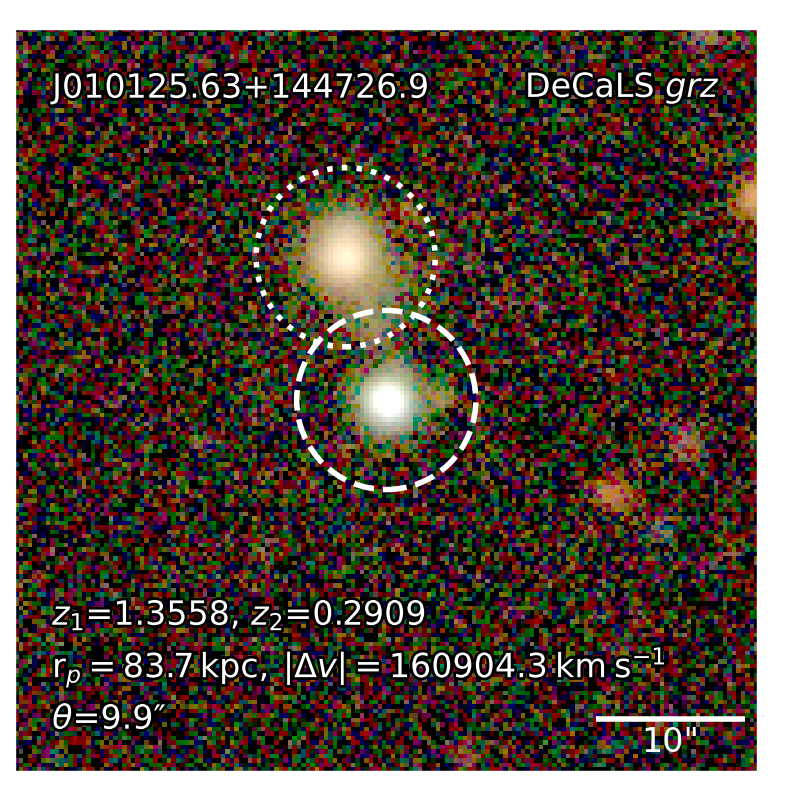}}\\
    \vspace{-4.5mm}
    \subfloat{\includegraphics[width=0.2\linewidth]{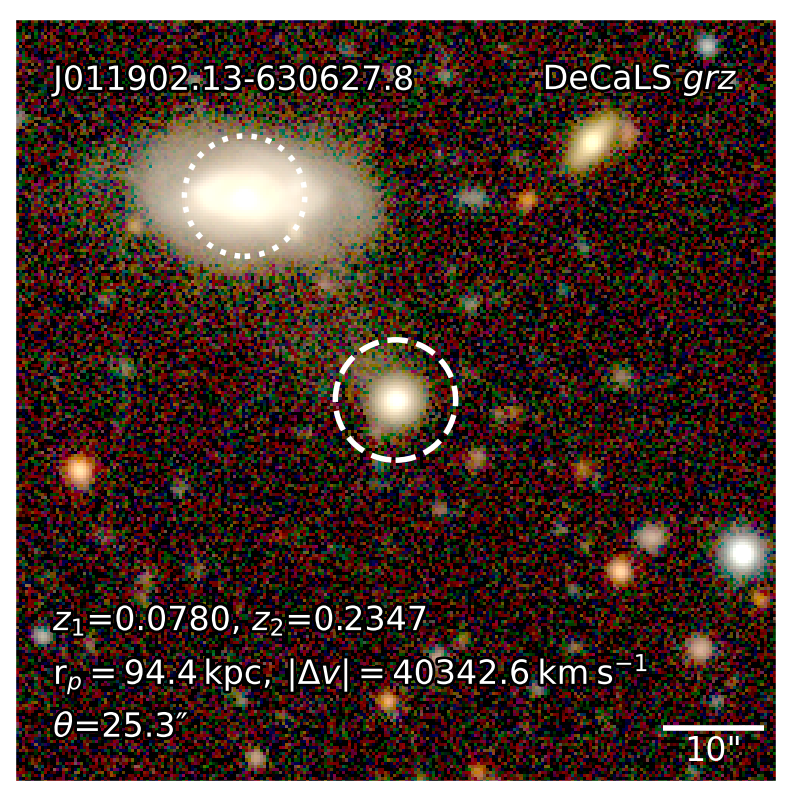}}
    \subfloat{\includegraphics[width=0.2\linewidth]{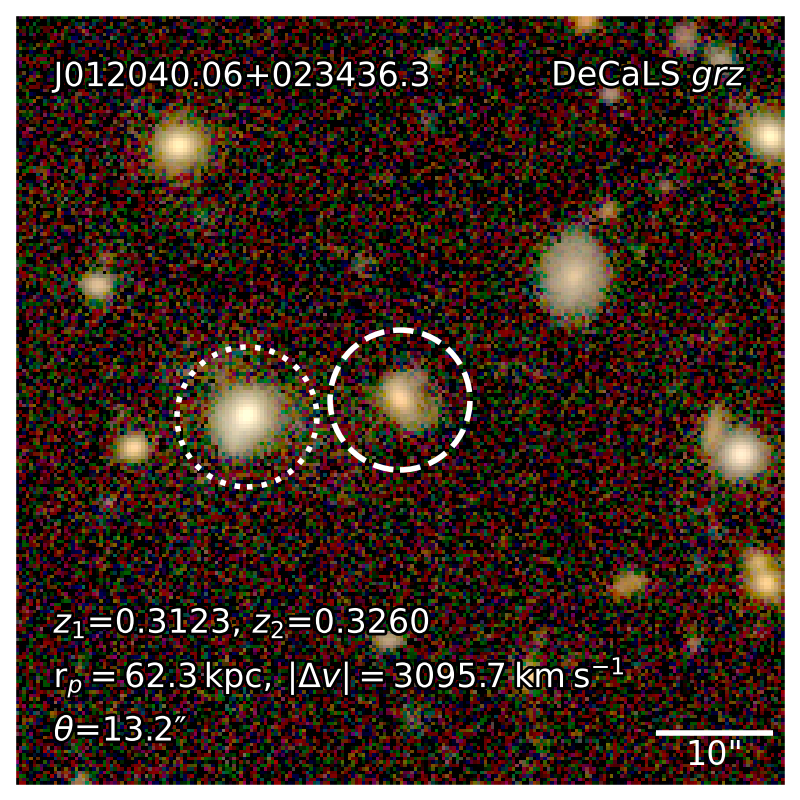}}
    \subfloat{\includegraphics[width=0.2\linewidth]{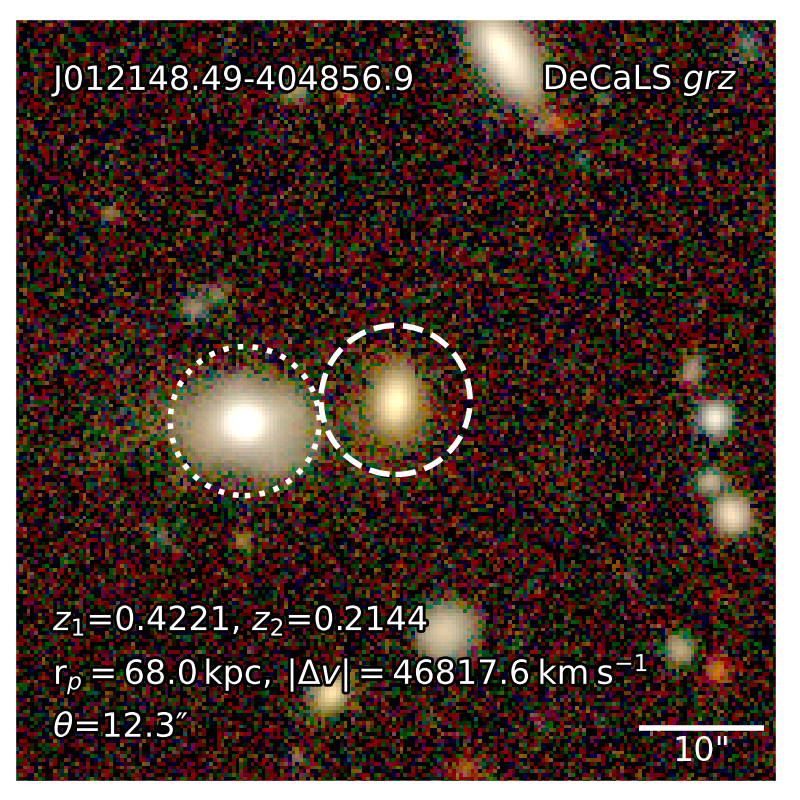}}
    \subfloat{\includegraphics[width=0.2\linewidth]{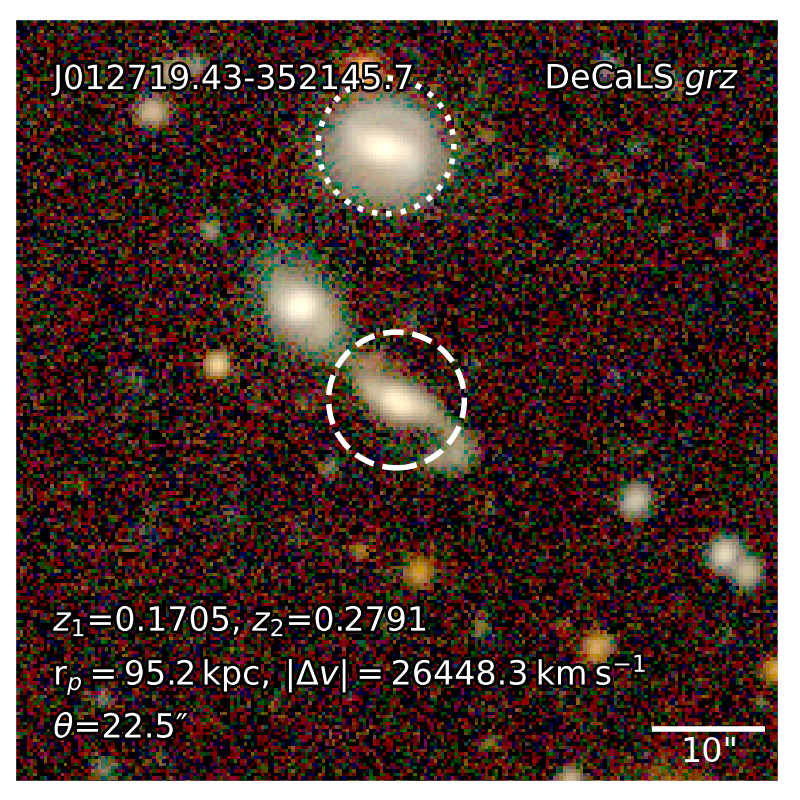}}
    \subfloat{\includegraphics[width=0.2\linewidth]{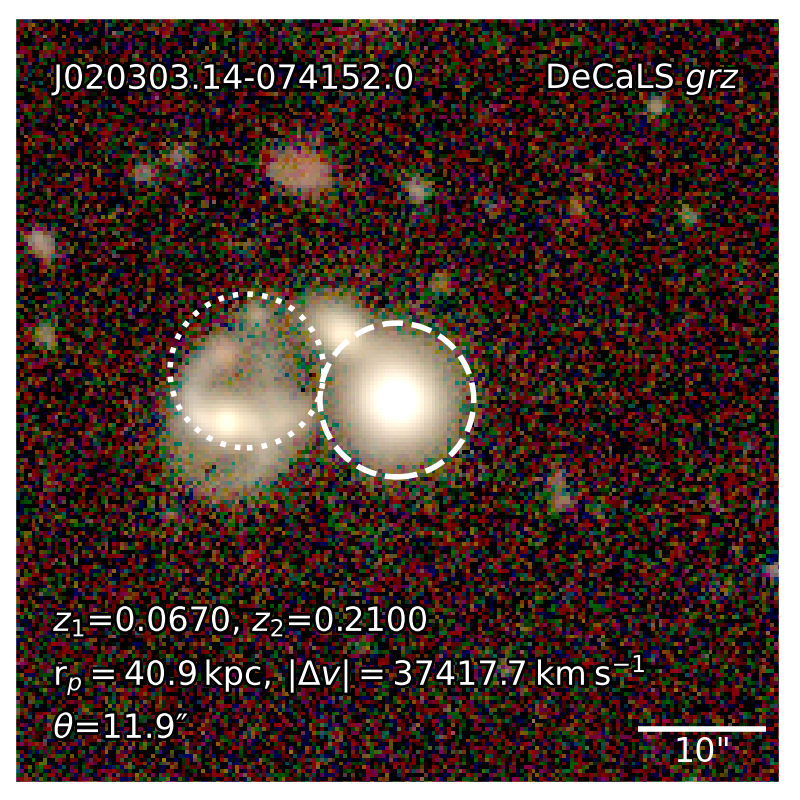}}\\
    \vspace{-4.5mm}
    \subfloat{\includegraphics[width=0.2\linewidth]{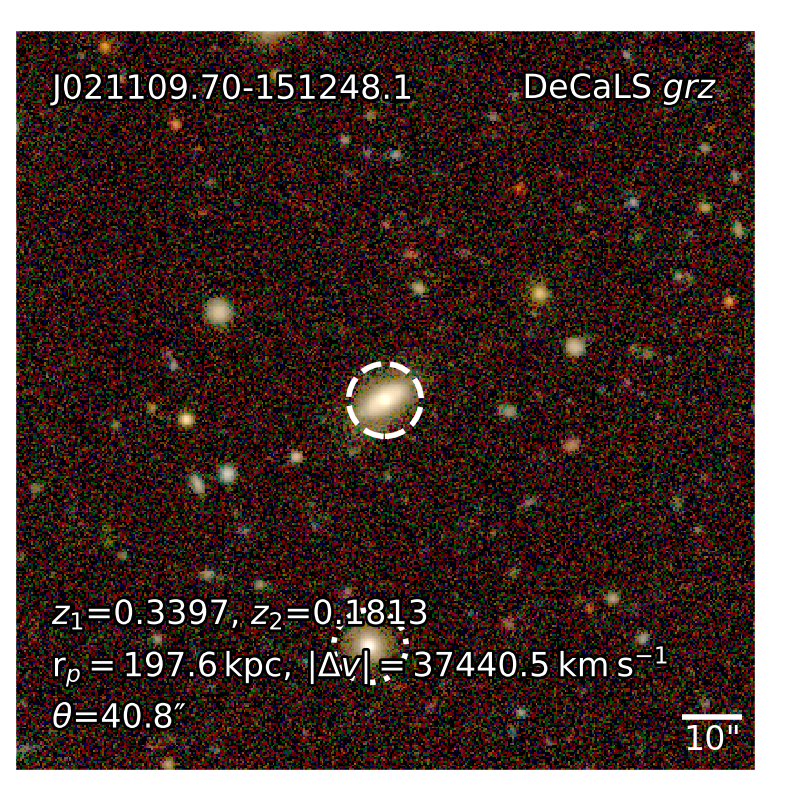}}
    \subfloat{\includegraphics[width=0.2\linewidth]{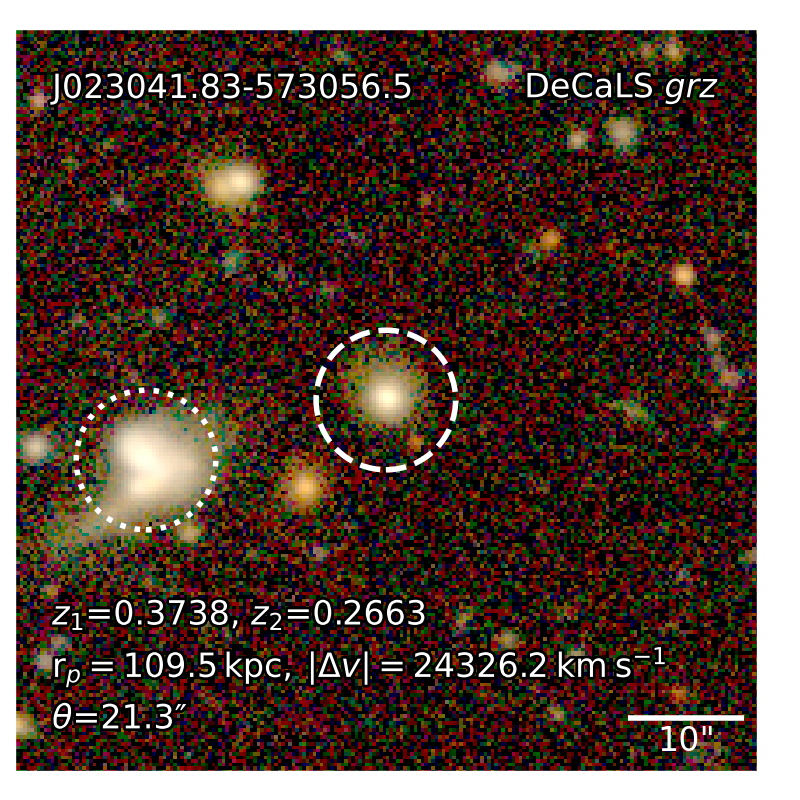}}
    \subfloat{\includegraphics[width=0.2\linewidth]{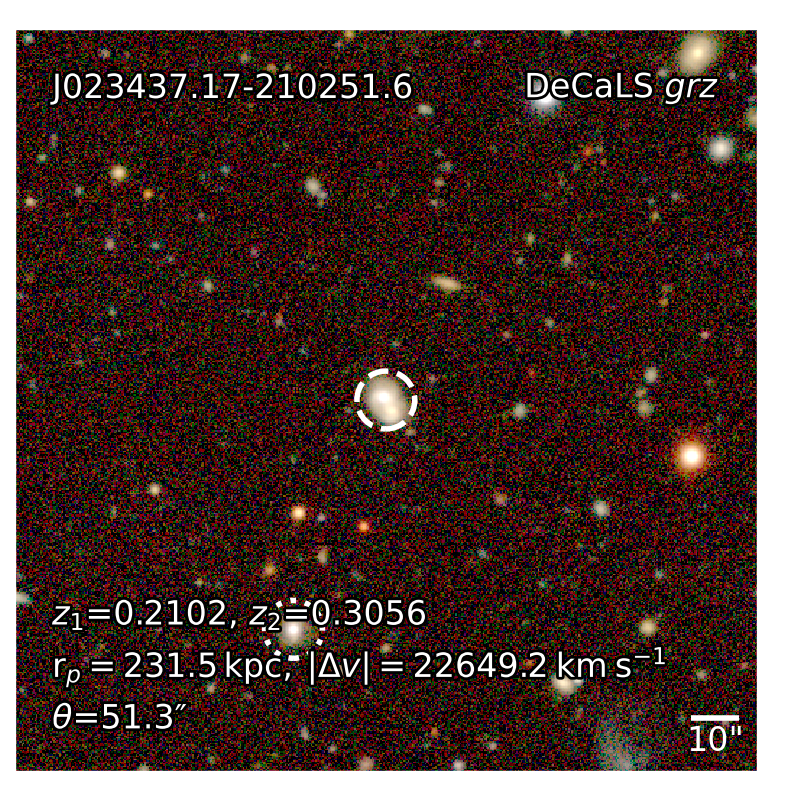}}
    \subfloat{\includegraphics[width=0.2\linewidth]{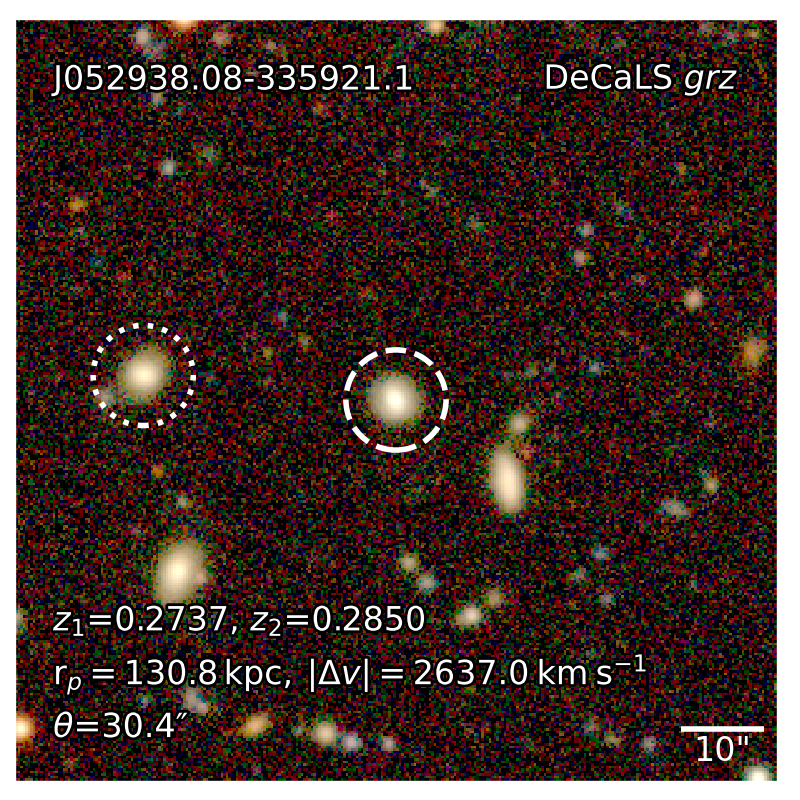}}
    \subfloat{\includegraphics[width=0.2\linewidth]{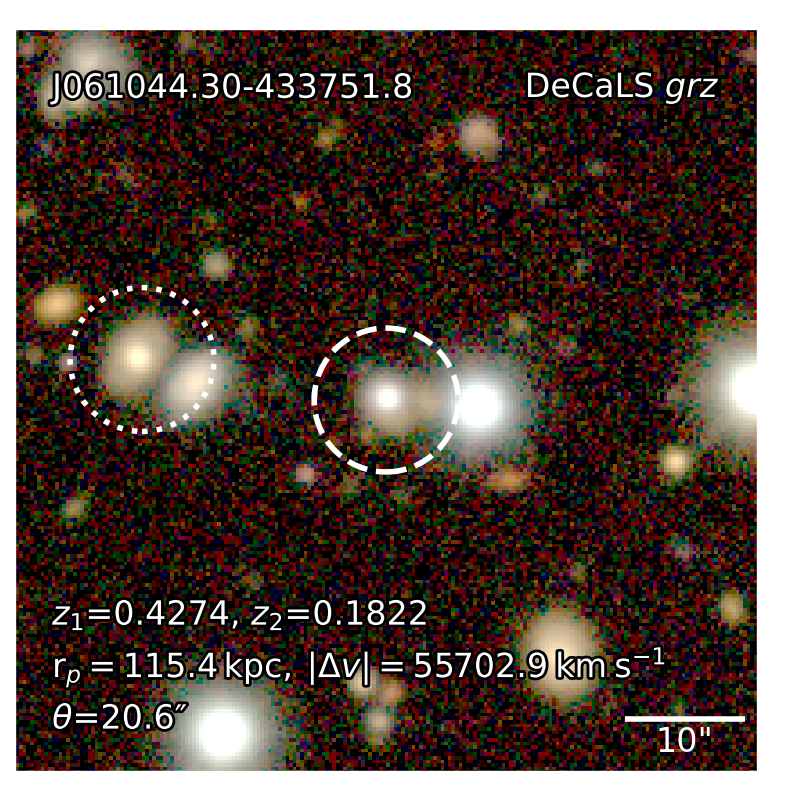}}\\
    \vspace{-4.5mm}
    \subfloat{\includegraphics[width=0.2\linewidth]{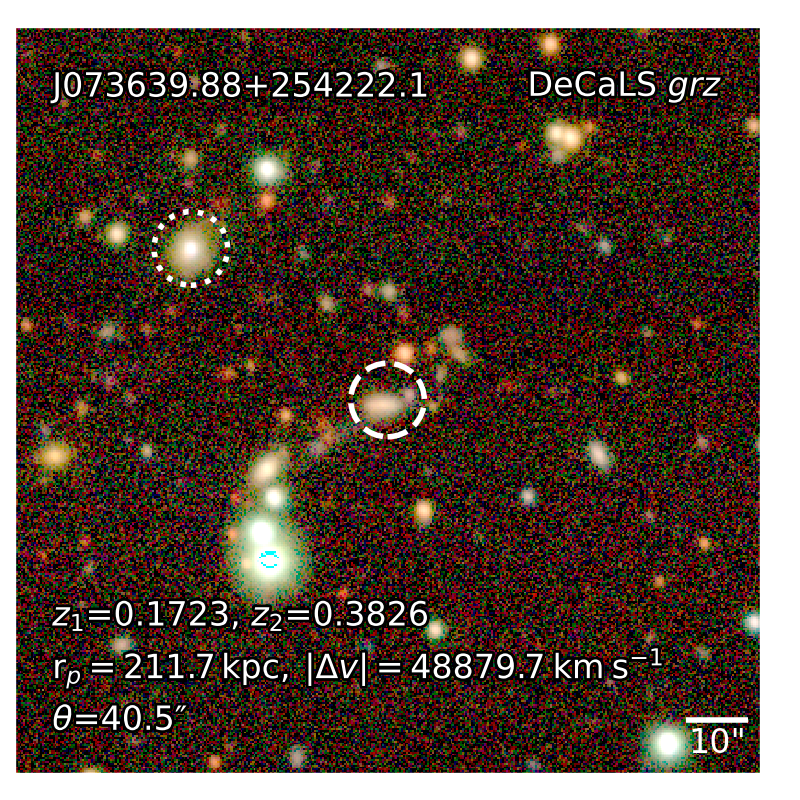}}
    \subfloat{\includegraphics[width=0.2\linewidth]{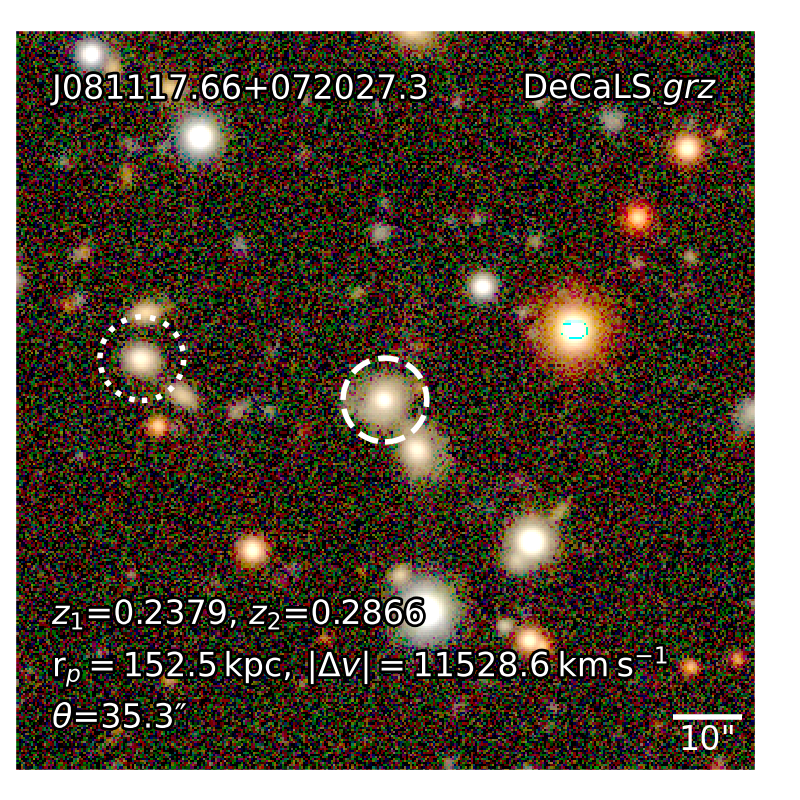}}
    \subfloat{\includegraphics[width=0.2\linewidth]{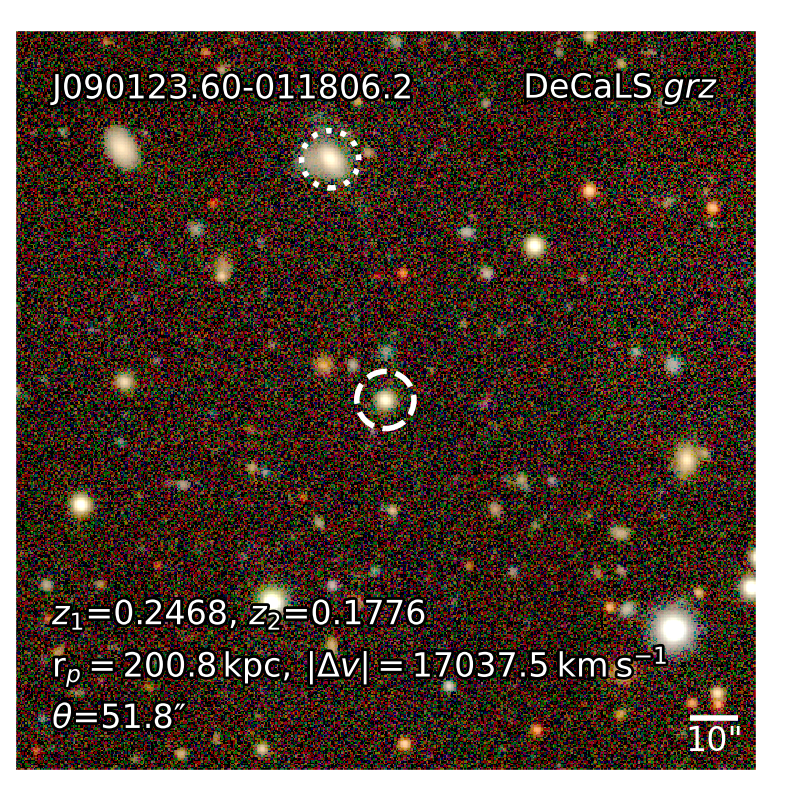}}
    \subfloat{\includegraphics[width=0.2\linewidth]{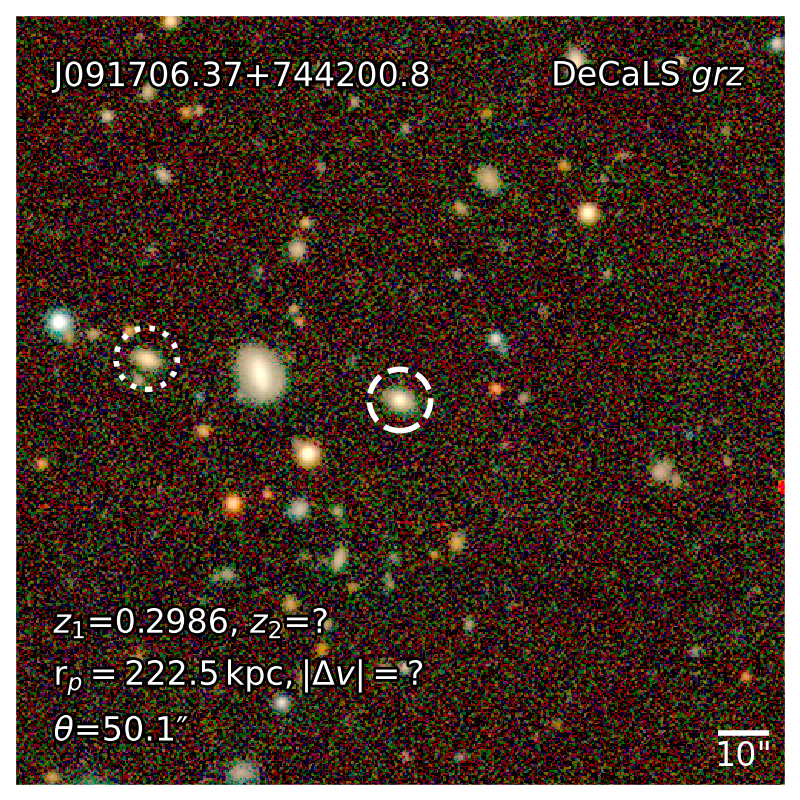}}
    \subfloat{\includegraphics[width=0.2\linewidth]{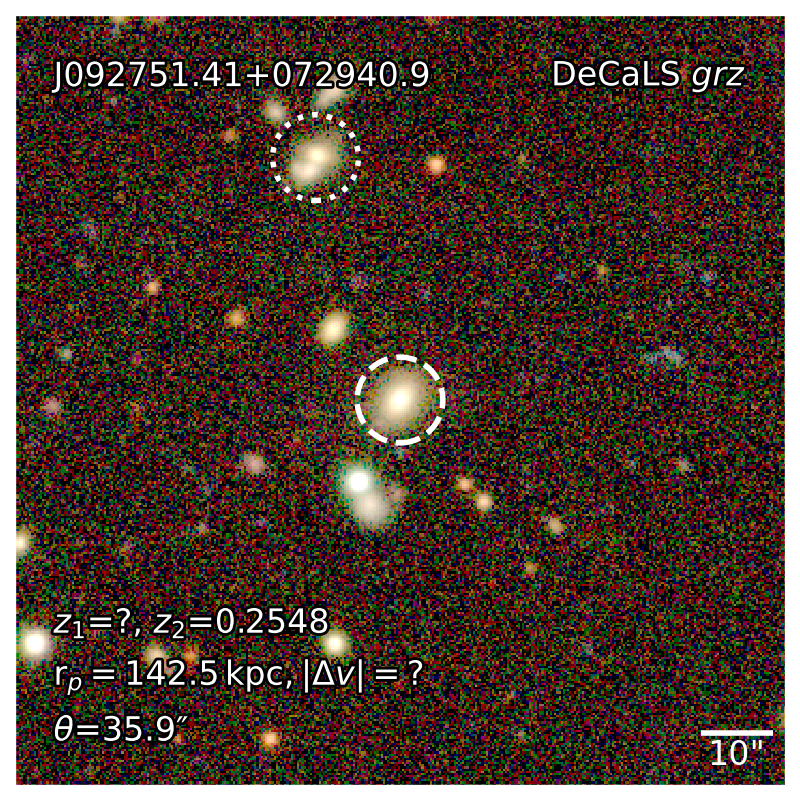}}\\
    \vspace{-4.5mm}
    \subfloat{\includegraphics[width=0.2\linewidth]{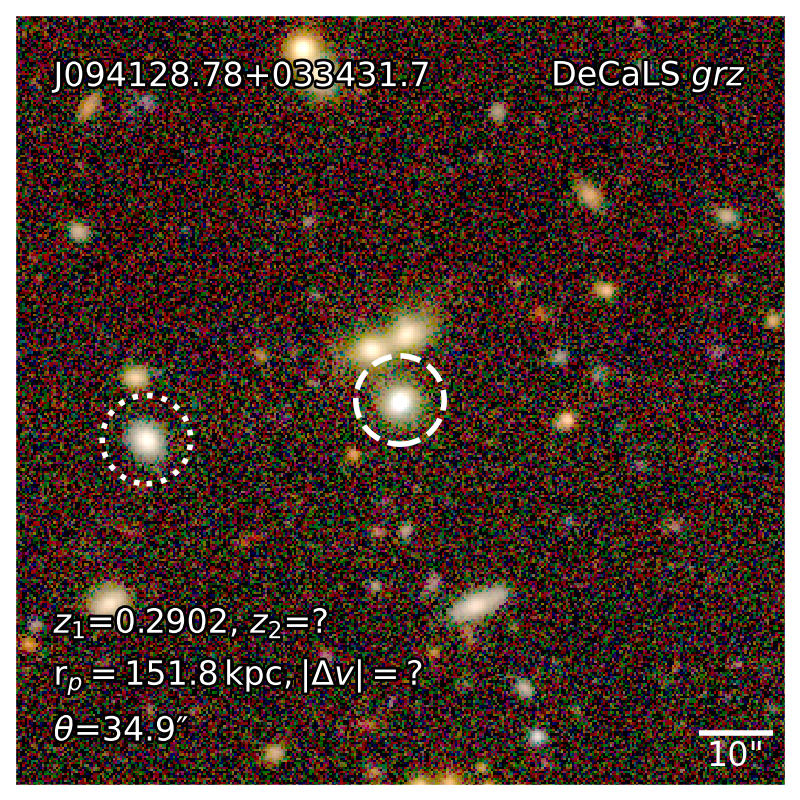}}
    \subfloat{\includegraphics[width=0.2\linewidth]{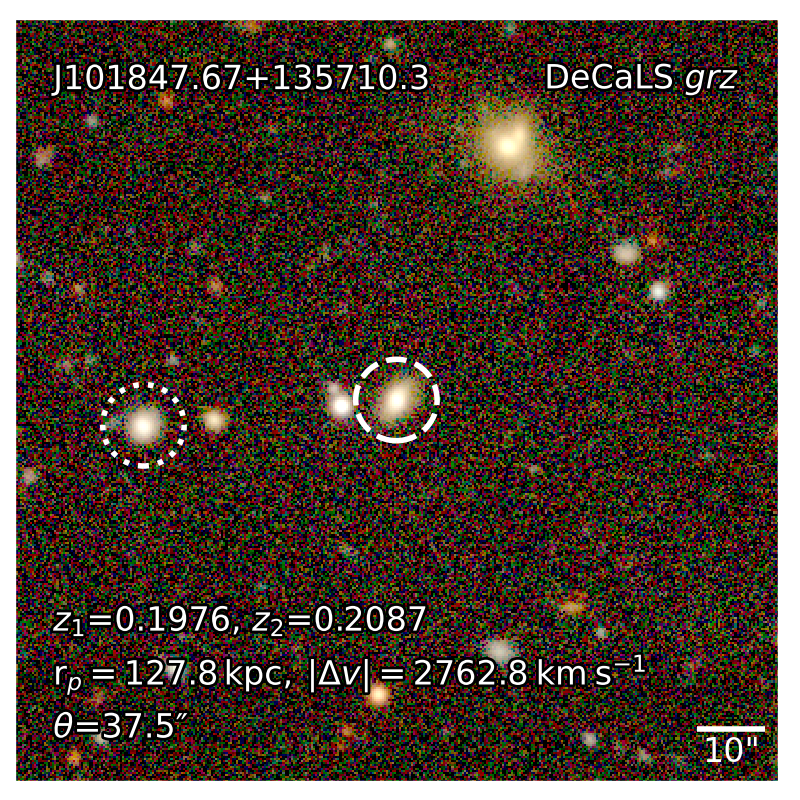}}
    \subfloat{\includegraphics[width=0.2\linewidth]{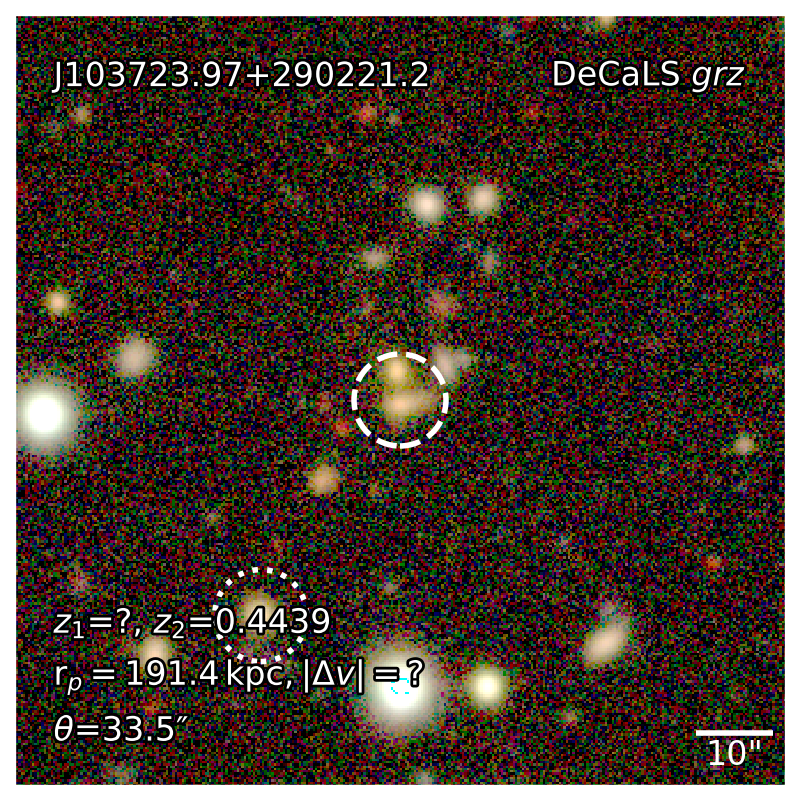}}
    \subfloat{\includegraphics[width=0.2\linewidth]{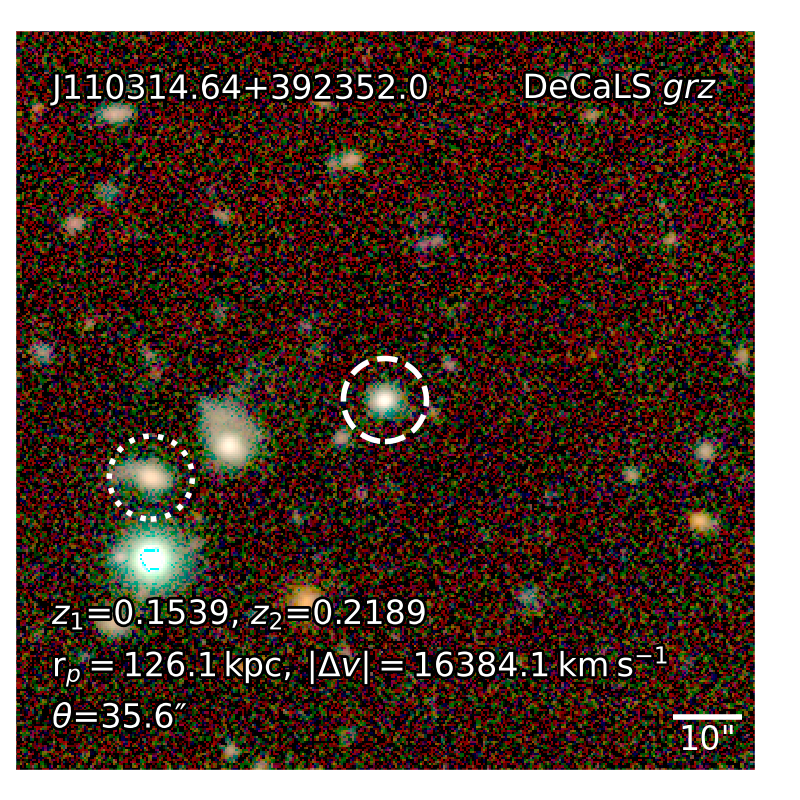}}
    \subfloat{\includegraphics[width=0.2\linewidth]{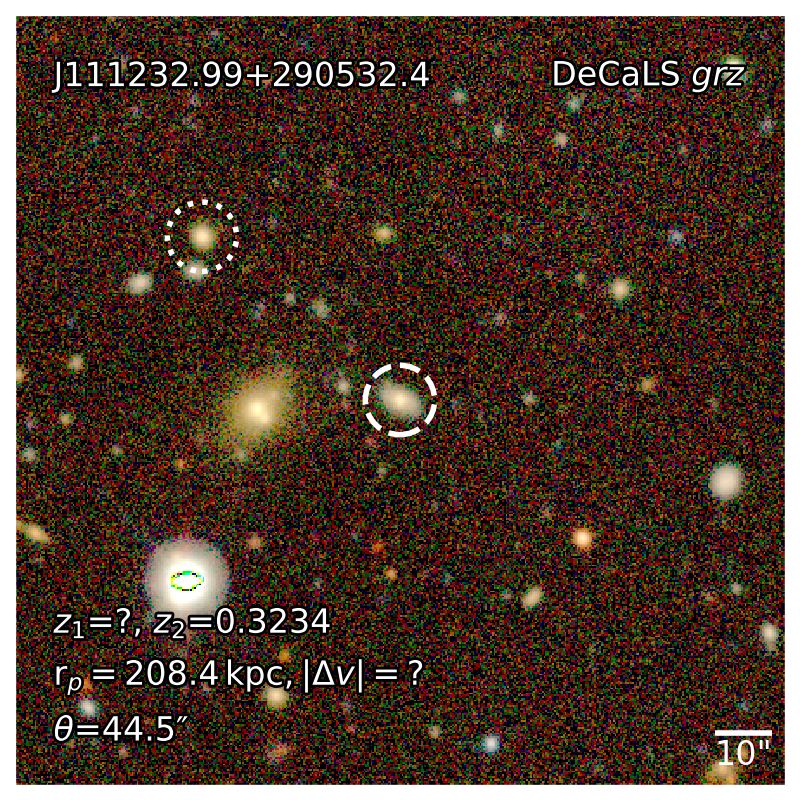}}\\
    \vspace{-4.5mm}
    \subfloat{\includegraphics[width=0.2\linewidth]{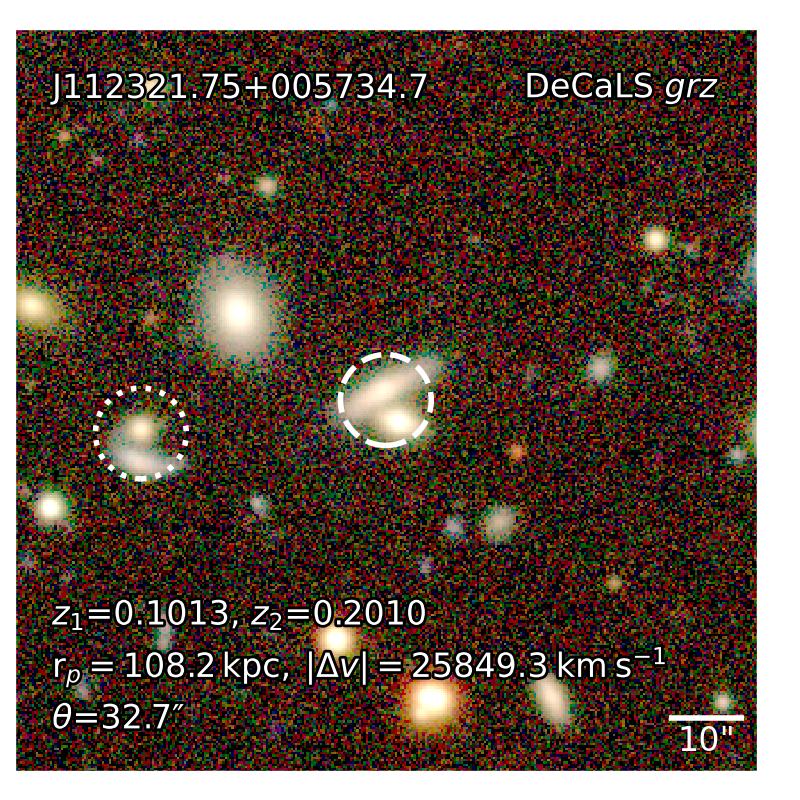}}
    \subfloat{\includegraphics[width=0.2\linewidth]{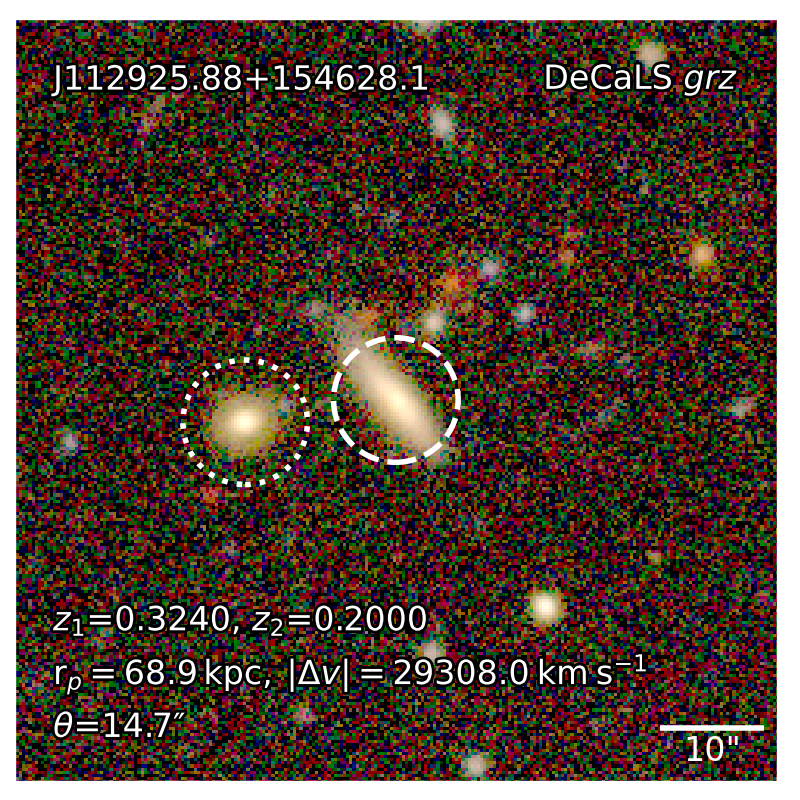}}
    \subfloat{\includegraphics[width=0.2\linewidth]{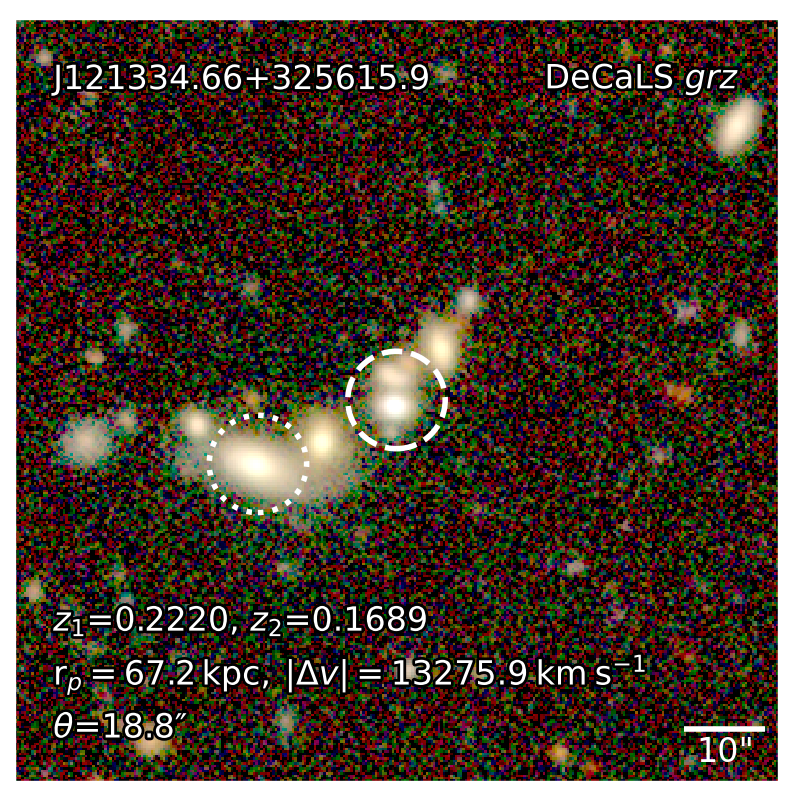}}
    \subfloat{\includegraphics[width=0.2\linewidth]{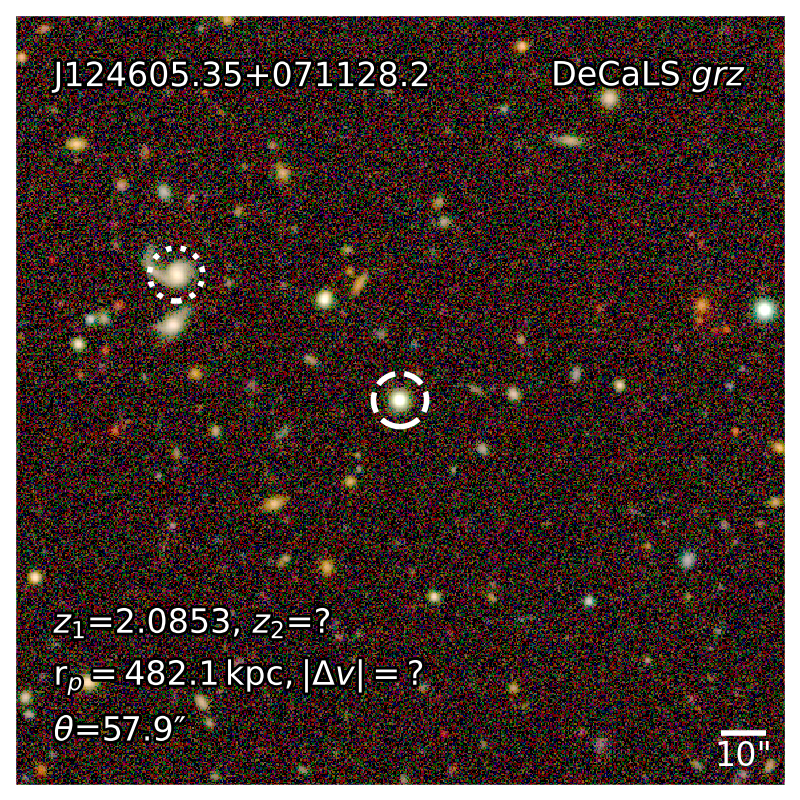}}
    \subfloat{\includegraphics[width=0.2\linewidth]{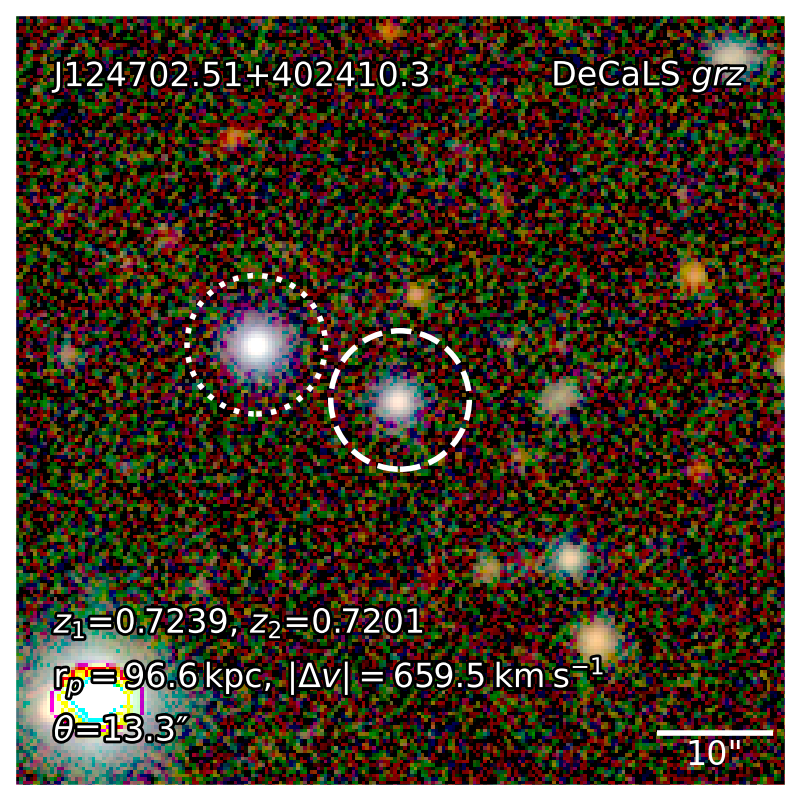}}\\
    \caption{Rejected mid-IR dual AGN candidates. Each panel is constructed in an identical fashion to those in Figure~\ref{fig:rank1_duals}. These dual AGN candidates were rejected based on the physical, projected separation and velocity difference criteria laid out in Section~\ref{sec:selection} or if they were identified as lenses in previous studies \citep[e.g.,][]{walsh1979}.}
    \label{fig:rejects}
\end{figure*}

\begin{figure*}
\ContinuedFloat
\centering
    \subfloat{\includegraphics[width=0.2\linewidth]{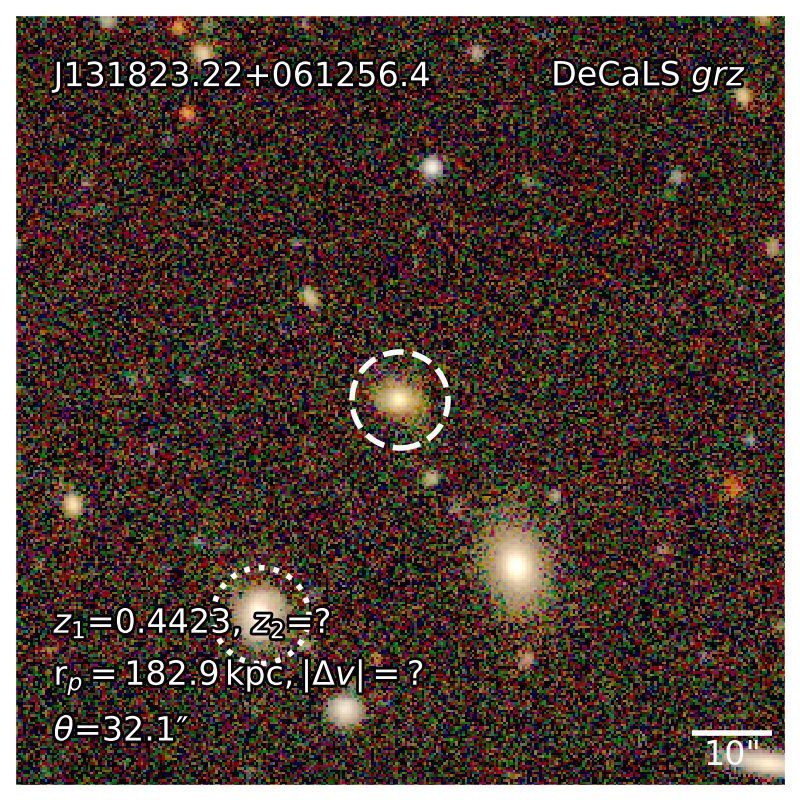}}
    \subfloat{\includegraphics[width=0.2\linewidth]{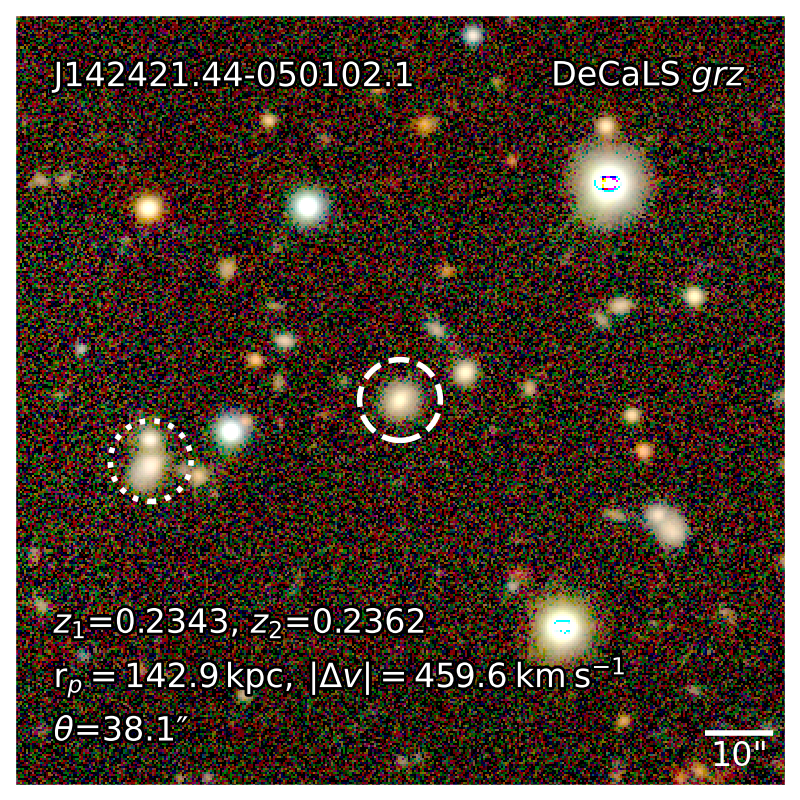}}
    \subfloat{\includegraphics[width=0.2\linewidth]{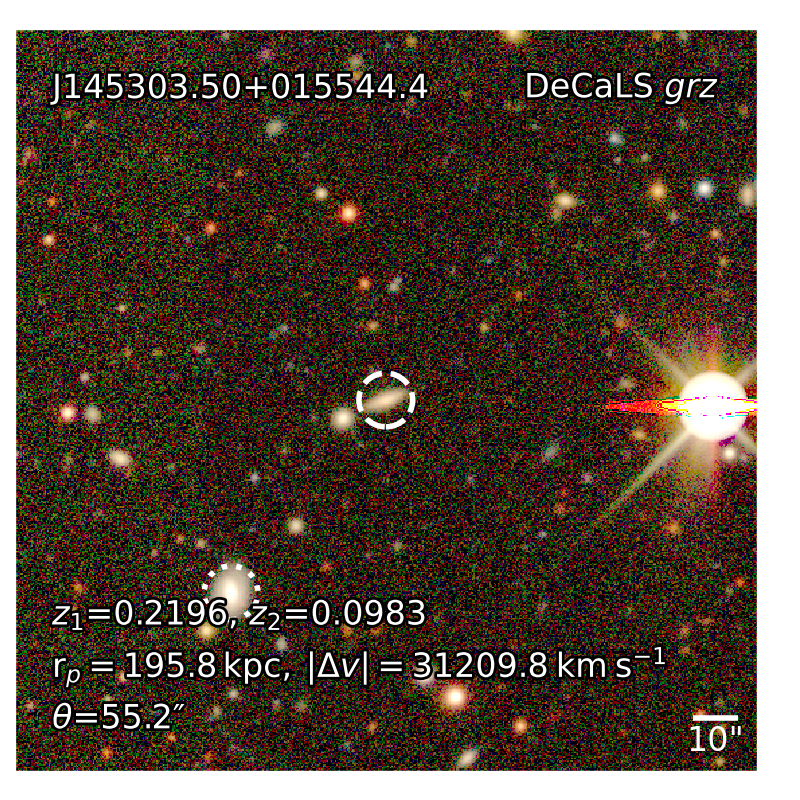}}
    \subfloat{\includegraphics[width=0.2\linewidth]{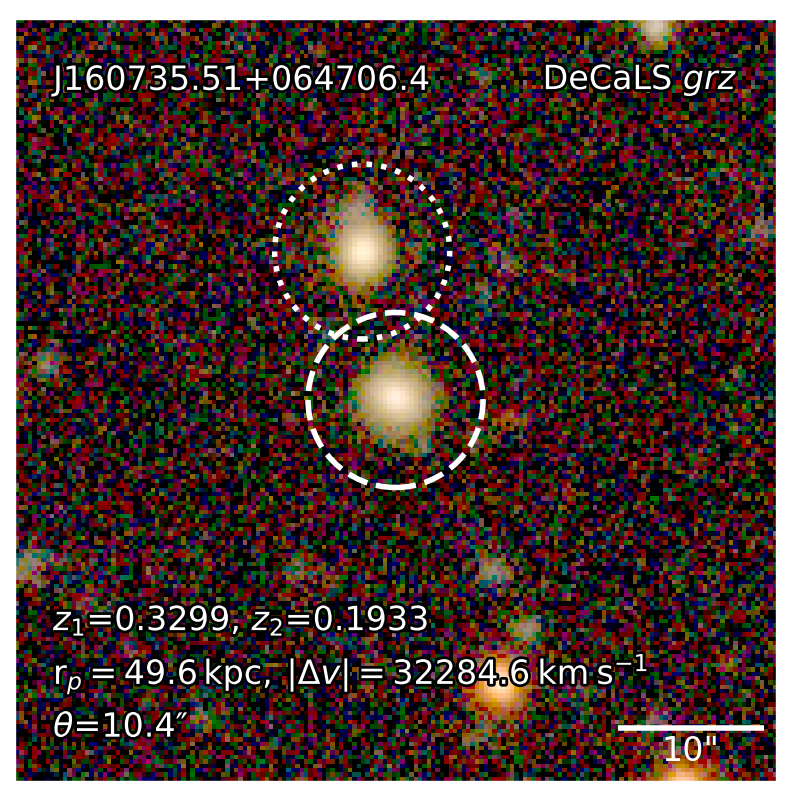}}
    \subfloat{\includegraphics[width=0.2\linewidth]{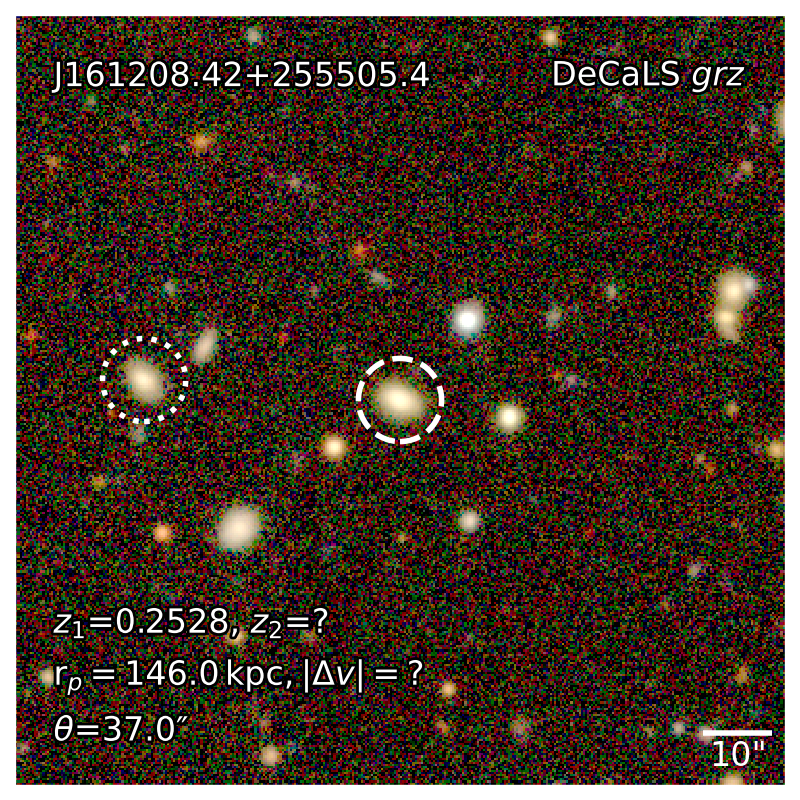}}\\
    \vspace{-4.5mm}
    \subfloat{\includegraphics[width=0.2\linewidth]{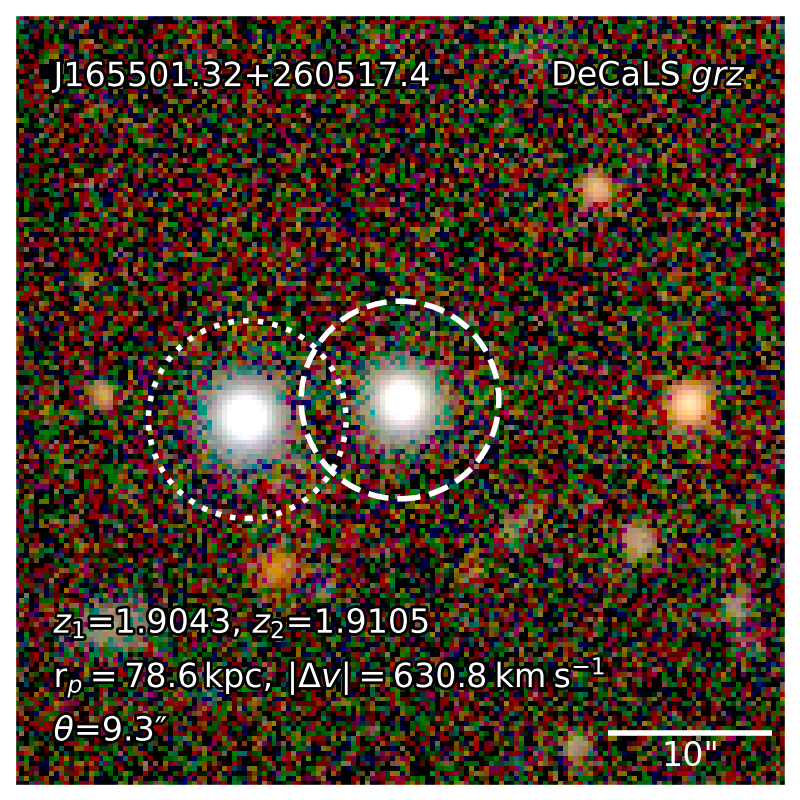}}
    \subfloat{\includegraphics[width=0.2\linewidth]{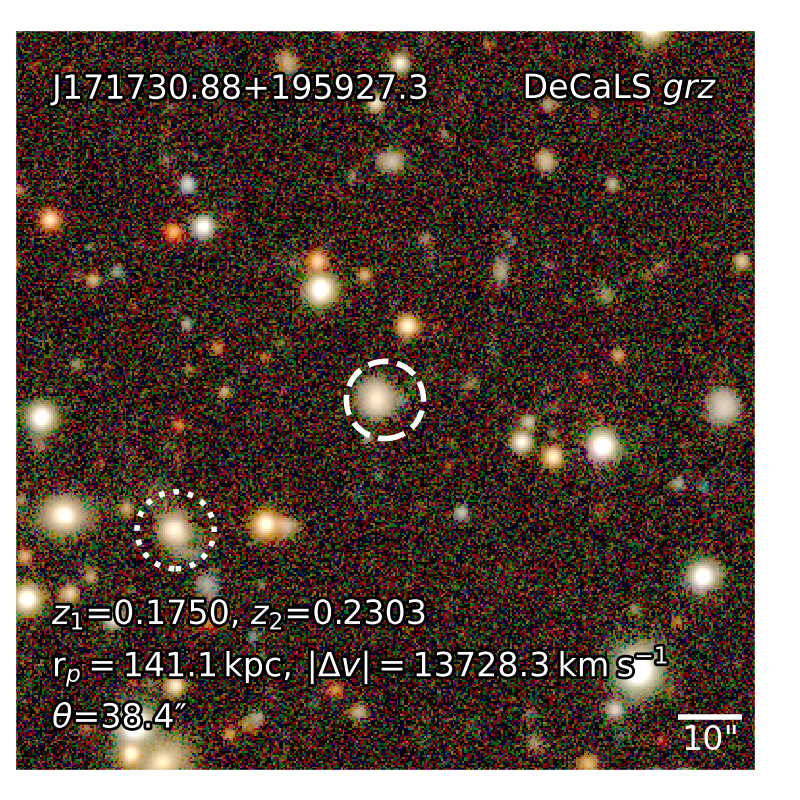}}
    \subfloat{\includegraphics[width=0.2\linewidth]{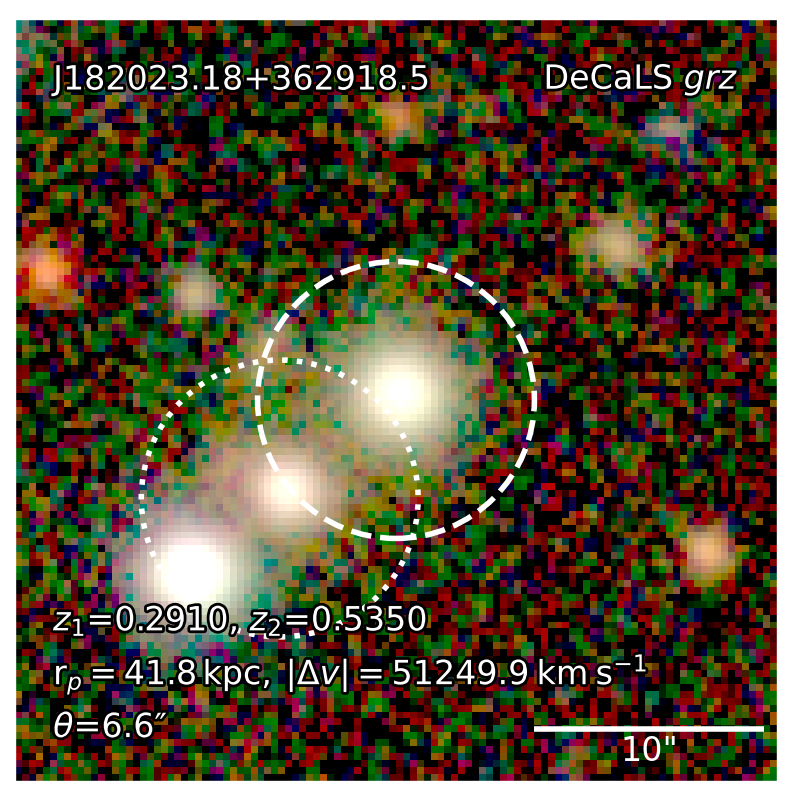}}
    \subfloat{\includegraphics[width=0.2\linewidth]{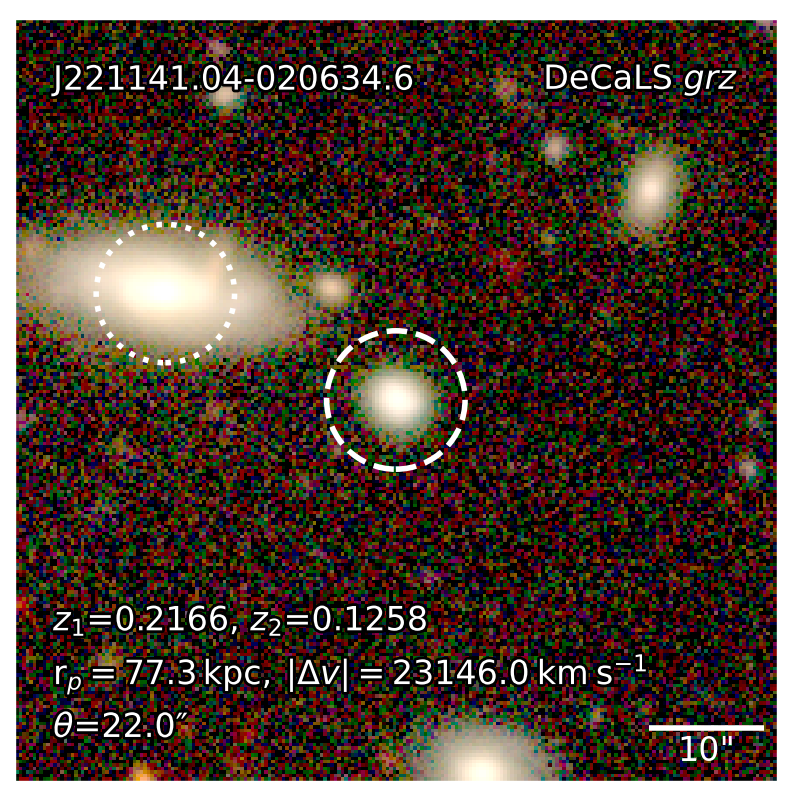}}
    \subfloat{\includegraphics[width=0.2\linewidth]{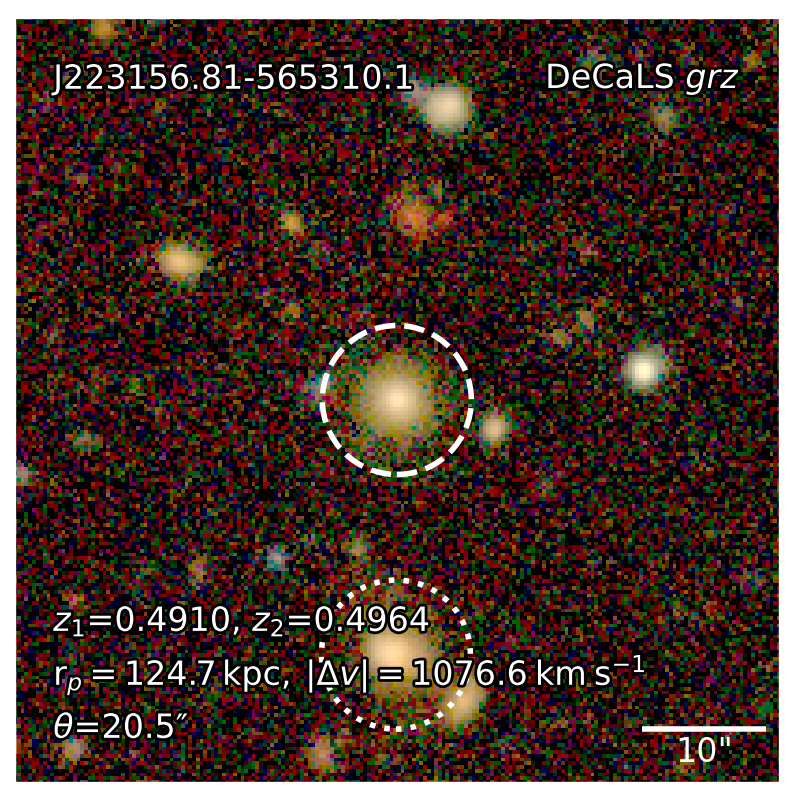}}\\
    \vspace{-4.5mm}
    \subfloat{\includegraphics[width=0.2\linewidth]{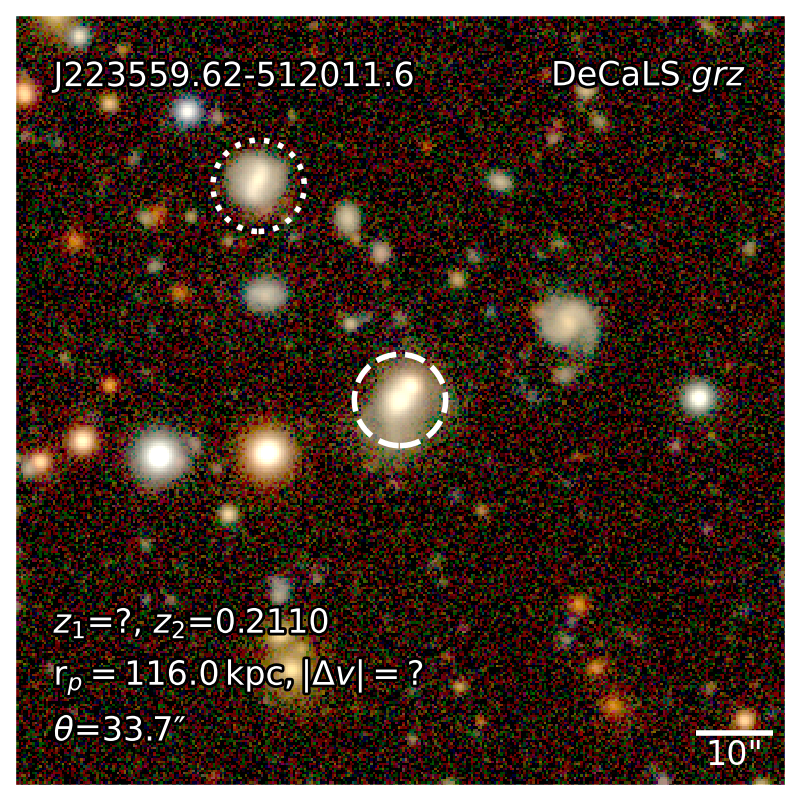}}
    \subfloat{\includegraphics[width=0.2\linewidth]{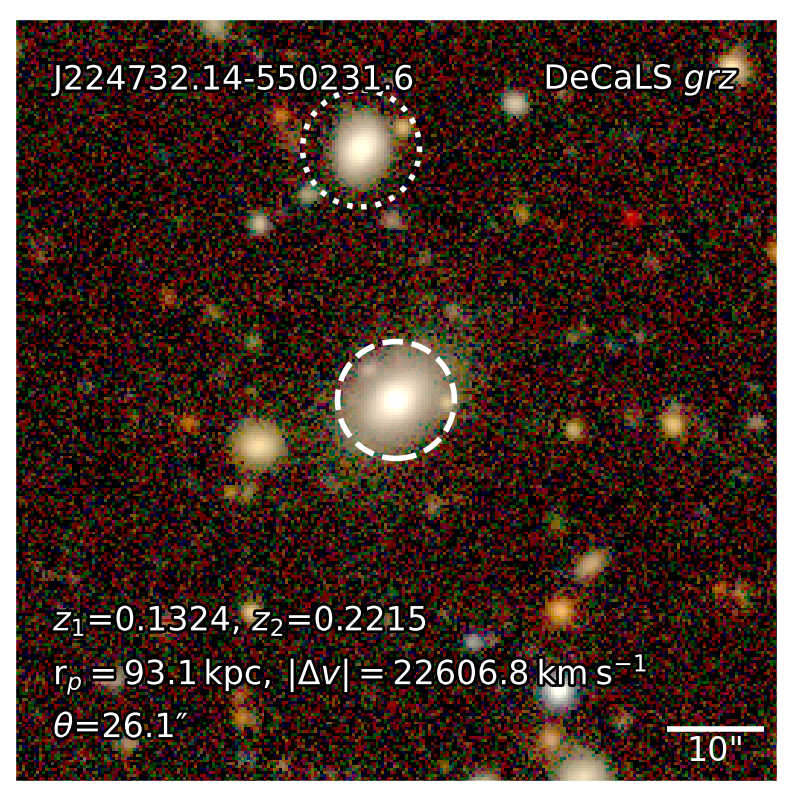}}
    \subfloat{\includegraphics[width=0.2\linewidth]{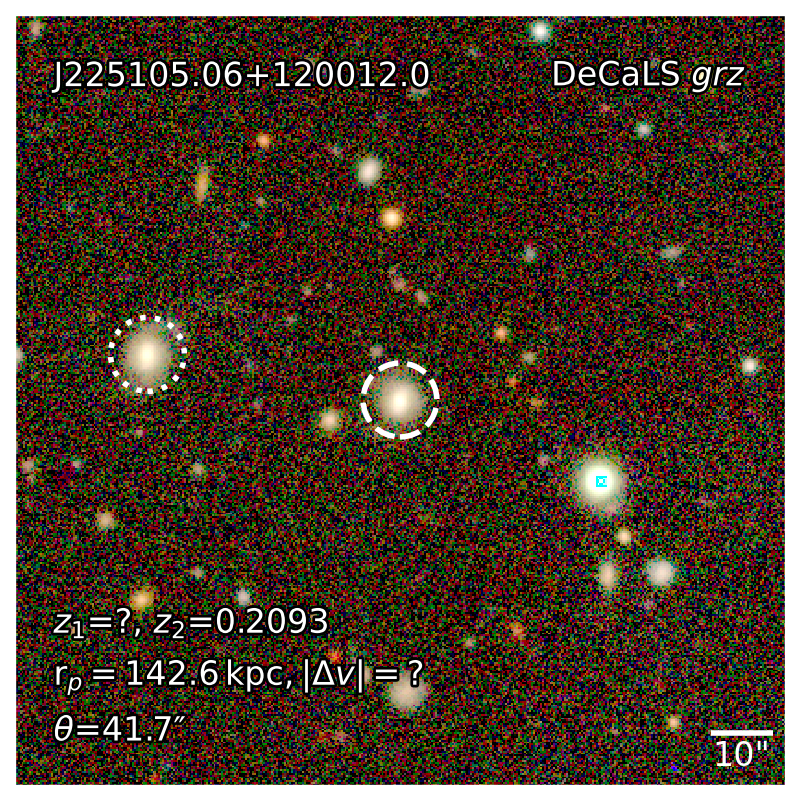}}
    \subfloat{\includegraphics[width=0.2\linewidth]{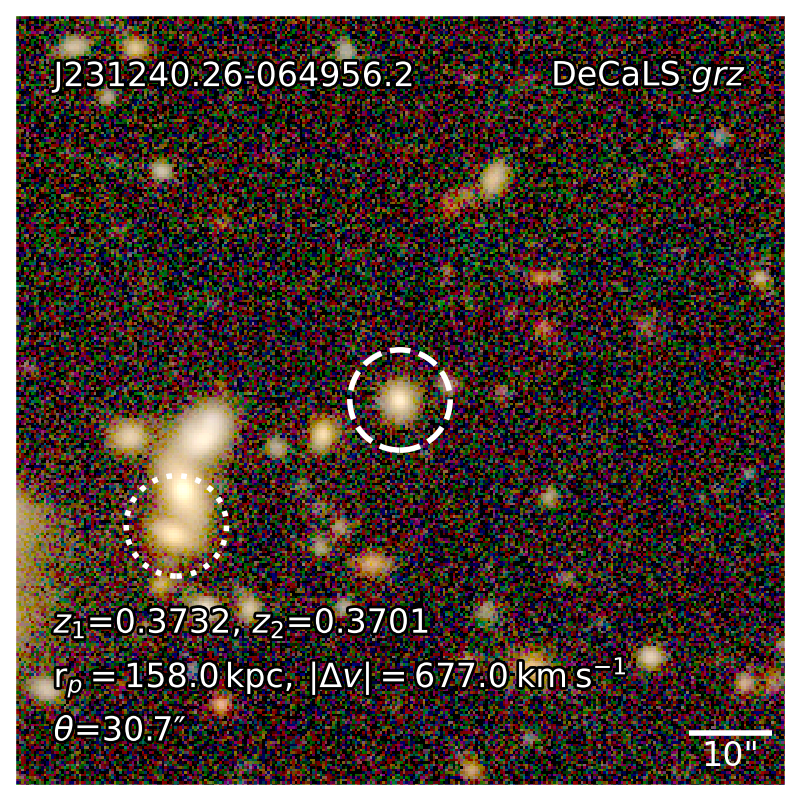}}
    \subfloat{\includegraphics[width=0.2\linewidth]{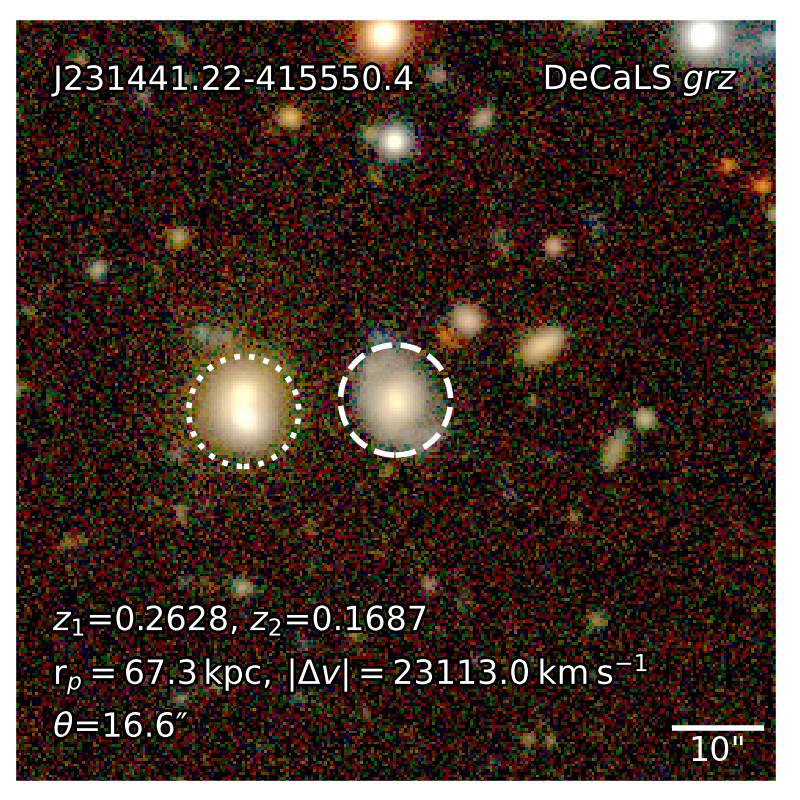}}\\
    \caption{Continued}
\end{figure*}

\section{Prioritizing Spectroscopic Redshifts Over Photometric Redshifts}
\label{sec:app_photz_vs_specz}

\begin{figure}
    \centering
    \includegraphics[width=0.99\linewidth]{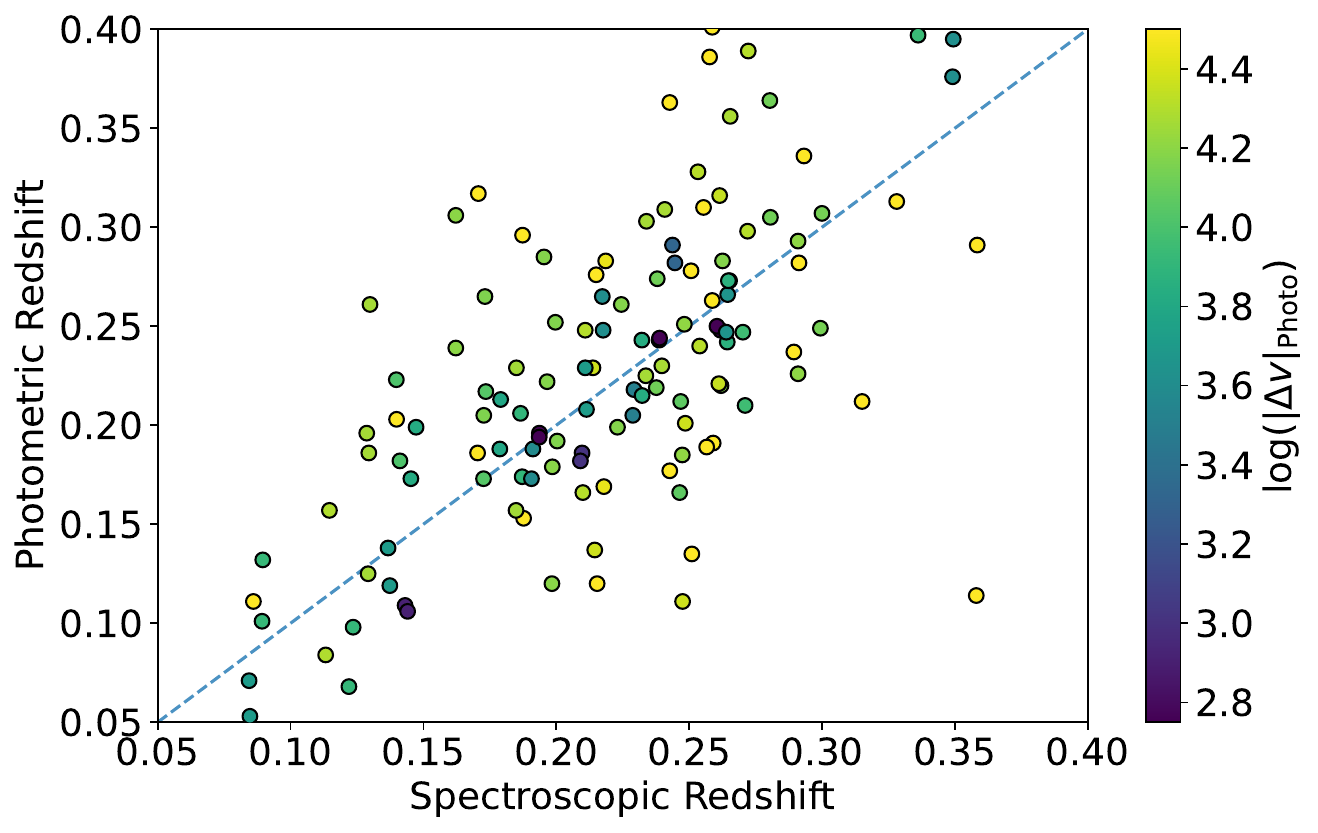}
    \caption{Spectroscopic vs. photometric redshifts for Rank 0.5 and 1 dual AGNs in this work, where both nuclei have measured spectroscopic redshifts. Spectroscopic redshifts are plotted on the x-axis, and photometric redshifts are plotted on the y-axis. The dashed line indicates unity between the phot- and spec-z's. The data points represent individual nuclei in the sample and are color coded by the logarithm of the absolute value of the velocity difference between the two nuclei based on their photometric redshifts ($\rm{log}(|\Delta v _{\rm{photo}}|)$. For scaling purposes, the color bar limits have been set to 2.78 and 4.5, which correspond to velocity differences of $\approx600$\,km\,s$^{-1}$ and $\approx31600$\,km\,s$^{-1}$, respectively. }
    \label{fig:photz_vs_specz}
\end{figure}

Throughout the course of this work, we relied strictly upon spectroscopic redshifts to remove non-dual AGN and non-merger contaminants during the selection and analysis process (see Sections~\ref{sec:contam_removal_phasei} and \ref{sec:contam_removal_phaseii}) and specifically avoided using photometric redshifts. To demonstrate the reasoning behind this decision, we plot in Figure~\ref{fig:photz_vs_specz} the spectroscopic and photometric redshifts for the dual AGN candidates (Ranks 0.5 and 1) selected in Section~\ref{sec:selection}. These photometric redshifts are the latest versions available on the DESI LS Sky Viewer (``DR 9.1'') and were computed using a random forest algorithm, where details on training and methodology can be found in \citet{zhou2023}. The spectroscopic redshifts come from archival measurements (e.g., SDSS, DESI, etc.), or were derived from our follow-up observations with Keck, Gemini, Palomar, and SOAR. Photometric redshifts are plotted on the y-axis, spectroscopic redshifts are plotted on the x-axis, and the dashed diagonal line indicates a photo-z/spec-z ratio of one (unity). Only systems where both nuclei have spectroscopic coverage are plotted here. The data points represent individual nuclei in the sample. The data are color coded by the logarithm of the absolute value of the velocity difference between the two nuclei based on their photometric redshifts ($\rm{log}(|\Delta v _{\rm{photo}}|$). For scaling purposes, the color bar limits have been set to 2.78 and 4.5, which correspond to velocity differences of $\approx600$\,km\,s$^{-1}$ and $\approx31600$\,km\,s$^{-1}$, respectively; recall from Section~\ref{sec:selection} that velocity differences above $>600-700$\,km\,s$^{-1}$ were flagged as contaminants and removed from the sample. When comparing photometric to spectroscopic redshifts for individual nuclei, approximately 95\% of all nuclei exhibit photometric-to-spectroscopic redshift velocity differences $|\Delta v _{\rm{photo-to-spec}}|>700$\,km\,s$^{-1}$. Turning to pairs of nuclei and focusing on photometric redshifts alone, nearly all systems ($\approx96\%$) exhibit velocity differences $|\Delta v _{\rm{photo}}|>700$\,km\,s$^{-1}$, and in fact $\approx93\%$ exhibit extremely large velocity differences ($>2000$\,km\,s$^{-1}$) that would in no way suggest a given pair of nuclei were interacting and would cast doubt on a merger scenario even when significant tidal features were present. The distribution of spectroscopic and photometric redshifts here -- along with the velocity difference information derived from the photometric redshifts alone -- demonstrates the clear dangers of relying upon photometric redshifts to assemble dual AGN samples like that assembled here. While on large statistical scales, photometric redshifts can be used as reliable proxies for spectroscopic redshifts \citep[e.g.,][]{zhou2023}, on the scales of smaller samples (such as that described in this work) pipeline-generated photometric redshifts are not sufficient for calculating accurate velocity differences and projected separations. Photometric redshifts therefore cannot be used to reliably remove contaminants without removing true merging systems and true dual AGNs{; the exception to this may be pair samples assembled using probabilistic distributions of photometric redshifts, such as that presented in \citet[][though there are some foreground-background contaminants in that sample, as noted in Section~\ref{sec:disc_overlapwbarrows} in this work.]{barrows2023}.}

\section{Morphological Misclassifications}
\label{app:morpho_mislabels}

As described in Section~\ref{sec:morphclass}, morphological classification based on visual inspection of optical imaging formed a primary component of our selection strategy for this mid-IR dual AGN program. There are several limitations for this approach however: (1) morphological classifications and visual identification of tidal features are limited by imaging depth, and low surface brightness tidal features become increasing difficult to observe with increasing redshift for a given observation depth. Thus, this technique is biased toward lower redshifts. (2) Not all galaxy mergers produce tidal features; the prominence and presence of tidally induced structures such as bridges, tidal arms, shells, ripples, and other debris depend upon the masses, structures, and orbital properties of the constituent galaxies \citep[e.g.,][]{mihos1998,dubinski1996,dimatteo2007,barnes2016,saleh2024}. This technique is therefore also biased against systems less prone to the development of tidal structures. 

For these reasons, we erred on the side of inclusion rather than exclusion during our morphological classification approach, incorporating subjective visual assessments such as optical color and size similarity, proximity of other similar-looking objects nearby (which could suggest group environments), and the presence of very low surface brightness fuzziness in the imaging that could be evidence of faint tidal debris or shells rather than image noise. Naturally, this more liberal approach increases the likelihood of contaminants (projected pairs residing at different redshifts, or pairs with separations too large to be considered a merger) in the overall sample of mid-IR dual AGN candidates. Figure~\ref{fig:rejects} shows the systems so far classified as rejected dual AGN candidates. While some of these rejected systems do --- by eye --- have more ambiguous morphologies in terms of interacting versus non-interacting (e.g., J021109.70-151248.1, J221141.04-020634.6), other systems show nuclei where both appear to exhibit tidal features or potential tidal features (e.g., J012040.06+023446.4, J012719.43-352145.7, J023041.83-573056.3) but the nuclei are in fact caught in projection and reside at entirely different redshifts. There are also a few cases where we have identified two mid-IR AGN candidates in a potential group of galaxies (and are therefore related to one another, e.g., J231240.26-064956.2), but the projected 2D separations or the nuclear velocity differences were slightly higher than our adopted definitions in Section~\ref{sec:defining_duals} and were therefore excluded. It is also worth noting that there are a few cases where contaminant mid-IR AGN pairs exhibit striking similarities to Rank 0.5 or Rank 1 mid-IR AGN candidates (in which the nuclei show consistent redshifts). A prime example is the Rank 1 dual AGN, J100030.84+553634.7, and the Rank -1 (rejected) candidates, J012040.06+023436.3 and J012148.49-404856.9: all three show similar 2D orbital configurations and fuzzy, tidal-like features (with J100030.84+553634.7 showing larger, faint features that could be mistaken for noise), yet the latter two are projected pairs where the nuclei are unrelated. While this morphological selection component was essential for avoiding the incompleteness of spectroscopic surveys (Section~\ref{sec:whystartfromimaging}), these rejected candidates reinforce the need for spectroscopic redshift verification in most cases to avoid erroneously classifying contaminants as bona fide dual AGNs based on perceived tidal features alone.

\bibliography{ref}{}
\bibliographystyle{aasjournal}



\end{document}